\documentclass[]{aa}

\usepackage{graphicx}
\usepackage{multirow}
\usepackage{amsmath}
\usepackage{siunitx}
\usepackage{lipsum}
\usepackage{placeins}
\usepackage{hyperref}
\usepackage{threeparttable}
\usepackage{tablefootnote}
\usepackage{mathtools}
\usepackage[thinc]{esdiff}
\usepackage{rotating,tabularx}
\usepackage{caption}
\usepackage{arydshln}

\DeclareRobustCommand{\rchi}{{\mathpalette\irchi\relax}}
\newcommand{\irchi}[2]{\raisebox{\depth}{$#1\chi$}} 

\usepackage[varg]{txfonts}

\def \bp{$\beta$\,Pic}
\def \betapictoris{$\beta$\,Pictoris}

\def\fei{Fe\,{\sc i}}
\def\feii{Fe\,{\sc ii}}
\def\feiii{Fe\,{\sc iii}}

\def\mnii{Mn\,{\sc ii}}

\def\znii{Zn\,{\sc ii}}

\def\mgi{Mg\,{\sc i}}
\def\mgii{Mg\,{\sc ii}}

\def\niii{Ni\,{\sc ii}}
\def\niiii{Ni\,{\sc iii}}

\def\cai{Ca\,{\sc i}}
\def\caii{Ca\,{\sc ii}}
\def\caiii{Ca\,{\sc iii}}

\def\crii{Cr\,{\sc ii}}

\def\coii{Co\,{\sc ii}}

\def\sii{Si\,{\sc i}}
\def\siii{Si\,{\sc ii}}
\def\siiii{Si\,{\sc iii}}
\def\siiv{Si\,{\sc iv}}
\def\Si{S\,{\sc i}}
\def\Sii{S\,{\sc ii}}
\def\Siii{S\,{\sc iii}}

\def\alii{Al\,{\sc ii}}
\def\aliii{Al\,{\sc iii}}

\def\ci{C\,{\sc i}}
\def\cii{C\,{\sc ii}}

\def\civ{C\,{\sc iv}}
\def\Xi{X\,{\sc i}}
\def\Xii{X\,{\sc ii}}

\def \A{$\si{\angstrom}$}
\def\feplus{Fe$^+$}
\def\mgplus{Mg$^+$}
\def\alplus{Al$^+$}
\def\aldeuxplus{Al$^{2+}$}

\def\mgplus{Mg$^+$}
\def\niplus{Ni$^+$}
\def\siplus{Si$^+$}
\def\splus{S$^+$}
\def\mnplus{Mn$^+$}
\def\coplus{Co$^+$}
\def\znplus{Zn$^+$}
\def\crplus{Cr$^+$}
\def\caplus{Ca$^+$}

\begin{document}
\title{Abundances of refractory ions in Beta Pictoris exocomets}
\titlerunning{Ion abundances in exocomets}

\author{
T.\ Vrignaud\inst{1}
\and
A. Lecavelier des Etangs\inst{1}
\and
P. A.~Str\o m \inst{2}
\and
F. Kiefer\inst{3}  
}

\institute{
Institut d'Astrophysique de Paris, CNRS, UMR 7095, Sorbonne Université, 98$^{\rm bis}$ boulevard Arago, 75014 Paris, France 
\and
Department of Physics, University of Warwick, Coventry CV4 7AL, UK
\and 
LESIA, Observatoire de Paris, Université PSL, CNRS, 5 Place Jules Janssen, 92190 Meudon, France
}

\abstract{

\betapictoris\ is a young, A5V star, known for harbouring a large number of cometary-like object (or exocomets) which frequently transit the star and create variable absorption signatures in its spectrum. The physical and chemical properties of these exocomets can be probed by the recently introduced curve of growth approach, which enables column densities measurements in cometary tails using absorption measurements in numerous spectral lines. Using this approach, we present a new study of archival spectra of \bp\ obtained with the \emph{Hubble} Space Telescope, the HARPS spectrograph, and at the Mont John University Observatory, aimed at constraining the abundance of refractory ions in \bp\ exocomets. 29 individual objects are studied, all observed in \feii\ lines (used as a reference ion) and at least one other species (\niii, \caii, \crii...). We find that the refractory composition of \bp\ exocomets is overall stable, especially for singly ionised species, and consistent with solar abundances. This validates the use of the curve of growth approach to study exocometary composition. We also show that some ions, such as \caii, are significantly depleted compared to solar abundances, which allows us to constrain the typical ionisation state in \bp\ exocomets. We find that most refractory elements (Mg, Ni, Fe...) are split in similar fractions between their first and second ionisation states, with the exception of Ca, mostly ionized twice. A strong correlation between the \aliii/\feii\ ratio and radial velocity is also found, showing that the most redshifted exocomets tend to be more ionised. These results open the way for further modelling of exocomets, in order to unveil their composition and the physical processes that affect their tails.

}


\maketitle

\section{Introduction}

\betapictoris\ (\bp) is a well-studied young \citep[20 Myr,][]{Miret-Roig_2020} planetary system, often considered a benchmark for studying planetary system formation and evolution. It has attracted significant attention due to its prominent debris disk \citep{Smith_1984}, which is rich in gas and dust \citep{Vidal-Majar_1986,Roberge_2006,Apai_2015} and extends over hundreds of au. The system is also known for hosting planets \citep{Lagrange_2010,Lagrange_2019,Nowak2020} and a large number of exocomets \citep{Ferlet_1987,Zieba_2019, Lecavelier_2022}. In fact, \bp\ is, by far, the star showing the most intense exocomet activity, with hundreds of objects detected so far using transit spectroscopy \citep{Kiefer_2014}. Other systems with reported exocomets (typically A-type stars, such as HD 172555 and 49 Cet) generally show much weaker activity \citep[][]{Kiefer_2014b,Montgomery2012,rebollido2020,Strom_2020}. Exocomets are of high scientific interest, as they provide valuable insights into the physical and dynamical processes occurring in the early stages of planetary system evolution. For instance, it has been proposed that the presence of exocomets in the inner region of the \bp\ system is linked to the gravitational perturbation of planets, which can elongate the orbits of small bodies formed at large stellar distances and bring them to the stellar vicinity \citep[][]{Beust1996, Beust2024}. Hints of collisions in the \bp\ system have also been reported \citep[][]{Rebollido2024, Chen2024}, providing an alternative scenario for the origin of the exocomets. Yet, despite extensive observational campaigns over the past decades, the intrinsic properties of the exocomets — such as their composition and production rates — have remained largely unknown. In particular, no abundance measurements have been conducted so far, depriving us of key information to understand their nature and origin.

A significant advancement in the study of exocomets came with the work of \cite{Vrignaud24}, who introduced a new approach for analysing spectroscopic transit data, termed the "exocomets curve of growth". This approach is based on the fact that the absorption depth of a transiting comet in a set of lines from a fixed species (eg. \feii) is generally a simple function of the lines parameters (oscillator strength, excitation energy). The comet's absorption depths can thus be easily fitted to extract the main physical properties of the comet (covering factor, column density) and to infer critical information about the excitation states of ions within the transiting gas. 

By applying this approach to a comet observed on December 6, 1997, the most recent study of \cite{Vrignaud24b} showed that the gas excitation in \bp\ exocomets is primarily controlled by the photon flux received from the star, rather than by electronic collisions. In this so-called radiative regime, the excitation temperature of the gas is set to the stellar effective temperature ($\sim 8000 \ \si{K}$); the excitation state is thus de-correlated from the local properties of the gas (density, kinetic temperature). This regime is typically associated with a low electronic density ($n_e \leq 10^{7} \ \si{cm^{-3}}$) and close transit distance ($d \leq 60 R_\star$). The study of \cite{Vrignaud24b} also showed that the various ions (e.g. \feii, \crii, \niii...) detected in the December 1997 comet are characterised by similar radial velocity profiles and excitation states, hinting that they are well-mixed. This shared behaviour among the observed ions is of great use for estimating their abundance ratios in exocomets. For instance, the \niplus/\feplus\ and \crplus/\feplus\ ratios in the December 6, 1997 comet were estimated to be $8.5 \pm 0.8 \cdot 10^{-2}$ and $1.04 \pm 0.15 \cdot 10^{-2}$ respectively, rather close to the solar Ni/Fe and Cr/Fe values \citep[][]{Vrignaud24b}.

In this paper, we aim to apply the exocomet curve of growth method to all the exocomets detected so far with the Space Telescope Imaging Spectrograph (STIS) onboard the Hubble Space Telescope (HST). By analysing the absorption features observed in an extended list of lines from a dozen of chemical species, we aim to place strong constraints on the composition of the cometary tails that routinely transit \bp. This comprehensive analysis will allow us to investigate the physical mechanisms that affect the composition of these tails (e.g. ionisation), and to further our understanding of the chemical diversity within the \bp\ system. 

The set of observations used in our analysis is presented in Sect.~\ref{Sect. Studied observations}. Sections \ref{Sect. A refined curve of growth model} and~\ref{Sect. Application to 30 comets} then describe our modified curve of growth model, aimed at analysing several species in a given exocomet simultaneously, and its application to a selection of 29~\bp\ exocomets observed with the HST. The discussion and conclusion are provided in Sect.~\ref{Sect. Discussion} and~\ref{Sect. Conclusion}.

\section{Studied observations}
\label{Sect. Studied observations}

\subsection{Raw data}
\label{Sect. Raw data}

Our study is primarily based on high resolution spectroscopic observations of \bp\ obtained with the STIS spectrograph onboard the HST. Obtained between 1997 and 2024, these spectra cover wavelength ranges between 1600 and 3000 \A, and include a large number of spectral lines from various ionised species (\feii, \niii, \siii, \crii...). These lines, particularly from \feii, have been shown to be a powerful probe of the physical properties of the transiting comets \citep{Vrignaud24,Vrignaud24b}.

Our dataset is completed with spectroscopic observations of \bp\ obtained with other instruments, very close in time to the STIS data (typically a few hours apart). These additional observations were collected using the Cosmic Origins Spectrograph (COS), also aboard the HST, as well as the High Accuracy Radial Velocity Planet Searcher (HARPS) on the European Southern Observation 3.6m telescope in La Silla, Chile, and the échelle spectrograph on the McLellan 1-m Telescope at the Mount John University Observatory (MJUO) in New Zealand. The COS spectra cover the far-ultraviolet (FUV) range ($\sim~1100~-1500$ \A), while the HARPS and MJUO spectra probe the visible domain. These complementary datasets provide access to spectral lines that were never observed with STIS, such as the 1250\,\A\ \Sii\ triplet (COS) and the 3900\,\A\ \caii\ doublet (HARPS, MJUO). 

The COS and STIS data were retrieved from the MAST archive, while the HARPS data was obtained from the ESO science portal. The MJUO spectra were obtained from \cite{Tobin2019}; only one night (December 6, 1997) was used in our study. A summary of the studied observations is provided in Table \ref{Tab. data}.

\subsection{Data Reduction}
\label{Sect. Data Reduction}

The HARPS and STIS data were reduced in a similar manner as that described in \cite{Vrignaud24}. First, the different spectral orders of each spectrum were weighted together and resampled on a common wavelength table. Then, to correct from flux calibration disparity from one observation to another, each spectrum was renormalised by a cubic spline fitted to stable spectral regions \citep[see Fig. 1 of][]{Vrignaud24}). Finally, each set of spectra collected during a short time frame (e.g. the December 6 and December 19, 1997 HST visits) was averaged together, to boost their Signal-to-Noise ratio (S/N) and facilitate their analysis. The COS spectra were simply renormalised by their mean value in the 1350-1380 \A\ range, which does not show spectral variations due to exocomets. Contrary to STIS and HARPS, no chromatic variation of flux calibration was found for the COS data. The MJUO spectra were not renormalized, as \cite{Tobin2019} already provide flux-calibrated data. Finally, all spectra were shifted to the rest frame of \bp, assuming an heliocentric radial velocity of 20 km/s \citep{Gontcharov2007}.

\subsection{Exocomet-free spectrum}
\label{Sect. The exocomet-free spectrum}

\begin{figure}[h!]
\centering
    \includegraphics[scale = 0.44,     trim = 0 0 0 40,clip]{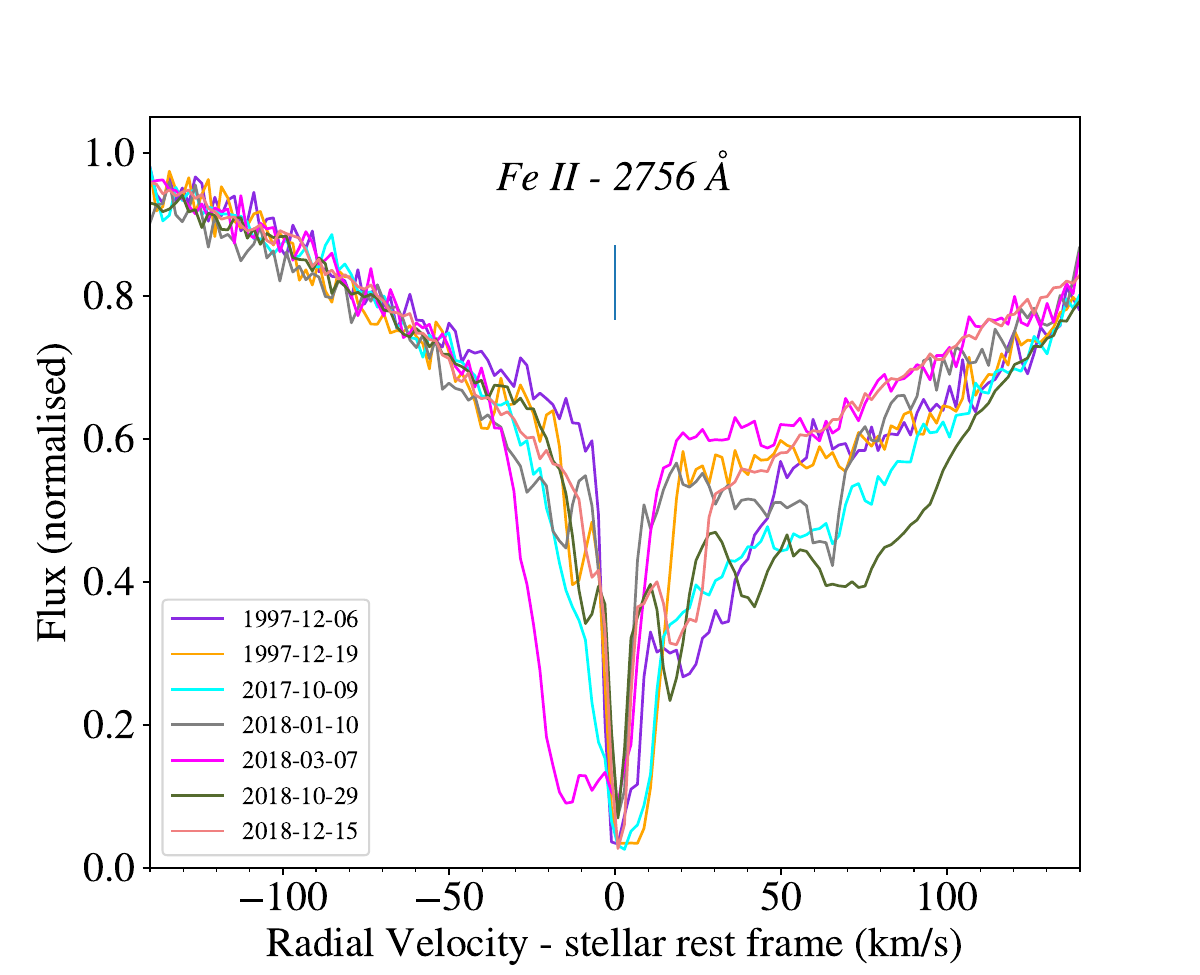}    
    \includegraphics[scale = 0.44,     trim = 0 0 0 40,clip]{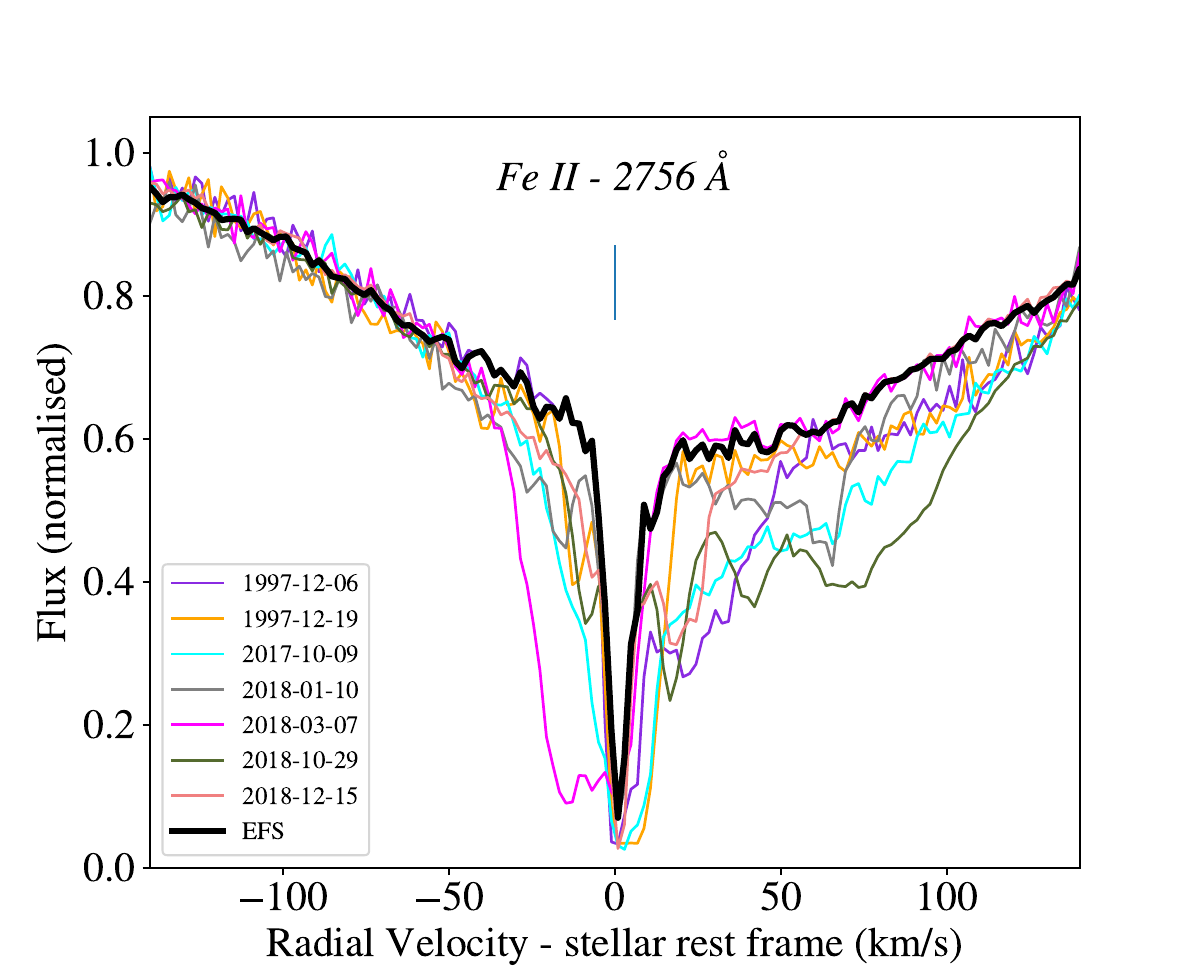}    
    \caption{\small Illustration of the EFS recovery. \textbf{Top}: 2728 \A\ \feii\ line, showing clear spectral variations from one visit to another. \textbf{Bottom}: same line, showing the calculated EFS.}
    \label{Fig. illustration EFS}
\end{figure}

Once the studied spectra have been renormalised to a common flux level, it becomes easy to visualise the spectral variations caused by the intense cometary activity around the star (see Fig.~\ref{Fig. illustration EFS} for \feii). However, to retrieve the absorption profile of a given exocomet, one needs to recover a reference stellar spectrum, free of any cometary absorption. 

This Exocomet-Free Spectrum (EFS) is obtained using a method similar to the one described in \cite{Vrignaud24}: First, we identify, for each spectrum, the set of lines affected by comet absorption, and the radial velocity ranges at which this absorption occurs (as an example, Table~\ref{Tab. data} provides the ranges used for the strongest \feii\ lines). We then average, for each wavelength pixel, all the spectra for which no comet absorption is expected (at that pixel).

This method was used for all lines, except for the 1850\,\A\ \aliii\ and 3900\,\A\ \caii\ doublets. In \aliii, the comets studied in the following (which are also detected in \feii, \niii...) are often blended with wider, highly variable absorption features, barely seen in singly ionised species. We thus did not calculate any EFS for this doublet; instead, for each exocomet, a continuum was fitted on nearby spectral regions using a cubic spline algorithm, and used as a reference spectrum. For the HARPS observations of the \caii\ doublet, we used the same method as introduced in \cite{Kiefer_2014}, which takes advantage of the very high number of available observations of \bp\ at these wavelengths (9\,000+ HARPS spectra obtained from 2003 to 2024). Finally, for the MJUO data, we used the reference spectrum provided in \cite{Tobin2019}.

\begin{figure}[h!]
\centering
    \includegraphics[scale = 0.44,     trim = 0 0 0 40,clip]{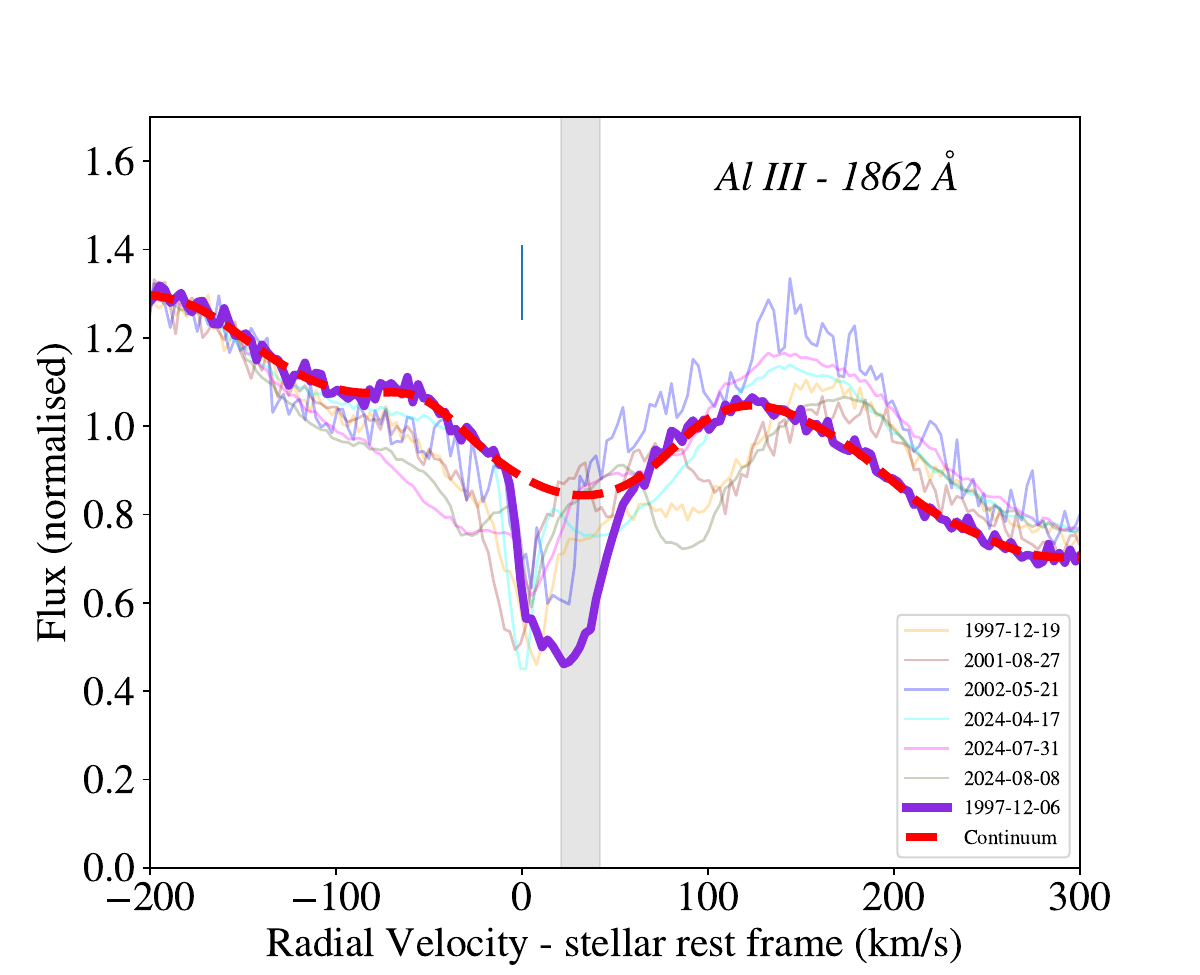}    
    \caption{\small Recovery of a reference spectrum for comet \#1 (1997-12-06, grey area) in the \aliii\ 1862 \A\ line. Contrary to other lines, we do not try to recover the EFS; instead, we fit a continuum to the observed spectrum using a cubic spline algorithm. This allows us to correct from the blending with other wide, shallow exocomets.}
    \label{Fig. illustration reference Al III}
\end{figure}

A sample of spectral lines showing clear comet absorption is provided on Fig. \ref{Fig. line examples}, along with the calculated EFS. For lines where very few observations are available, the recovery of the EFS might be incomplete (e.g. the \feii\ 2399 \A\ line).

\subsection{Correlation between cometary features}
\label{Sect. Correlation between cometary features}

\begin{figure}[h!]
\centering
    \includegraphics[scale = 0.44,     trim = 0 0 0 40,clip]{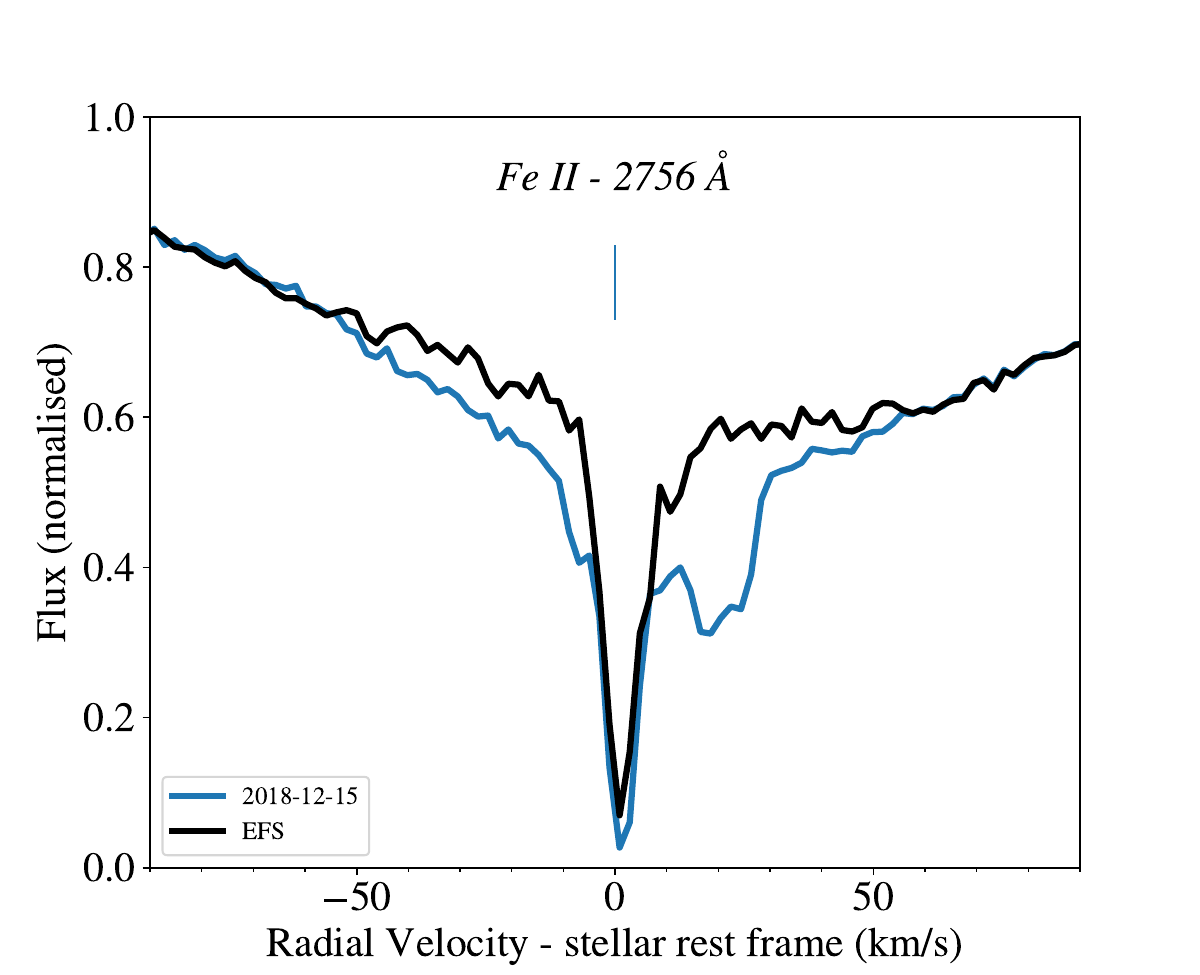}    
    \includegraphics[scale = 0.44,     trim = 0 0 0 40,clip]{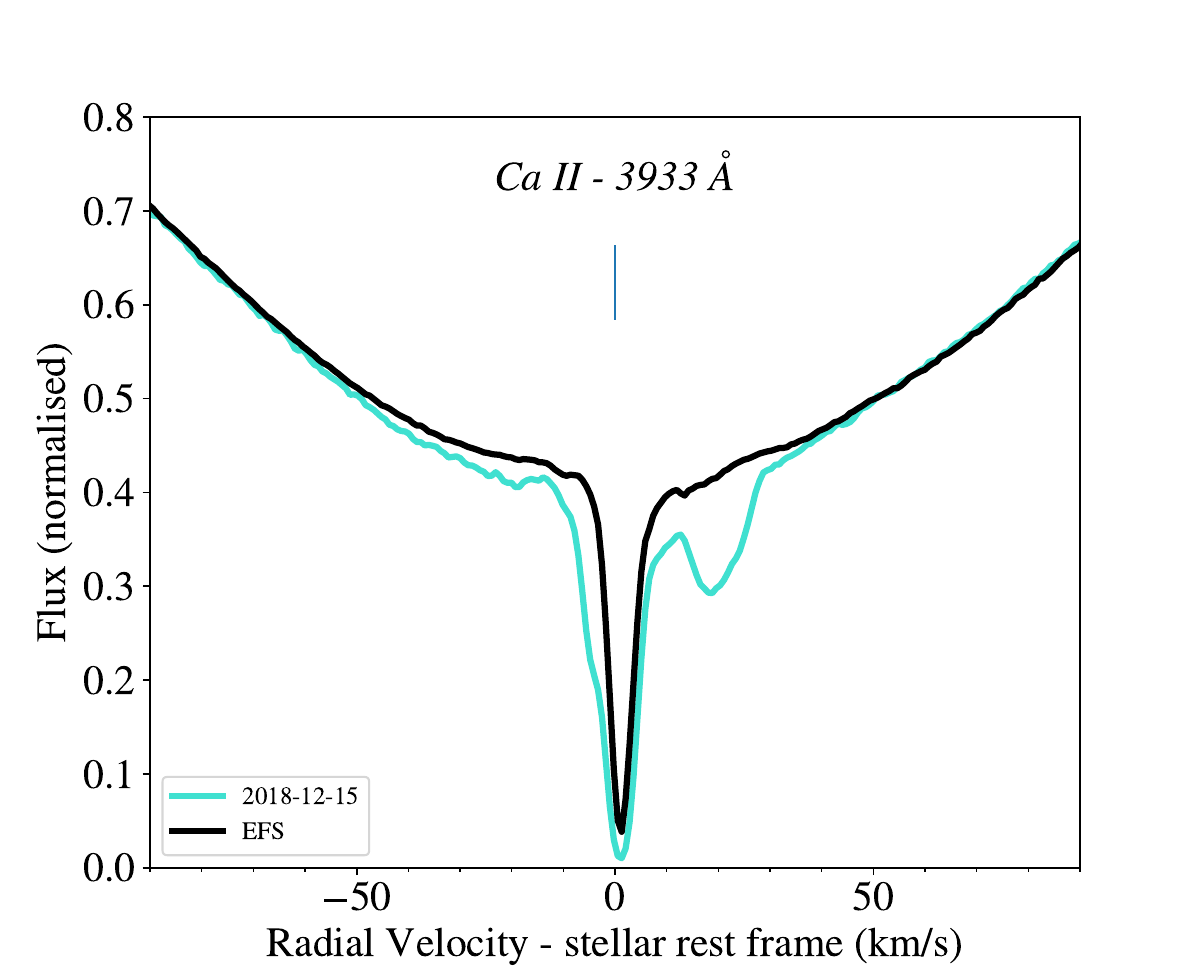}     
    \includegraphics[scale = 0.44,     trim = 0 0 0 40,clip]{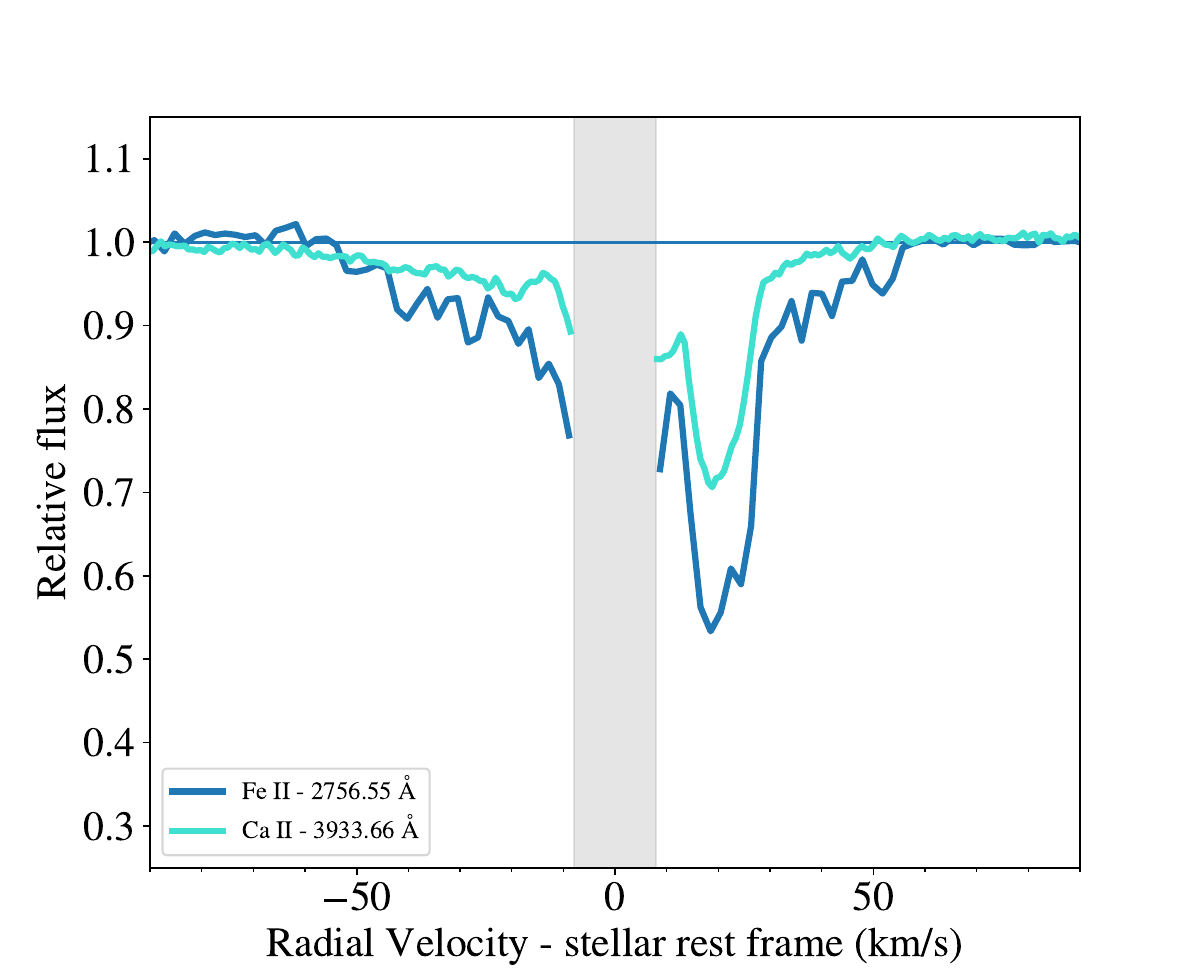}    
    \caption{\small Cometary absorption observed in \feii\ (top) and \caii\ lines (middle, one hour later) on December 15, 2018. The EFS is plotted with a thick black line. The comparison of the absorption after division by the reference spectrum is given in the bottom panel (the region near the circumstellar line was removed, due to a poorer EFS determination in \feii\ lines).  
    }
    \label{Fig. Compare Fe/Ca}
\end{figure}

As pointed out in \cite{Vrignaud24b}, the cometary absorptions observed in \feii\ lines correlate well with signatures observed in other species (\niii, \crii). As an example, Fig. \ref{Fig. Compare Fe/Ca} shows two observations of \bp\ in \feii\ (STIS) and \caii\ (HARPS), obtained one hour apart on December 15, 2018. The similarity of the absorption features is striking; in particular, a strong redshifted comet (designated below as comet 18, see table \ref{Tab. comets}) is clearly visible in both spectra, roughly between 15 and 30 km/s. 

\FloatBarrier

The strong correlation between exocometary features observed in lines from different species shows that ions in exocometary tails usually stay mixed, thanks to efficient momentum exchange via Coulomb scattering \citep{Vrignaud24b}. This ion mixing is a key feature of \bp\ exocomets, as it ensures that the signatures of a comet in a set of lines from different species are not influenced by the spatial distribution of each species individually, but rather by the comet geometry as a whole (e.g., covering factor), by the intrinsic spectral properties of each species (e.g., oscillator strengths), and by the global composition of the tail. Combined with the curve of growth approach, this ion mixing greatly facilitates the estimate of abundance ratios in \bp\ exocomets.

\section{A multi-species curve of growth model}
\label{Sect. A refined curve of growth model}

\subsection{The curve of growth approach}

Introduced in \cite{Vrignaud24}, the curve of growth approach aims at constraining the physical properties of a transiting exocomet observed in spectroscopy. Instead of focusing on a single spectral line, this approach relies on comparing the exocomet absorption depth in many (typically 10-100) lines, with diverse oscillator strengths and excitation levels. The dependency of the comet's absorption depth with the line oscillator strength and excitation energy can provide valuable estimates of the comet's physical properties (column density, excitation temperature, and size given by the covering factor of the stellar disk).

The study of \cite{Vrignaud24} presented a first curve of growth model, where a transiting comet is approximated by a cloud with an homogeneous column density, covering a limited fraction of the stellar disk, and characterised by a uniform excitation temperature. In this model, the average absorption depth $\overline{\rm abs}_{lu}$ \citep[see definition in][]{Vrignaud24} of the comet in any line of a fixed species (such as \feii) is given by: 
\begin{equation}
    \hspace{3 cm}\overline{\rm abs}_{lu}  = \overline{\alpha}  \cdot \left(  1 - e^{- \tau_{lu}} \right),
\label{Eq. Model 1 zone 1 species}
\end{equation}
with $\tau_{lu} = \gamma \cdot \frac{\lambda_{lu}}{\lambda_0} g_l f_{lu} \cdot e^{-E_l/k_B T}$ the optical depth of the considered line. Here, $l$ and $u$ denote the lower and upper levels of the transition, and $\lambda_{lu}$, $f_{lu}$, $g_l$ and $E_l$ its wavelength, oscillator strength, lower level multiplicity, and lower level energy (respectively). The model is characterised by three parameters: $\overline{\alpha}$, the comet's average covering factor; $\gamma$, the typical optical thickness of the studied species (linked to its column density, see Sect. \ref{Sect. Estimating abundance ratios} below); and $T$, its excitation temperature. These parameters are, a priori, specific to the radial velocity range where the comet's average absorption depths are measured, and to the considered species (e.g., \feii). Finally, $\lambda_0$ is an arbitrary wavelength. As in previous studies, we will use $\lambda_0 = 2756.551$ \A\ (which is the wavelength of a strong \feii\ line), regardless of the studied species.

Despite its simplicity, this one-zone model happens to fit most of the observed variable absorptions in \bp\ spectrum well (see Sect. \ref{Sect. Application to 30 comets}). However, \cite{Vrignaud24b} noted that some cometary signatures can slightly deviate from this model, and are better described by the transit of several gaseous components with various sizes and optical depths. 

\cite{Vrignaud24b} thus introduced a more sophisticated curve of growth model, in which a transiting comet is represented with two gaseous components: a dense core and a surrounding, thinner envelope. This new model has now five parameters: $\overline{\alpha_t}$, the total comet's covering factor, $\overline{\alpha_c}$, the covering factor of the core component, $\gamma_c$, the typical optical thickness in the core component, $\gamma_e/\gamma_c$, the optical thickness ratio between the external and core components, and $T$, the excitation temperature throughout the comet. The average absorption depth $\overline{\rm abs}_{lu}$ of the comet in any line of the studied species is now expressed as: 
\begin{equation}
    \hspace{1.4 cm} \overline{\rm abs}_{lu}  = \overline{\alpha}_{\rm c}  \cdot \left(  1 - e^{- \tau_{\textrm{c, } lu}} \right) + \ \overline{\alpha}_{\rm e} \cdot \left(  1 - e^{- \tau_{\textrm{e, } lu}} \right),
\label{Eq. Model 2 zones 1 species}
\end{equation}
with: 
$$\tau_{\textrm{c, } lu} = \gamma_{\rm c} \cdot \frac{\lambda_{lu}}{\lambda_0} g_l f_{lu} e^{-E_l/k_B T},$$
$$\tau_{\textrm{e, } lu} = \gamma_{\rm e} \cdot \frac{\lambda_{lu}}{\lambda_0} g_l f_{lu} e^{-E_l/k_B T},$$
and $\overline{\alpha}_{\rm e}$ and $\gamma_{\rm e}$ the covering factor and typical optical thickness of the outer envelope, given by:
$$
\begin{array}{ll}
     \overline{\alpha}_{\rm e} = \overline{\alpha}_{\rm t} - \overline{\alpha}_{\rm c},  \\
     \gamma_{\rm e} = \gamma_{\rm c} \cdot \frac{\gamma_{\rm e}}{\gamma_{\rm c}}. 
\end{array}
$$

Again, the five model parameters are - a priori - specific to the studied species (\feii, \niii...) and to the radial velocity range where the comet's average absorption depth is measured.

\subsection{Fitting several species simultaneously}

In previous studies, these models were mostly applied to \feii\ lines, which are very numerous and show strong comet absorption. However, \cite{Vrignaud24} showed that the absorption signatures of individual exocomets in lines from different species (\niii, \crii) tend to have the same shape and to follow the same curve of growth as \feii; in other words, the comet geometry ($\overline{\alpha}$, $\overline{\alpha}_c$...) and excitation temperature ($T$) constrained by \feii\ lines also apply to other species. 

In the following, we will thus assume that exocometary gaseous tails always remain well-mixed, regardless of their overall shape. This hypothesis allows us to introduce a multi-species curve of growth model, able to fit several species simultaneously. Going back to Eq.~\ref{Eq. Model 1 zone 1 species} and~\ref{Eq. Model 2 zones 1 species}, we now assume that the geometric parameters ($\overline{\alpha}$ for Eq.~\ref{Eq. Model 1 zone 1 species}, $\overline{\alpha}_t$, $\overline{\alpha}_c$, $\gamma_e/\gamma_c$ for Eq.~\ref{Eq. Model 2 zones 1 species}) and the excitation temperature ($T$) are the same for all species; only the $\gamma$ (Eq. \ref{Eq. Model 1 zone 1 species}) and $\gamma_c$ (Eq. \ref{Eq. Model 2 zones 1 species}) parameters remain specific to each atom or ion. These assumptions allow us to refine Eqs.~\ref{Eq. Model 1 zone 1 species} and~\ref{Eq. Model 2 zones 1 species} into Eqs.~\ref{Eq. Model 1 zone n species} and~\ref{Eq. Model 2 zones n species}:

\begin{equation}
    \hspace{2.7 cm}\overline{\rm abs}_{lu, \ i}  = \overline{\alpha}  \cdot \left(  1 - e^{- \tau_{lu, \ i}} \right) 
\label{Eq. Model 1 zone n species}
\end{equation}
with:
$$
\tau_{lu, \ i} = \gamma_i \cdot \frac{\lambda_{lu}}{\lambda_0} g_l f_{lu} \cdot e^{-E_l/k_B T},$$ 
and:
\begin{equation}
    \hspace{1.2 cm} \overline{\rm abs}_{lu, \ i}  = \overline{\alpha}_{\rm c}  \cdot \left(  1 - e^{- \tau_{\textrm{c, } lu, \ i}} \right) + \ \overline{\alpha}_{\rm e} \cdot \left(  1 - e^{- \tau_{\textrm{e, } lu, \ i}} \right)
\label{Eq. Model 2 zones n species}
\end{equation}
with: 
$$
\tau_{\textrm{c, } lu, \ i} = \gamma_{{\rm c},\ i} \cdot \frac{\lambda_{lu}}{\lambda_0} g_l f_{lu} e^{-E_l/k_B T},
$$
$$
\tau_{\textrm{e, } lu, \ i} = \gamma_{{\rm c},\ i} \cdot \frac{\gamma_{\rm e}}{\gamma_{\rm c}} \cdot \frac{\lambda_{lu}}{\lambda_0} g_l f_{lu} e^{-E_l/k_B T}.
$$
In the above equations, we now use a letter $i$ to specify the species associated with the line $(l,u)$. The two models have now respectively $n+2$ and $n+4$ parameters, with n the number of fitted species.

\subsection{Estimating abundance ratios}

Equations~\ref{Eq. Model 1 zone n species} and \ref{Eq. Model 2 zones n species} can be fitted to the absorption depths of a given comet in a set of lines from several species, in order to constrain the comet geometry, excitation temperature, and typical optical thicknesses of each species ($\gamma_i$ for Eq.~\ref{Eq. Model 1 zone n species}, $\gamma_{{\rm c}, \ i}$ for Eq.~\ref{Eq. Model 2 zones n species}). Once constrained, these parameters allow for the estimate of the total column density of each species, through \citep[see][]{Vrignaud24,Vrignaud24b}:

$$
{\rm N}_{{\rm tot},\ i} = \frac{4 \varepsilon_0 m_e c}{e^2 \lambda_0} \Delta v \ Z_i\big(T\big) \ \gamma_i \ \overline{\alpha}
$$
for the one-zone model (Eq.~\ref{Eq. Model 1 zone n species}), and:
$$
{\rm N}_{{\rm tot},\ i} = \frac{4 \varepsilon_0 m_e c}{e^2 \lambda_0} \Delta v \ Z_i\big(T\big) \ \gamma_{{\rm c},\ i} \ \Big (\overline{\alpha}_{\rm c} + \left( \gamma_{\rm e}/\gamma_{\rm c} \right) \overline{\alpha}_{\rm e}\Big)
$$
for the two-component model (Eq.~\ref{Eq. Model 2 zones n species}). Here, $\Delta v$ is the radial velocity width of the comet, $Z_i(T)$ is the partition function of the considered species, and $\varepsilon_0$, $m_e$, and $c$ are respectively the vacuum permittivity, electron mass and light velocity. Using the assumption that the geometric parameters are identical for all the studied species, the abundance ratio of two species $i$, $j$ then writes:

\begin{equation}
    \hspace{2.4 cm} \left[ \frac{i}{j} \right] = \frac{{\rm N}_{{\rm tot},\ i}}{{\rm N}_{{\rm tot},\ j}} = \frac{Z_i\big(T\big) \ \gamma_i}{Z_j\big(T\big) \ \gamma_j},
\label{Eq. abundance ratio}
\end{equation}
where the $\gamma_i$ and  $\gamma_j$ parameters are replaced by $\gamma_{c,i}$ and $\gamma_{c,j}$ in the case of a two-component model.

\section{Application to a sample of comets}
\label{Sect. Application to 30 comets}

\begin{table*}[h]
    \centering 
    \begin{threeparttable}
    \renewcommand{\arraystretch}{1.2}
    \caption{\small List of the studied chemical species sorted by ionisation energy.}
    \begin{tabular}{c c c c c c c c c c c c c}          
            
            \cline{1-13}  
            \noalign{\smallskip}
            \cline{1-13}  
            
            Species & \Si & \caii  & \mgii & \mnii & \feii & \siii & \crii & \coii & \znii & \niii & \Sii & \aliii \\
            Ionisation energy (eV) & 10.36   & 11.87 &  15.04    & 15.64      &  16.19 & 16.35 & 16.49 & 17.08 & 17.96 & 18.17 & 23.34 & 28.45         \\

            \cline{1-13}

        \end{tabular} 
        \label{Tab. List ions}
\end{threeparttable}
\end{table*}

\subsection{The sample of exocomets}

Our study aims at measuring the abundance ratio of several atoms and ions (\Si, \feii, \niii...) in \bp\ exocomets, using the available spectroscopic data. To carry out this analysis, we selected a set of 29 comets that appear to be well suited for abundance measurement. All these comets produce significant, well-individualised absorption in a large number of \feii\ lines (used as a reference species), as well as in, at least, one other refractory species (\crii, \niii, \caii...). Basic information on the 29 selected comets is provided in Tab. \ref{Tab. comets}. A unique identifier was attributed to each object (e.g., "\bp\ C19971206 a" for comet 1), according to the upcoming nomenclature of Lecavelier des Etangs et al. (in prep). A snapshot of each comet in one strong \feii\ line is also provided in Fig. \ref{Fig. curves of growth all}.

\subsection{Studied species and line selection}

Throughout our cometary sample, we were able to probe a total of 12 refractory species, listed in Table~\ref{Tab. List ions}, reflecting most of the abundant exocometary species with optically-allowed spectral lines in the UV/visible. Most of the studied species are ions, except for \Si, which stands out by its fairly high ionisation potential compared to most neutral species (10.36 eV). Other atoms (\fei, \mgi, \sii, \cai...) with lower ionisation potentials ($\leq 8$ eV) were not detected in any exocomet, and thus not included in the study. This non-detection is probably the result of efficient photo-ionisation. Despite noticeable spectral variations, \alii, \siiii\ and \siiv\ were also not included in the study, due to a very poor Signal-to-noise ratio (S/N) in the 1670~\A\ \alii\ and 1206~\A\ \siiii\ lines, and to highly blended absorption in the 1400 \A\ \siiv\ doublet.

For each individual exocomet, it was never possible to probe the abundance of all 12 species simultaneously, either because the appropriate lines were not observed, or because the lines were too faint or too saturated to be exploited. As an example, the \crii/\feii\ could be constrained only for the strongest comets (typically when N$_{\text{tot, \feii}} /\Delta v > 10^{13}$ cm$^{-2}$\,/\,km\,s$^{-1}$), as cometary signatures in \crii\ are usually very faint (a few \% at most). On the other hand,  the \mgii/\feii\ ratio was estimated only for the faintest comets (N$_{\text{tot, \feii}}/\Delta v \sim 10^{12}$ cm$^{-2}$\,/\,km\,s$^{-1}$), as \mgii\ signatures are easily saturated. 

\begin{figure}[h!]
\centering
    \includegraphics[scale = 0.44,     trim = 0 0 0 50,clip]{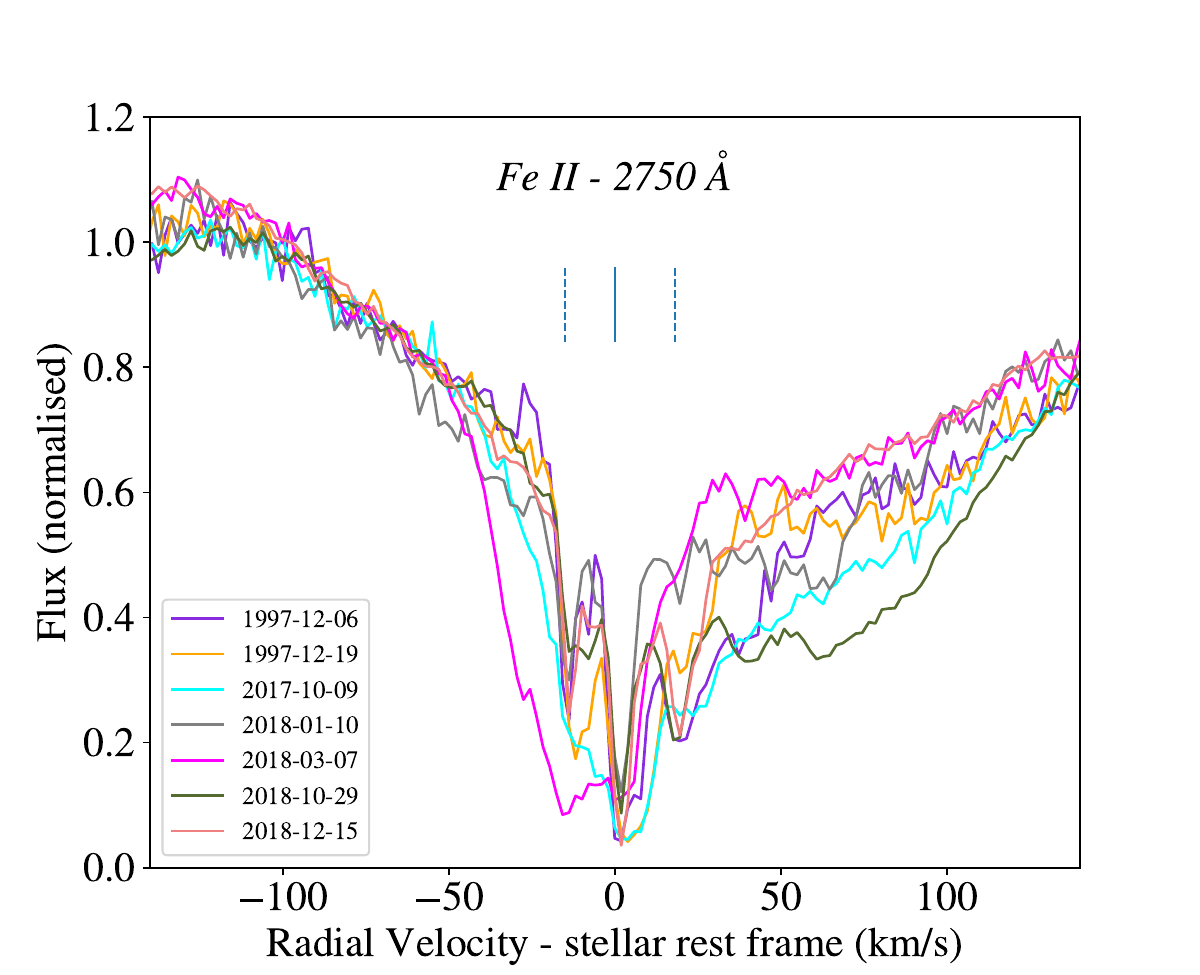}    
    \caption{\small Zoomed-in view of the 2750.13\,\A\ \feii\ line (indicated with a solid, blue marker). Even if cometary signatures are detected, they are blended with absorptions from other close \feii\ lines, at 2749.99 and 2750.30~\A\ (dotted markers). Due to this blending, the triplet was not included in our curve of growth analysis.}
    \label{Fig. Fe II 2750}
\end{figure}

The set of lines used to fit each exocomet was selected carefully. In particular, we did not include lines blended in close multiplets (i.e., when the lines separation is lower than the typical width of cometary signatures; see for instance Fig. \ref{Fig. Fe II 2750}). Lines for which the S/N on the cometary absorption depth is below unity were also not included in the fit, as well as those showing spurious features. Finally, we removed the lines for which no reliable EFS can be determined, generally due to an insufficient number of observations. The complete list of lines used throughout the study is given in 
Table~\ref{Tab. list lines}. Most of the line parameters ($\lambda_{lu}, A_{ul}$,...) were obtained from the NIST database \citep{NIST_ASD}. Table~\ref{Tab. comets} also provides, for each comet, the list of line series considered in the fit. 

\subsection{Fit procedure}
\label{Sect. Fit procedure}

Each comet was fitted using a similar procedure: 
\begin{itemize}
    \item We first measured the average absorption depth (relative to the EFS) of each comet in each line of its selected set. As in \cite{Vrignaud24}, uncertainties on absorption measurements were estimated by taking into account photon noise (generally between $0.5$ and $3 \%$) and systematic errors due to imperfect flux calibration ($0.5 - 1 \%$, estimated by comparing our spectra on stable spectral domains). For each comet, the RV range used to measure its absorption depth was chosen as the range where the absorption is the deepest, to limit the impact of systematic uncertainties. The RV ranges used in the fits are provided in Table \ref{Tab. comets}. 
    
    \item We then fitted the measured absorption depths of each comet using one of the multi-species curve of growth models (Eq.~\ref{Eq. Model 1 zone n species} or~\ref{Eq. Model 2 zones n species}) and a MCMC algorithm from the {\tt emcee} python library \citep{emcee2013}. These fits provided constrains on the geometry and excitation temperature of each comet, as well as on the $\gamma$ values of each fitted species. 
    \item We finally estimated the abundance of each fitted species, using Eq. \ref{Eq. abundance ratio}. To properly estimate uncertainties, this ratio is calculated at each step of the Markov chains.
\end{itemize}

Comets 22-29 (HST program \#17421) were all observed during a rather long time (3-4 HST orbits), enough to observe variation in their absorption profile. Each of these comets was analysed once, using the HST orbit showing the more individualised and deep absorption (e.g. for comet \#22, the second orbit of the 2024-04-17 visit). The specific HST orbits used for these comets are provided in Table~\ref{Tab. comets}.

\begin{figure*}[t]
    \centering    
    \includegraphics[clip,  trim = 220 65 0 40,  scale = 0.49]{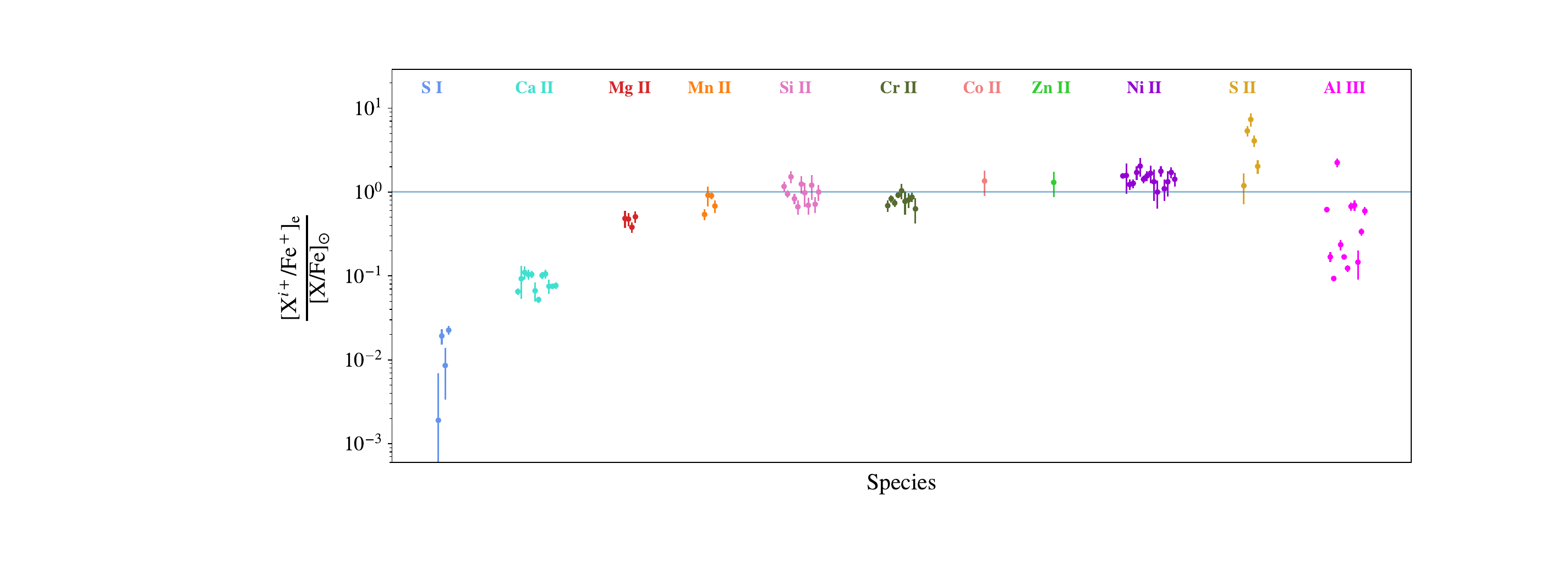}
    \caption[]{\small Overview of all abundance measurements performed throughout the 29 exocomet sample, grouped by chemical species. For each species (e.g. \Si), the measured ratios with \feii\ were divided by the solar ratio of the corresponding elements (e.g. S/Fe), as given by \cite{Asplund2009}. The ordering of the species reflect their ionisation energy; each group of measurements is ordered according to the comet indexes (Tab. \ref{Tab. comets}).}
    \label{Fig. recap all comets}
\end{figure*}

The one-zone curve of growth model was used for most comets, as it generally provides satisfactory fits. The only exceptions are comets \#1, \#2 and \#6, which appeared to be significantly better fitted with a two-component model (based on the comparison of the Bayesian Information Criteria (BIC) of the two models). 

For comets observed in a set of \feii\ lines covering a wide energy range 
($E/k_b$ typically in the range $0-12000 \, \si{K}$), the excitation temperature was set as a free parameter. For other comets, we used a Gaussian prior ($T = 9000 \pm 500 \,\si{K}$), based on values found for exocomets where the excitation temperature could be directly measured. This prior enabled us to estimate column densities in all exocomets, even those for which no direct estimate of the excitation temperature could be performed.

\subsection{Results}

The curves of growth of the fitted comets are shown on Fig.~\ref{Fig. curves of growth all}. The physical properties of the comets (covering factor, total \feii\ column density, excitation temperature) and their measured compositions (abundance ratios) are provided in Table~\ref{Tab. comets properties} and~\ref{Tab. abundances}, respectively. All abundance ratios are expressed relatively to \feii, which serves as a reference species. A visual recapitulation of all abundances measurements is also provided in Fig.~\ref{Fig. recap all comets}, allowing to visualise the dispersion of each species.

\section{Discussion}
\label{Sect. Discussion}

\subsection{Excitation temperatures}
\label{Sect. Stable excitation temperatures}

Before digging into the composition of our cometary sample, it is interesting to look at the temperatures that has been measured in comets \#1 to \#6 and \#8 (Table \ref{Tab. comets properties}).  In all these cases, the excitation temperature of the transiting comets remains close to $9000 \, \si{K}$ . This result validates the conclusion of \cite{Vrignaud24b}, which studied in details the excitation state of \feii\ in comet \#1: the density in the gaseous cloud surrounding the exocomets is generally sufficiently low for the gas excitation to be controlled by the stellar flux rather than by electronic collisions (radiative regime). In this context, the excitation temperature of the gas is roughly set to the stellar effective temperature \citep[$\sim$8052 \si{K},][]{Gray2006}. The slight discrepancy observed between the typical excitation temperature found in exocomets ($\sim$$9000$ \si{K}) and the stellar effective temperature could be explained by the presence of absorption lines in the stellar spectrum \citep[as proposed for solar system comets,][]{Manfroid2021} or by the self-opacity of the transiting gas, which distorts the stellar spectrum received in the innermost regions of the transiting comets.

This consistency of excitation temperature led us to assume that all \bp\ exocomets follow a radiative regime, even those for which the gas excitation cannot be directly probed because of an insufficient wavelength coverage. For these comets, we thus applied a gaussian prior on $T$ of $9000 \pm 500 \ \si{K}$, based on the values found for comets 1-6 and 8 (particularly comets 1, 4, 5 and 8, for which the most accurate values of $T$ could be obtained).

\subsection{Ion abundance}
\label{Sect. Consistency of ion abundance}

\begin{figure}[h!]
    \centering    
    \includegraphics[clip,  trim = 15 0 0 40,  scale = 0.38]{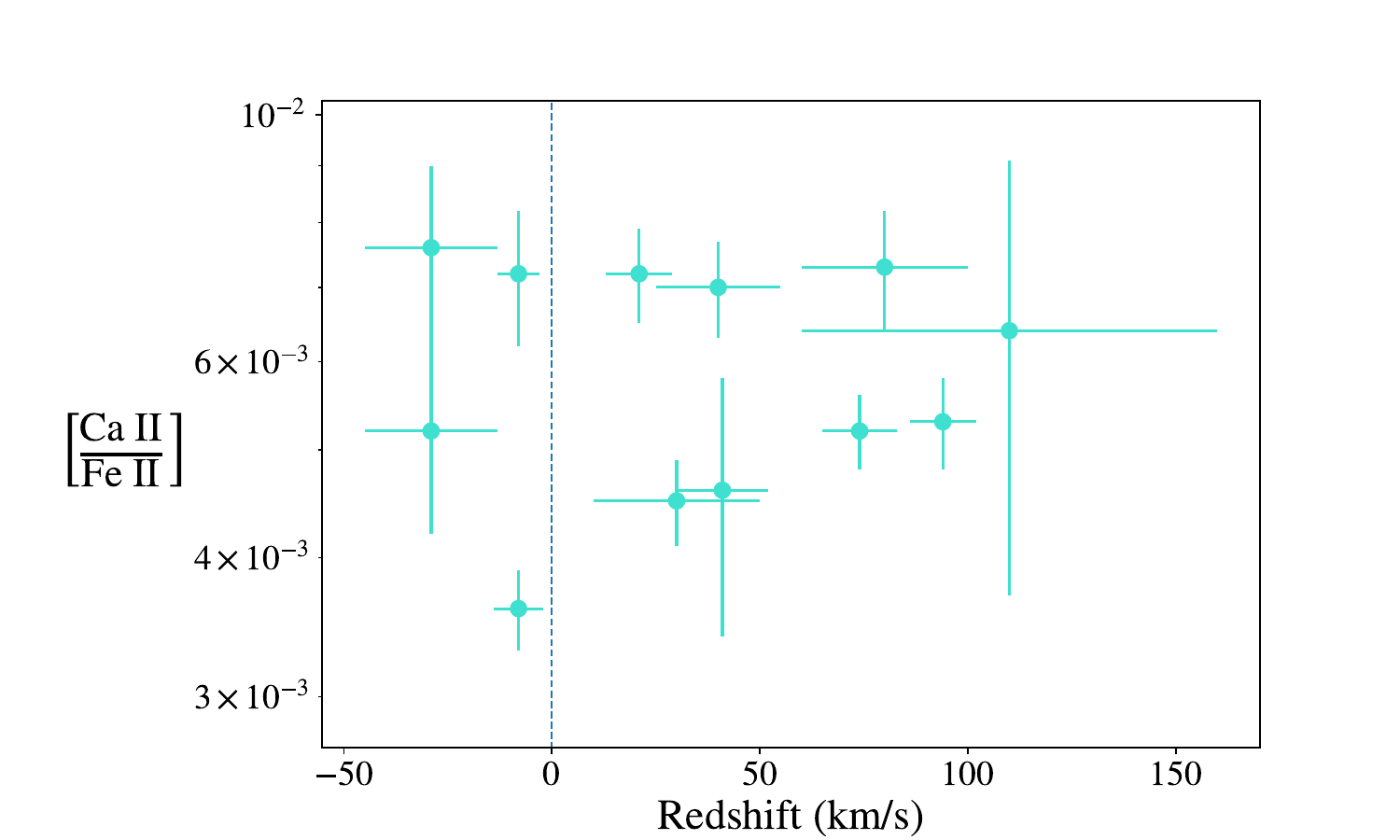}
    \caption[]{\small All \caii/\feii\ ratios measured in our exocometary sample, as a function of the comet redshift. Horizontal error bars reflect the radial velocity extent of the cometary signatures. }
    \label{Fig. Ca2 RV}
\end{figure}

The abundances of most of the singly ionised species (\caii, \niii, \siii, \crii, \mnii, \mgii) relative to \feii\ appear to be fairly consistent throughout our cometary sample, with values typically spanning over a factor 2-3 at most. As an example, the measured \caii/\feii\ ratio spans from 3$\times 10^{-3}$ to 8$\times 10^{-3}$ (mean: $5.8 \times 10^{-3}$), the \niii/\feii\ ratio from 5$\times 10^{-2}$ to 10 $\times 10^{-2}$ (mean: $7.6 \times 10^{-2}$), and the \siii/\feii\ ratio from 0.7 to 1.5 (mean: $0.95$). For \caii, the spread of abundance measurements is wider than the typical uncertainty on each measurement, indicating that the \caii/\feii\ ratio can vary slightly from one comet to another, up to a factor of $\sim$$3$. The origin of this diversity is however unclear; no correlation of the \caii/\feii\ ratio with radial velocity is observed (Fig. \ref{Fig. Ca2 RV}), which could have been expected given that low-velocity and high-velocity comets usually belong to two separated families with different transit distance \citep{Kiefer_2014}. For \niii, \siii, \crii, \mnii\ and \mgii, the observed abundance distributions are consistent with the estimated error bars; it is therefore difficult to conclude on any diversity in the composition of our comet sample. In addition, the abundance of these species does not seem to be correlated with radial velocity.

\begin{figure}[h!]
    \centering    
    \includegraphics[clip,  trim = 20 0 0 40,  scale = 0.38]{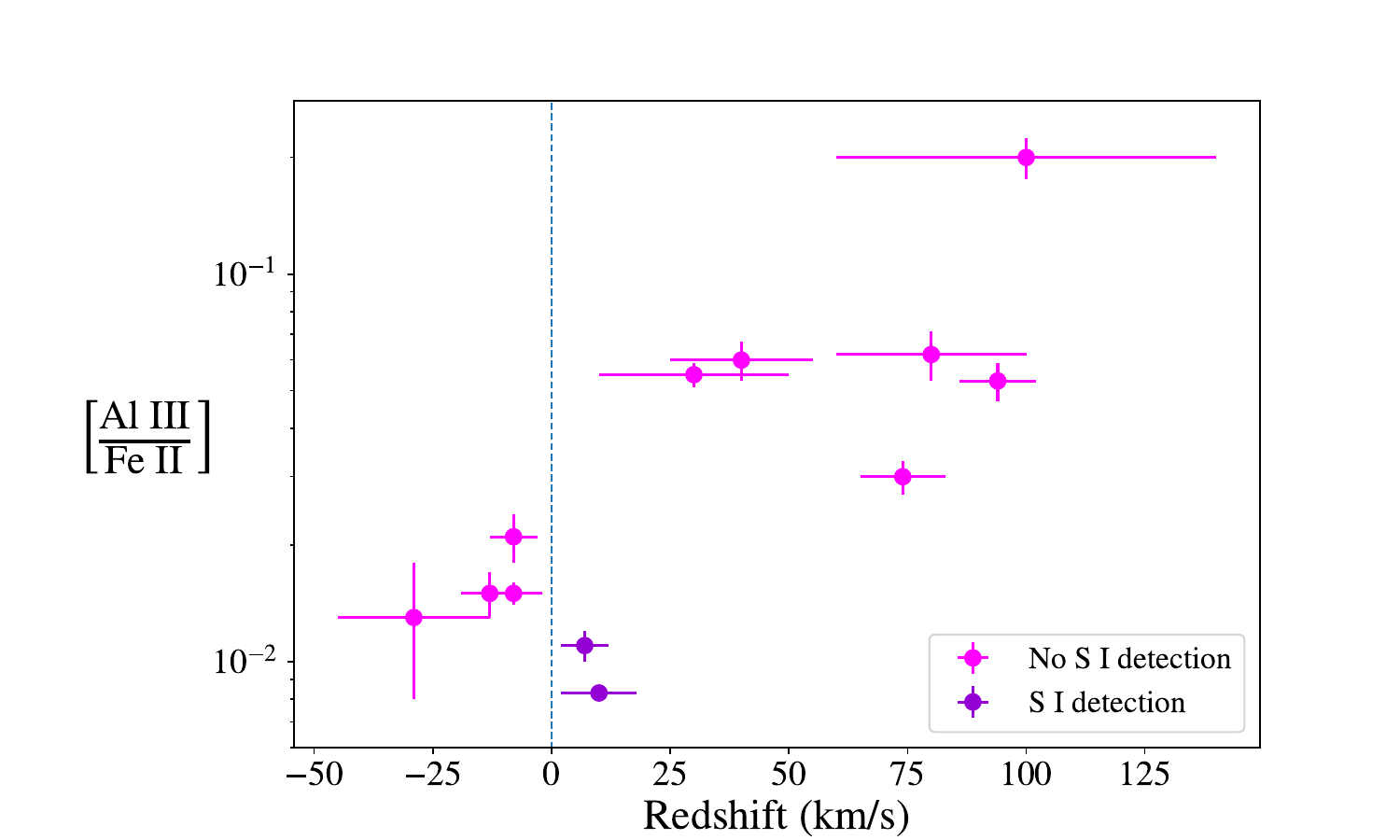}
    \caption[]{\small All \aliii/\feii\ ratios measured in our cometary sample, as a function of the comet redshift. Horizontal error bars reflect the radial velocity extent of the cometary signatures. The two violet markers indicate the two comets with clear \Si\ detections (see text).}
    \label{Fig. Al3 RV}
\end{figure}

A few species, however, exhibit a much wider abundance distribution (Fig.~\ref{Fig. recap all comets}). The most obvious one is \aliii, with a \aliii/\feii\ ratio spanning a factor of 20 and ranging from $\sim$$0.01$ (comets \#5, \#21) to $0.2$ (comet \#6). The \aliii\ abundance is found to be strongly correlated with redshift: the most redshifted comets are usually those who exhibit the strongest \aliii\ absorptions (Fig.~\ref{Fig. Al3 RV}). The \Si\ abundance in \bp\ exocomets also seems fairly diverse: this atom is clearly detected in comets \#5 and \#21 (\Si/\feii\ $\sim9\cdot 10^{-3}$, see Table~\ref{Tab. abundances}), but barely noticed in comets \#1 and \#20 (abundance measurements for other comets were not included in the study, due to much larger uncertainties). Here, it is interesting to note that the only two comets found to harbour a significant amount of \Si\ are those who display the weakest \aliii\ signatures (Fig.~\ref{Fig. Al3 RV}): contrary to \aliii, the \Si\ abundance in \bp\ exocomets seems to be anti-correlated with redshift, with low-velocity comets showing higher \Si/\feii\ ratio. These correlations (or anti-correlations) are  very likely linked to the transit distance of the comets: highly redshifted comets are generally closer to the star, meaning that they receive a greater irradiation. According to the model proposed in \cite{Beust_1993}, this would lead to a higher ionisation degree in cometary gas, due to a denser and hotter bow shock at the front of the coma. On the contrary, some low-redshift comets appear to be sufficiently far away from the star to allow \Si\ to survive ionisation (and prevent the formation of large amounts of \aliii), indicating that the compression mechanisms in their comae are much softer.

Finally, important variations were found for the \Sii/\feii\ ratio (estimated in five exocomets; see Fig.~\ref{Fig. recap all comets}), with values ranging from 0.5 to 3. However, these measurements should be taken with cautious: \feii\ and \Sii\ are observed with two different HST instruments (STIS and COS), at different resolutions, and during separate HST orbits, necessarily separated by 96~minutes. Further observations may thus be needed to confirm this diversity in \Sii/\feii\ ratios. For \coii\ and \znii, only comet \#1 (1997-12-06) could provide constraints on their abundances, with rather large uncertainties. It is therefore impossible to characterise the diversity of exocomet composition regarding these two species.

\begin{figure*}[t]
    \centering    
    \includegraphics[clip,  trim = 220 50 0 40,  scale = 0.49]{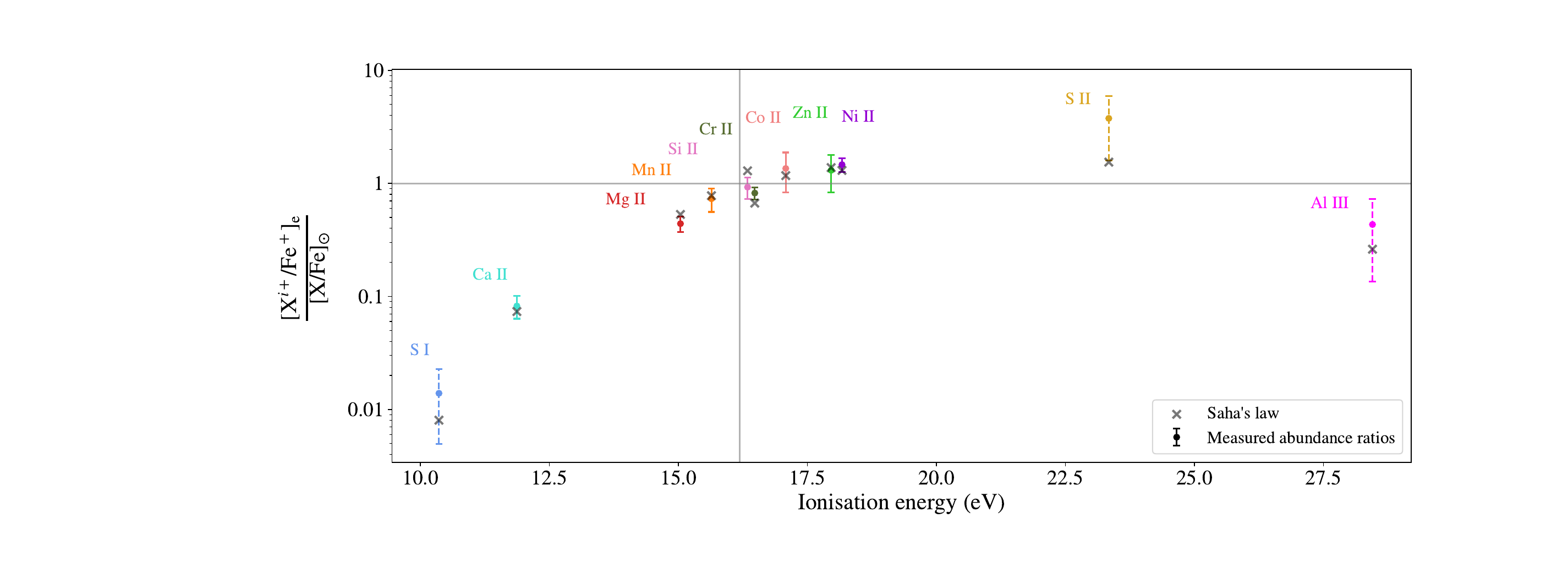}
    \caption[]{\small Typical abundances of 11 atom and ions in \bp\ exocomets, relative to \feii\ and normalised by solar abundances. The numerical values used to build this figures are provided in Tab. \ref{Tab. Average abundances}. Grey crosses indicate the best-fit model with Saha's equation (see Sect. \ref{Sect. Saha}.}
    \label{Fig. abondance vs ion pot}
\end{figure*}

\subsection{Link between abundances and ionisation energies}
\label{Sect. Ionisation energy}

Figure~\ref{Fig. recap all comets} shows that the typical abundances of the studied species in \bp\ exocomets are generally consistent with solar abundances. For instance, the exocometary \siii/\feii\ ratio is close to the solar Si/Fe ratio, $\simeq 1$. These quasi-solar values hint that the first ionisation degree is probably a major form of the studied elements. However, we also note some discrepancies: \Si, \caii\ and \mgii\ appear to be always depleted (by factors of $\sim$100, 10 and 2, respectively), while \niii\ and \Sii\ seem to be over-abundant (by a factor 1.5 for \niii, and 2 to 6 for \Sii).

\begin{table}[b!]
    \centering 
    \begin{threeparttable}
    \renewcommand{\arraystretch}{1.1}
    \caption{\small Average abundances of 11 refractory species in \bp\ exocomets, compared to solar ratios. 
    }
    \begin{tabular}{c c c c}          
            
            \cline{1-4}  
            \noalign{\smallskip}
            \cline{1-4}  
            \noalign{\smallskip}

            X$^{i+}$ & [X$^{i+}$/\feplus]$_e$ & [X/Fe]$_\odot$\tnote{a}  & $\eta_{\rm X} =  \frac{[\text{X}^{i+}/\text{\feplus}]_e}{[\text{X/Fe}]_\odot}$ \\
                  
    \noalign{\smallskip}
    \cline{1-4}  
    \noalign{\smallskip}

            S$^0$ & $5.8 \pm 3.7$        &   $420 \pm 50$   &  $0.014 \pm 0.009$\tnote{b} \\
                  & ($\cdot10^{-3}$) & ($\cdot10^{-3}$) &  \\

    \noalign{\medskip}

            \caplus & $0.57 \pm 0.11$  & $6.9 \pm 0.9$    &  $0.082 \pm 0.19$ \\
                    & ($\cdot10^{-2}$) & ($\cdot10^{-2}$) &   \\

    \noalign{\medskip}

            \mgplus & $0.55 \pm 0.05$ & $1.26 \pm 0.16$ &  $0.43 \pm 0.07$ \\

    \noalign{\medskip}

            \mnplus &  $6.2 \pm 1.2$     &  $8.5 \pm 1.1$    &   $0.73 \pm 0.17$ \\
                    &   ($\cdot10^{-3}$) &  ($\cdot10^{-3}$) &   \\

    \noalign{\medskip}

            \siplus & $0.95 \pm 0.17$ & $1.02 \pm 0.12$ &  $0.93 \pm 0.19$ \\

    \noalign{\medskip}

            \crplus & $1.13 \pm 0.05$    & $1.38 \pm 0.16$    &   $0.82 \pm 0.10$ \\
                    &   ($\cdot10^{-2}$) &   ($\cdot10^{-2}$) &   \\

    \noalign{\medskip}

            \coplus & $4.1 \pm 1.4$      &    $3.1 \pm 0.6$   &   $1.35 \pm 0.52$ \\
                    &  ($\cdot10^{-3}$)  &   ($\cdot10^{-3}$) &    \\

    \noalign{\medskip}

            \znplus & $1.5 \pm 0.5$     & $1.15 \pm 0.17$  &   $1.30 \pm 0.47$ \\
                    &  ($\cdot10^{-3}$) & ($\cdot10^{-3}$) &    \\

    \noalign{\medskip}

            \niplus & $7.61 \pm 0.28$   &  $5.2 \pm 0.7$    &  $1.46 \pm 0.20$ \\
                    &  ($\cdot10^{-2}$) &  ($\cdot10^{-2}$) &   \\

    \noalign{\medskip}

            \splus & $1.58 \pm 0.90$ &  $0.42 \pm 0.05$ & $3.8 \pm 2.1$ \\

    \noalign{\medskip}

            \aldeuxplus & $3.9 \pm 2.6$      & $8.9 \pm 1.0$      &   $0.43 \pm 0.29$ \\
                        &   ($\cdot10^{-2}$) &   ($\cdot10^{-2}$) &   \\

    \noalign{\smallskip}
    \cline{1-4}  

        \end{tabular} 
        
    \begin{tablenotes}
      \item[a] \small \citep[taken from][]{Asplund2009}. 
      \item[a] \small Error bars on [X$^{i+}$/\feplus]$_e$ reflect both the uncertainties on our measurements, and their dispersion around their average value. 
    \end{tablenotes}
    
    \label{Tab. Average abundances}
\end{threeparttable}
\end{table}

To understand this trend, it is useful to plot the average abundances (normalised to the solar values) as a function of their ionisation potentials (Fig.~\ref{Fig. abondance vs ion pot}; see the numerical values in Table~\ref{Tab. Average abundances}). A clear correlation is visible: the more difficult to ionise a species is, the more abundant it is in \bp\ exocomets. The observed deviations from solar ratios are thus probably due to a diversity of ionisation states among the studied species. For instance, the depletion of \caii\ is probably due to the higher ionisation degree of Ca than other species, as \caii\ is much more easily ionised into \caiii\ (11.87~eV, Table~\ref{Tab. List ions}) than \feii\ into \feiii\ or \niii\ into \niiii\ (note that, for all these species, the neutral fraction is negligible, due to photo-ionisation). Similarly, the high ratio of \Sii /\feii\ compared to the solar S/Fe ratio is likely caused by the very difficult ionisation of \Sii\ into \Siii\ (23.34~eV), which leads a higher fraction of sulphur to remain as S$^+$ in the cometary tails.

In between, the nearly solar ratios found for \mnii, \siii\ and \crii\ relatively to \feii\ are consistent with these four species having similar ionisation energies ($\sim 16$ eV). These solar ratios suggest that the dust sublimation is efficient enough for the composition of the gaseous tail to reflect the dust composition itself, which is probably solar for refractory elements. This result is very different from solar system comets, for which the composition of the gas is primarily influenced by the varying sublimation rates of the cometary materials. For instance, in some solar system comets the Ni/Fe ratio has been found to be strongly super-solar \citep{Manfroid2021}, due to faster sublimation rates for nickel carbonyls (e.g. Ni(CO)$_4$) compared to iron carbonyls (e.g.\ Fe(CO)$_5$).

The \aliii/\feii\ ratio is slightly more difficult to interpret, since \aliii\ is ionised twice: a significant fraction of Al is probably retained as \alii, while, for neutral or singly ionised species (\Si, \caii...), the fraction in lower ionisation states is either null or negligible. This explains why the \aliii/\feii\ ratio is generally sub-solar, even if \aliii\ is very difficult to ionise (28.4 eV).

\subsection{The ionisation state}

\subsubsection{Using the \aliii/\feii\ ratio}
\label{Sect. Ion state Al III}

The \aliii/\feii\ ratios can be used as probes of their ionisation states. Let us consider a given exocomet, and some chemical element X (e.g. Fe, Ca...). Assuming that the [X/Fe] and [Al/Fe] ratios in the studied exocomet are solar, the [\Xii/X] ratio can be written as (see App.~\ref{App. Ionisation state}):

\begin{equation}
    \hspace{2.2 cm}\left[ \frac{\text{\Xii}}{\text{X}} \right]_{\rm e} = \left(
    \frac{\eta_{\text{Al}}}{\eta_{\text{X}}}  +   \frac{1}{\eta_{\text{X}} s_{\text{Al}}} \left[\frac{\text{\aliii}}{\text{\feii}}\right]_{\rm e} \right)^{-1},
    \label{Eq. ionisation ratio}
\end{equation}
with:
$$s_{\text{Al}} = \left[ \text{Al}/\text{Fe} \right]_\odot,$$
$$\eta_{\text{X}} = \frac{\left[ \text{\Xii}/\text{\feii} \right]_\text{e}}{\left[ \text{X}/\text{Fe} \right]_\odot}, \ \ \ \ \ \ \ \ \ \ \ \ \ \ \eta_{\text{Al}} = \frac{\left[ \text{\alii}/\text{\feii} \right]_\text{e}}{\left[ \text{Al}/\text{Fe} \right]_\odot},$$
where the index {\tt e} denotes ratios in the considered exocomet and $\eta_\text{X}$ is the abundance of the single ionized state of the species $\text{X}$ relative to \feii\ in solar units.

Our study showed that the abundances of singly ionised species (\caii, \crii, \niii...) relative to \feii\ in \bp\ exocomets are fairly stable. We can thus make the assumption that $\eta_\text{X}$ and $\eta_{\rm Al}$ are fixed parameters, independent of the considered exocomet. The value of $\eta_\text{X}$ can be estimated from our measurements (e.g. $\eta_\text{Ca} = 0.082 \pm 0.019$, $\eta_\text{Fe} = 1$, $\eta_\text{Ni} = 1.46 \pm 0.20$; Table \ref{Tab. Average abundances}). However, the \alii/\feii\ ratios of the studied exocomets could not be measured due to the very low stellar flux in the \alii\ 1670 \A\ line, preventing us from directly estimating $\eta_\text{Al}$. To mitigate this problem, we can note that the $\eta$ value of a given element is primarily controlled by its ionisation potential (Fig. \ref{Fig. abondance vs ion pot}). Since the ionisation potential of \alplus (18.8 eV) is rather close to that of \niplus (18.2 eV), we an assume that $\eta_\text{Al} \approx \eta_{\rm Ni} = 1.46$. We will thus assume in the following $\eta_{Al} = 1.5 \pm 0.3$.

Under these assumptions, Eq. \ref{Eq. ionisation ratio} provides a direct link between the \aliii/\feii\ ratio of a given exocomet and its \Xii/X ratio, provided that the $\eta$ value of X can be constrained. For instance, Fig. \ref{Fig. ionisation state - Al, Fe, Ca} provides estimates of the \caii/Ca, \feii/Fe and \alii/Al ratios of the comets for which the \aliii/\feii\ ratio could be measured, using the $\eta$ values given in Table \ref{Tab. Average abundances}. These three elements appear to have contrasted ionisation states: while most of Calcium is ionised at least twice (\caii/Ca\ $\sim 0.04$), about half of iron remains singly-ionised (\feii/Fe $\sim 0.5$), as well as most of aluminium (\alii/Al\ $\sim 0.6-0.9$). Overall, the ionisation state among the studied exocomets appears to be rather stable, with only a slightly larger fraction of Al being doubly-ionised in the most redshifted comets.

\subsubsection{Using the Saha equation}
\label{Sect. Saha}

The measured cometary abundances corrected from solar composition (Fig. \ref{Fig. abondance vs ion pot}) can also be fitted using Saha's law. This law writes, for any element X: 

\begin{equation}
    \left[ \frac{\text{X}^{(i+1)+}}{\text{X}^{i+}} \right] = \frac{2}{n_e \Lambda(T)^3} \frac{Z_{X^{(i+1)+}}(T)}{Z_{X^{i+}}(T)} \exp\left( - \frac{E_{i+1}}{k_B T}\right),
\label{Saha}     
\end{equation}
with $\Lambda(T) = \sqrt{\frac{h^2}{2 \pi m_e k_B T}}$. The parameters of this law are $T$, the electronic temperature in the ionisation region, and $n_e$, the electronic density.

The typical composition of \bp\ exocomets appears to match rather well this simple law (see the grey crosses in Fig.~\ref{Fig. abondance vs ion pot}). The fitted temperature is $T = 12\,200 \pm 300$ K, which may be considered as the typical temperature in the ionising part of \bp\ exocomets. However, to fit the measurements with the Saha's law, the leading factor in Eq.~\ref{Saha} yields an equivalent electronic density of $\log\left(n_e\ (\si{cm^{-3}})\right) = 21.33 \pm 0.16$, which is, of course, much too large to be realistic \citep[geometric estimate provide values in the range $10^5$ - $10^6$ cm$^{-3}$,][]{Vrignaud24b}. Saha's equation is thus probably not the right tool for modelling the ionisation mechanisms in \bp\ exocomets: a complete description of the ionisation and recombination rates is probably needed to fully understand these mechanisms.

\begin{figure}[h]
    \centering    
    \includegraphics[clip,  trim = 0 0 0 50,  scale = 0.36]{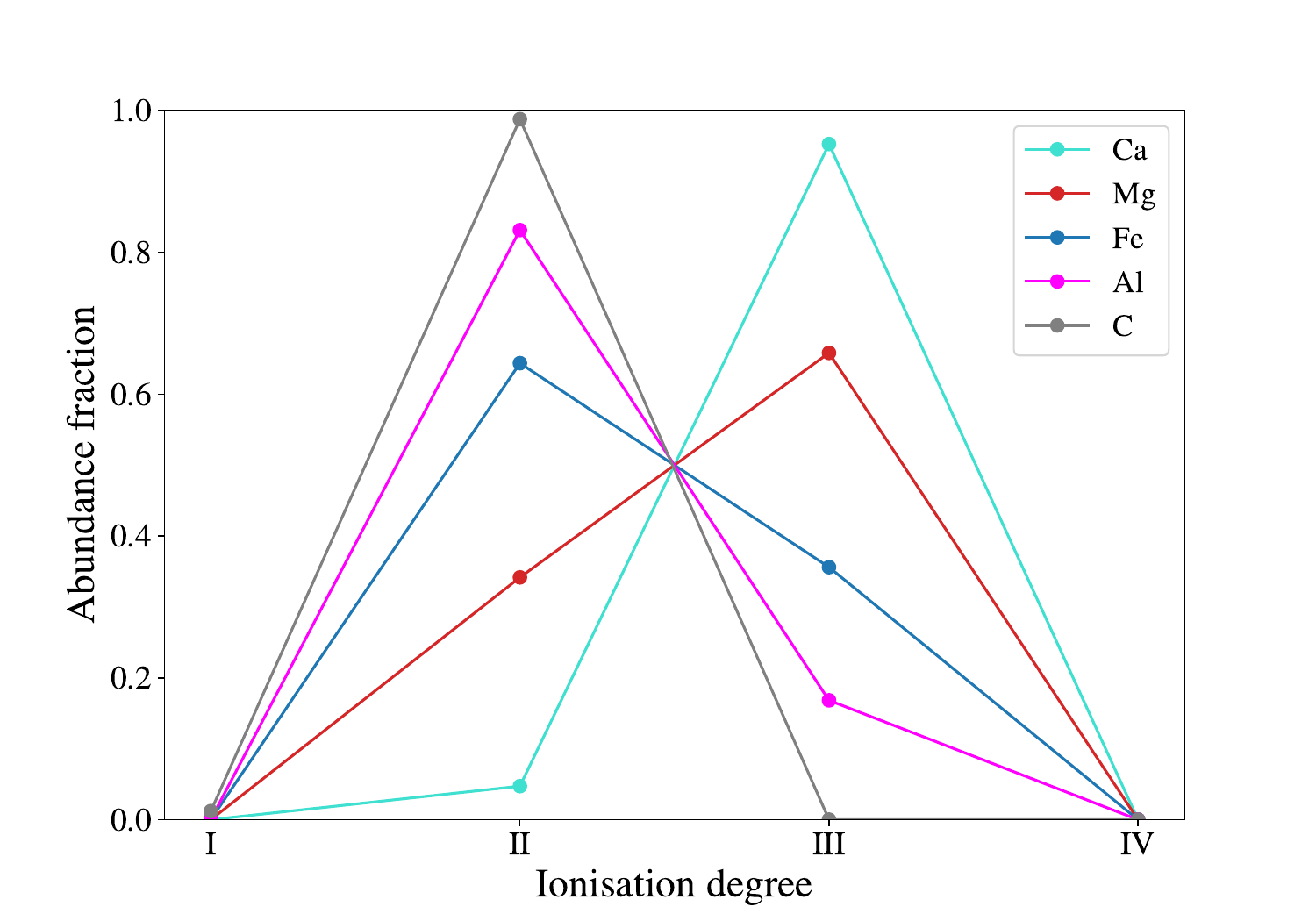}
    \caption[]{\small Typical ionisation state of five elements in \bp\ exocomets, as constrained by the fit of our abundances measurements (Fig. \ref{Fig. abondance vs ion pot}) with Saha's equation.}
    \label{Fig. ion state}
\end{figure}

Nevertheless, our fit provides a first quantitative description of the ionisation state in \bp\ exocomets. For instance, Fig. \ref{Fig. ion state} shows the ionisation distributions of a few elements, as constrained by Saha's law. While Fe, Mg and Al appear to be split in similar fractions between their singly and doubly ionised forms, Ca is, for the most part, doubly ionised. This is fully consistent with direct estimates from Sect. \ref{Sect. Ion state Al III}. With the constrained parameters ($T$, $n_e$), one can also predict the expected ionisation state of other elements. For instance, the main form of C in \bp\ exocomets should be \cii\ (Fig. \ref{Fig. ion state}), with only a small fraction in neutral form (about 1-2\%). However, \civ\ \citep[also detected in \bp\ exocomets,][]{Deleuil1993,Vidal-Madjar_1994} is not produced in the model, showing that this first description of the ionisation state in \bp\ exocomets is perfectible and other mechanisms should play an important role \citep[see, for instance,][]{Beust_1993}.

Finally, the dispersion of \aliii/\feii\ ratios in the cometary sample can be fairly well reproduced by Saha's equation, by letting $T$ vary from $10\,800$ K (comets 5 and 21, with the lowest \feii/\aliii\ ratio) to $14\,900$ K (comet 6, showing the highest ratio). It thus appears that the physical conditions in the ionising tails of the studied exocomets are rather uniform, despite the apparent diversity of these objects (e.g., they cover a wide range of redshift and line width).

\subsection{Limitations}
\label{Sect. Measurement limitations}

The main limitation of our abundance measurements is that they are very fragmented: even though the number of studied species is rather large (12, including \feii), these species were never observed altogether in one individual exocomet. In fact, most of the studied comets are generally observed in less than 5~different species simultaneously. This considerably limits our ability to study the variety of composition and ionisation states among \bp\ exocomets, as we are still unable to build diagrams similar to Fig.~\ref{Fig. abondance vs ion pot} but for individual comets. In Sect.~\ref{Sect. Saha}, Saha's equation was thus fitted to the average abundance ratios measured across our comet sample (Table~\ref{Tab. Average abundances}), providing rough constraints on the typical ionisation state of \bp\ exocomets rather than on the properties of any individual object.

Some of our abundances measurements should also be taken with caution, due to observing limitations. This is in particular the case for abundance ratios derived from non-simultaneous observations (e.g. the \caii/\feii\ ratios), which may be affected by time-variation of the total amount of gas transiting the star. The different spectral resolutions of COS ($R$ = 16\,000 - 24\,000) and STIS ($R$ = 115\,000) also make the estimate of the \Sii/\feii\ ratio slightly hazardous, as the cometary features observed in \Sii\ (with COS) are often blended. The curve of growth approach may thus not be the best tool to study exocomet absorption in \Sii; here a complete modelling of the cometary profile may be required.

It should also be noted that the assumption that the gas follows a Boltzmann distribution at $T \sim 9000 \ \si{K}$ may not be valid for all the studied species. In particular, due to the very low stellar flux in the bottom of the \caii\ H\&K lines, the excitation of \caii\ in the transiting comets may be fairly different than assumed by our model. Modelling the level population of \caii\ in a similar way to that of \cite{Vrignaud24b} and using radiative and collision coefficients from \cite{Melendez_2007_CaII}, we calculated that $\sim 87 \%$ of \caii\ should remain in its ground state, compared to $\sim 62 \%$ if the gas follows a Boltzmann distribution at 9000~K. Since our measurements are based on observations of the ground state of \caii\ (through the H\&K lines), the \caii/\feii\ ratios of the studied exocomets might be overestimated by as much as $\sim 40 \%$. Although the case of \caii\ is probably special due to the extreme depth of the photospheric H\&K lines, abundance measurements in other species may also be biased for this reason. Overcoming this issue would require a complete modelling of the energy distribution of all the studied species, which is beyond the scope of the present work.

\section{Conclusion}
\label{Sect. Conclusion}

Our study provided the first abundance measurements of 12 ionised species in a large sample of exocomets that transited \bp\ between 1997 and 2024. We found that the composition of the gaseous tails is fairly stable from one comet to another, particularly for singly-ionised species (\caii, \niii, \crii...), and overall consistent with solar abundances. This validates the use of the curve of growth approach to study the composition of \bp\ exocometary tails. 

We also showed that the typical abundances of refractory ions in \bp\ exocomets is strongly correlated with ionisation energy, hinting that ionisation is playing a prominent role in the composition of these gaseous tails. For instance, \caii, fairly easy to ionise (11.9 eV), is systematically depleted by a factor $\sim 10$ compared to \feii\ (16.2 eV). A closer study of the measured abundances showed that Fe is generally split in similar fraction between \feii\ and \feiii, while only $\sim 4$ \% of Ca is in the form of \caii, the rest being \caiii. These estimates were confirmed by the fit of our abundance measurements with the Saha equation. A clear correlation between the \aliii/\feii\ ratio and the radial velocity was also noted, showing that the most redshifted comets are generally more ionised, likely due to their closer distance to \bp. These results could enable further studies of the ionization processes in \bp\ exocomets, in order to quantify the relative contributions of potential mechanisms such as photoionization, electron impact ionization, or stellar wind charge exchange.

This first characterization of the typical ionization state in \bp\ exocomets now allows us to predict the ionisation state of other elements, including tracers of volatiles (C, N, O). For instance, extrapolating the fitted Saha's law to carbon, we can estimate that the typical neutral fraction of C in \bp\ exocomets is around 1-2 \%. This value, coupled with observations of exocomets in \ci\ and \feii\ lines (both accessible with STIS), could enable the determination of the C/Fe ratio of \bp\ exocomets, yielding valuable insight into their formation mechanisms.

However, our abundance measurements remain fragmented: the number of detected species in each individual comet is generally too low to allow a case-by-case study of their ionisation state and composition. To overcome this lack, new observations of \bp\ on a broader wavelength domain are required, to allow the detection of new exocomets in a large number of species simultaneously. Such observations will be obtained by 2025 with HST program \#17790 \citep{Vrignaud24_HST}, which will target the whole wavelength domain between 1200 and 2900 \A\ at four different epochs. This program will allow in-depth studies of the ionisation state in individual comets, thanks to simultaneous observations in a large number of species with diverse ionisation potentials (\Si, \feii, \niii, \Sii, \mgii...). The C/Fe of the detected exocomets will also be inferred, using measurements in \ci\ and \feii\ lines and estimates of the ionisation state in each object. The present paper thus serves as a benchmark for these future studies, demonstrating the relevance of abundance ratios measurements to probe the composition of \bp\ exocomets and identify the physical processes that ionize the gas.

\begin{acknowledgements}
T.V. \& A.L. acknowledge funding from the Centre National d’\'Etudes Spatiales (CNES).
\end{acknowledgements}

\bibliographystyle{aa}
\bibliography{bibliography}

\newpage

\begin{appendix}

\section{Ionisation state}
\label{App. Ionisation state}
Let X be a given chemical element (Fe, Al, Ca...). Assuming that X and Al are present in solar abundances in exocometary tails, and that most of the Aluminium gas is either singly or doubly ionised, we have:

$$
\frac{s_{\text{X}}}{s_{\text{Al}}} = 
\left[ \frac{\text{X}}{\text{Al}} \right]_\odot = 
\frac{\text{[X]}_{\rm e}}{\text{[\alii]}_{\rm e} + \text{[\aliii]}_{\rm e}} = 
\frac{\left[ \frac{\text{X}}{\text{\Xii}} \right]_{\rm e}}{\left[\frac{\text{\alii}}{\text{\Xii}}\right]_{\rm e} + \left[\frac{\text{\aliii}}{\text{\Xii}}\right]_{\rm e}},
$$
or equivalently: 
$$
\left[ \frac{\text{\Xii}}{\text{X}} \right]_{\rm e} = 
\frac{s_{\text{Al}}}{s_{\text{X}}} \cdot \left( \ \left[\frac{\text{\alii}}{\text{\Xii}}\right]_{\rm e} + \left[\frac{\text{\aliii}}{\text{\Xii}}\right]_{\rm e} \ \right)^{-1},
$$
where $s_{\text{Al}} = \left[ \text{Al}/\text{Fe} \right]_\odot$ and $s_{\text{X}} = \left[ \text{X}/\text{Fe} \right]_\odot$, and where the index {\tt e} refers to exocometary abundances. Assuming that $\left[ \text{\Xii}/\text{\feii} \right]_{\rm e} = \eta_{\text{X}} s_{\text{X}}$ and $\left[ \text{\alii}/\text{\feii} \right]_{\rm e} = \eta_{\text{Al}} s_{\text{Al}}$, with $\eta_{\text{X}}$ and $\eta_{\text{Al}}$ fixed values common to all exocomets (as found for singly ionised species, Fig. \ref{Fig. recap all comets}), we have:
$$
\left[\frac{\text{\alii}}{\text{\Xii}}\right]_{\rm e} = \frac{\eta_{\text{Al}} s_{\text{Al}}}{\eta_{\text{X}} s_{\text{X}}} \ , \ \ \ \ \ \ 
\left[\frac{\text{\aliii}}{\text{\Xii}}\right]_{\rm e} = \frac{1}{\eta_{\text{X}} s_{\text{X}}} \left[\frac{\text{\aliii}}{\text{\feii}}\right]_{\rm e},
$$
yielding: 

\begin{equation}
    \hspace{2.2 cm}\left[ \frac{\text{\Xii}}{\text{X}} \right]_{\rm e} = \left(
    \frac{\eta_{\text{Al}}}{\eta_{\text{X}}}  +   \frac{1}{\eta_{\text{X}} s_{\text{Al}}} \left[\frac{\text{\aliii}}{\text{\feii}}\right]_{\rm e} \right)^{-1}.
    \label{Eq. ionisation ratio (App)}
\end{equation}

Provided that $\eta_{\text{X}}$ can be estimated from measurements in many exocomets, Eq. \ref{Eq. ionisation ratio (App)} allows us to convert the measured \aliii/\feii\ ratio of a given exocomet into its \Xii/X ratio. As an example, Fig. \ref{Fig. ionisation state - Al, Fe, Ca} provides the \caii/Ca, \feii/Fe and \alii/Al ratios of our cometary exocomet sample (when the \aliii/\feii\ ratio could be measured), using $\eta$ values of 0.081 for Ca, 1 for Fe and 1.5 for Al (Table \ref{Tab. Average abundances}).

\begin{figure}[h]
    \centering    
    \includegraphics[clip,  trim = 10 0 0 40,  scale = 0.38]{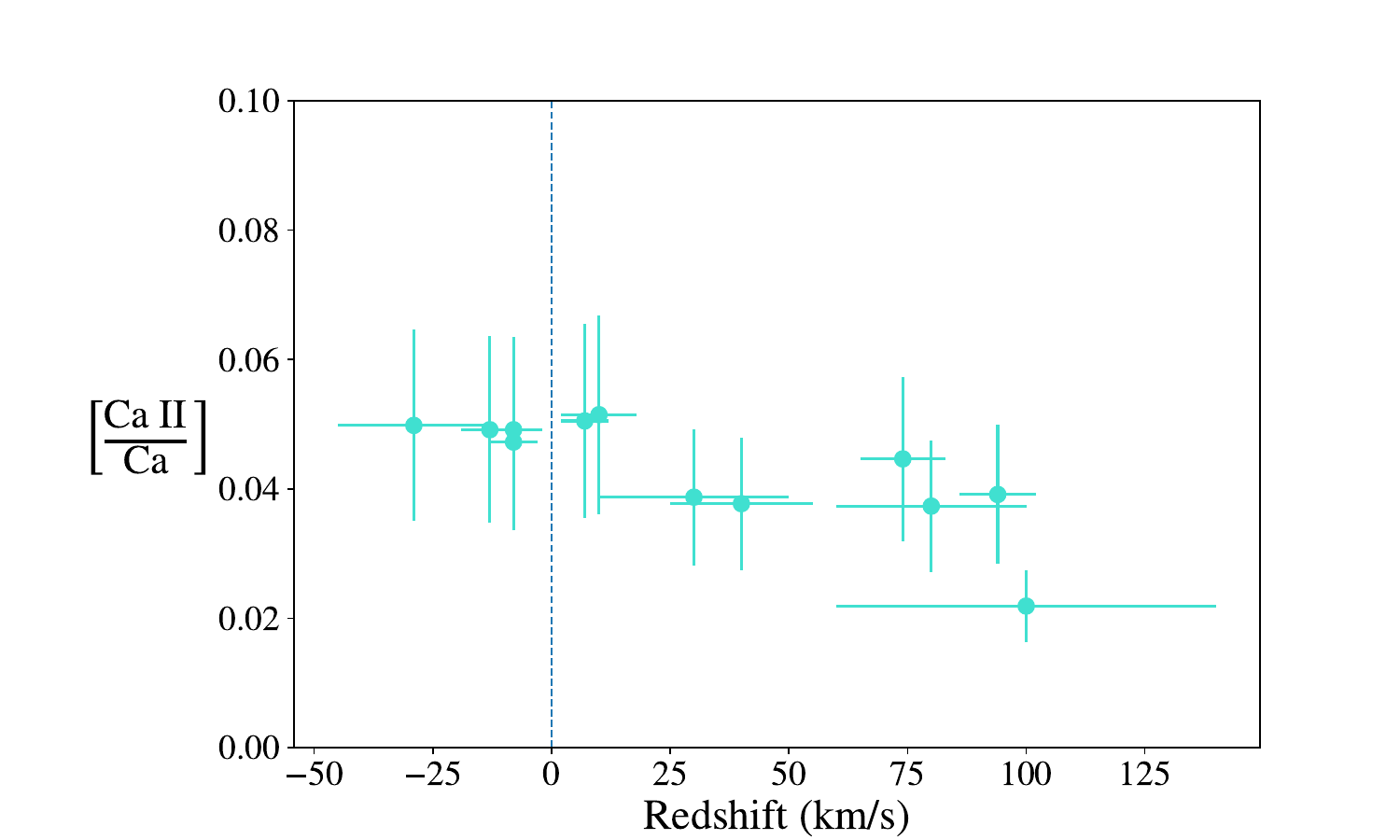}
    \includegraphics[clip,  trim = 10 0 0 40,  scale = 0.38]{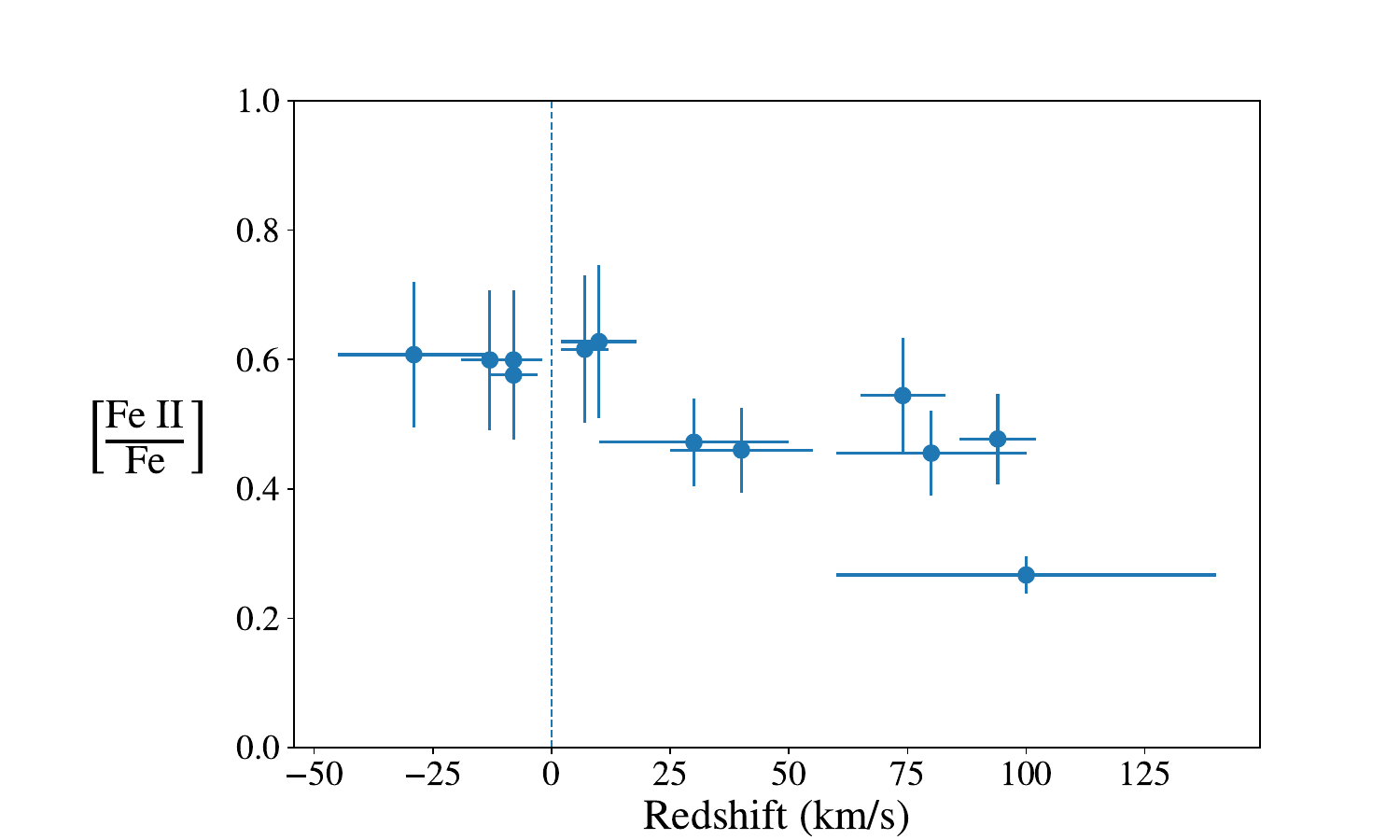}
    \includegraphics[clip,  trim = 10 0 0 40,  scale = 0.38]{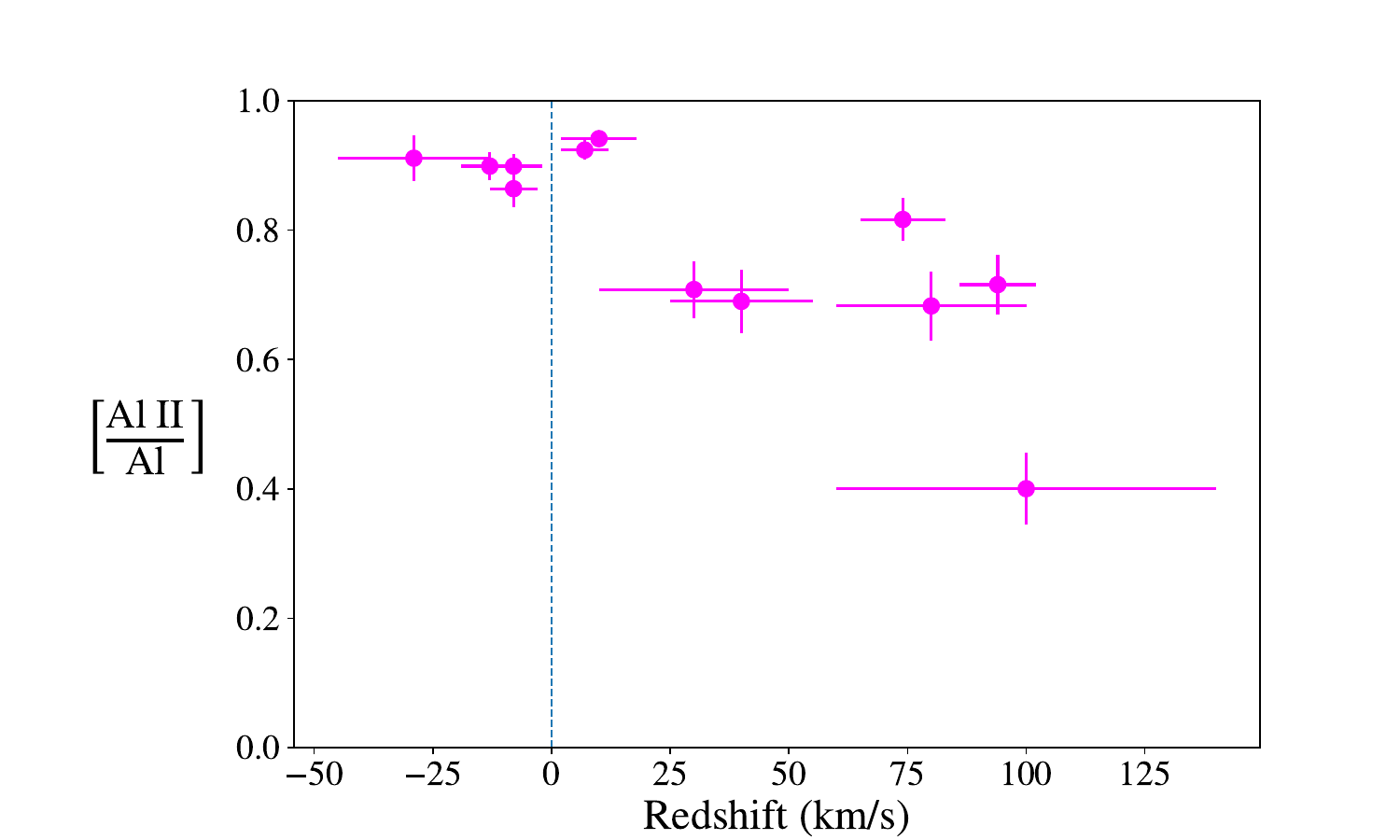}
    \caption[]{\small Estimates of the \caii/\caii (top), \feii/\feii\ (middle) and \alii/\alii (bottom) ratios of the studied exocomets, using Eq. \ref{Eq. ionisation ratio (App)} and the measured \aliii/\feii\ ratios. Error bars reflect uncertainties on the measured \aliii/\feii\ ratios, and on the fixed parameters of Eq. \ref{Eq. ionisation ratio (App)} (see Table \ref{Tab. Average abundances}). }
    \label{Fig. ionisation state - Al, Fe, Ca}
\end{figure}

\FloatBarrier
\newpage
\onecolumn

\section{Observations}

\begin{table}[h!]
    \centering 
    \begin{threeparttable}
    \caption{\small Log of the 11 observations of \bp\ from which our cometary sample was extracted. Other observations, not listed below, were used to reconstruct the EFS.}
    \renewcommand{\arraystretch}{1.25} \begin{tabular}{ c c c c c c c c } 		   
    \cline{1-8}
    \noalign{\smallskip}
    \cline{1-8}                 
     
     Date & Instrument & Program ID  & Wavelength & Start time & Exposure time & RV ranges showing \\
          &            &             &    (\A)    &    (UT)    &     (s)  & comet absorption (km/s)\tnote{a}    \\
    		  
    \cline{1-8}      
    \noalign{\smallskip}

    \multirow{6}{*}{1997-12-06} & \multirow{6}{*}{STIS} & \multirow{6}{*}{7512} & 1463 - 1661 & 07:18 & 900 &  \\ 
    & & & 1628 - 1902 & 07:48 & 679 & \\ 
    & & & 1879 - 2150 & 07:42 & 80 & \\ 
    & & & 2128 - 2396 & 09:16 & 288 & \multirow{1.5}{*}{[-4;+200]} \\ 
    & & & 2377 - 2650 & 09:02 & 360 & \\ 
    & & & 2620 - 2888 & 08:48 & 360 & \\ 
    \noalign{\smallskip}
    \cdashline{2-6}
    \noalign{\smallskip}
    & MJUO &  & 3900 - 4000  & 09:24 & 1800 & \\   
    
    \noalign{\smallskip}
    \cline{1-8} 
    \noalign{\smallskip}
    
    \multirow{6}{*}{1997-12-19} & \multirow{6}{*}{STIS} & \multirow{6}{*}{7512} & 1463 - 1661 & 19:51 & 900 & \multirow{6}{*}{[-17;+19], [+50;+150]} \\ 
    & & & 1628 - 1902 & 20:22 & 679 & \\ 
    & & & 1879 - 2150 & 20:14 & 80 & \\ 
    & & & 2128 - 2396 & 21:53 & 288 & \\ 
    & & & 2377 - 2650 & 21:40 & 360 & \\ 
    & & & 2620 - 2888 & 21:26 & 360 & \\ 
    
    \noalign{\smallskip}
    \cline{1-8} 
    \noalign{\smallskip}

    2001-08-27 & STIS & 9154 & 1628 - 1897 & 01:15 & 210 & [-17;-3] \\ 
    
    \noalign{\smallskip}
    \cline{1-8} 
    \noalign{\smallskip}
        
    2002-05-22 & STIS & 9154 & 2327 - 2606 & 17:09 & 200 & [-8,40] \\ 
    
    \noalign{\smallskip}
    \cline{1-8} 
    \noalign{\smallskip}
        
    2017-10-09 & STIS & 14735 & 2665 - 2931 & 23:40 & 475 & [-35;+120] \\ 
    
    \noalign{\smallskip}
    \cline{1-8} 
    \noalign{\smallskip}

    2018-03-07 & STIS & 14735 & 2665 - 2931 & 02:33 & 475 & [-40;+13] \\ 
    
    \noalign{\smallskip}
    \cline{1-8} 
    \noalign{\smallskip}
        
    \multirow{2.55}{*}{2018-10-29} & COS & 15479 & 1163 - 1480 & 23:15 & 2012 & \multirow{2.55}{*}{[-30;-4], [+8;+140]} \\ 
    \noalign{\smallskip}
    \cdashline{2-6}      
    \noalign{\smallskip}
    & STIS & 15479 & 2665 - 2931 & 21:49 & 2101 & \\
    
    \noalign{\smallskip}
    \cline{1-8} 
    \noalign{\smallskip}
    
    & COS & 15479 & 1163 - 1480 & 05:50 & 2101 &  \\ 
    \noalign{\smallskip}
    \cdashline{2-6}      
    \noalign{\smallskip}
    2018-12-15 & STIS & 15479 & 2665 - 2931 & 04:22 & 2101 & [-40;+50] \\ 
    \noalign{\smallskip}
    \cdashline{2-6}      
    \noalign{\smallskip}
    & HARPS & 0102.C-0584 & 3783 - 6913 & 05:38 & 540 & \\ 
    
    \noalign{\smallskip}
    \cline{1-8} 
    \noalign{\smallskip}

    & & & 1629 - 1897 & 00:09 & 1886 &  \\ 
    \multirow{2.6}{*}{2024-04-17} & STIS & 17421 & 1629 - 1897 & 01:36 & 2392 & \multirow{2.6}{*}{[-17;+100]} \\ 
    & & & 1629 - 1897 & 03:11 & 2392 & \\
    \noalign{\smallskip}
    \cdashline{2-6}
    \noalign{\smallskip}               
    & HARPS & 112.26ZF & 3783 - 6913 & 23:12\tnote{b} & 10800 & \\ 

    \noalign{\smallskip}
    \cline{1-7}      
    \noalign{\smallskip}

    & & & 1629 - 1897 & 11:42 & 1886 &  \\ 
    2024-07-31 & STIS & 17421 & 1629 - 1897 & 13:09 & 2392 & [-65;-5], [+3;+100] \\ 
    & & & 1629 - 1897 & 14:44 & 2392 & \\  
    
    \noalign{\smallskip}
    \cline{1-7}     
    \noalign{\smallskip}
                    
    & \multirow{4}{*}{STIS} & \multirow{4}{*}{17421} & 1629 - 1897 & 11:03 & 1886 & \multirow{4}{*}{ } \\ 
    & & & 1629 - 1897 & 12:17 & 1886 & \multirow{2.5}{*}{[-55,-15], [+2,+35],}\\ 
    \multirow{1.7}{*}{2024-08-08}  & & & 1629 - 1897 & 13:51 & 2392 & \multirow{2.5}{*}{[+62,+140]} \\ 
    & & & 1629 - 1897 & 15:26 & 2392 & \\
    \noalign{\smallskip}
    \cdashline{2-6}
    \noalign{\smallskip}
    & HARPS & 112.26ZF & 3783 - 6913 & 09:51 & 2400 & \\   

    \noalign{\smallskip}
    \cline{1-8}                 
\end{tabular}

    \begin{tablenotes}
      \item[a] \small The RV ranges apply to all exposures of each visit.
      \item[b] \small On April 16, 2024
    \end{tablenotes}
    
\label{Tab. data}

\end{threeparttable}
\end{table}

\newpage
\section{Line examples}

\begin{figure*}[h!]
\centering
    \includegraphics[scale = 0.4,     trim = 10 40 20 40,clip]{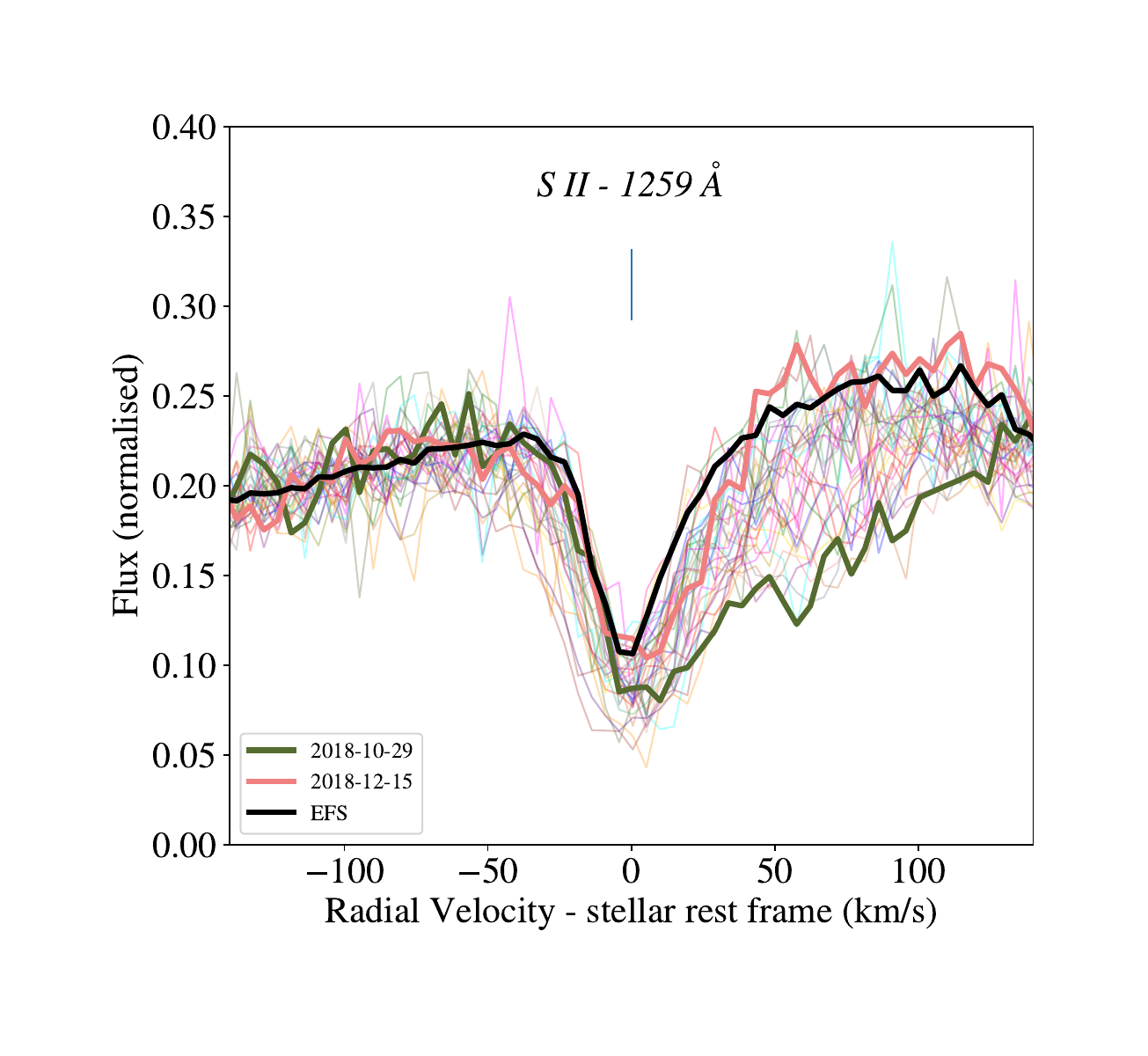}    
    \includegraphics[scale = 0.4,     trim = 10 40 40 40,clip]{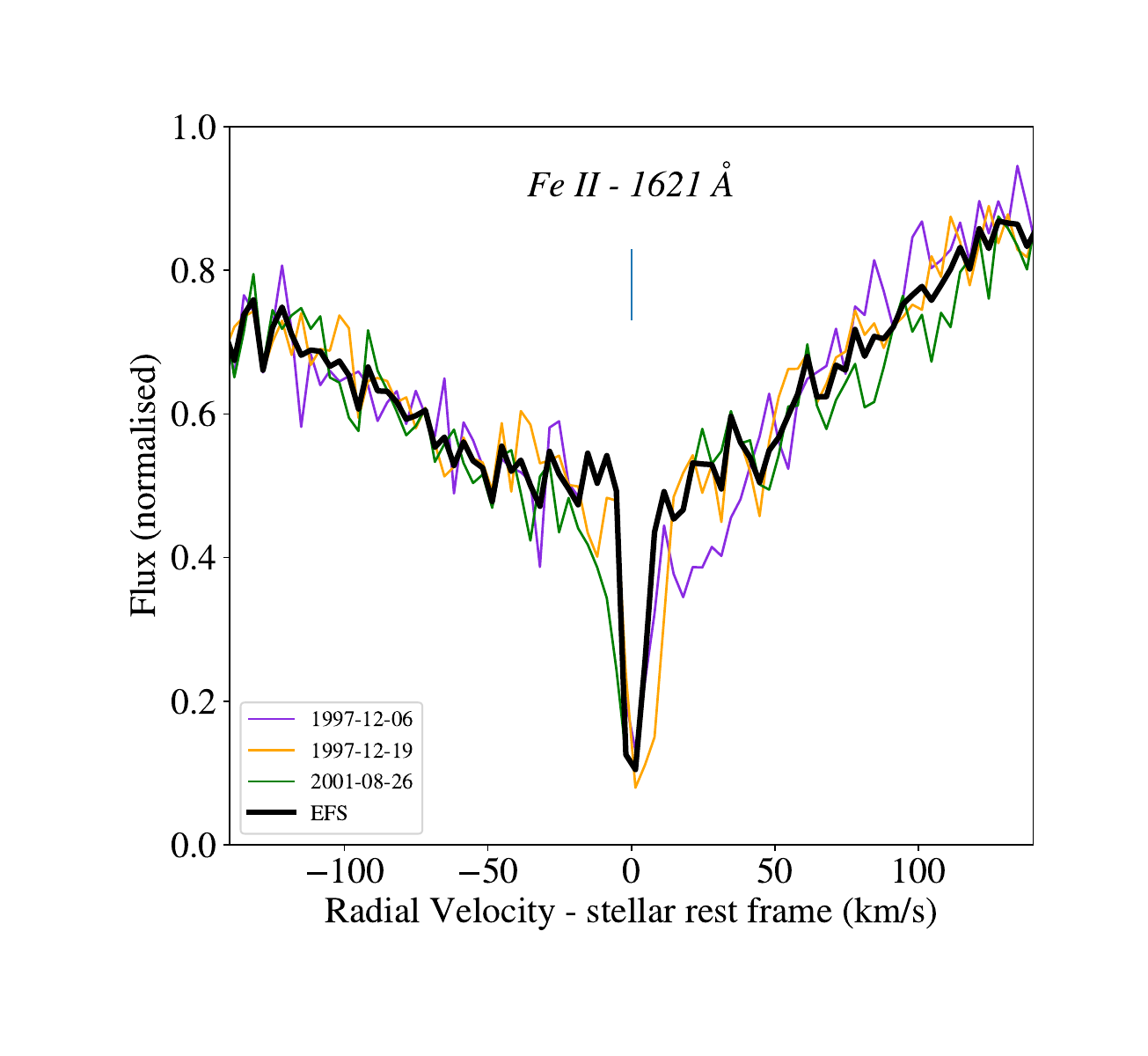}    
    
    \includegraphics[scale = 0.4,     trim = 10 40 20 40,clip]{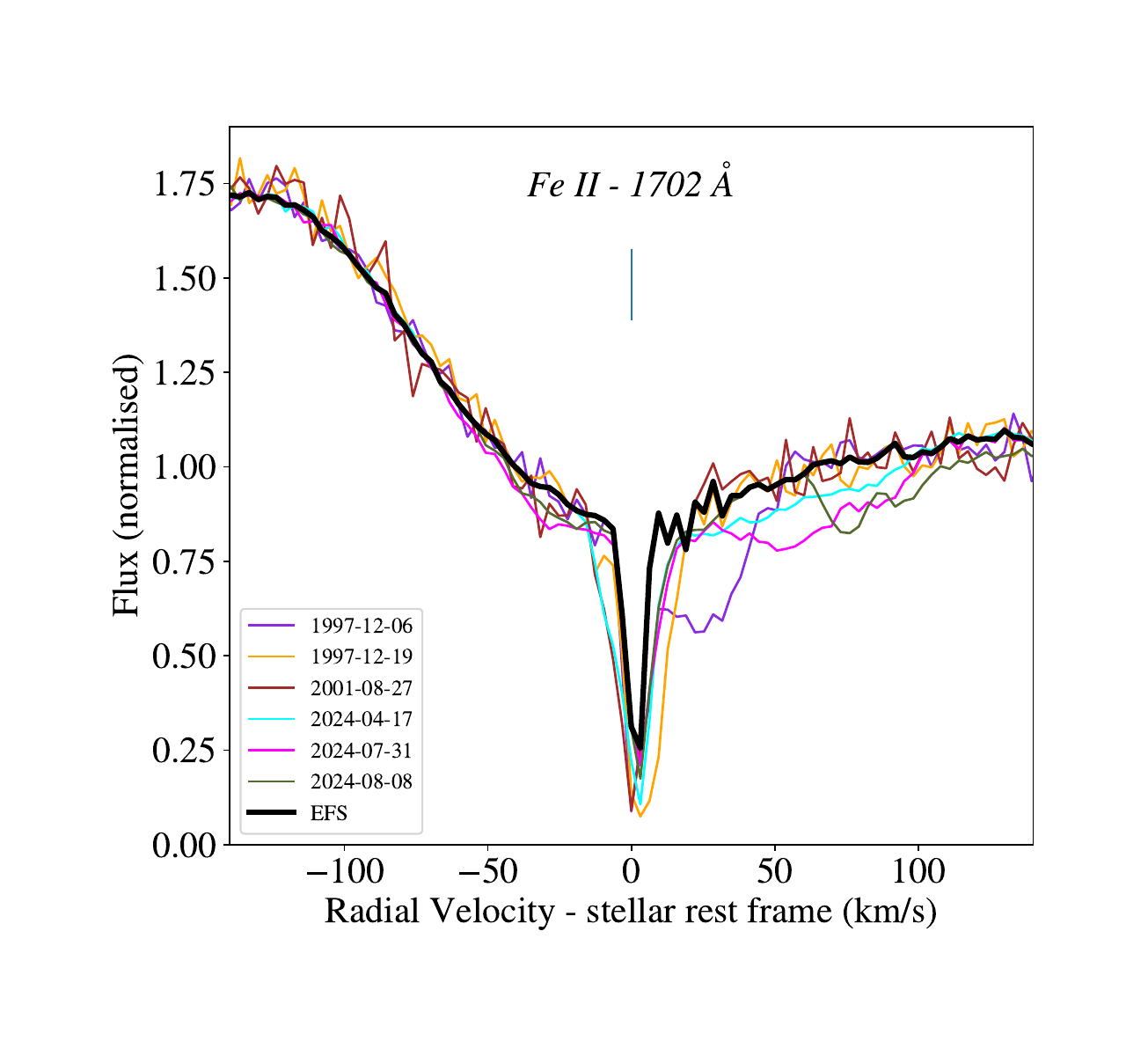}    
    \includegraphics[scale = 0.4,     trim = 10 40 40 40,clip]{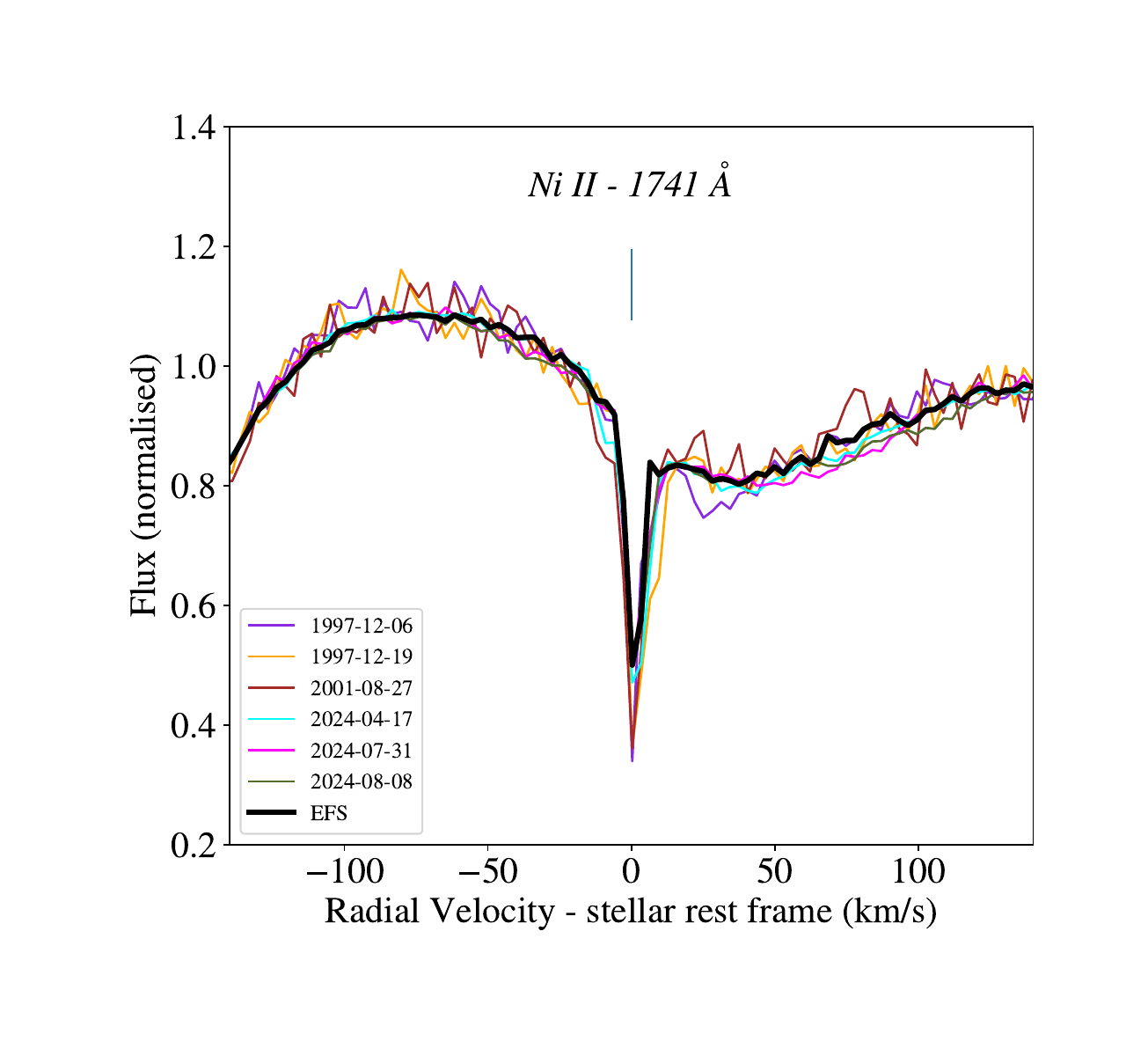}
    
    \includegraphics[scale = 0.4,     trim = 10 40 20 40,clip]{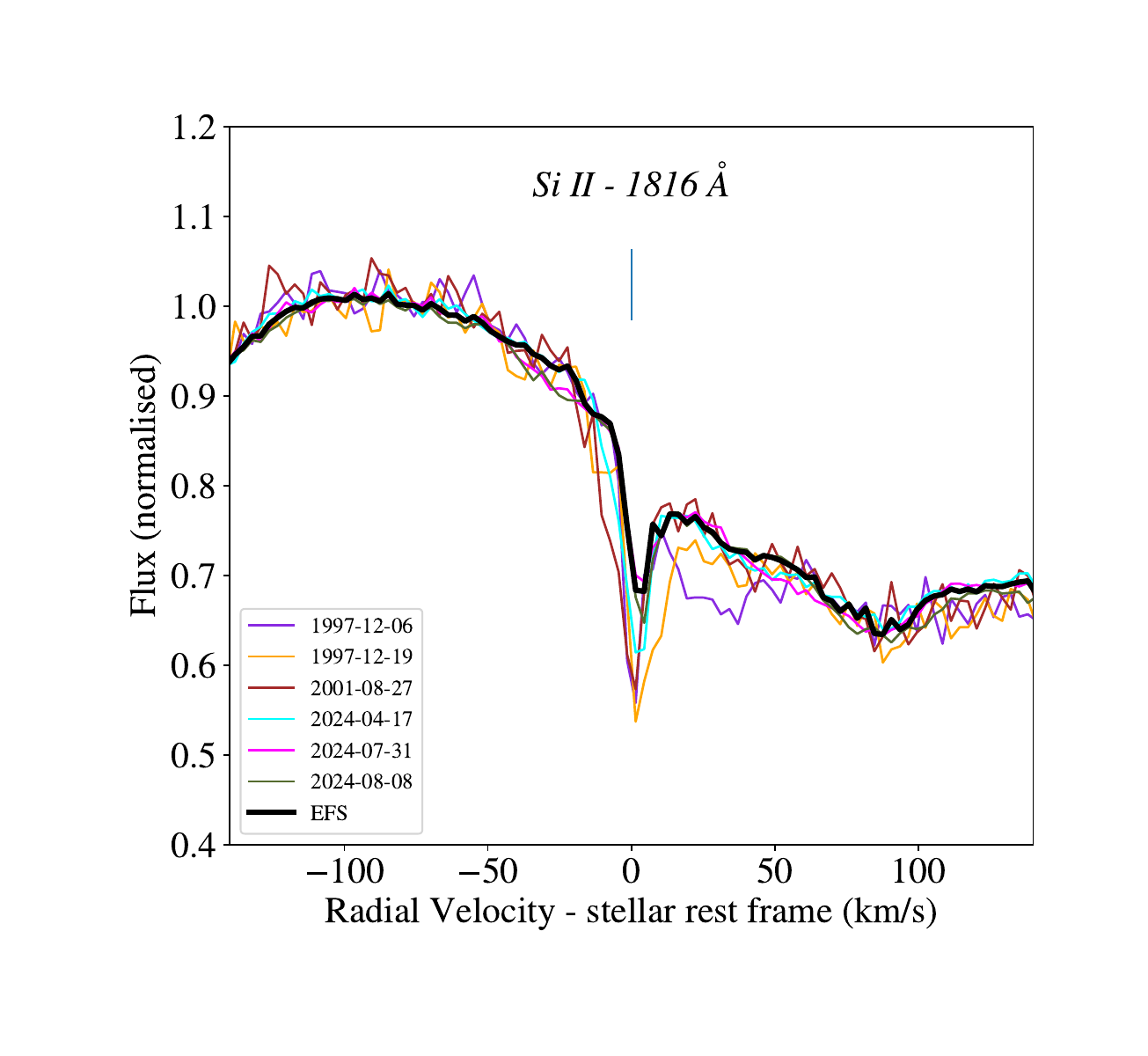}    
    \includegraphics[scale = 0.4,     trim = 10 40 40 40,clip]{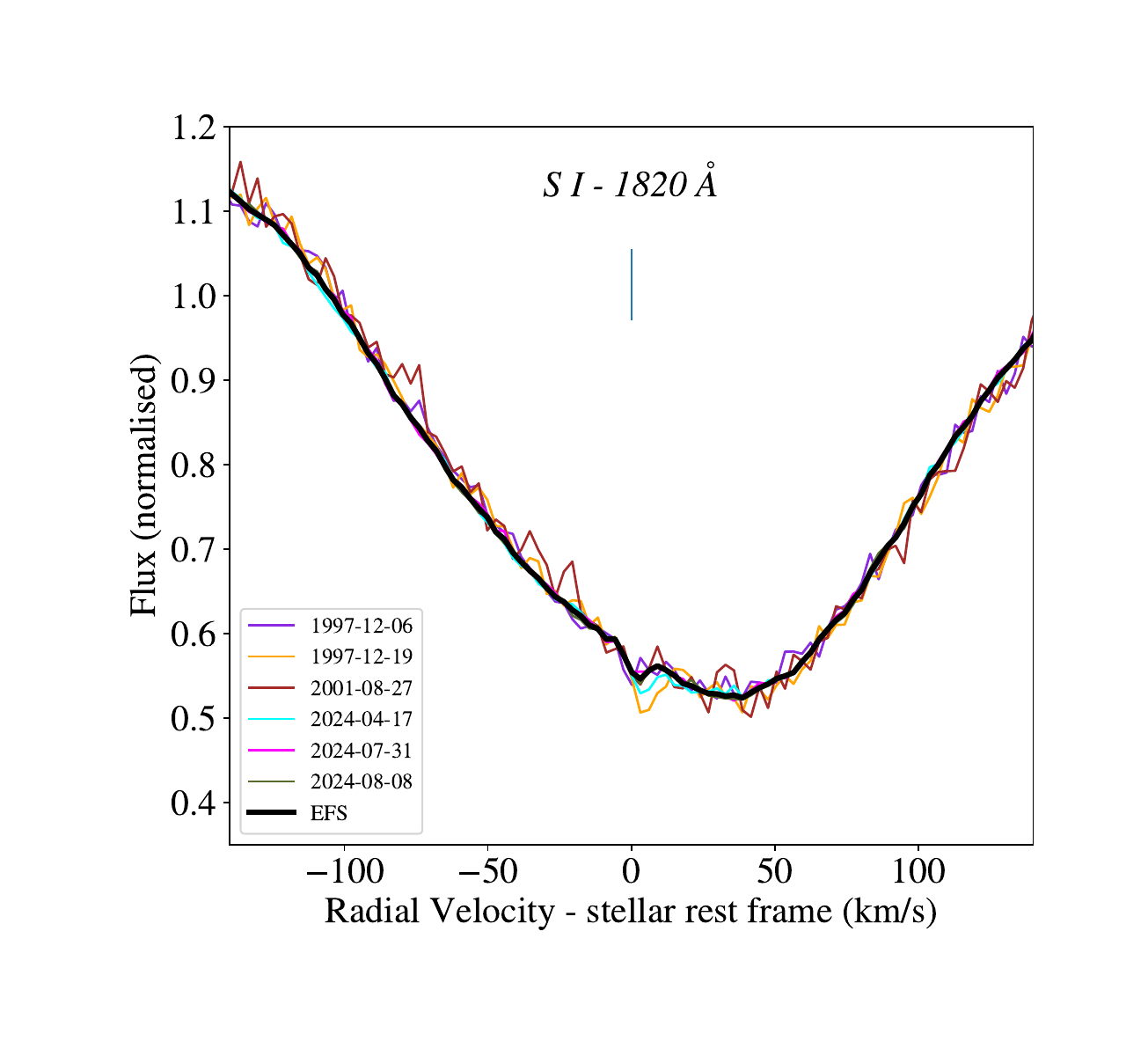}    
    \caption{\small Examples of spectral lines used in our study. For conciseness, we show only one line for each series provided in Tab. \ref{Tab. list lines}.}
    \label{Fig. line examples}
\end{figure*}

\newpage

\begin{figure*}[h!]
\centering
    \includegraphics[scale = 0.4,     trim = 10 40 20 30,clip]{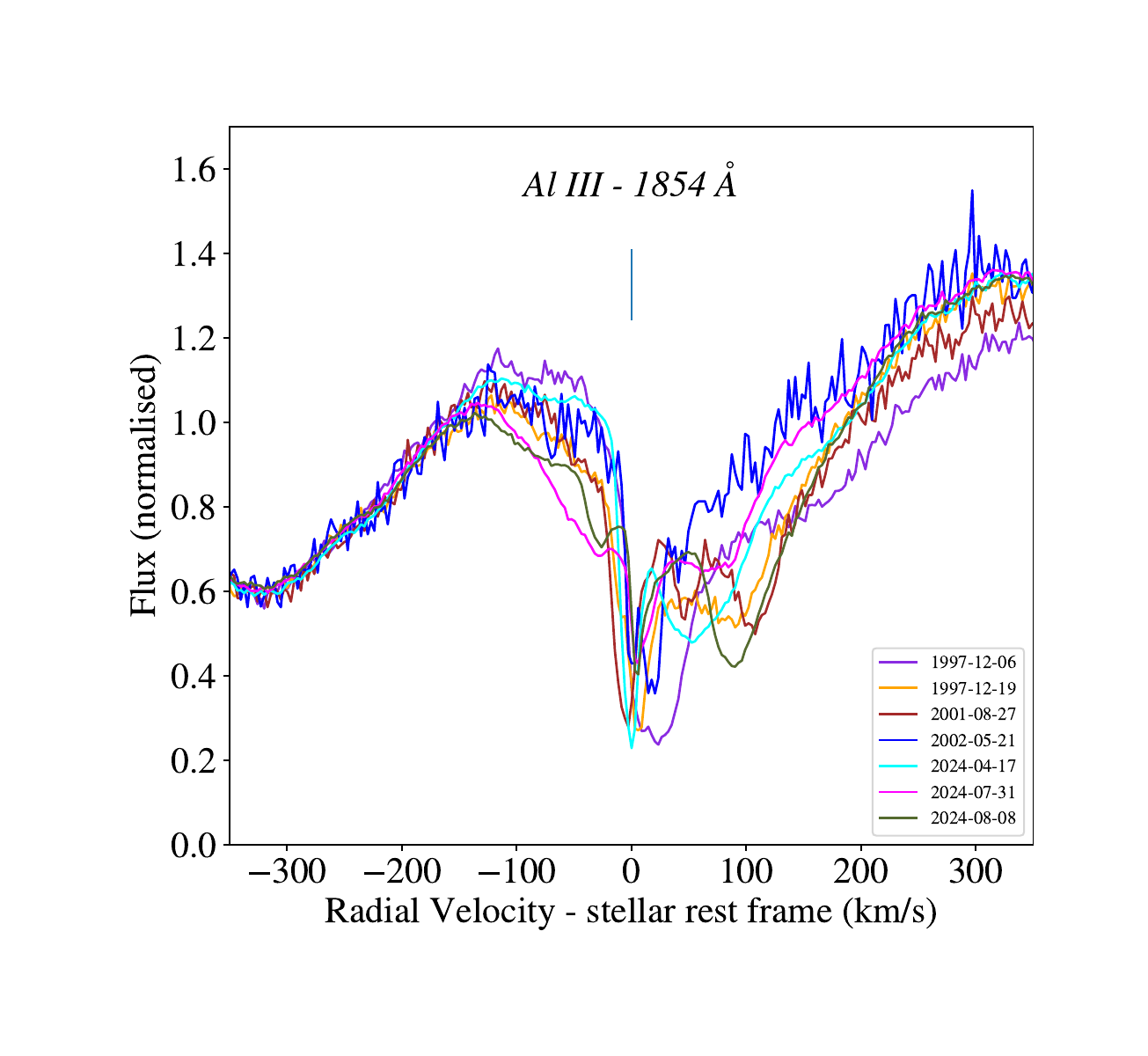}    
    \includegraphics[scale = 0.4,     trim = 10 40 40 30,clip]{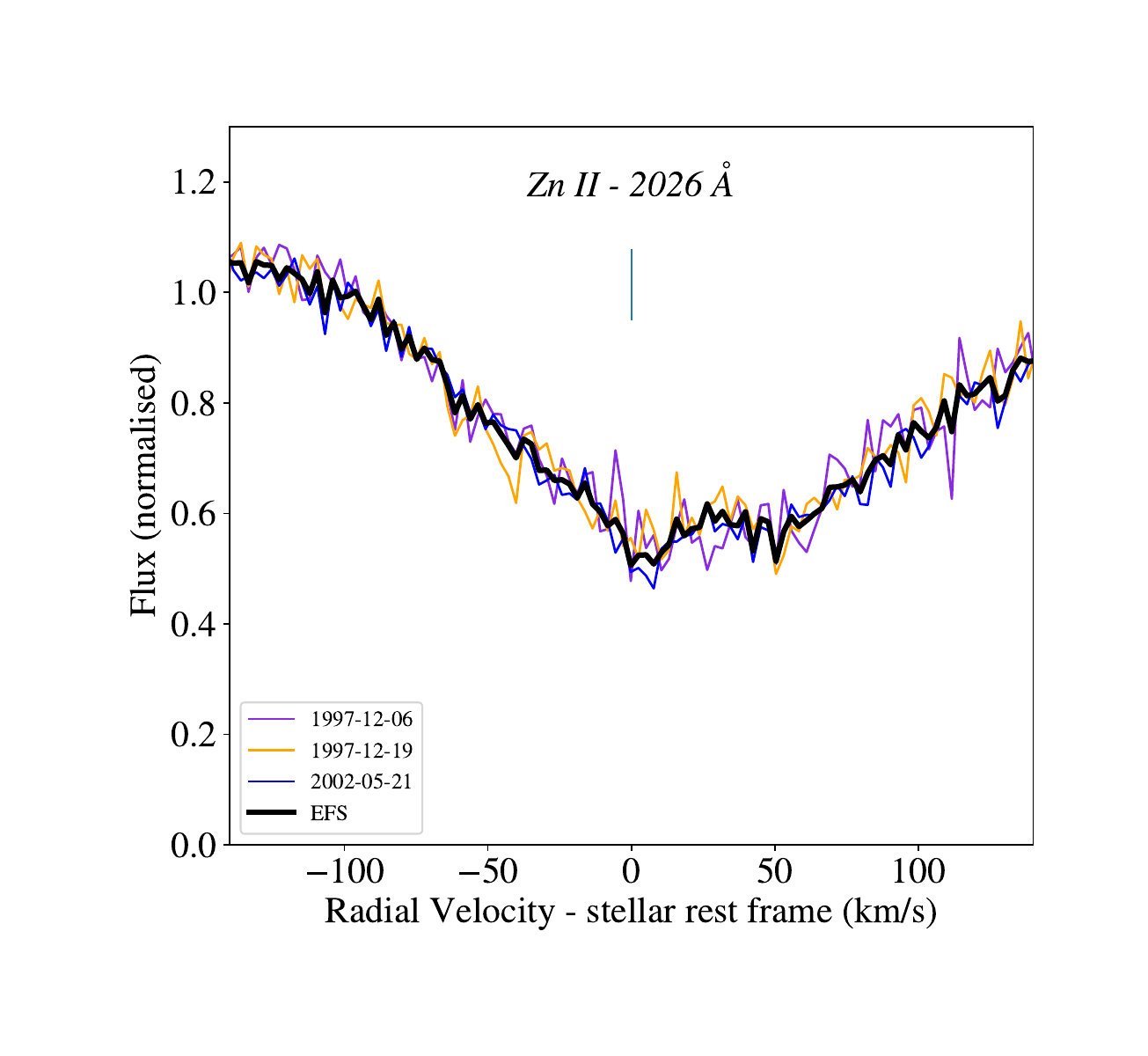}    
    
    \includegraphics[scale = 0.4,     trim = 10 40 20 40,clip]{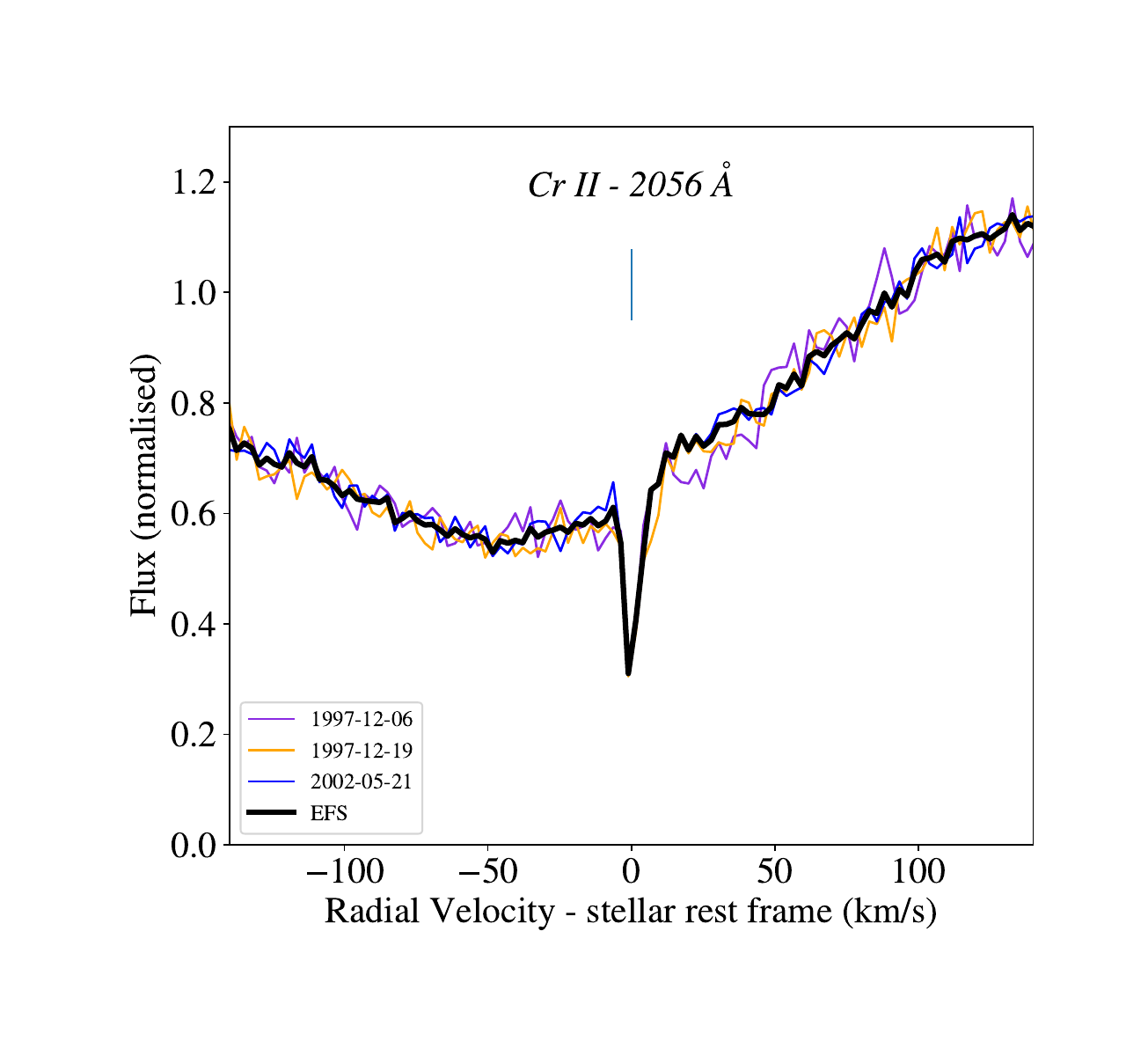}    
    \includegraphics[scale = 0.4,     trim = 10 40 40 40,clip]{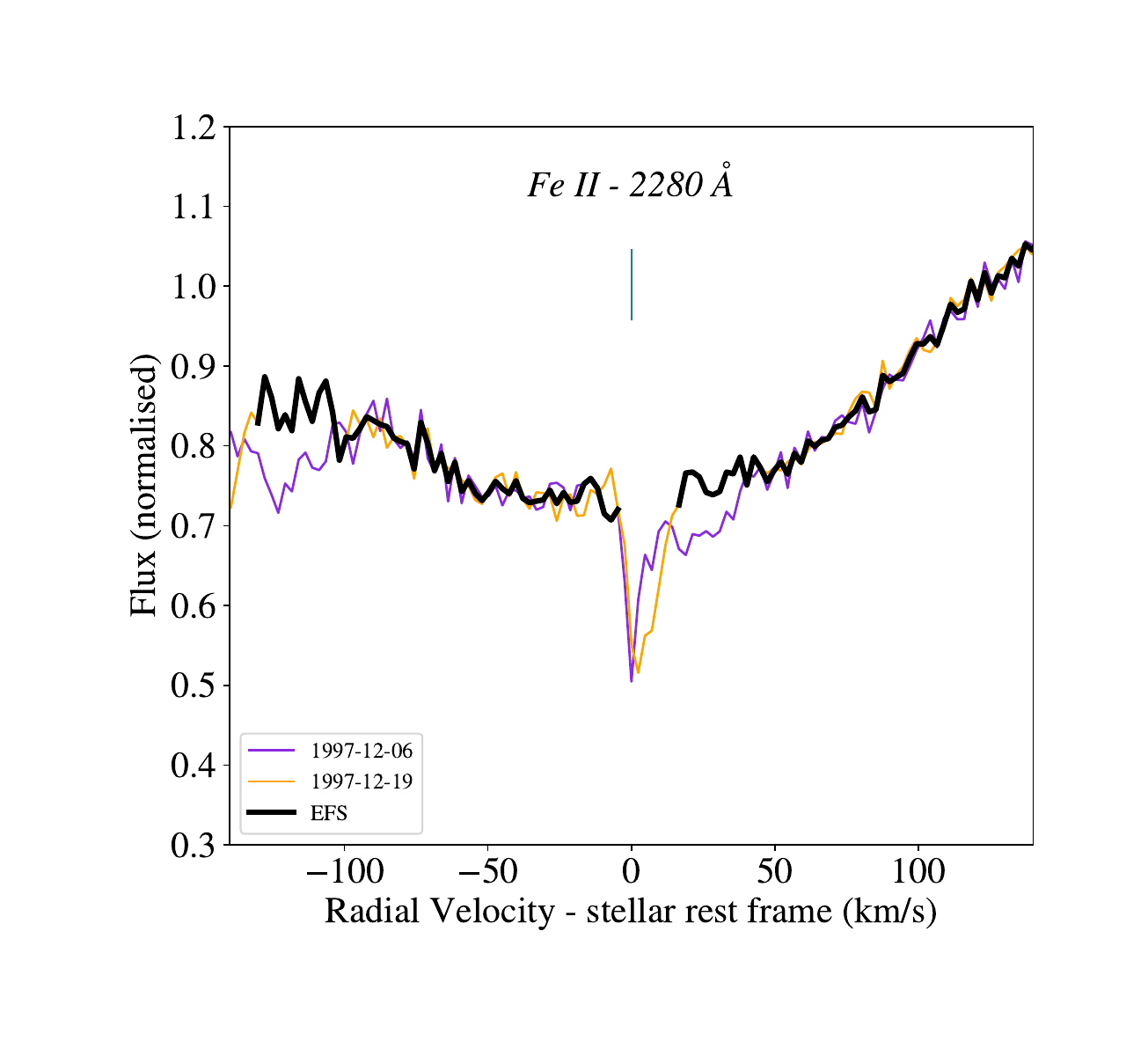}
    
    \includegraphics[scale = 0.4,     trim = 10 40 20 40,clip]{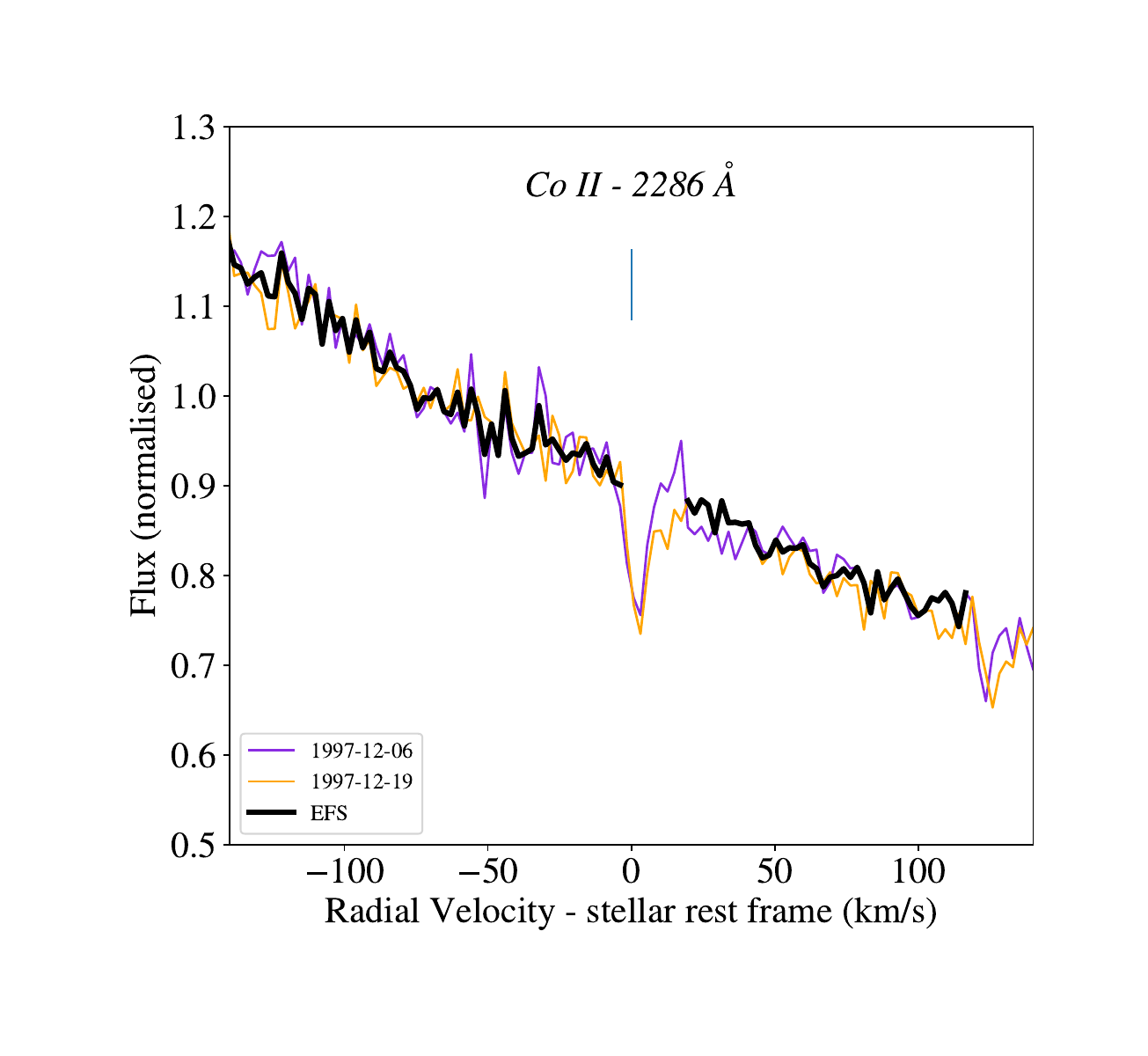}    
    \includegraphics[scale = 0.4,     trim = 10 40 40 40,clip]{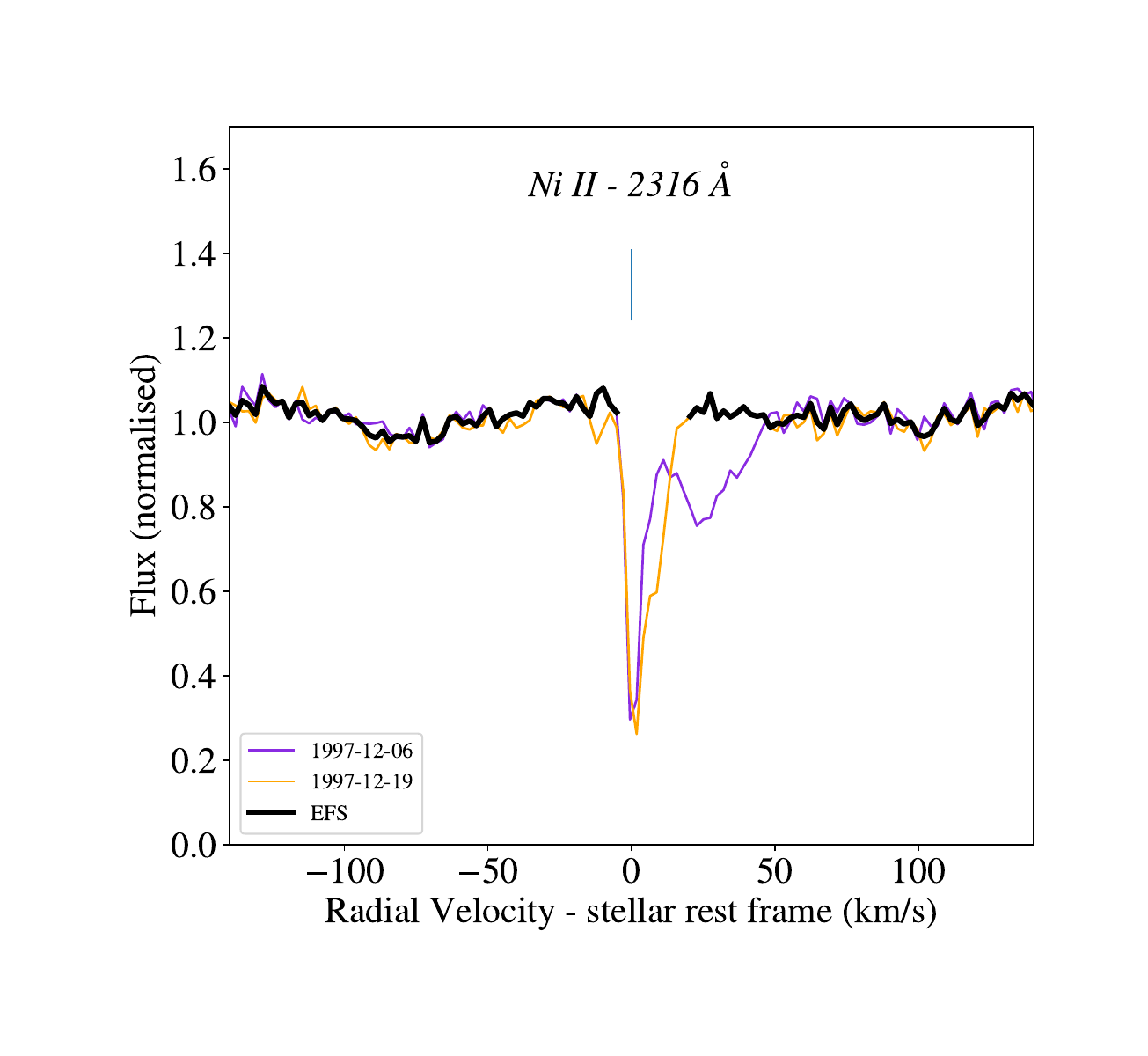}    
\end{figure*}
\centering \small \textbf{Figure \ref{Fig. line examples}}, continued.

\newpage

\begin{figure*}[h!]
\centering
    \includegraphics[scale = 0.4,     trim = 10 40 20 30,clip]{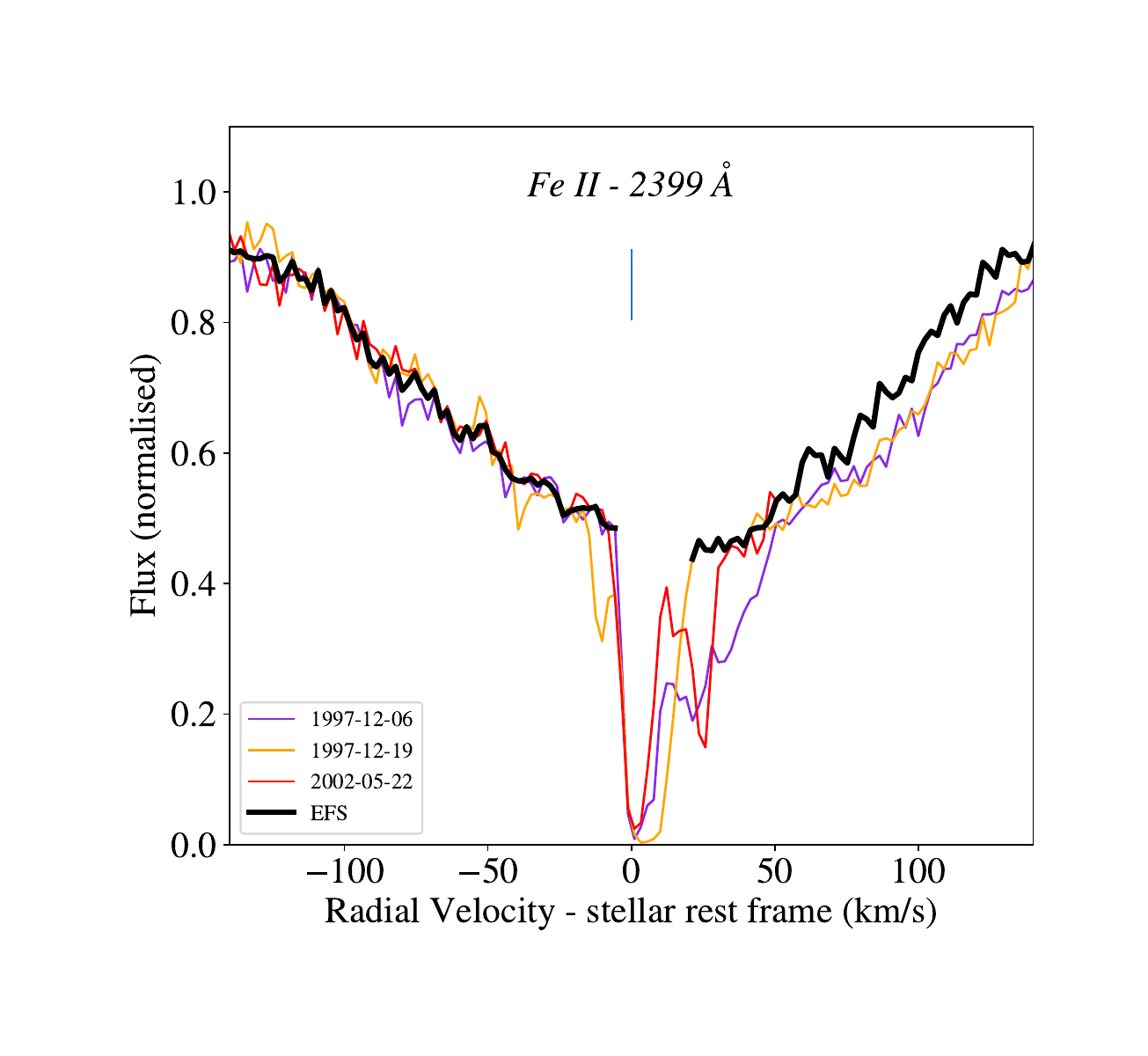}    
    \includegraphics[scale = 0.4,     trim = 10 40 40 30,clip]{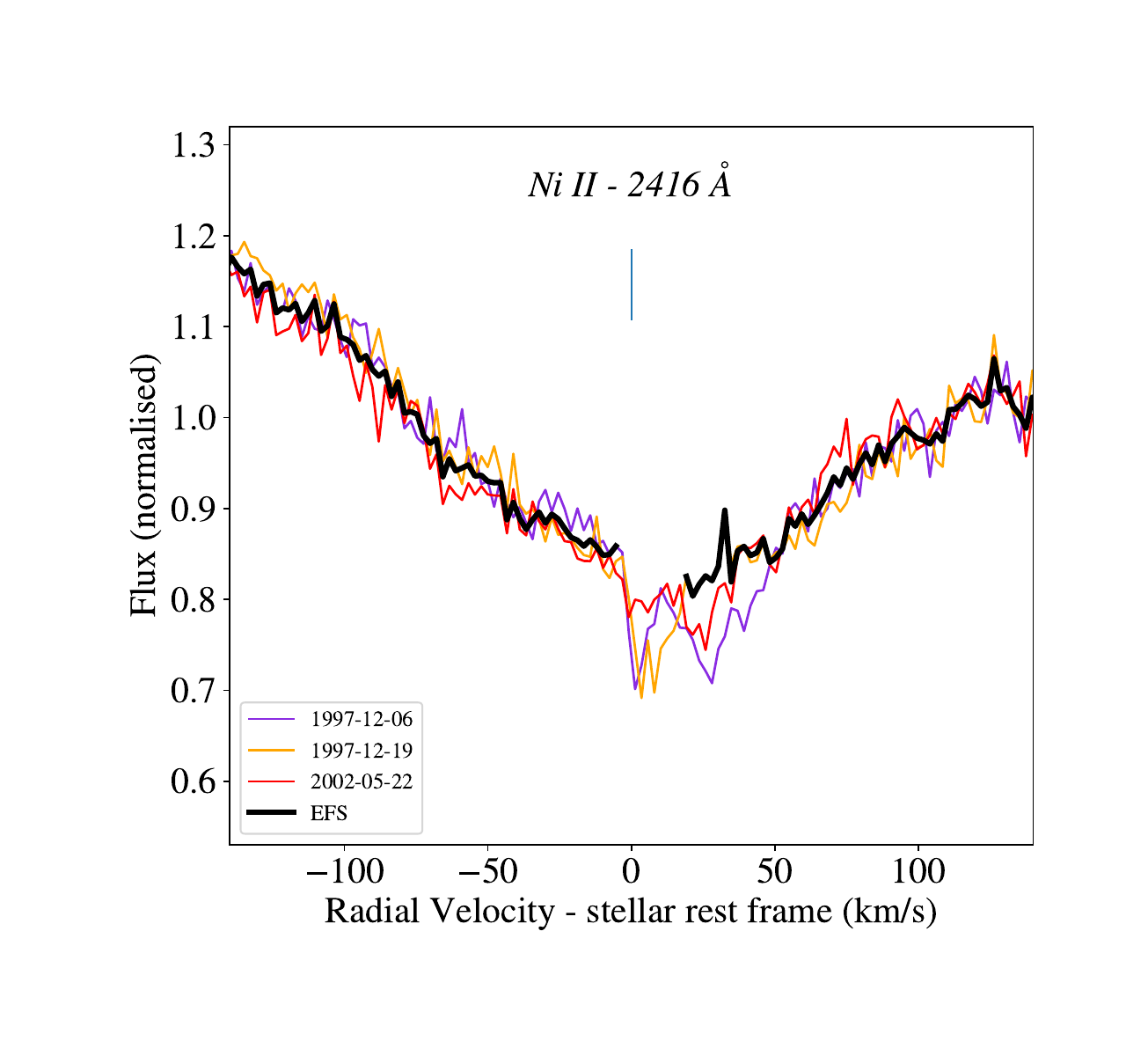}    
    
    \includegraphics[scale = 0.4,     trim = 10 40 20 40,clip]{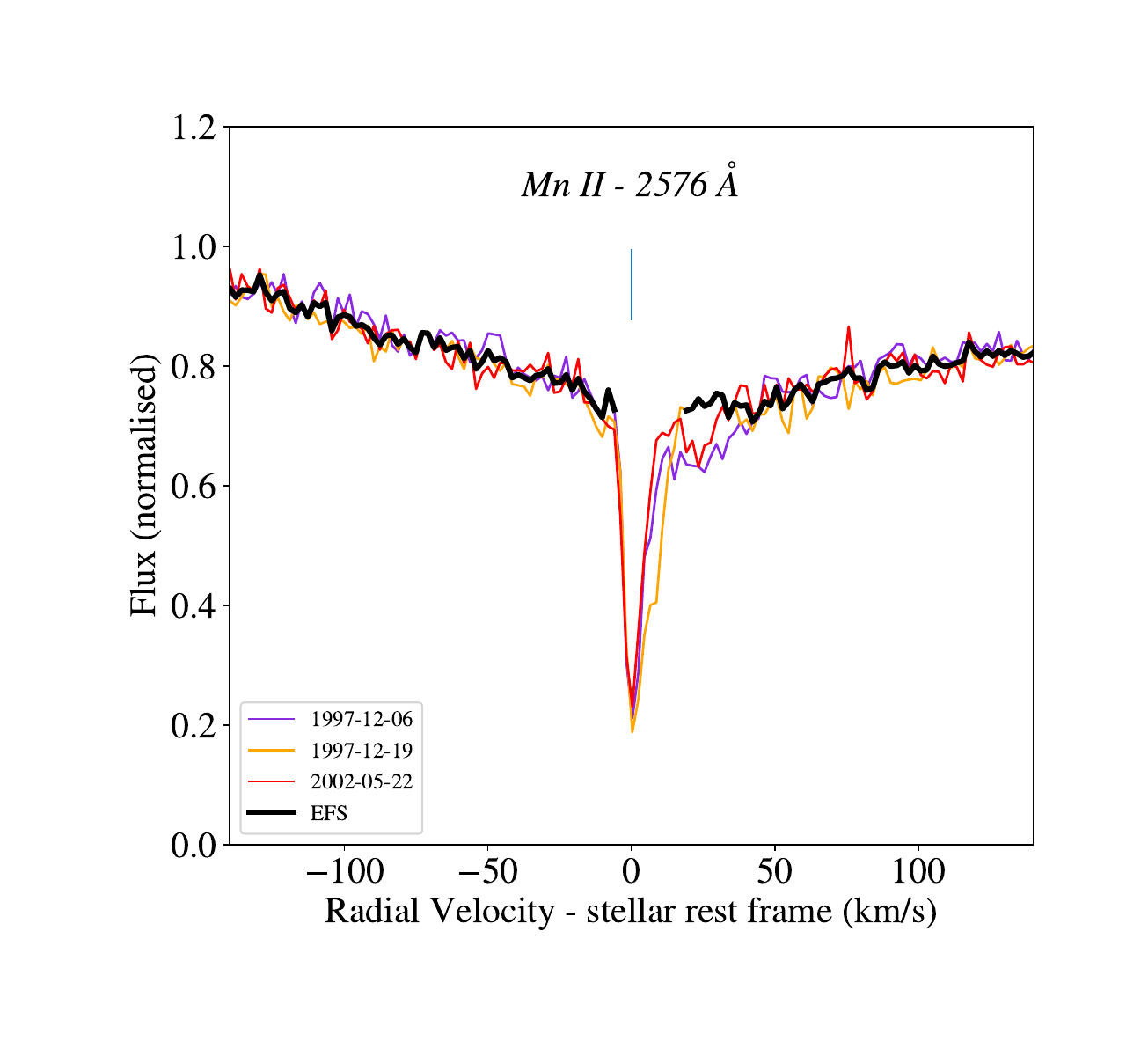}    
    \includegraphics[scale = 0.4,     trim = 10 40 40 40,clip]{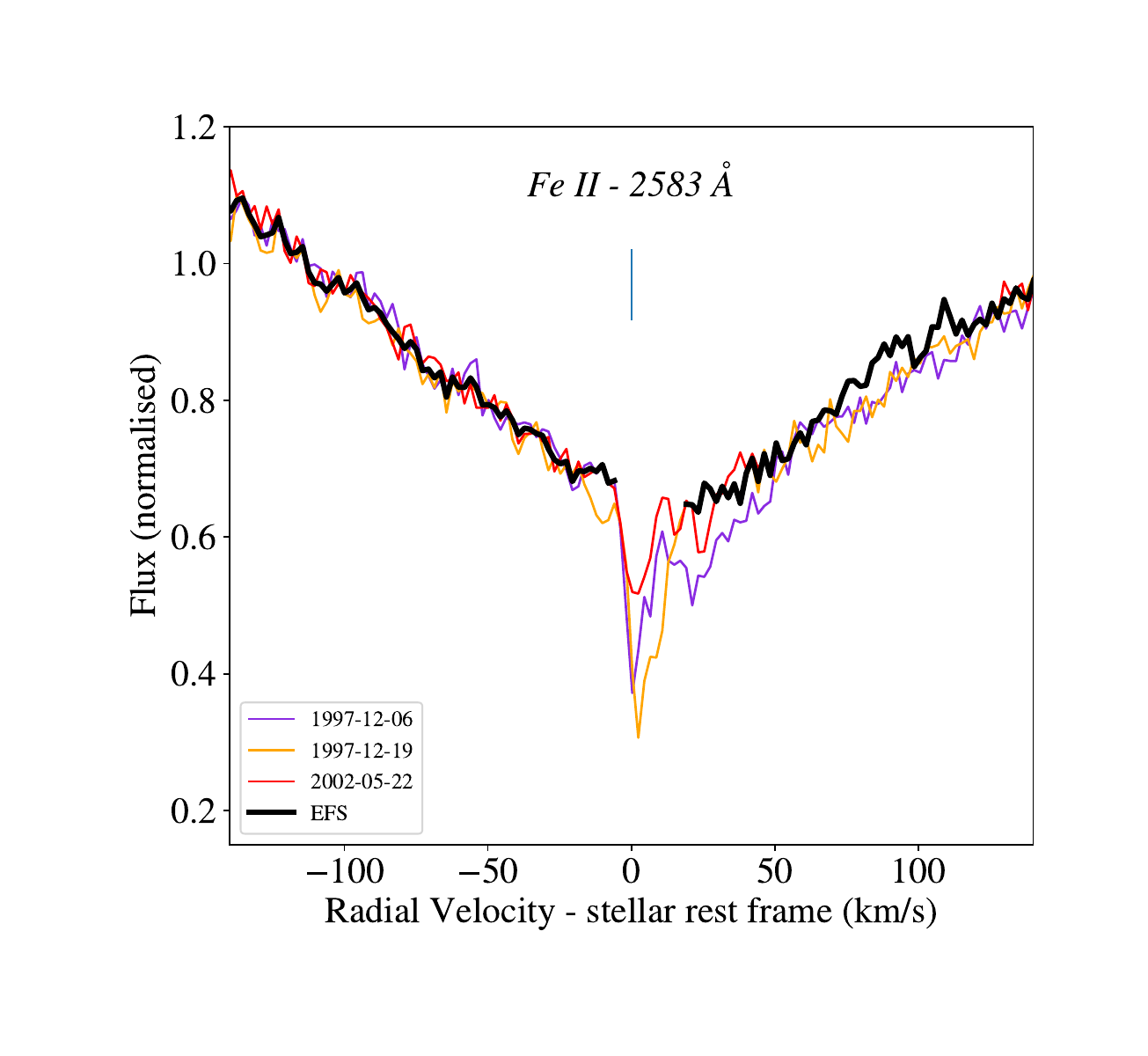}
    
    \includegraphics[scale = 0.4,     trim = 10 40 20 40,clip]{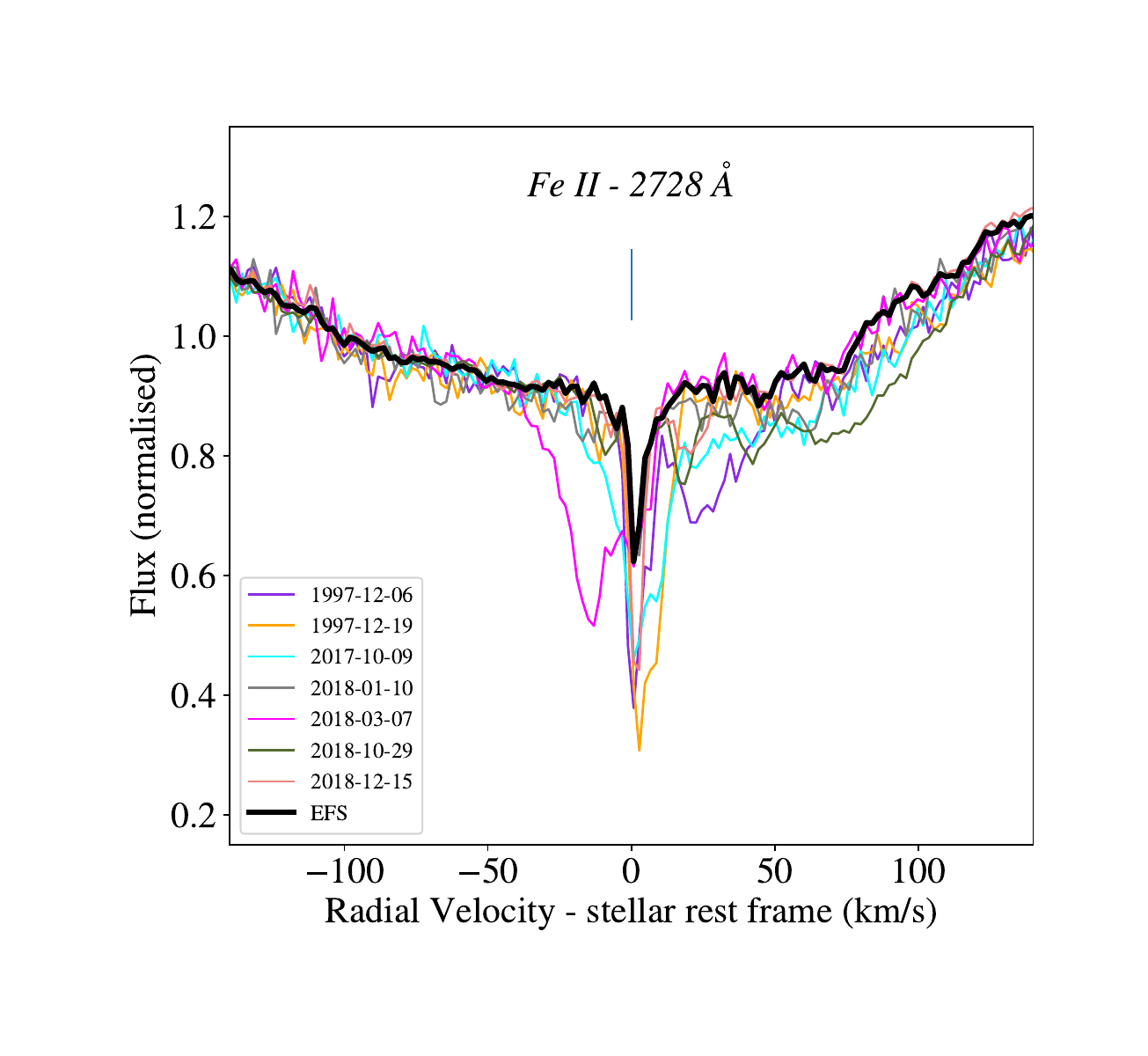}    
    \includegraphics[scale = 0.4,     trim = 10 40 40 40,clip]{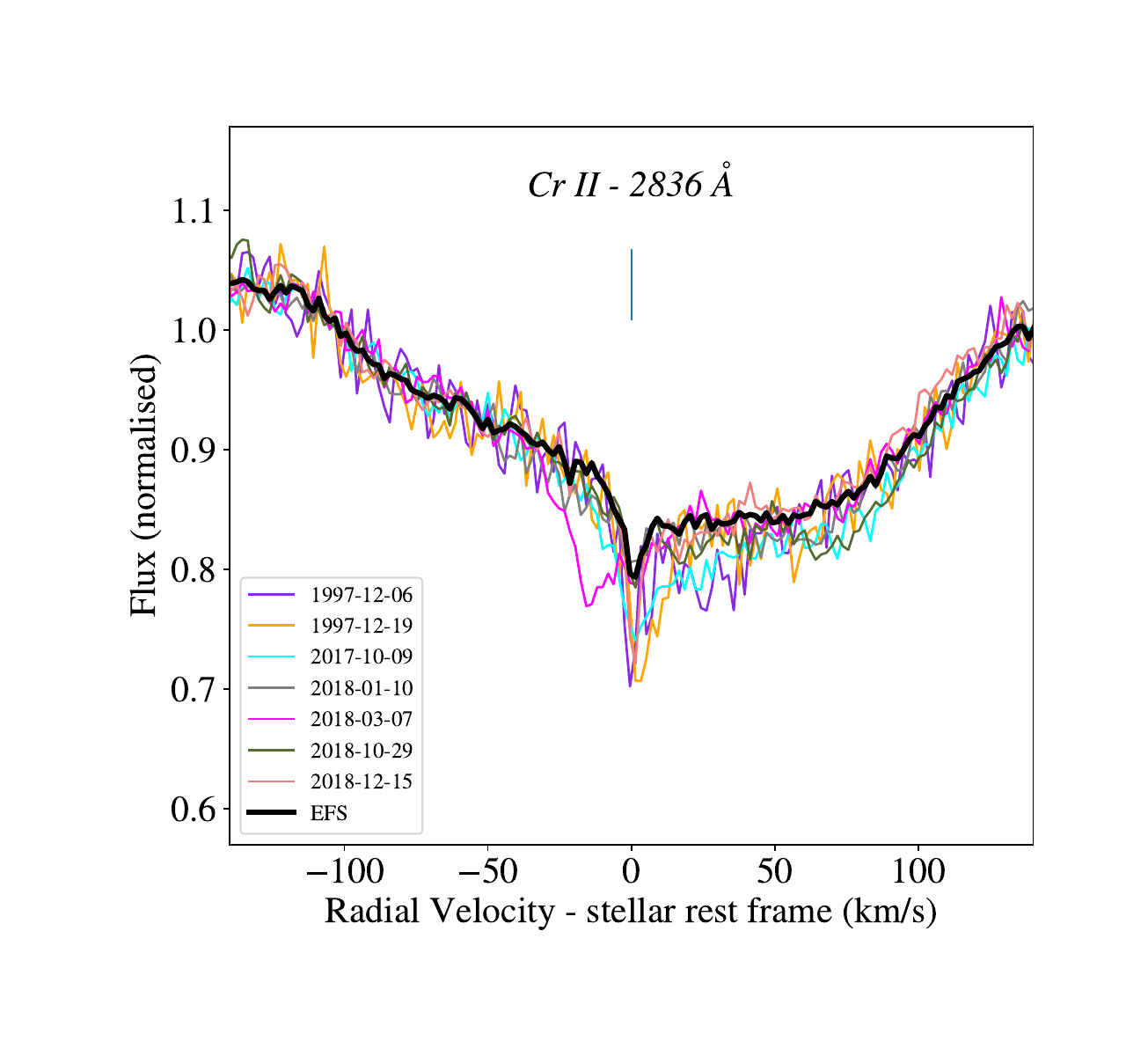}    
\end{figure*}
\small \textbf{Figure \ref{Fig. line examples}}, continued.

\newpage

\begin{figure*}[h!]
\centering
    \includegraphics[scale = 0.4,     trim = 10 40 20 30,clip]{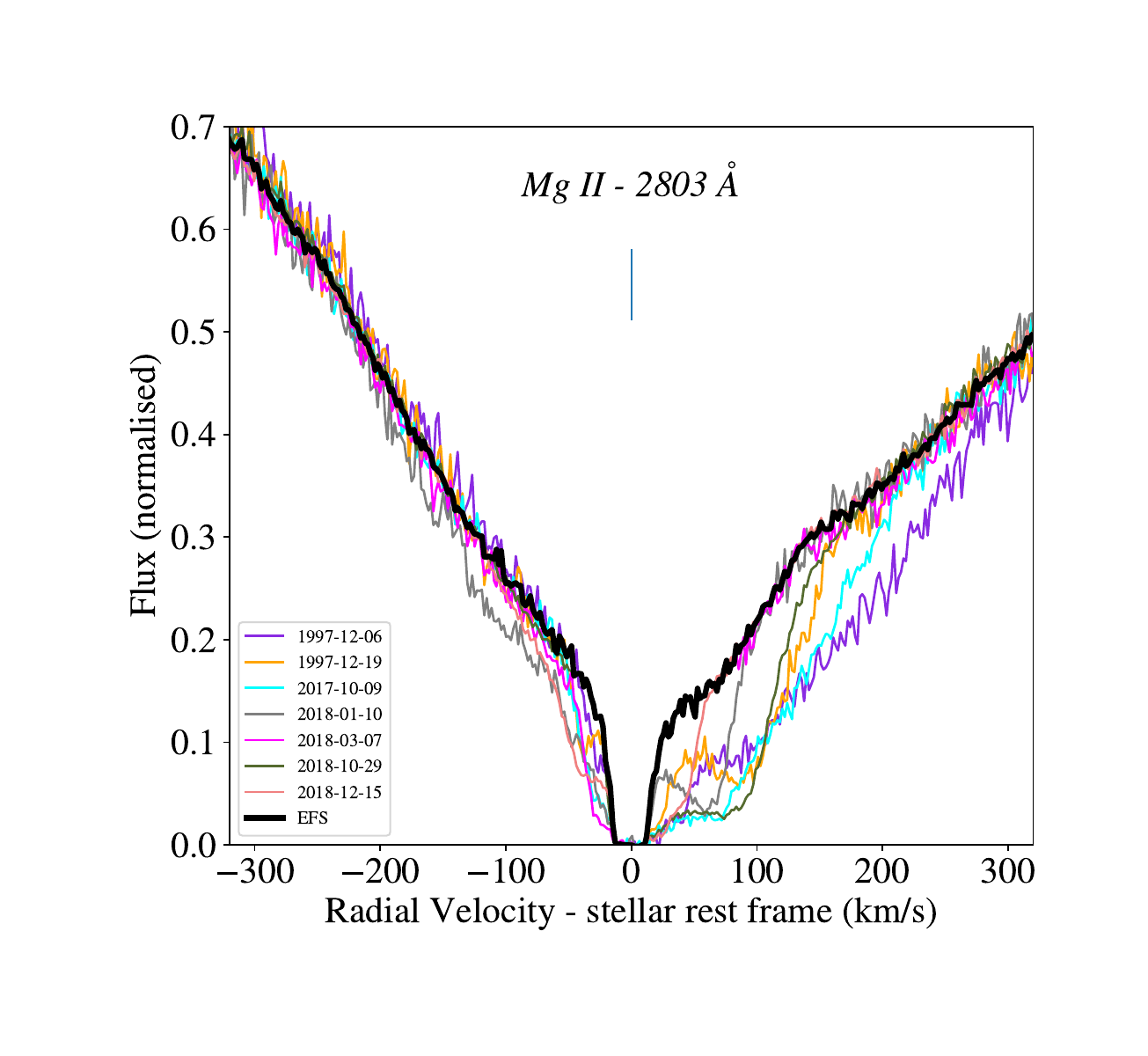}    
    \includegraphics[scale = 0.4,     trim = 10 40 40 30,clip]{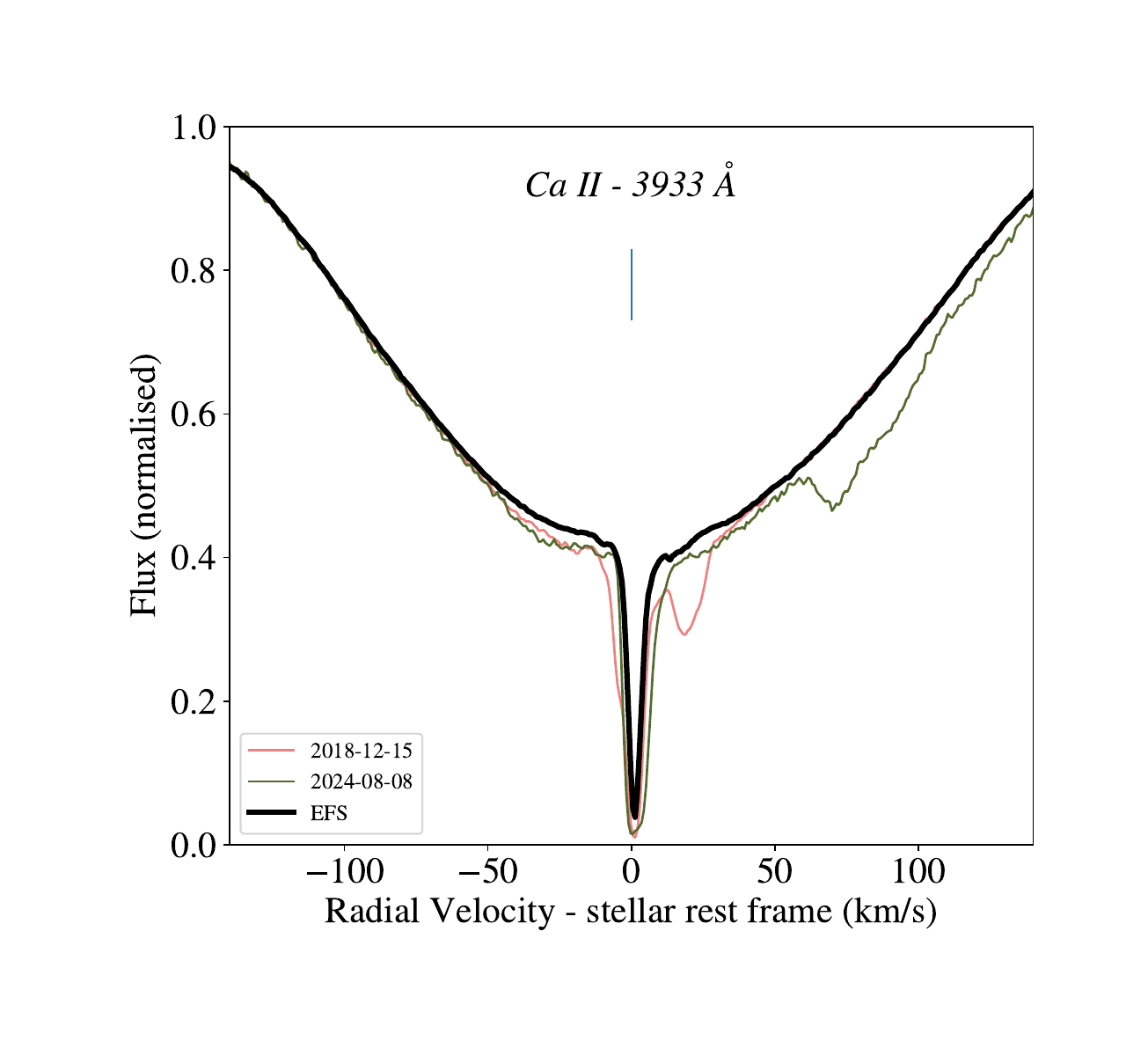}    
    \includegraphics[scale = 0.4,     trim = 10 40 40 30,clip]{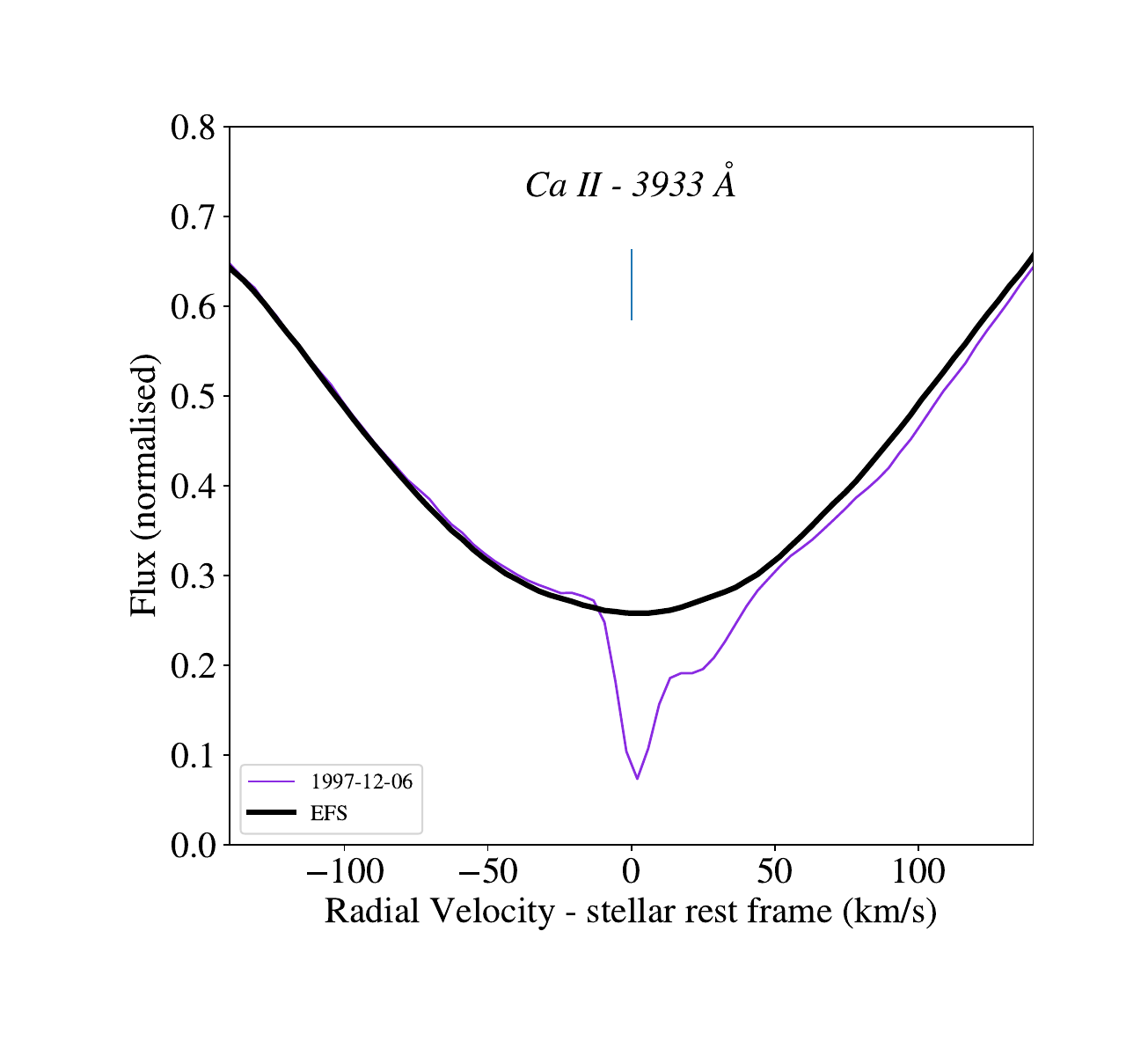}    
       
\end{figure*}
\small \textbf{Figure \ref{Fig. line examples}}, continued Note that the \caii\ doublet (3900 \A) was observed both by with HARPS (upper right) and MJUO (lower).

\newpage
\section{Comet list}

\begin{table}[h!]
    \centering 
    \begin{threeparttable}
    \renewcommand{\arraystretch}{1.15}
    \caption{\small List of the 29 comets studied in the present paper.}
    \begin{tabular}{ c c c c c c }          
            
            \cline{1-6}  
            \noalign{\smallskip}
            \cline{1-6}  
            
            Label & Date (orbit\tnote{a} ) & Nomenclature\tnote{b} & Total RV range\tnote{c} & Fitted RV range\tnote{d} & Studied lines\tnote{e} \\
                  &            &  &      (km/s)    &      (km/s)    &             \\
            
            \cline{1-6}
            \noalign{\smallskip}

            & & & & & \feii\ $\lambda$2250, $\lambda$2400,  \\
            & & & & & $\lambda$2600, $\lambda$2750 \A   \\
            & & & & &\niii\  $\lambda$1750, $\lambda$2200, \\
            & & & & &$\lambda$2400 \A \\
            & & & & &\crii\  $\lambda$2000,  $\lambda$2700 \A \\
            \multirow{2}{*}{1\tnote{f}} & \multirow{2}{*}{1997-12-06} & \multirow{2}{*}{C19971206 a} & \multirow{2}{*}{[+10,+50]} & \multirow{2}{*}{[+21,+42]} &\siii\  $\lambda$1800 \A \\
            & & & & & \coii\  $\lambda$2300 \A \\
            & & & & & \mnii\  $\lambda$2600 \A  \\
            & & & & & \Si\    $\lambda$1800 \A  \\
            & & & & & \aliii\ $\lambda$1850 \A  \\
            & & & & & \znii\  $\lambda$2000 \A  \\
            & & & & & \caii\  $\lambda$3900 \A  \\

    \noalign{\smallskip}
    \cline{1-6}
    \noalign{\smallskip}

            & & & & & \feii\ $\lambda$2400, $\lambda$2600, \\    
            \multirow{2}{*}{2} & \multirow{2}{*}{1997-12-06} & \multirow{2}{*}{C19971206 b}  & \multirow{2}{*}{[+60,+200]} & \multirow{2}{*}{[+70,+150]} & $\lambda$2750 \A \\
            & & & & &\mgii\  $\lambda$2800 \A \\
            & & & & &\caii\  $\lambda$3900 \A \\

    \noalign{\smallskip}
    \cline{1-6}
    \noalign{\smallskip}

            & & & & & \feii\ $\lambda$2400, $\lambda$2600,  \\
            \multirow{2}{*}{3} & \multirow{2}{*}{1997-12-19} & \multirow{2}{*}{C19971219 a} & \multirow{2}{*}{[-45,-31]} & \multirow{2}{*}{[-43,-33]} &  $\lambda$2750 \A  \\
            & & & & &\niii\  $\lambda$1700, $\lambda$2200, \\
            & & & & &\niii\ $\lambda$2400 \A  \\

    \noalign{\smallskip}
    \cline{1-6}
    \noalign{\smallskip}

            & & & & &\feii\ $\lambda$2250,  $\lambda$2400, \\
            & & & & &  $\lambda$2600, $\lambda$2750 \A  \\
            \multirow{2}{*}{4} & \multirow{2}{*}{1997-12-19} & \multirow{2}{*}{C19971219 b}  & \multirow{2}{*}{[-17,-6]} & \multirow{2}{*}{[-16,-7]} & \niii\  $\lambda$1700, $\lambda$2200, \\
            & & & & & $\lambda$2400 \A \\
            & & & & & \mnii\  $\lambda$2600 \A  \\
            & & & & & \aliii\  $\lambda$1850 \A  \\
            
    \noalign{\smallskip}
    \cline{1-6}
    \noalign{\smallskip}

            & & & & &\feii\ $\lambda$1600,  $\lambda$1700,   \\ 
            & & & & & $\lambda$2750 \A   \\ 
            & & & & &\mnii\  $\lambda$2600 \A  \\
            \multirow{2}{*}{5\tnote{g}} & \multirow{2}{*}{1997-12-19} & \multirow{2}{*}{C19971219 c} & \multirow{2}{*}{[0,+19]} & \multirow{2}{*}{[+3,+15]} & \niii\  $\lambda$1700\A  \\
            & & & & &\crii\  $\lambda$2700 \A  \\
            & & & & &\siii\  $\lambda$1800 \A  \\
            & & & & &\Si\    $\lambda$1800 \A  \\
            & & & & &\aliii\ $\lambda$1850 \A \\

    \noalign{\smallskip}
    \cline{1-6}
    \noalign{\smallskip}

            & & & & &\feii\ $\lambda$2400,  $\lambda$2600,    \\ 
            \multirow{2}{*}{6} & \multirow{2}{*}{1997-12-19} & \multirow{2}{*}{C19971219 d}  & \multirow{2}{*}{[+50,+150]} & \multirow{2}{*}{[+60,+120]} & $\lambda$2750 \A   \\ 
            & & & & & \aliii\  $\lambda$1850 \A   \\
            & & & & &\mgii\  $\lambda$2800 \A  \\

    \noalign{\smallskip}
    \cline{1-6}
    \noalign{\smallskip}

    \end{tabular} 
        
    \begin{tablenotes}
      \item[a] \small Specific HST orbits used to analyse comets 20-29 (see Sect. \ref{Sect. Fit procedure}.
      \item[b] \small The full nomenclature should also contain the stellar name, e.g. "\bp\ C19971206 a" for comet 1, as proposed by Lecavelier des Etangs et al. (in prep).
      \item[c] \small RV range where each comet is detected.
      \item[d] \small RV range used to measure the comets absorption depths. 
      \item[e] \small Fitted line series for each comet.
      \item[f] \small Comet 1 was already studied in \cite{Vrignaud24}.      
      \item[g] \small For comet 5, the \mnii/\feii\ ratio was estimated from the $\lambda$2600 \A\ \mnii\ and $\lambda$2400-2600 \A\ \feii\ line series, using the 2002-05-22 spectrum as a reference.
    \end{tablenotes}
    
\label{Tab. comets}

\end{threeparttable}
\end{table}

\newpage

\small \centering \textbf{Table \ref{Tab. comets}}, continued.

\begin{table}[h!]
    \centering 
    \begin{threeparttable}
    \renewcommand{\arraystretch}{1.15}
    \begin{tabular}{ c c c c c c}          
            
            \cline{1-6}  
            \noalign{\smallskip}
            \cline{1-6}  
            
            Label & Date (orbit) & Nomenclature & Total RV range & Fitted RV range & Studied lines \\
                  &            &  &      (km/s)    &      (km/s)    &             \\
            
            \cline{1-6}  
            \noalign{\smallskip}

            & & & & &\feii\ $\lambda$1700,  $\lambda$2600, \A   \\ 
            \multirow{2}{*}{7} & \multirow{2}{*}{2001-08-27} & \multirow{2}{*}{C20010827 a} & \multirow{2}{*}{[-15,0]} & \multirow{2}{*}{[-13,-6]} & $\lambda$2750 \A   \\ 
            & & & & & \aliii\  $\lambda$1850 \A  \\
            & & & & &\mgii\  $\lambda$2800 \A  \\

    \noalign{\smallskip}
    \cline{1-6}
    \noalign{\smallskip}

            & & & & &\feii\ $\lambda$2400,  $\lambda$2600 \A   \\ 
            8 & 2002-05-22 & C20020522 a & [+18,+31] & [+21,+29] & \niii\  $\lambda$2400 \A  \\
            & & & & &\mnii\  $\lambda$2600 \A  \\

        \noalign{\smallskip}
        \cline{1-6}
        \noalign{\smallskip}
    
            \multirow{2}{*}{9} & \multirow{2}{*}{2017-10-09} & C20171009 a & \multirow{2}{*}{[+2,+15]} & \multirow{2}{*}{[+4,+12]} & \feii\ $\lambda$2750\A  \\ 
            & & & & &\crii\  $\lambda$2700 \A  \\

        \noalign{\smallskip}
        \cline{1-6}
        \noalign{\smallskip}

            \multirow{2}{*}{10} & \multirow{2}{*}{2017-03-07} & C20170307 a & \multirow{2}{*}{[-20,-6]} & \multirow{2}{*}{[-35,0]} & \feii\ $\lambda$2750\A  \\ 
            & & & & &\crii\  $\lambda$2700 \A  \\

        \noalign{\smallskip}
        \cline{1-6}
        \noalign{\smallskip}
    
            & & & & &  \feii\ $\lambda$2750\A   \\ 
            11 & 2018-10-29 & C20181029 a &[-22,-5] & [-22,-5] & \Sii\  $\lambda$1250\A  \\
            & & & & & \crii\  $\lambda$2700\A  \\
            
    \noalign{\smallskip}
        \cline{1-6}
    \noalign{\smallskip}
            
            & & & & & \feii\ $\lambda$2750\A \\ 
            12 & 2018-10-29 & C20181029 b & [+11,+28] & [+11,+28] & \Sii\  $\lambda$1250\A  \\
            & & & & & \crii\  $\lambda$2700\A \\

    \noalign{\smallskip}
        \cline{1-6}
    \noalign{\smallskip}
            
            & & & &  &\feii\ $\lambda$2750\A \\ 
            13 & 2018-10-29 &C20181029 c &  [+30,+50] & [+30,+50] & \Sii\  $\lambda$1250\A  \\
            & & & &  &\crii\  $\lambda$2700\A \\

    \noalign{\smallskip}
        \cline{1-6}
    \noalign{\smallskip}
            
            & & & & & \feii\ $\lambda$2750\A  \\ 
            14\tnote{h} & 2018-10-29 & C20181029 d &  [+50,+150] & [+60,+90] & \Sii\  $\lambda$1250\A  \\
            & & & & & \crii\  $\lambda$2700\A  \\

    \noalign{\smallskip}
        \cline{1-6}
    \noalign{\smallskip}
            
            & & & & & \feii\ $\lambda$2750\A  \\ 
            15 & 2018-12-15 & C20181215 a & [-50,-20] & [-50,-30] & \mgii\  $\lambda$2800\A   \\
            & & & & & \caii\  $\lambda$3900\A  \\

    \noalign{\smallskip}
        \cline{1-6}
    \noalign{\smallskip}
            
            \multirow{2}{*}{16} & \multirow{2}{*}{2018-12-15} & \multirow{2}{*}{C20181215 b} & \multirow{2}{*}{[-11,-3]} & \multirow{2}{*}{[-12,-6]} & \feii\ $\lambda$2750\A  \\ 
            & & & & & \caii\  $\lambda$3900\A  \\

    \noalign{\smallskip}
        \cline{1-6}
    \noalign{\smallskip}
            
            \multirow{2}{*}{17\tnote{i}} & \multirow{2}{*}{2018-12-15} & \multirow{2}{*}{C20181215 c} & \multirow{2}{*}{[-2,+5]}  & \multirow{2}{*}{[-2,+5]} & \feii\ $\lambda$2750\A \\ 
            & & & & & \crii\  $\lambda$2700\A \\

    \noalign{\smallskip}
        \cline{1-6}
    \noalign{\smallskip}
            
            & & & & & \feii\ $\lambda$2750\A  \\ 
            \multirow{2}{*}{18} & \multirow{2}{*}{2018-12-15} & \multirow{2}{*}{C20181215 d} & \multirow{2}{*}{[+12,+32]}  & \multirow{2}{*}{[+13,+28]} & \Sii\  $\lambda$1250\A   \\
            & & & & & \crii\  $\lambda$2700\A  \\
            & & & & & \caii\  $\lambda$3900\A  \\

    \noalign{\smallskip}
        \cline{1-6}
    \noalign{\smallskip}
            
            & & & & & \feii\ $\lambda$2750\A  \\ 
            19 & 2018-12-15 & C20181215 e & [+32,+55] & [+32,+46] & \mgii\  $\lambda$2800\A  \\
            & & & & & \caii\  $\lambda$3900\A  \\

    \noalign{\smallskip}
        \cline{1-6}
    \noalign{\smallskip}

        \end{tabular} 
        
    \begin{tablenotes}
      \item[h] Comet 14 was also studied in \cite{Vrignaud24}.
      \item[i] For comets 17 and 22, the much poorer EFS determination of \feii\ lines compared to \caii\ makes the retrieval of the \caii/\feii\ ratio hazardous.
    \end{tablenotes}
\end{threeparttable}
\end{table}

\newpage

\small \textbf{Table \ref{Tab. comets}}, continued.

\begin{table}[h!]
    \centering 
    \begin{threeparttable}
    \renewcommand{\arraystretch}{1.15}
    \begin{tabular}{ c c c c c c}          
            
            \cline{1-6}  
            \noalign{\smallskip}
            \cline{1-6}

            Label & Date (orbit) & Nomenclature & Total RV range & Fitted RV range & Studied lines \\
                  &            &  &      (km/s)    &      (km/s)    &             \\
            
            \cline{1-6}  
    \noalign{\smallskip}

            & & & & & \feii\ $\lambda$1700\A  \\ 
            & & & & & \niii\  $\lambda$1700\A  \\
            \multirow{2}{*}{20} & \multirow{2}{*}{2024-04-17 \ (1-3) }  & \multirow{2}{*}{C20240417 a} & \multirow{2}{*}{[-14,-5]} & \multirow{2}{*}{[-12,-6]} & \siii\  $\lambda$1800\A  \\
            & & & & & \Si\  $\lambda$1800\A \\
            & & & & & \aliii\  $\lambda$1850\A \\
            & & & & & \caii\  $\lambda$3900\A \\

    \noalign{\smallskip}
        \cline{1-6}
    \noalign{\smallskip}
            
            & & & & & \feii\ $\lambda$1700\A  \\ 
            & & & & & \niii\  $\lambda$1700\A \\
            21\tnote{f} & 2024-04-17  (1-3)& C20240417 b & \ [+2,+11] & [+2,+11] & \siii\  $\lambda$1800\A  \\
            & & & & & \Si\  $\lambda$1800\A  \\
            & & & & & \aliii\  $\lambda$1850\A  \\

    \noalign{\smallskip}
            \cline{1-6}  
    \noalign{\smallskip}

            & & & & & \feii\ $\lambda$1700\A   \\ 
            & & & & & \niii\  $\lambda$1700\A  \\
            22 & 2024-04-17 \ (2) & C20240417 c & [+24,+58] & [+25,+55] & \siii\  $\lambda$1800\A  \\
            & & & & & \aliii\  $\lambda$1850\A  \\
            & & & & & \caii\  $\lambda$3900\A  \\
            
    \noalign{\smallskip}
            \cline{1-6}  
    \noalign{\smallskip}
                        
            & & & & & \feii\ $\lambda$1700\A  \\ 
            23 & 2024-04-17 \ (1) & C20240417 d  & [+60,+104] & [+60,+95] & \aliii\  $\lambda$1850\A   \\
            & & & & & \caii\  $\lambda$3900\A  \\

    \noalign{\smallskip}
            \cline{1-6}  
    \noalign{\smallskip}
            
            & & & & & \feii\ $\lambda$1700\A  \\ 
            24\tnote{j} & 2024-07-31 \ (1-3) & C20240731 a  & [-60,-10] & [-40,-15] & \niii\  $\lambda$1700\A \\
            & & & & & \siii\  $\lambda$1800\A \\

    \noalign{\smallskip}
            \cline{1-6}  
    \noalign{\smallskip}
            
            & & & & & \feii\ $\lambda$1700\A  \\ 
            25\tnote{j} & 2024-07-31 \ (2) & C20240731 b & [+40,+75] & [+45,+70] & \niii\  $\lambda$1700\A\\
            & & & & & \siii\  $\lambda$1800\A  \\

    \noalign{\smallskip}
            \cline{1-6}  
    \noalign{\smallskip}
            
            \multirow{2}{*}{26\tnote{j}}& \multirow{2}{*}{2024-07-31 \ (3)} & \multirow{2}{*}{C20240731 c} & \multirow{2}{*}{[+70,+105]} &\multirow{2}{*}{[+78,+102]} & \feii\ $\lambda$1700\A\\ 
            & & & & & \niii\  $\lambda$1700\A  \\

    \noalign{\smallskip}
            \cline{1-6}  
    \noalign{\smallskip}
            
            & & & & & \feii\ $\lambda$1700\A   \\ 
            & & & & & \niii\  $\lambda$1700\A  \\
            27 & 2024-08-08 \ (1-3) & C20240808 a & [-50,-10]  & [-40,-15] & \siii\  $\lambda$1800\A   \\
            & & & & & \aliii\  $\lambda$1850\A  \\
            & & & & & \caii\  $\lambda$3900\A  \\

    \noalign{\smallskip}
            \cline{1-6}  
    \noalign{\smallskip}
            
            & & & & & \feii\ $\lambda$1700\A   \\ 
            & & & & & \niii\  $\lambda$1700\A  \\
            28 & 2024-08-08 \  (1-3) & C20240808 b &  [+60,+85] & [+67,+83] & \siii\  $\lambda$1800\A \\
            & & & & & \aliii\  $\lambda$1850\A  \\
            & & & & & \caii\  $\lambda$3900\A  \\

    \noalign{\smallskip}
            \cline{1-6}  
    \noalign{\smallskip}

            & & & & & \feii\ $\lambda$1700\A  \\ 
            & & & & & \niii\  $\lambda$1700\A  \\
            29 & 2024-08-08 \  (1-3) &  C20240808 c &  [+85,+105] & [+85,+100] & \siii\  $\lambda$1800\A   \\
            & & & & & \aliii\  $\lambda$1850\A  \\
            & & & & & \caii\  $\lambda$3900\A  \\

    \noalign{\smallskip}
            \cline{1-6}  
    \noalign{\smallskip}

        \end{tabular} 
        
    \begin{tablenotes}
      \item[j] For comets 24 to 26, the \aliii\ doublet was removed from the analysis, due to many overlapping absorption features.
    \end{tablenotes}
\end{threeparttable}
\end{table}

\FloatBarrier
\newpage

\section{Comet plots and curves of growth}

\begin{figure*}[h!]
\centering
    \includegraphics[scale = 0.38,     trim = 10 30 30 40,clip]{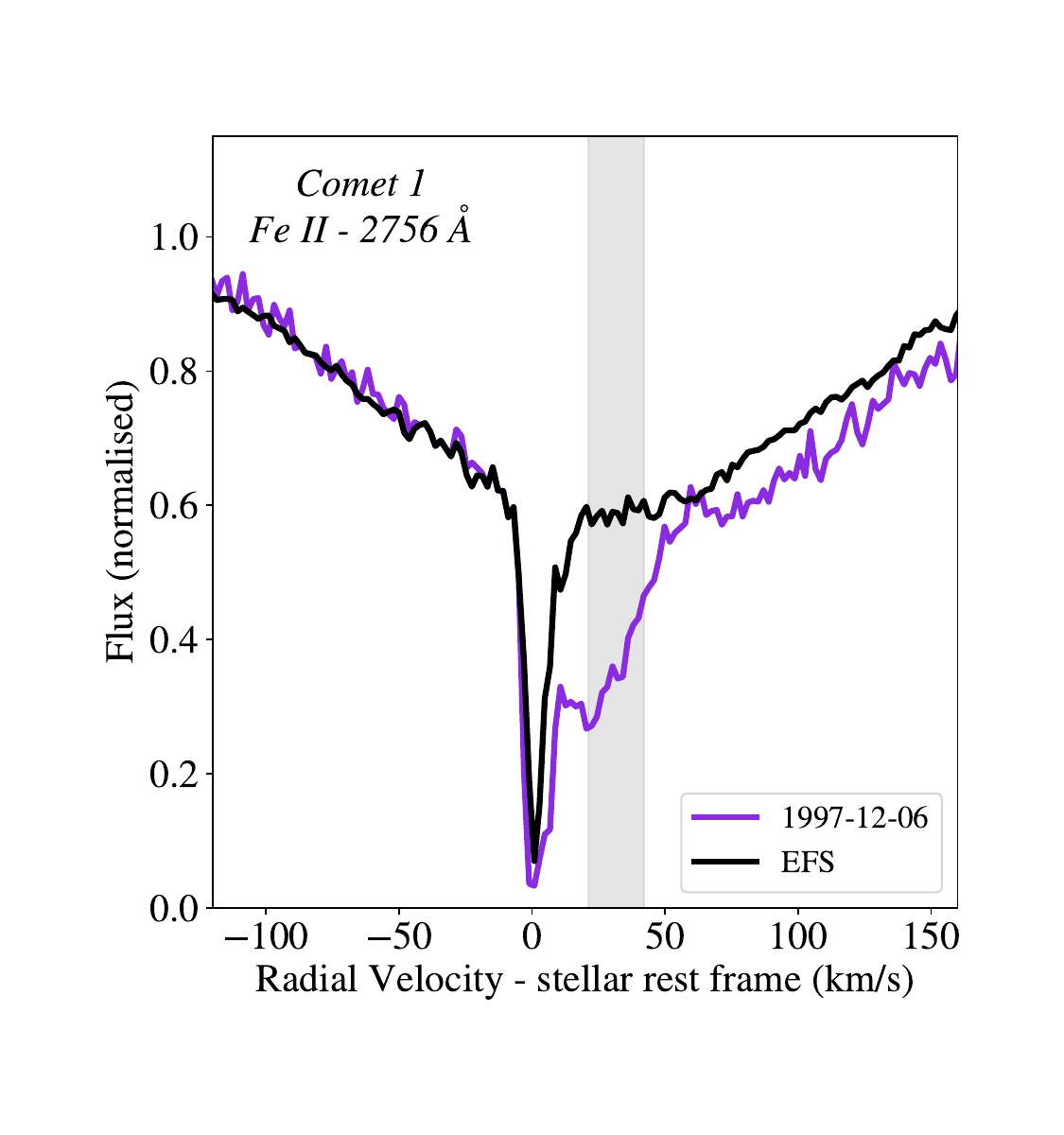}    
    \includegraphics[scale = 0.38,     trim = 85 30 80 40,clip]{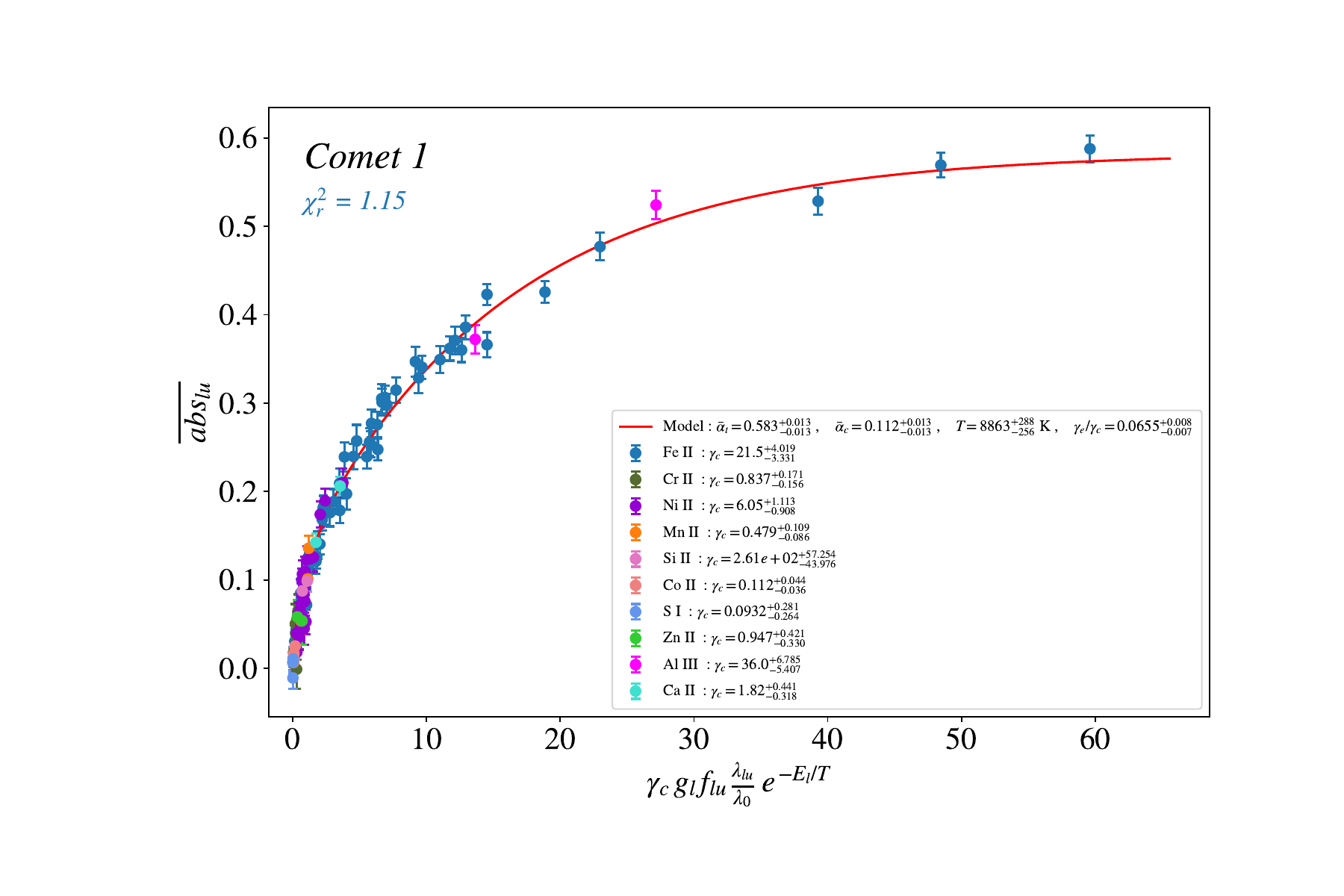}    
    
    \includegraphics[scale = 0.38,     trim = 10 30 30 40,clip]{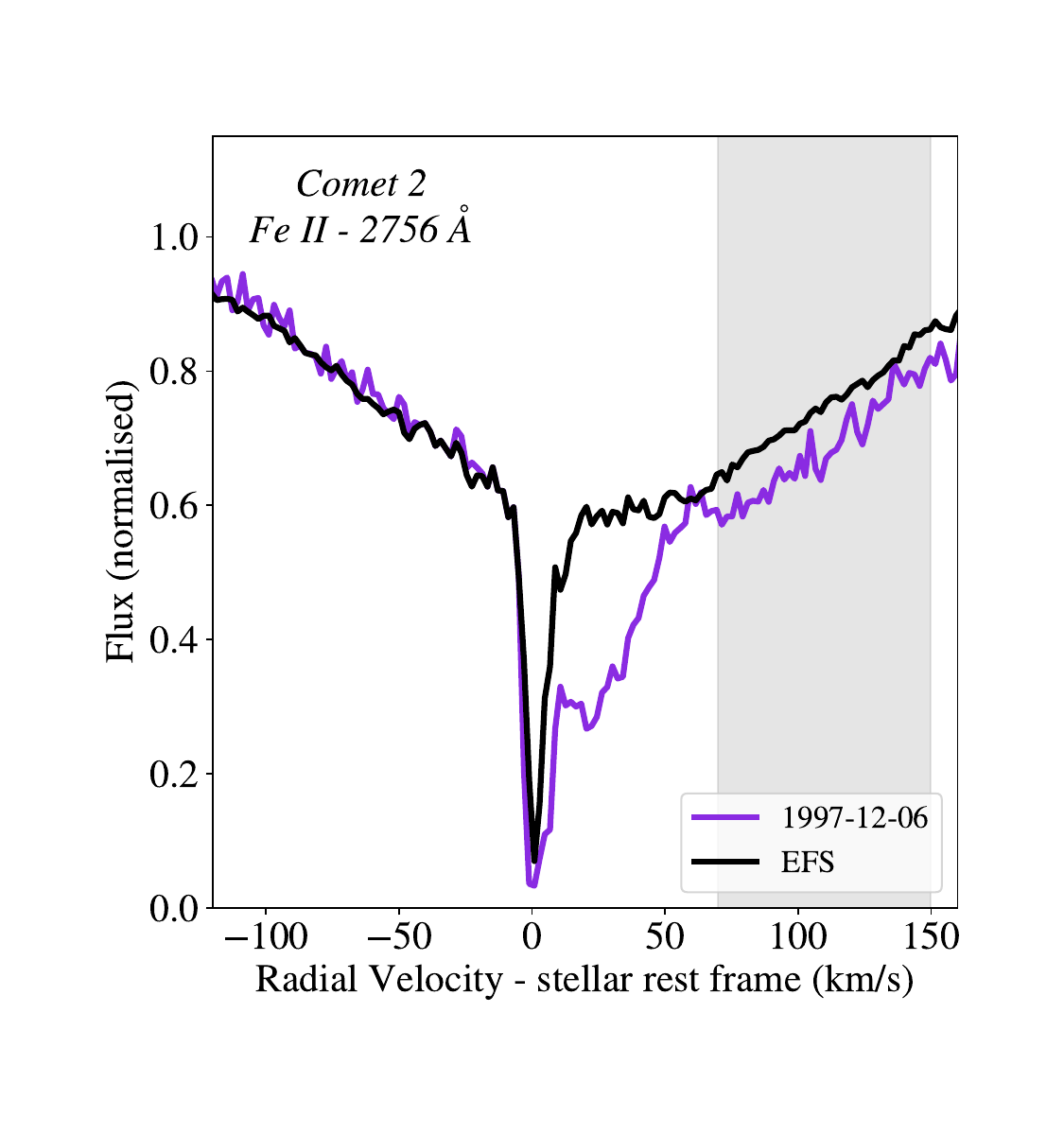}    
    \includegraphics[scale = 0.38,     trim = 85 30 80 40,clip]{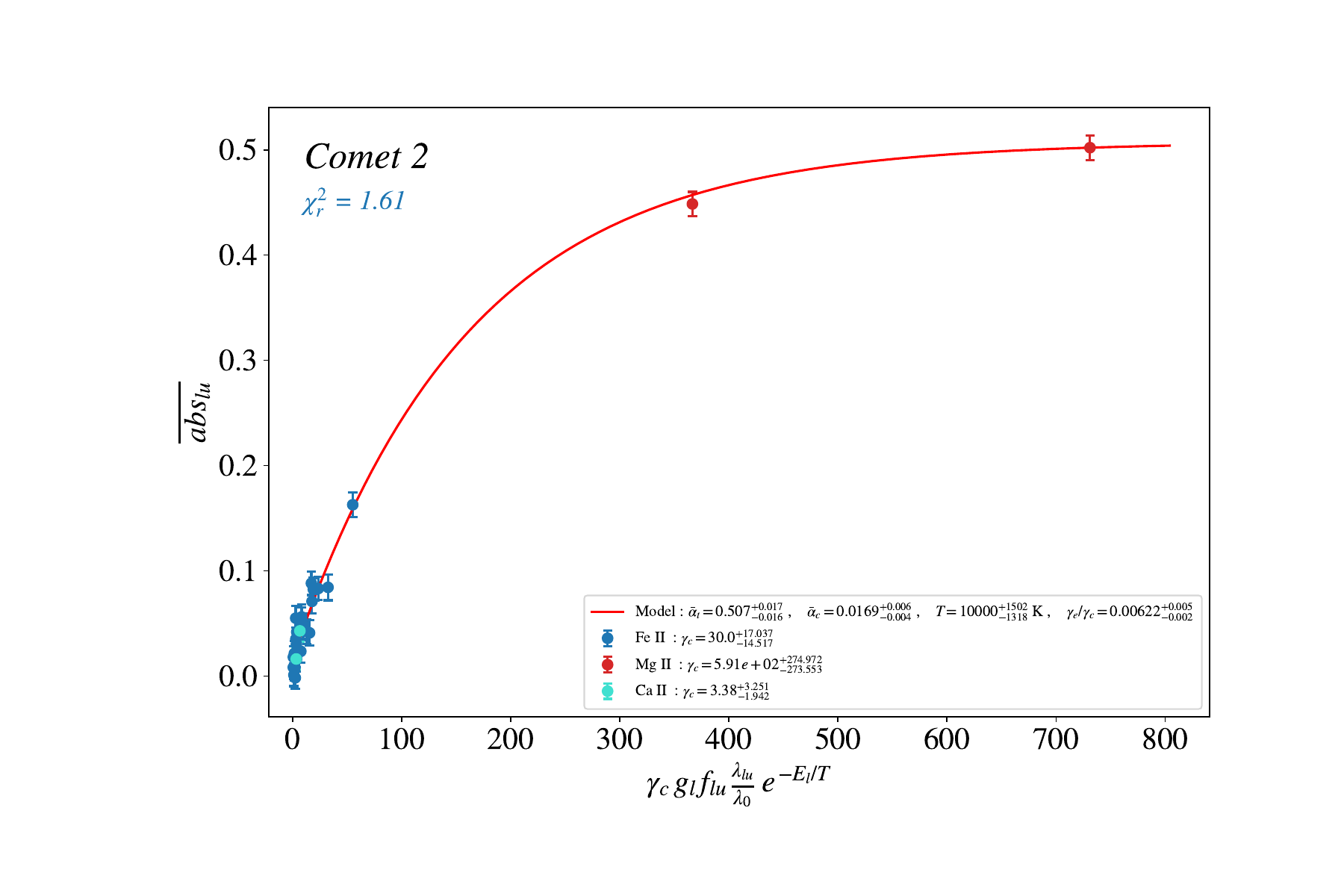}
    
    \includegraphics[scale = 0.38,     trim = 10 30 30 40,clip]{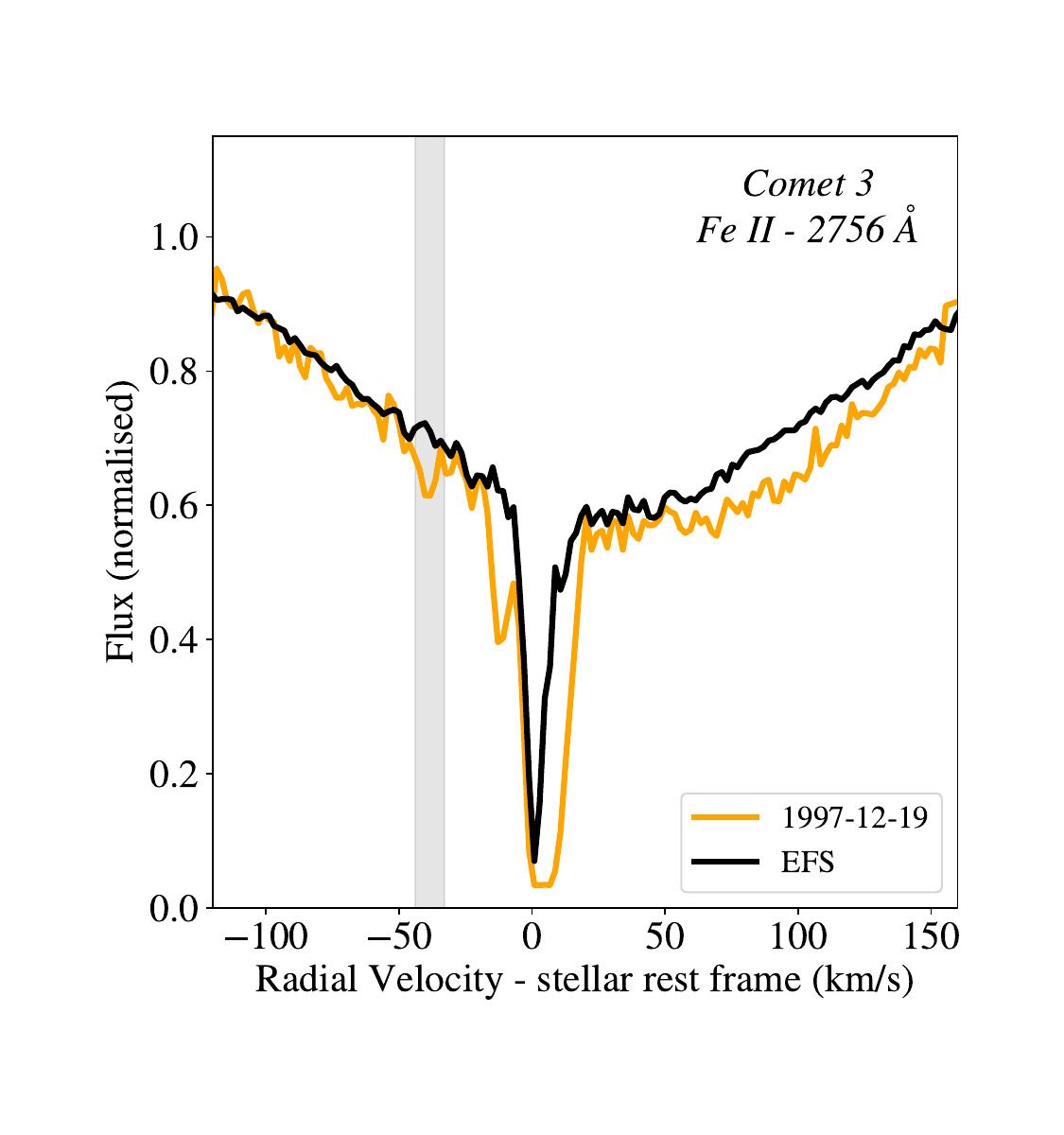}    
    \includegraphics[scale = 0.38,     trim = 85 30 80 40,clip]{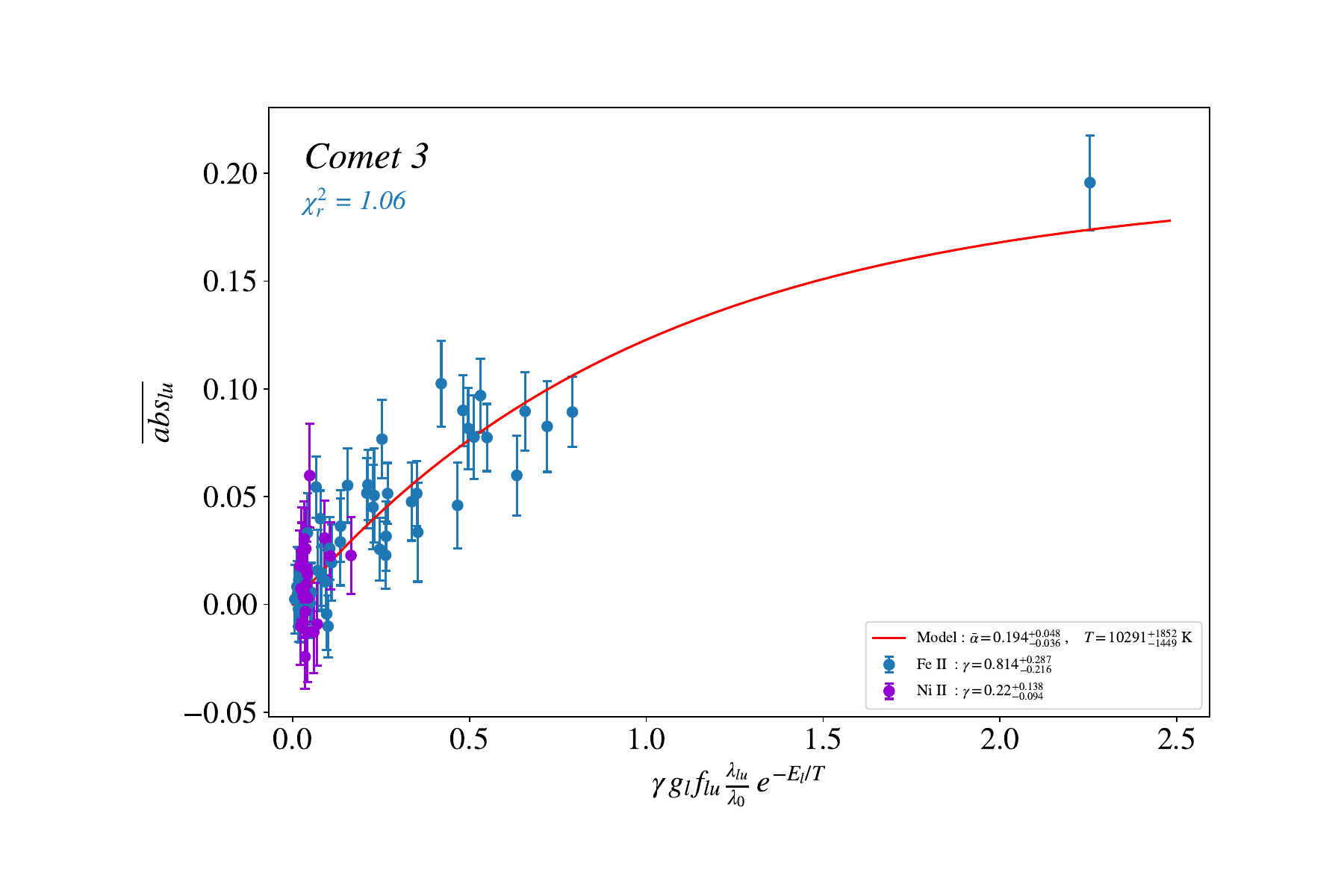}    

    \caption{\small \textbf{Left : } plots of the 29 studied comets, in one strong \feii\ line. The grey areas indicate the radial velocity range used to measure the absorption depth of each comet. For comets studied using a specific HST orbit (e.g. comets 23), the orbit number is indicated in parenthesis (e.g. '2024-04-17 (2)').  \textbf{Right} : curves of growth of the 29 studied comets. Different species are shown with different colours; the best-fit model is plotted with a red, solid line. The corresponding reduced $\rchi^2$ is also provided.}
    \label{Fig. curves of growth all}
\end{figure*}

\newpage

\begin{figure*}[h!]
\centering
    \includegraphics[scale = 0.38,     trim = 10 30 30 40,clip]{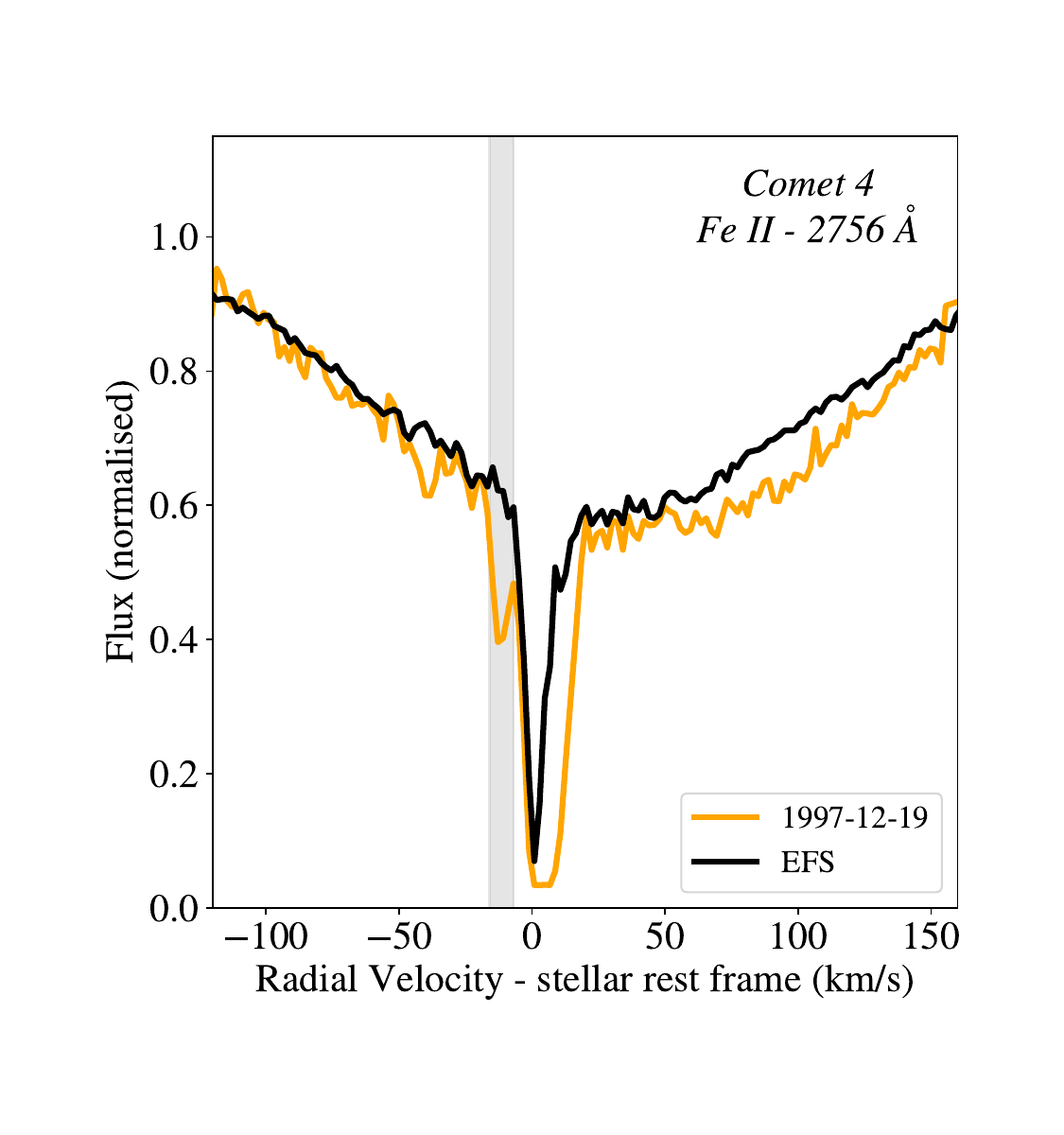}    
    \includegraphics[scale = 0.38,     trim = 85 30 80 40,clip]{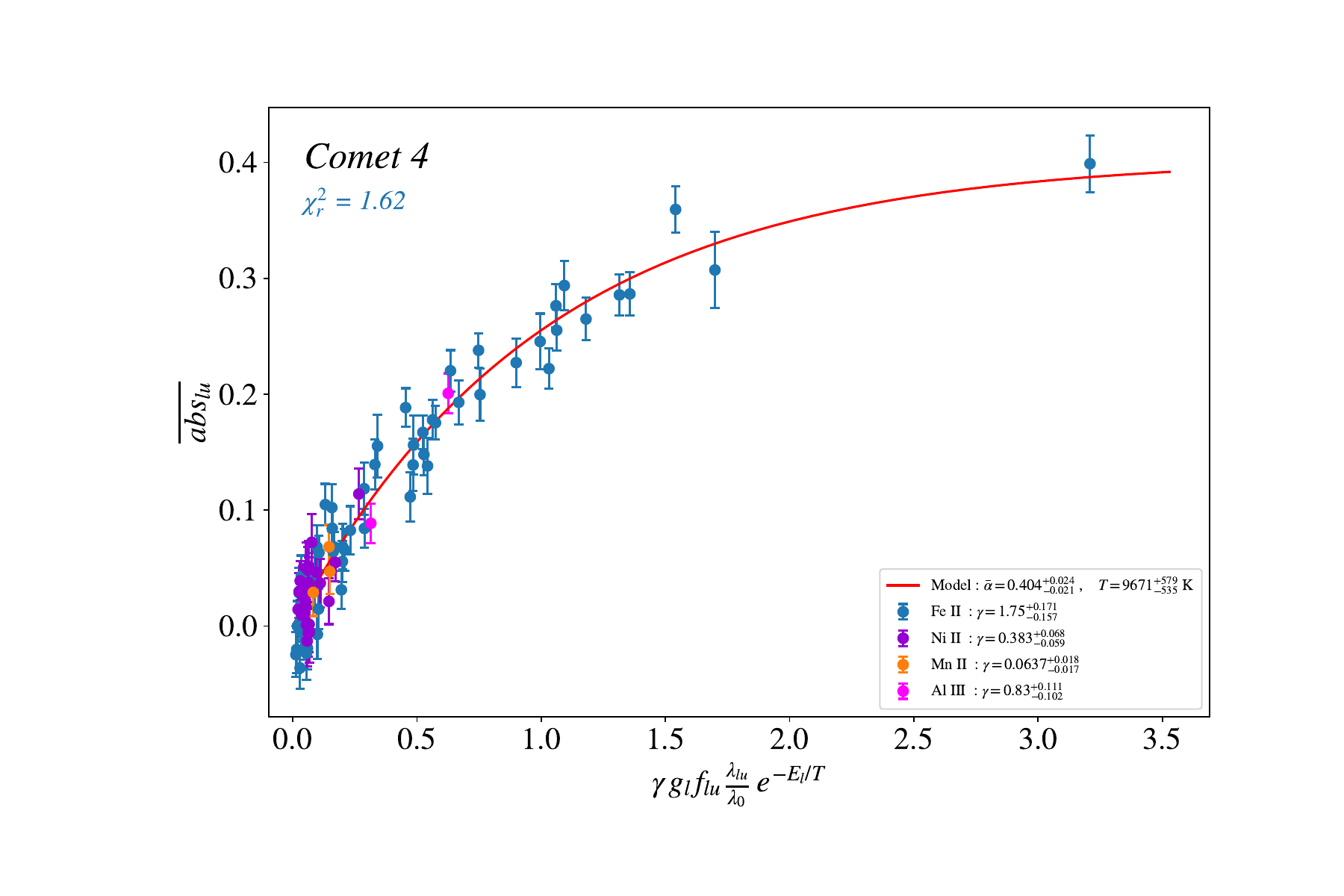}    
    
    \includegraphics[scale = 0.38,     trim = 10 30 30 40,clip]{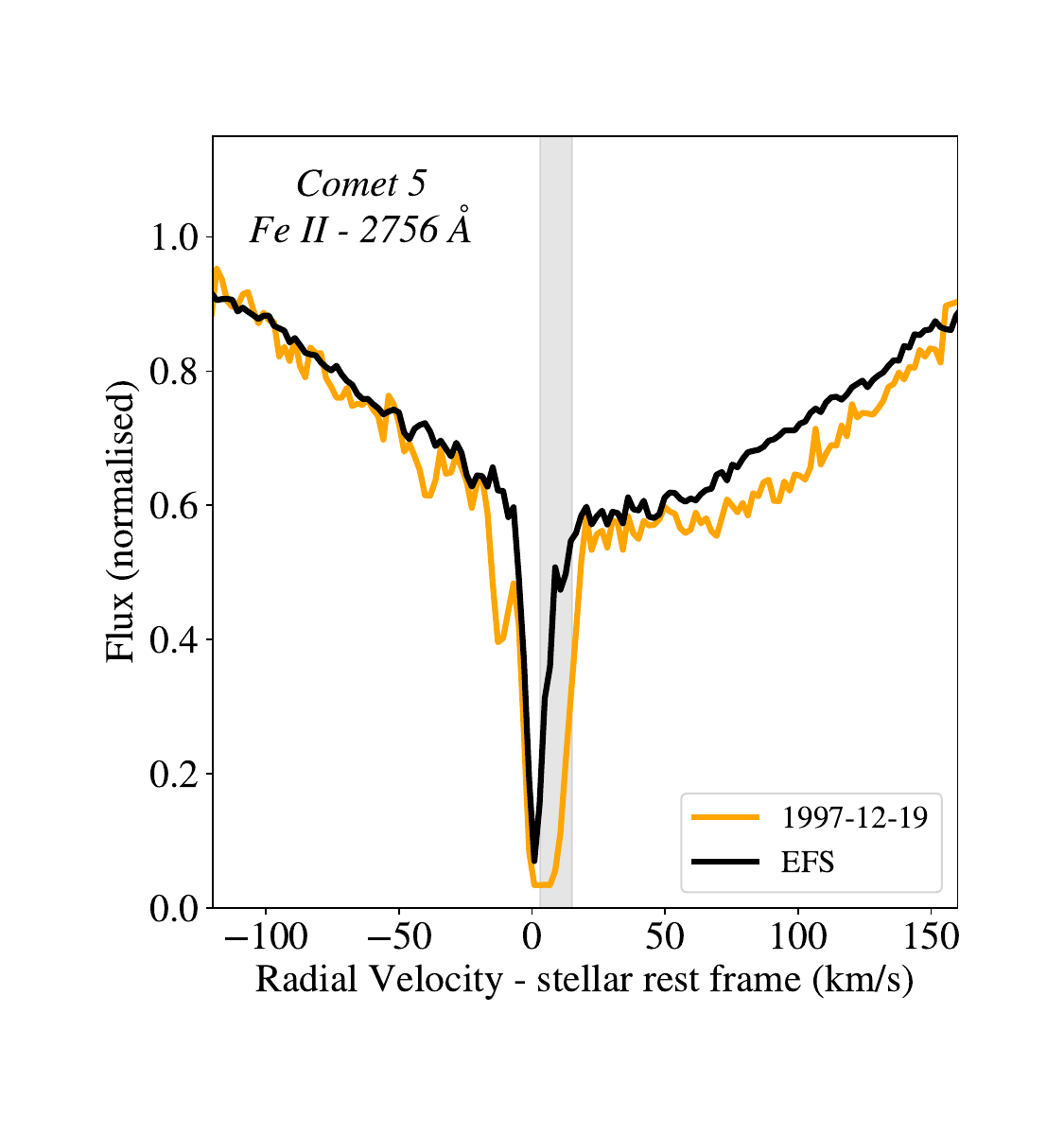}    
    \includegraphics[scale = 0.38,     trim = 85 30 80 40,clip]{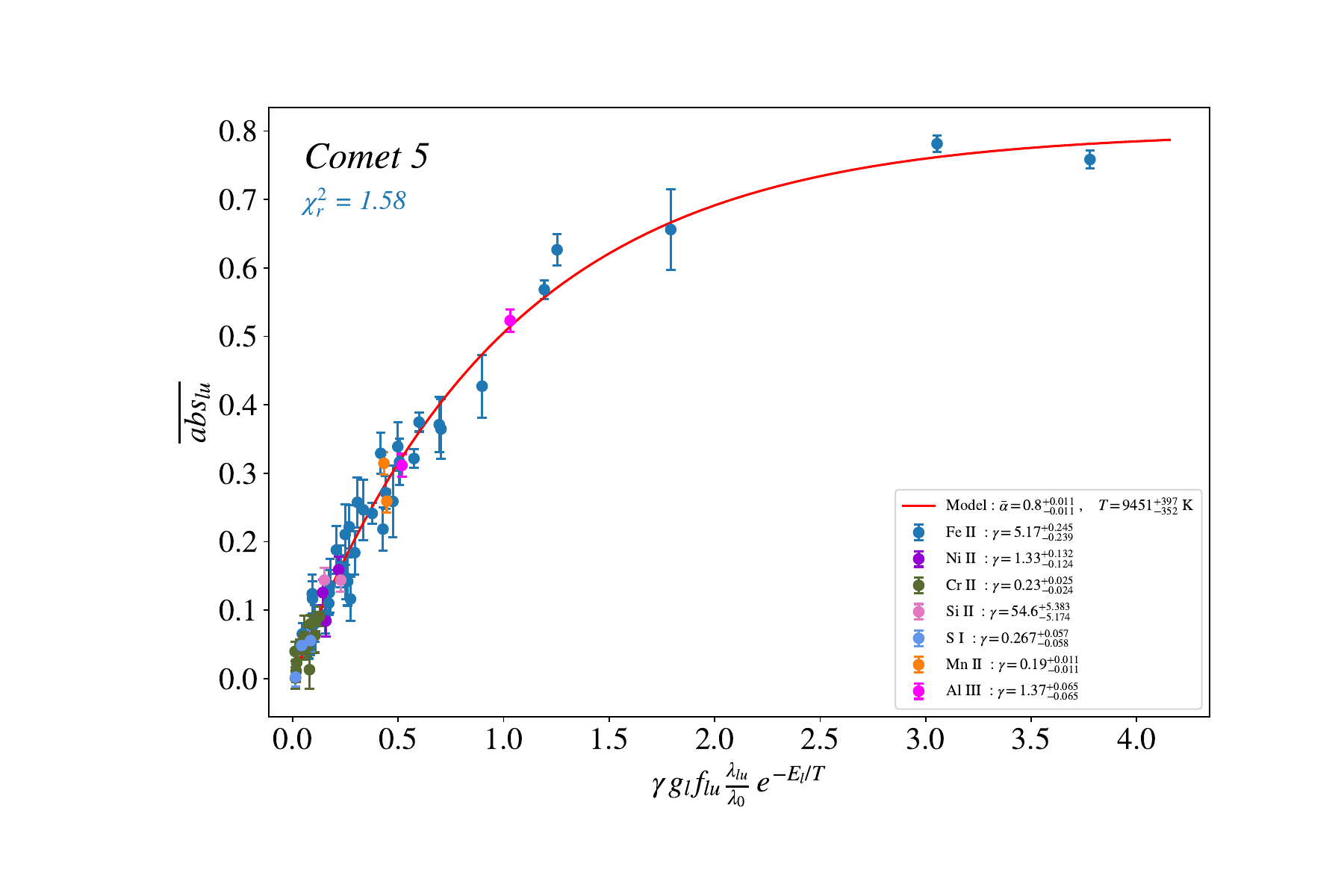}
    
    \includegraphics[scale = 0.38,     trim = 10 30 30 40,clip]{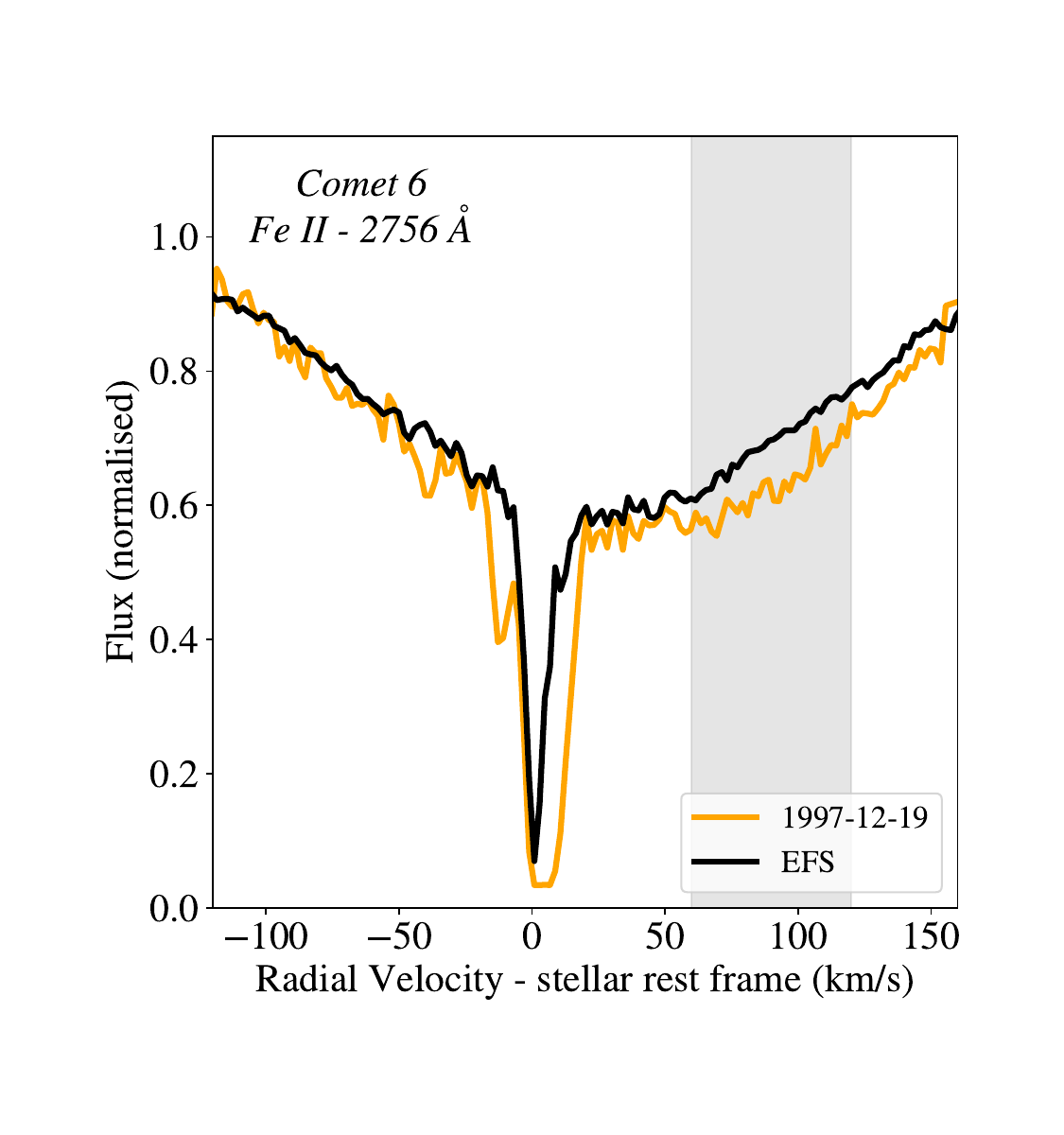}    
    \includegraphics[scale = 0.38,     trim = 85 30 80 40,clip]{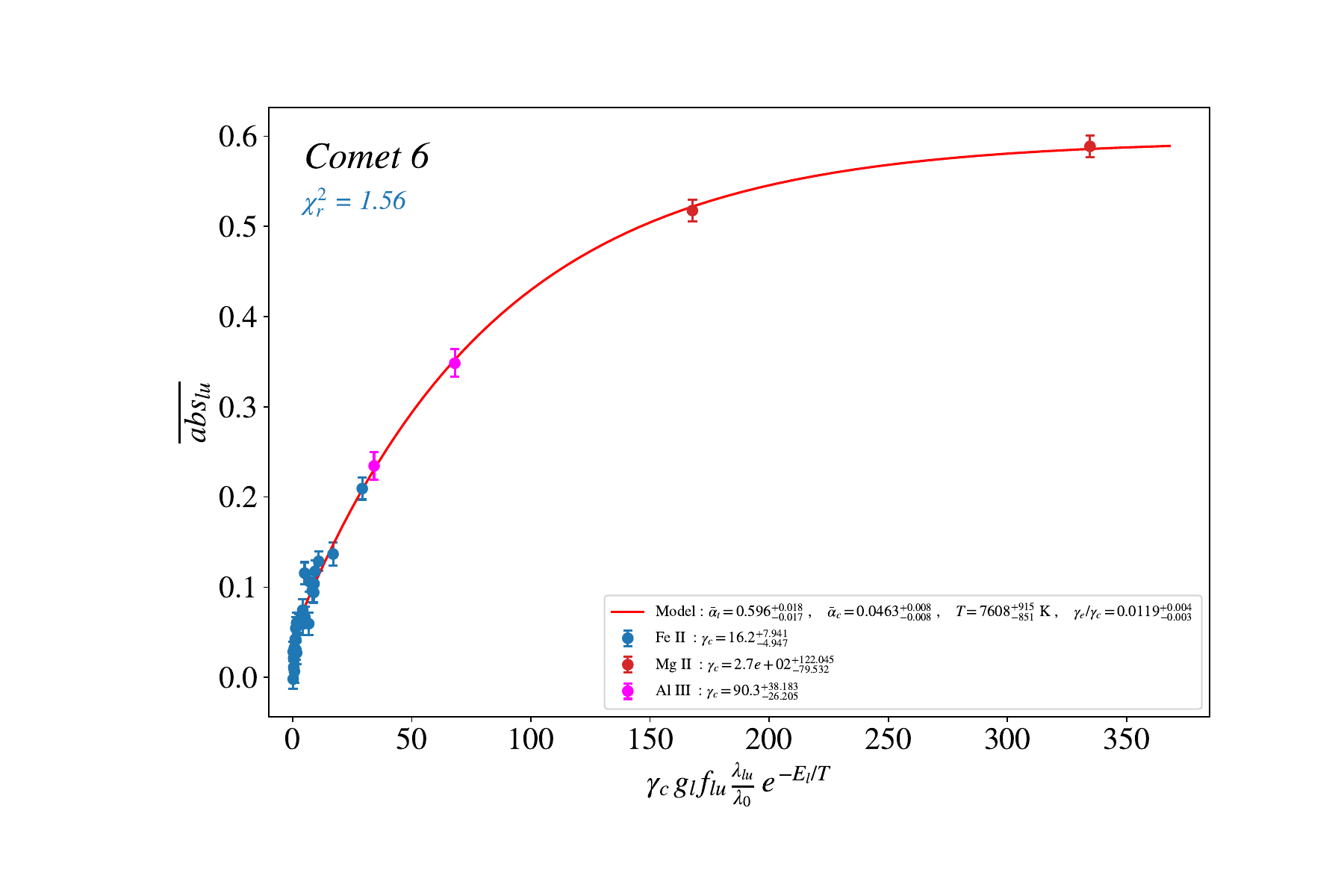}      
\end{figure*}
\small \centering \textbf{Figure \ref{Fig. curves of growth all}}, continued.

\newpage

\begin{figure*}[h!]
\centering
    \includegraphics[scale = 0.38,     trim = 10 30 30 40,clip]{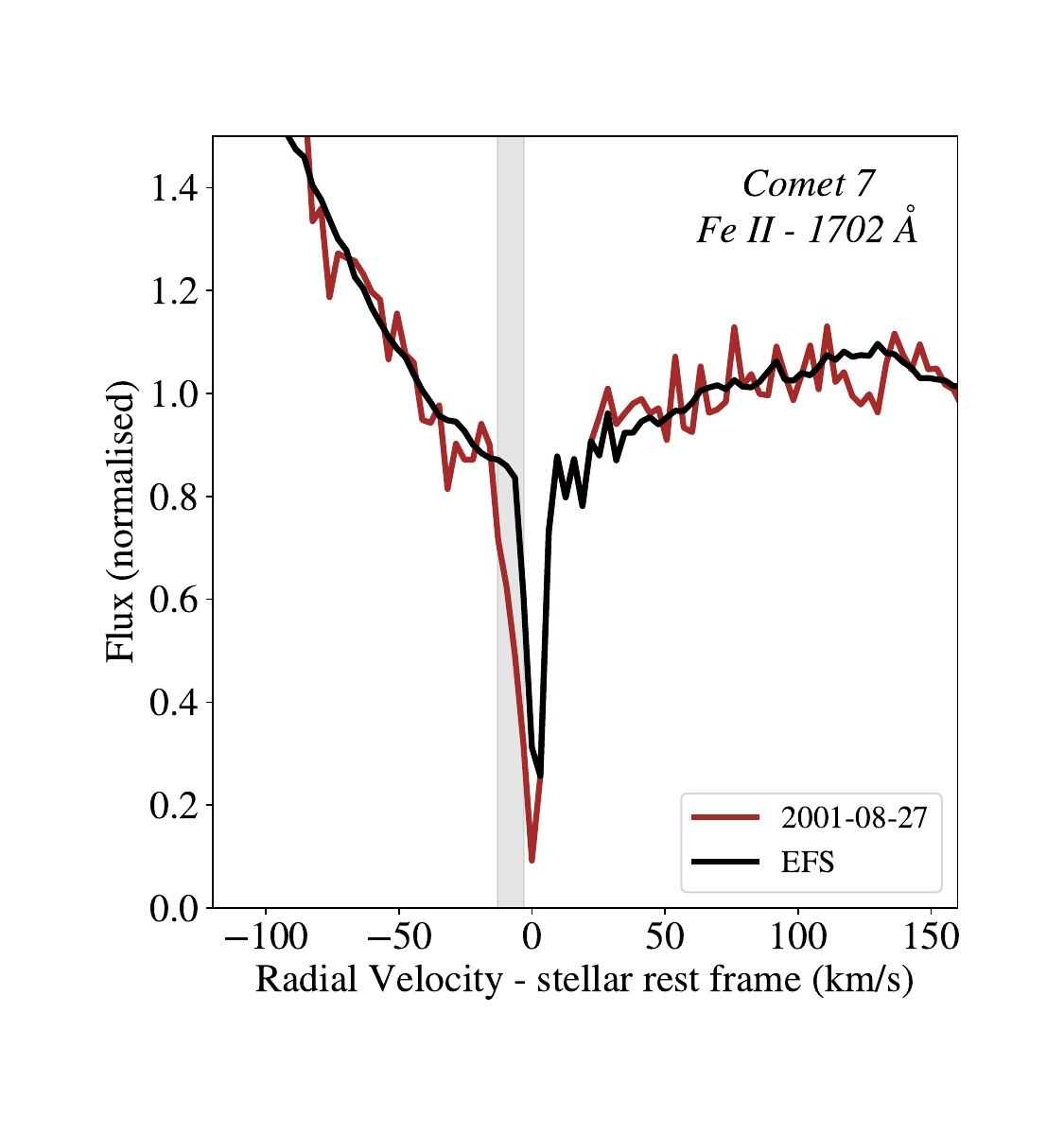}    
    \includegraphics[scale = 0.38,     trim = 85 30 80 40,clip]{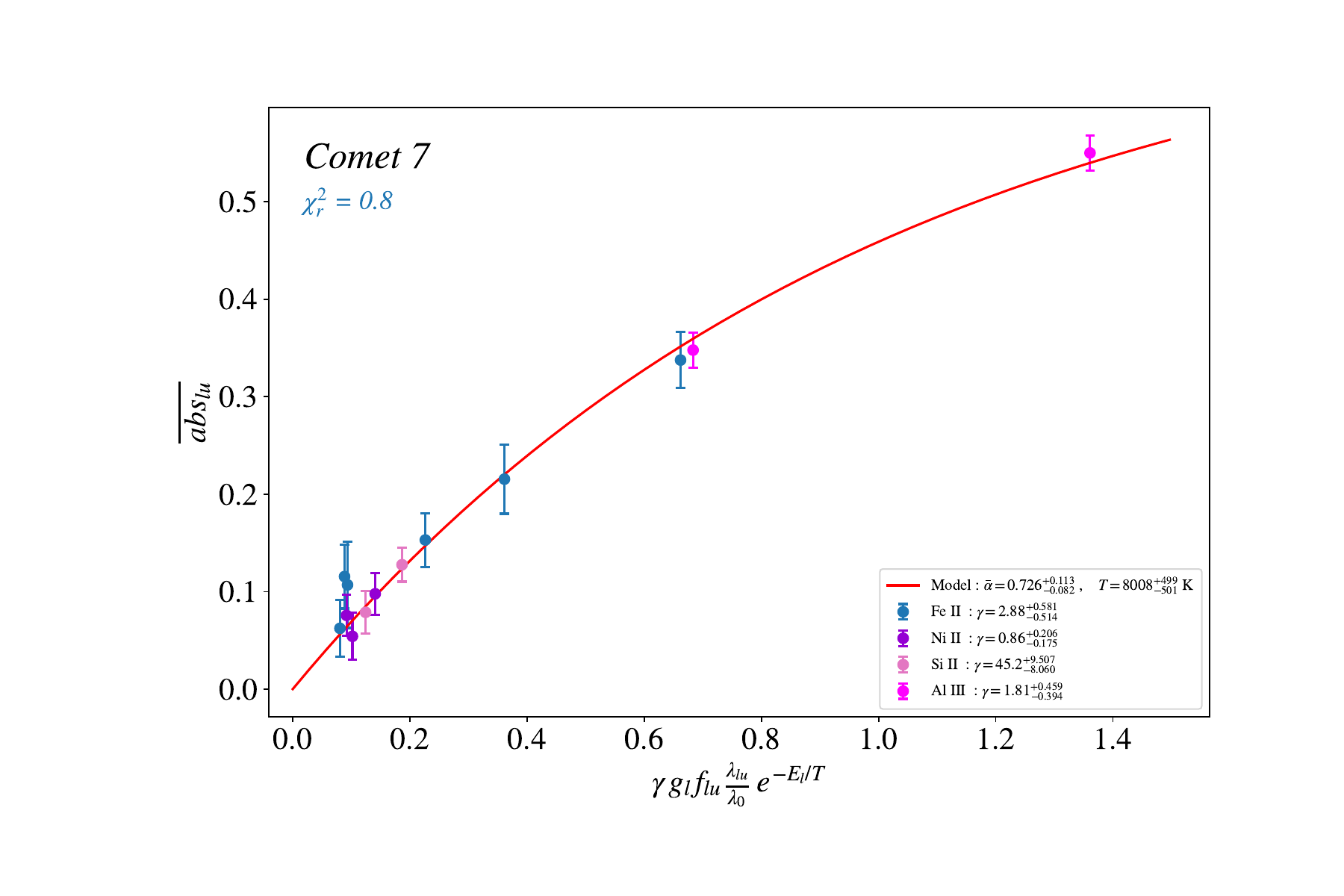}    
    
    \includegraphics[scale = 0.38,     trim = 10 30 30 40,clip]{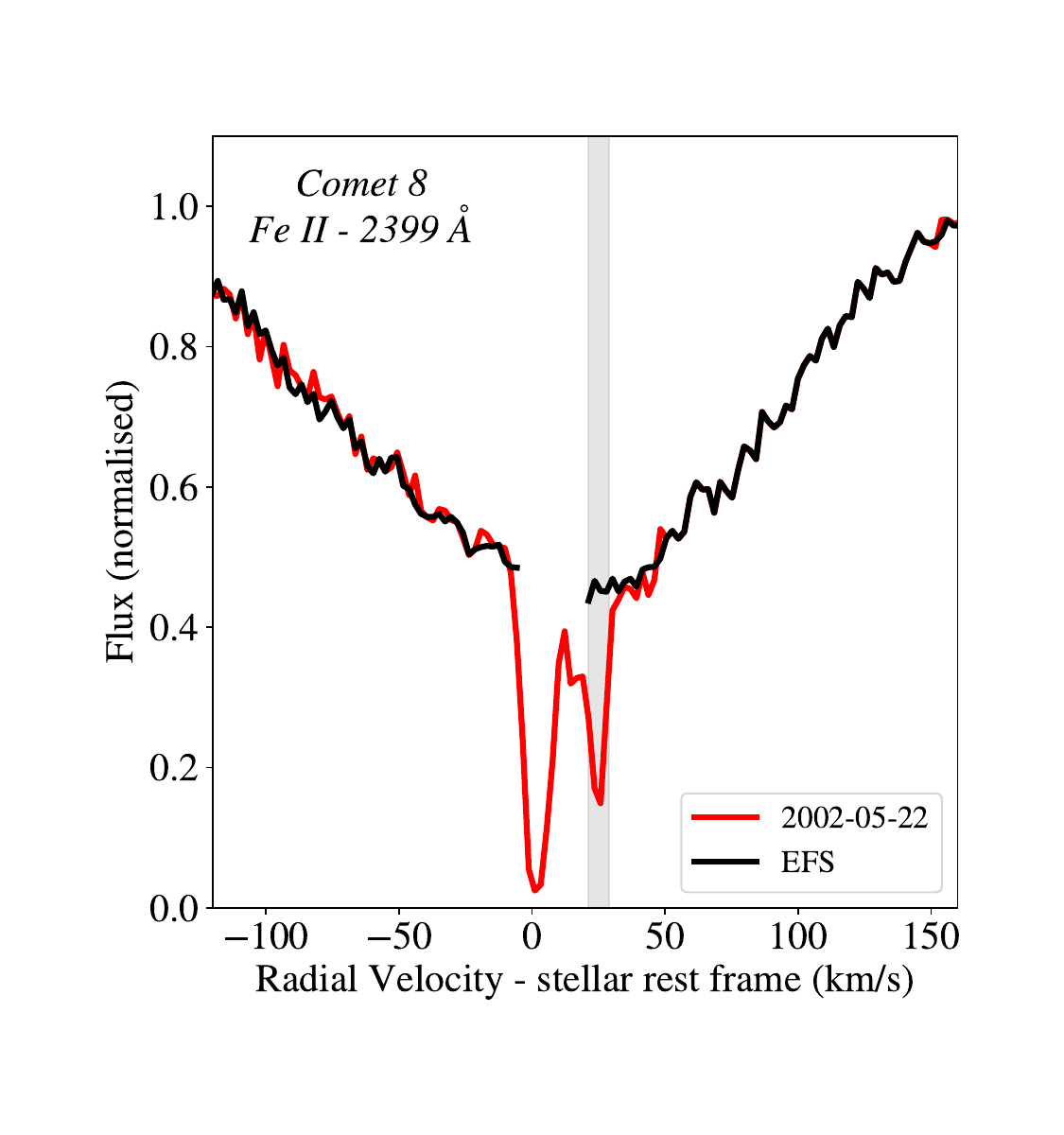}    
    \includegraphics[scale = 0.38,     trim = 85 30 80 40,clip]{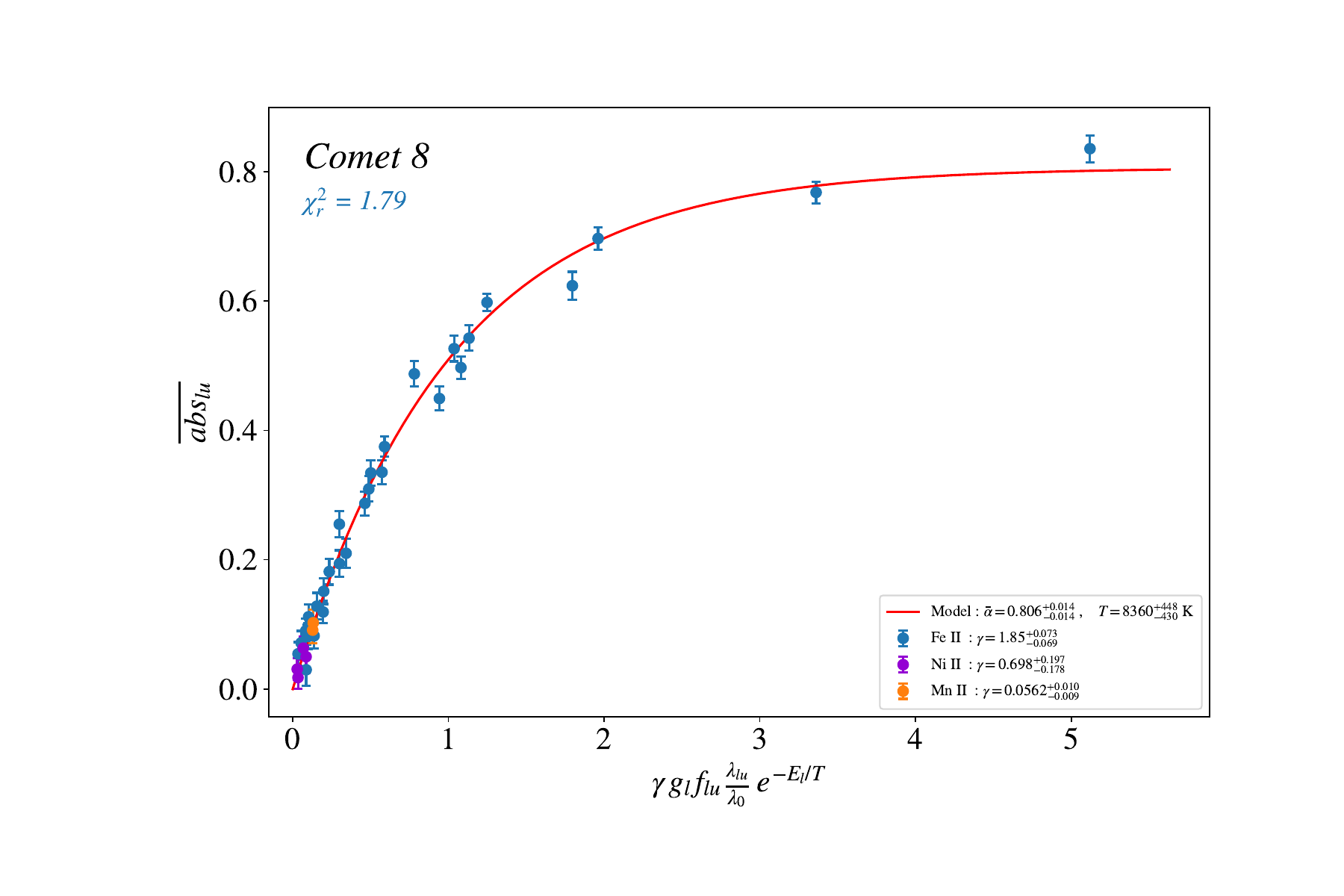}
    
    \includegraphics[scale = 0.38,     trim = 10 30 30 40,clip]{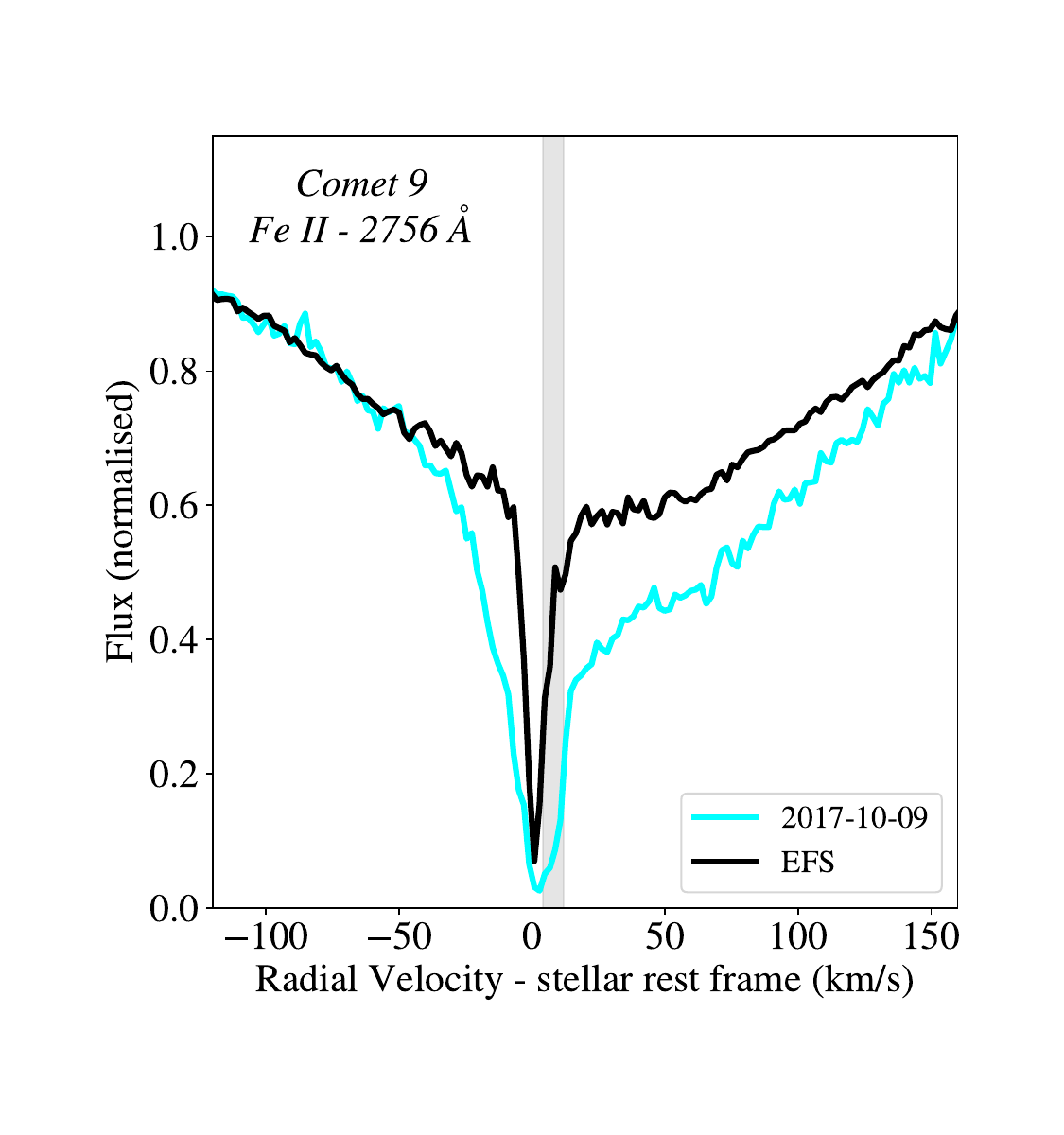}    
    \includegraphics[scale = 0.38,     trim = 85 30 80 40,clip]{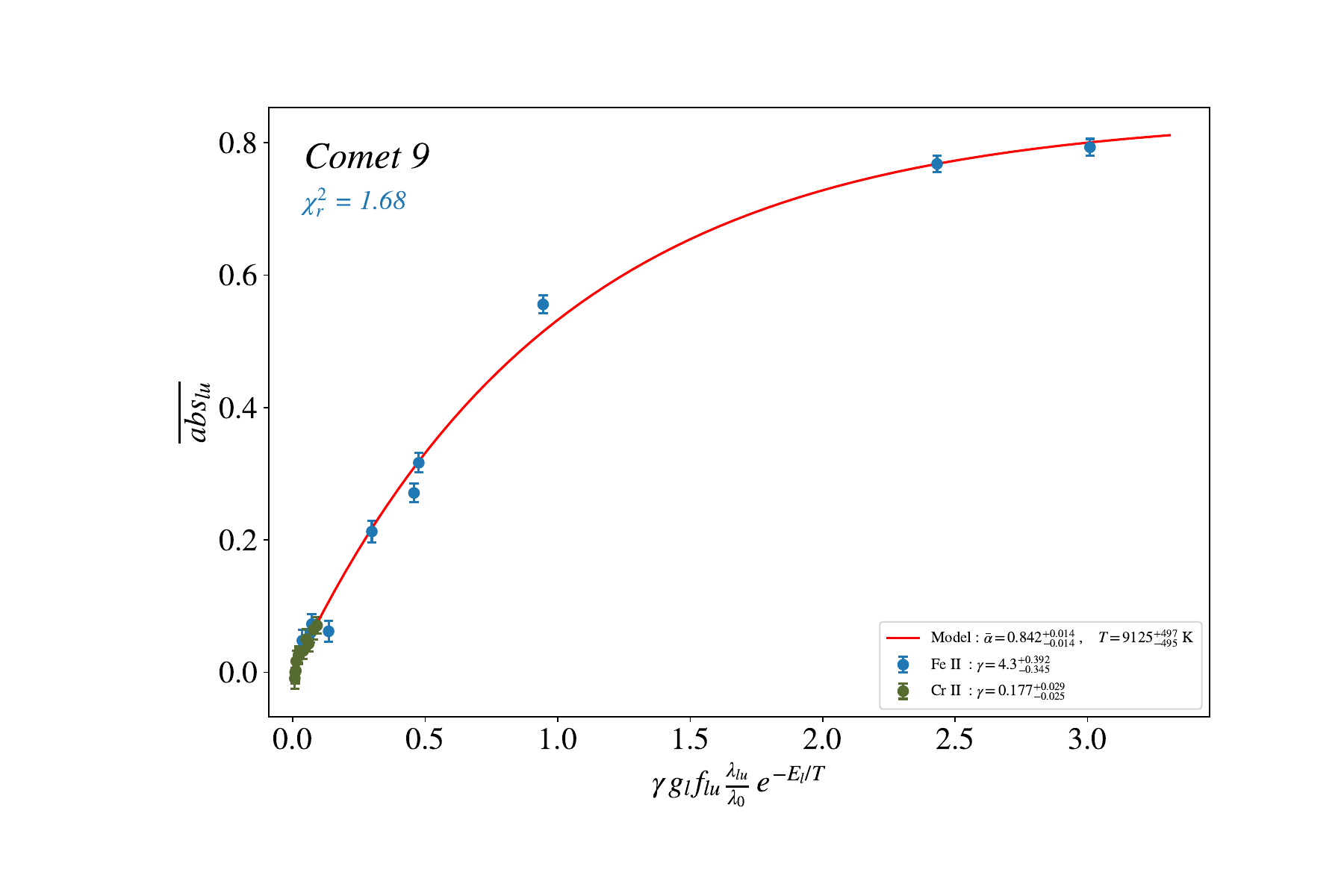}   
\end{figure*}
\small \textbf{Figure \ref{Fig. curves of growth all}}, continued.

\newpage

\begin{figure*}[h!]
\centering
    \includegraphics[scale = 0.38,     trim = 10 30 30 40,clip]{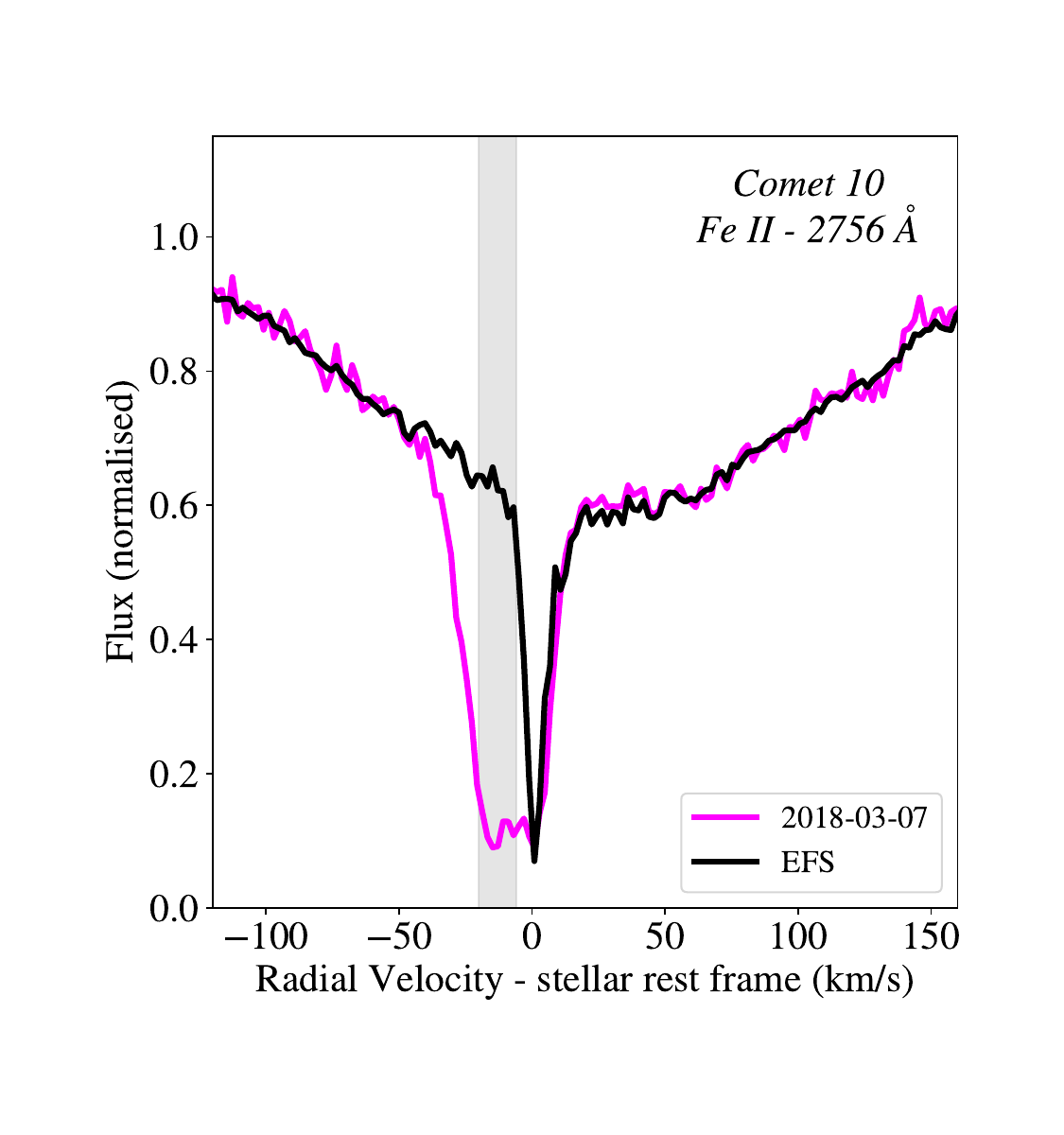}    
    \includegraphics[scale = 0.38,     trim = 85 30 80 40,clip]{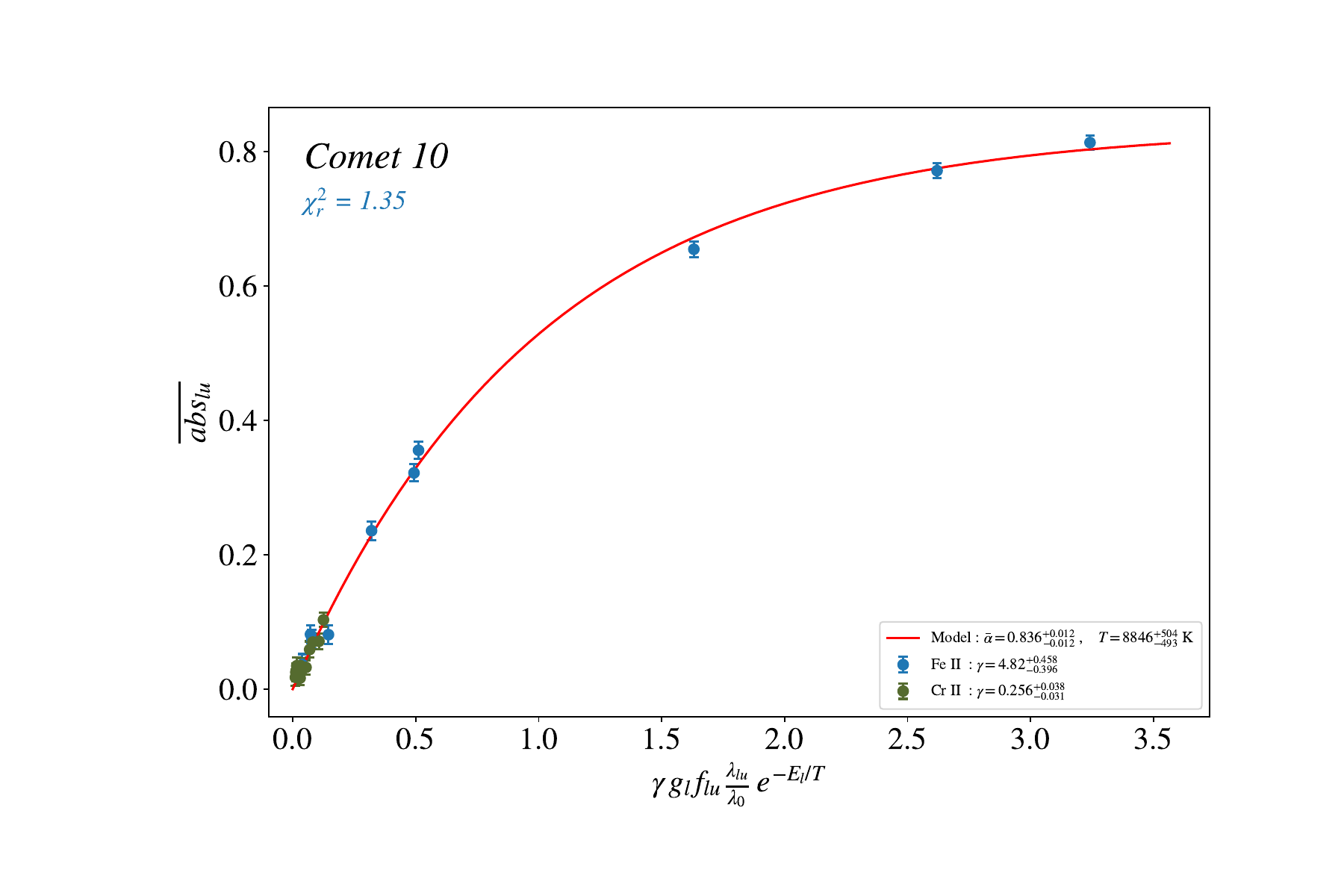}    
    
    \includegraphics[scale = 0.38,     trim = 10 30 30 40,clip]{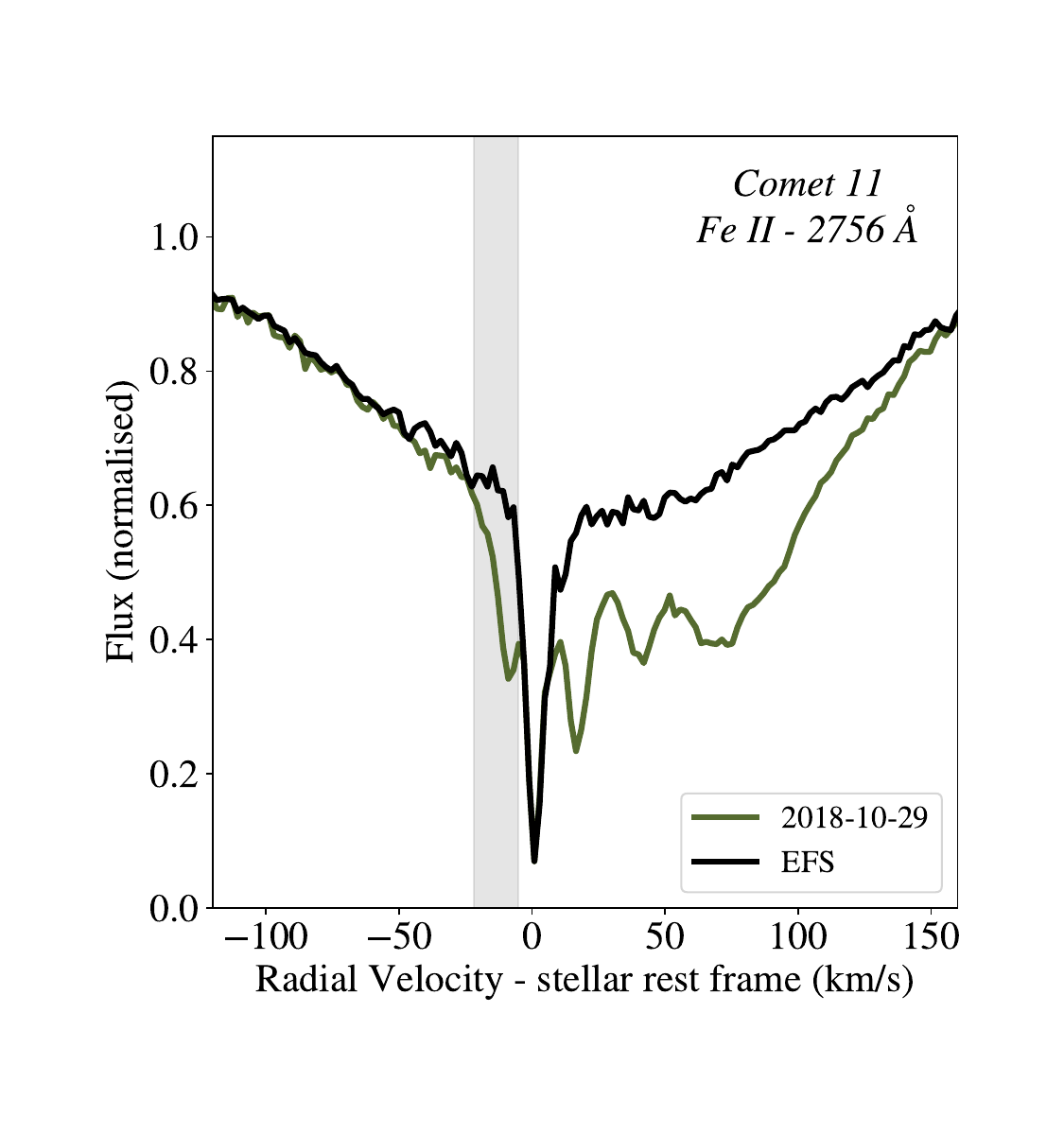}    
    \includegraphics[scale = 0.38,     trim = 85 30 80 40,clip]{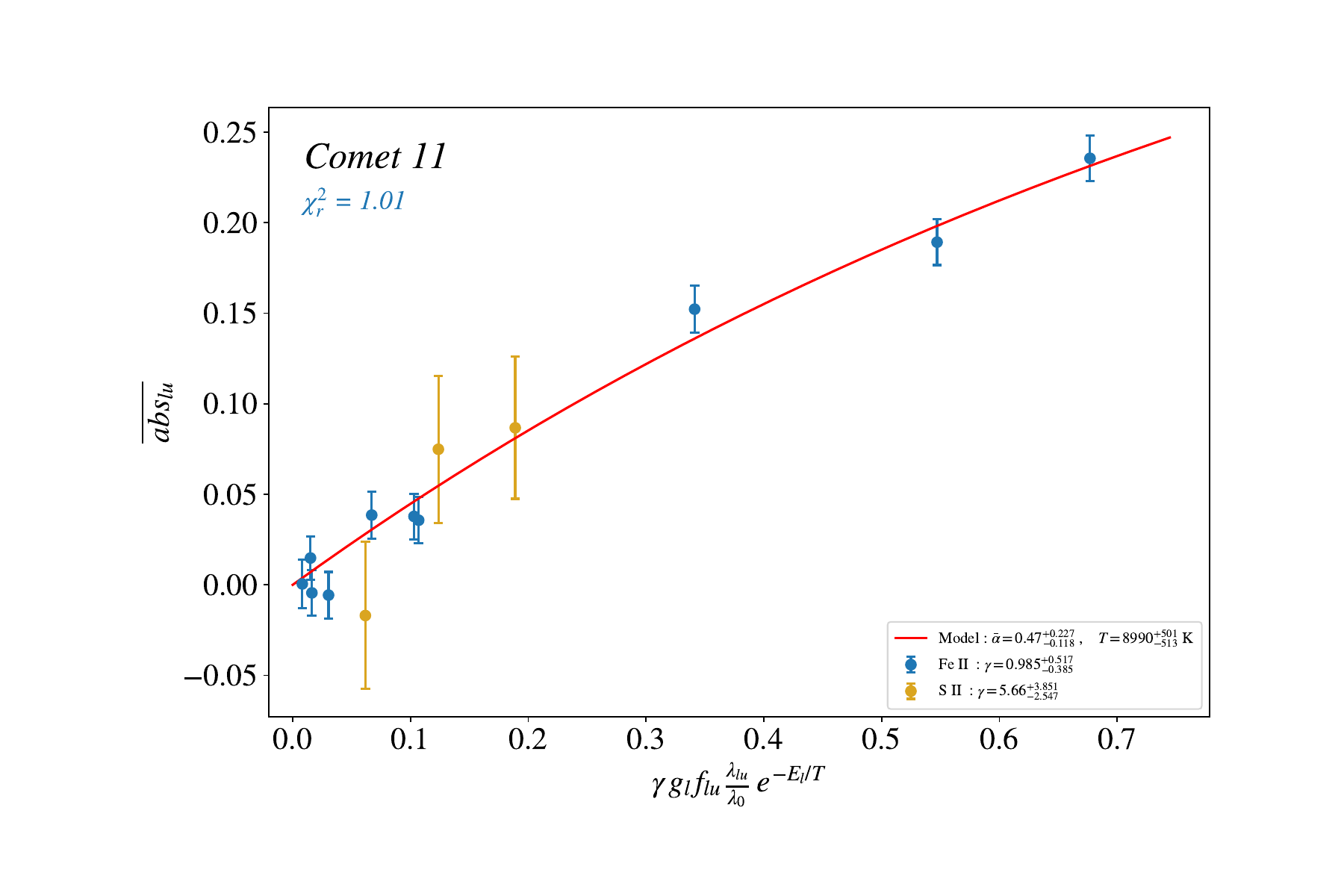}
    
    \includegraphics[scale = 0.38,     trim = 10 30 30 40,clip]{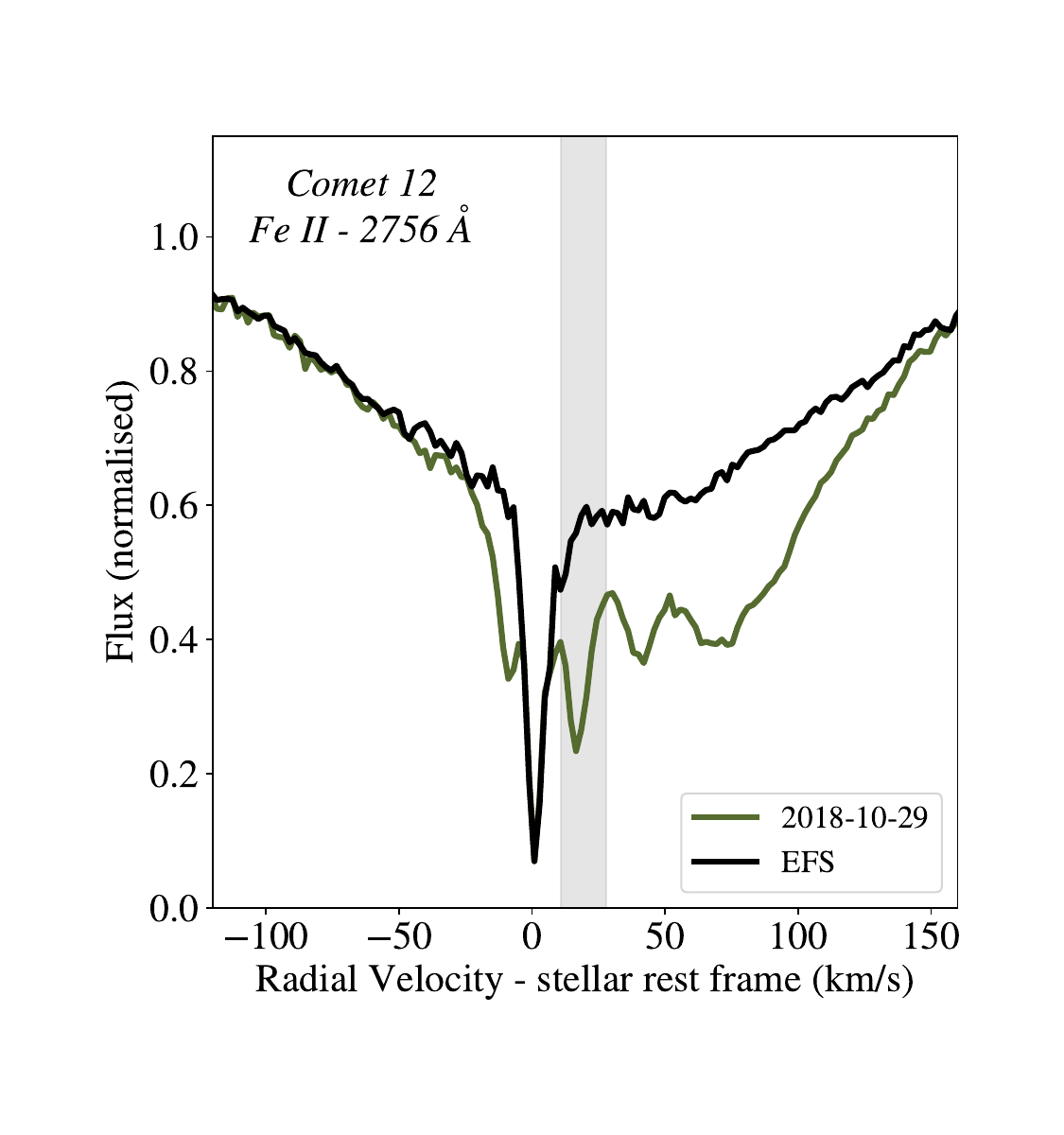}    
    \includegraphics[scale = 0.38,     trim = 85 30 80 40,clip]{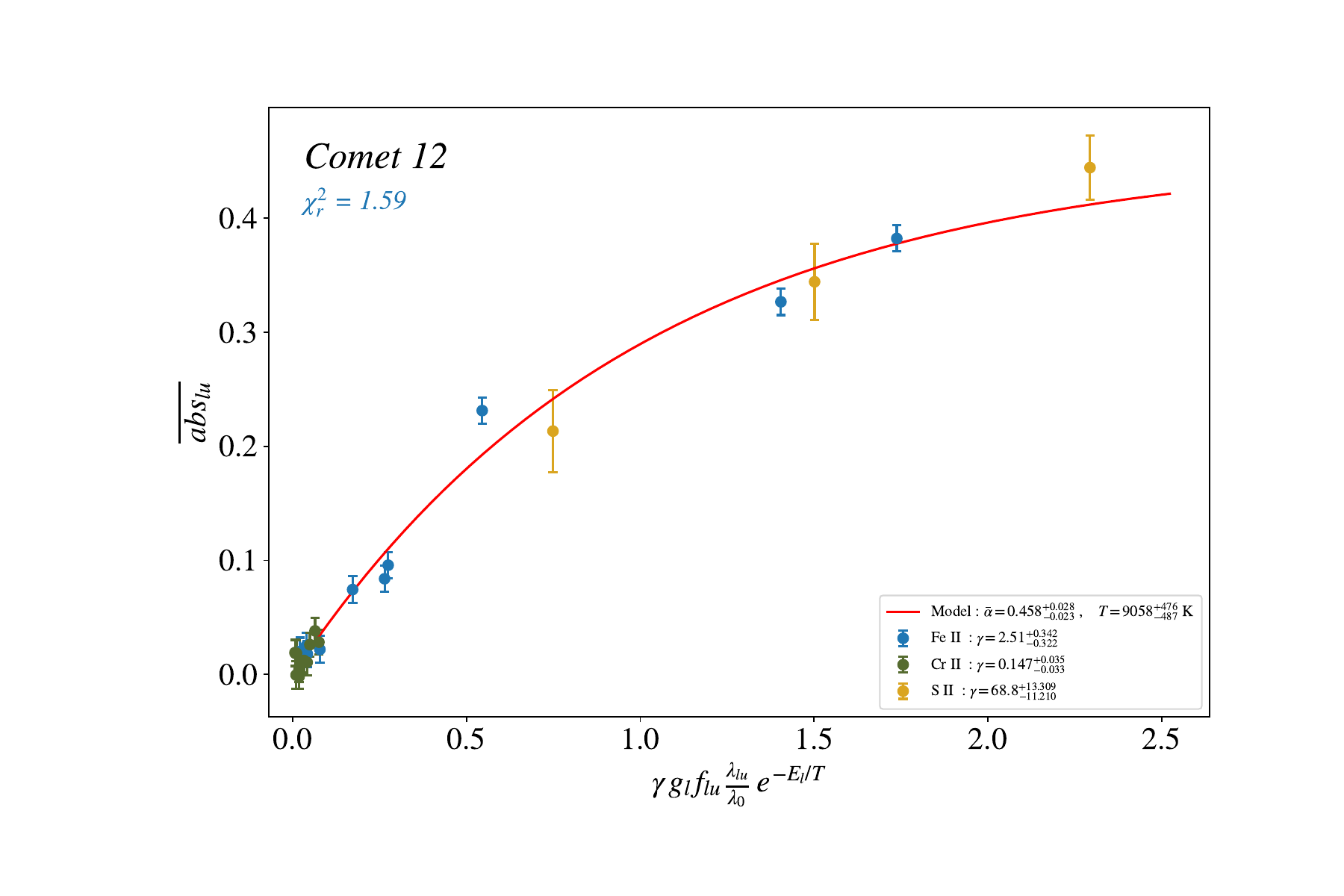}    
\end{figure*}
\small \textbf{Figure \ref{Fig. curves of growth all}}, continued.

\newpage

\begin{figure*}[h!]
\centering
    \includegraphics[scale = 0.38,     trim = 10 30 30 40,clip]{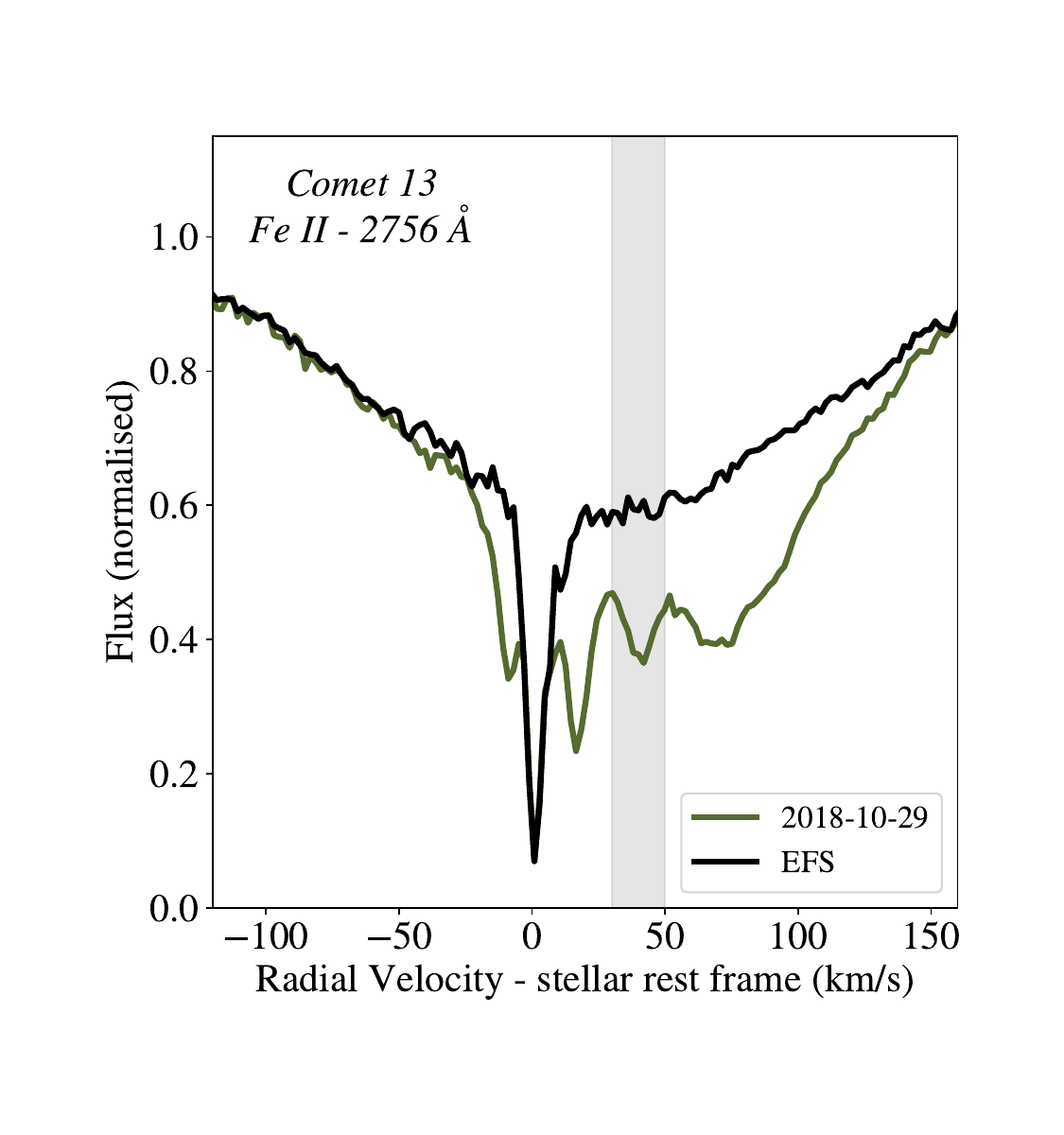}    
    \includegraphics[scale = 0.38,     trim = 85 30 80 40,clip]{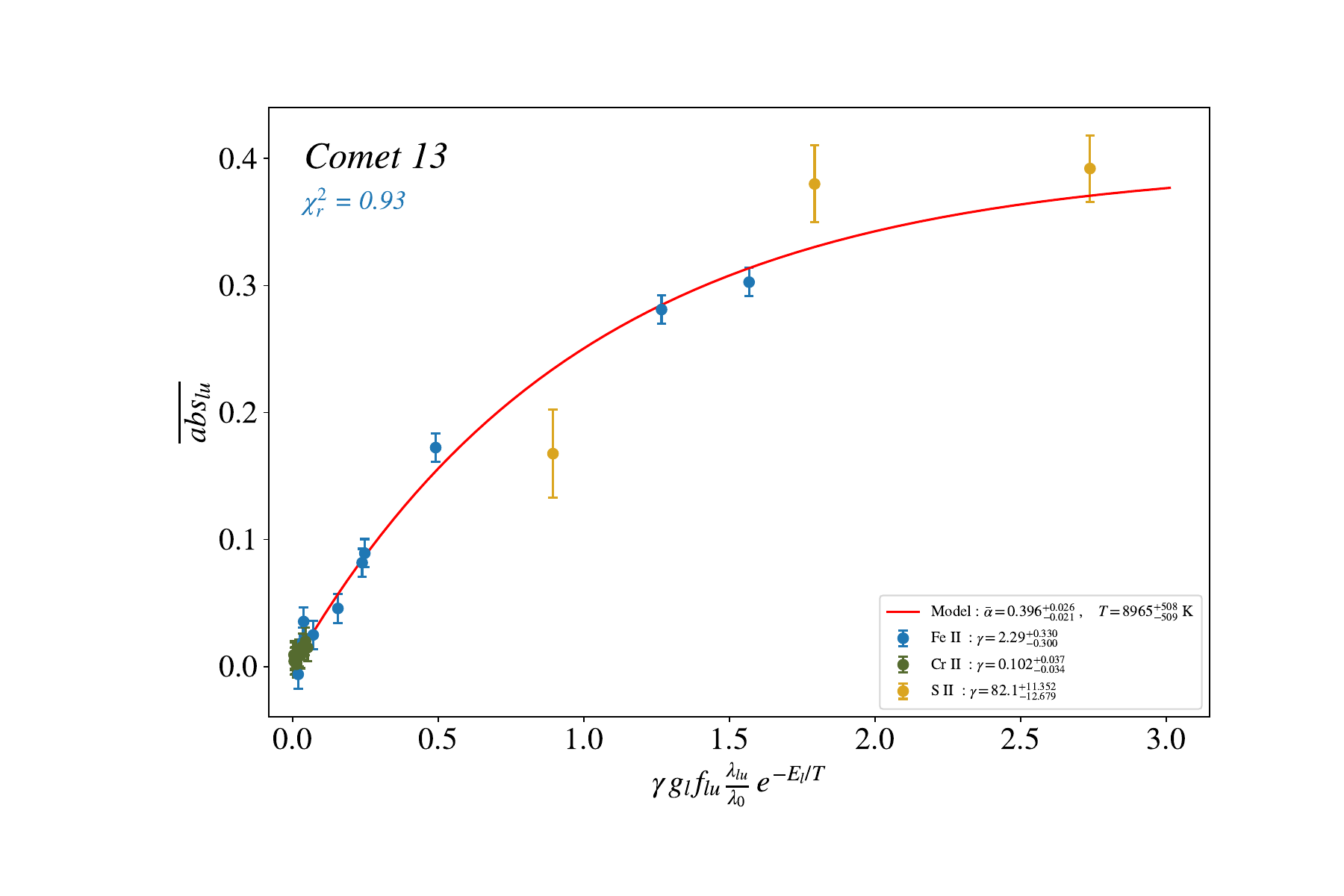}    
    
    \includegraphics[scale = 0.38,     trim = 10 30 30 40,clip]{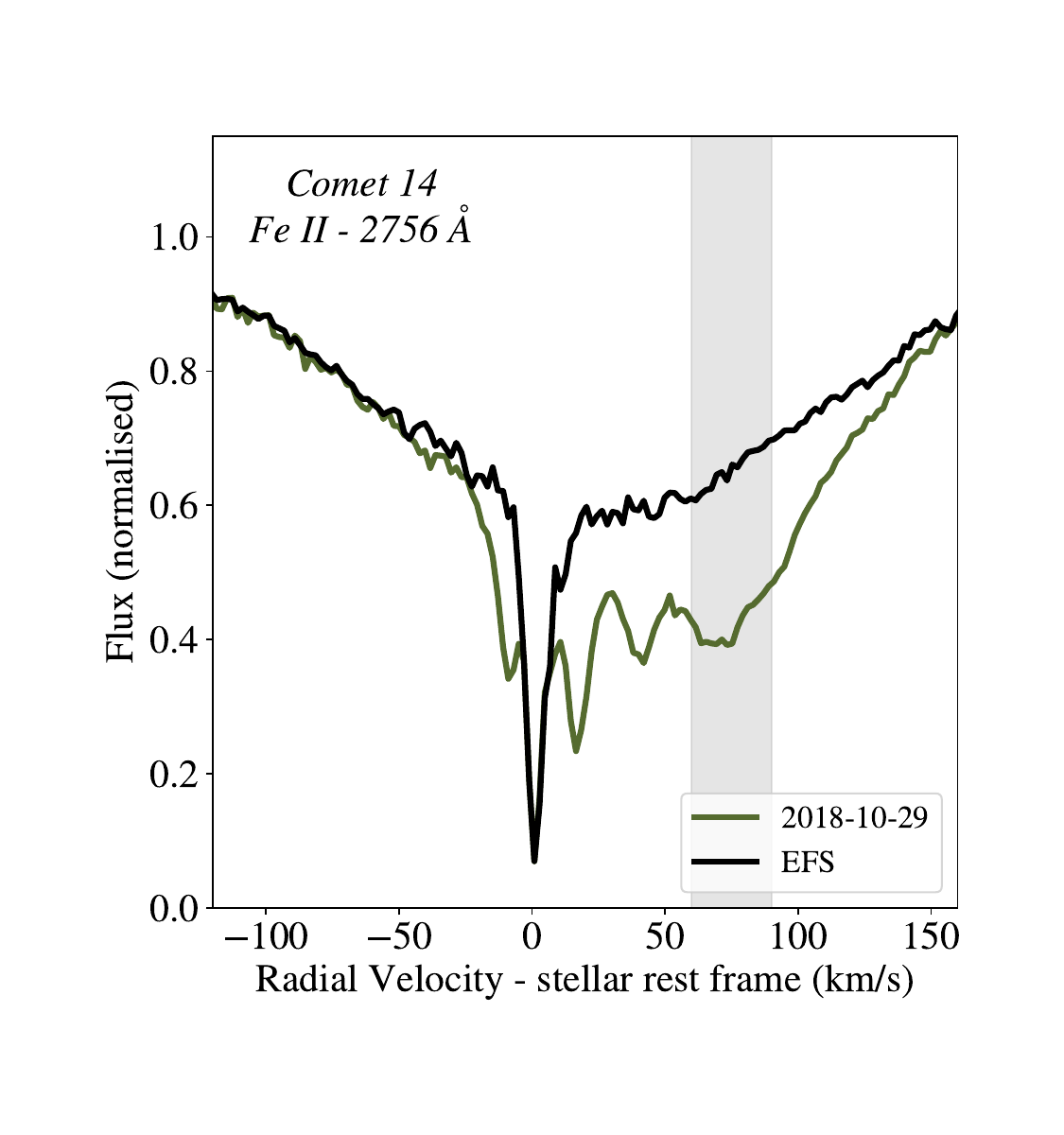}    
    \includegraphics[scale = 0.38,     trim = 85 30 80 40,clip]{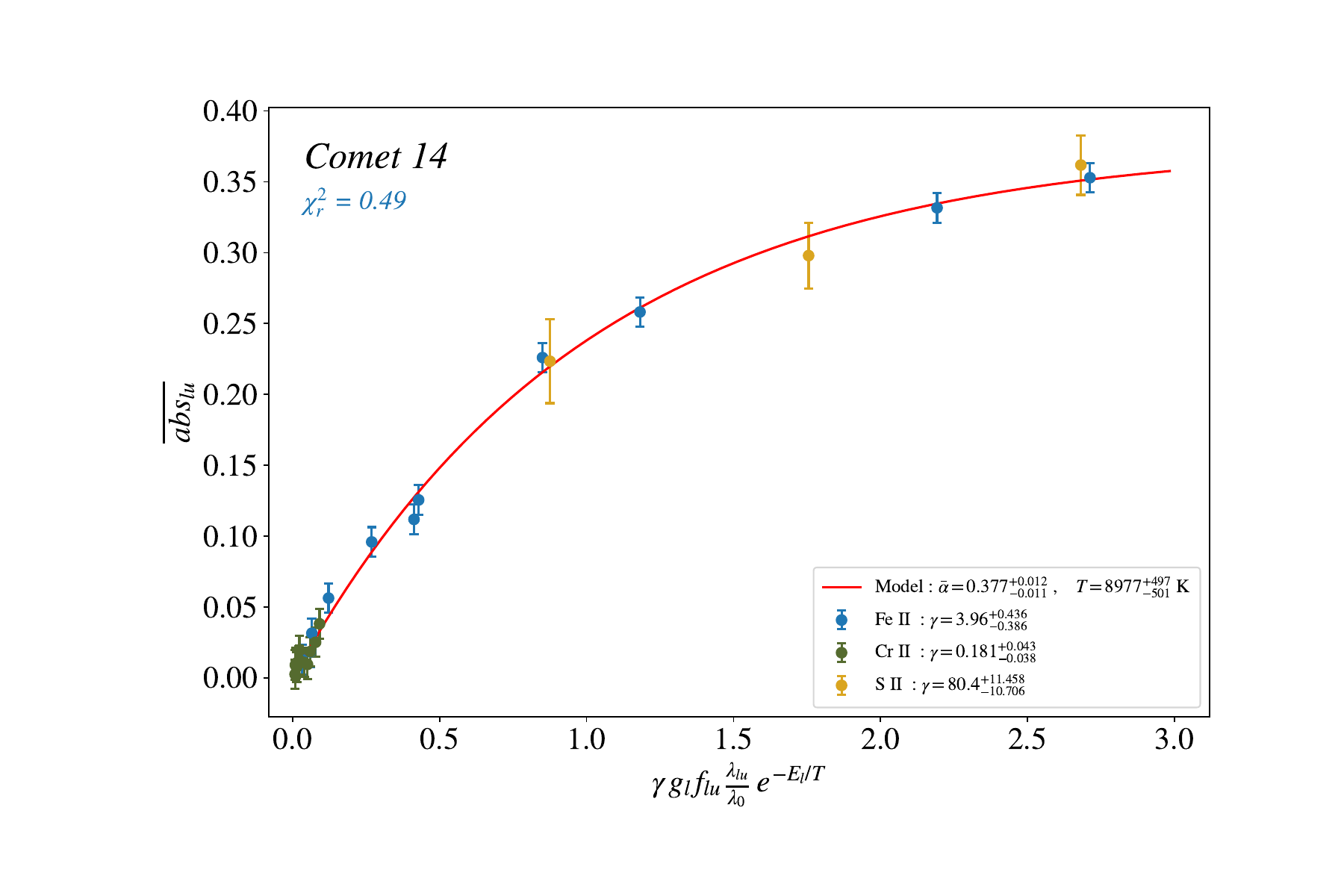}
    
    \includegraphics[scale = 0.38,     trim = 10 30 30 40,clip]{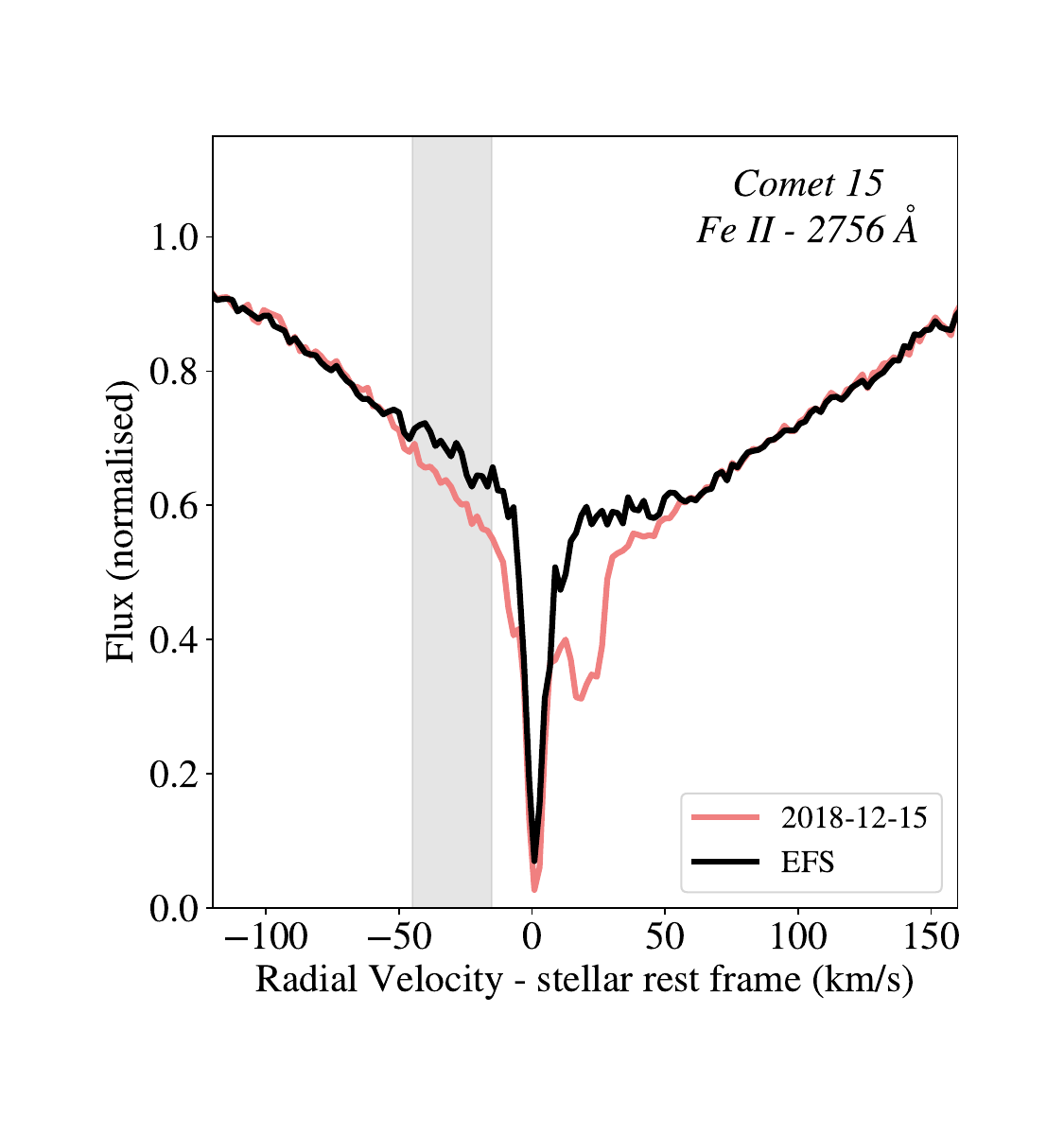}    
    \includegraphics[scale = 0.38,     trim = 85 30 80 40,clip]{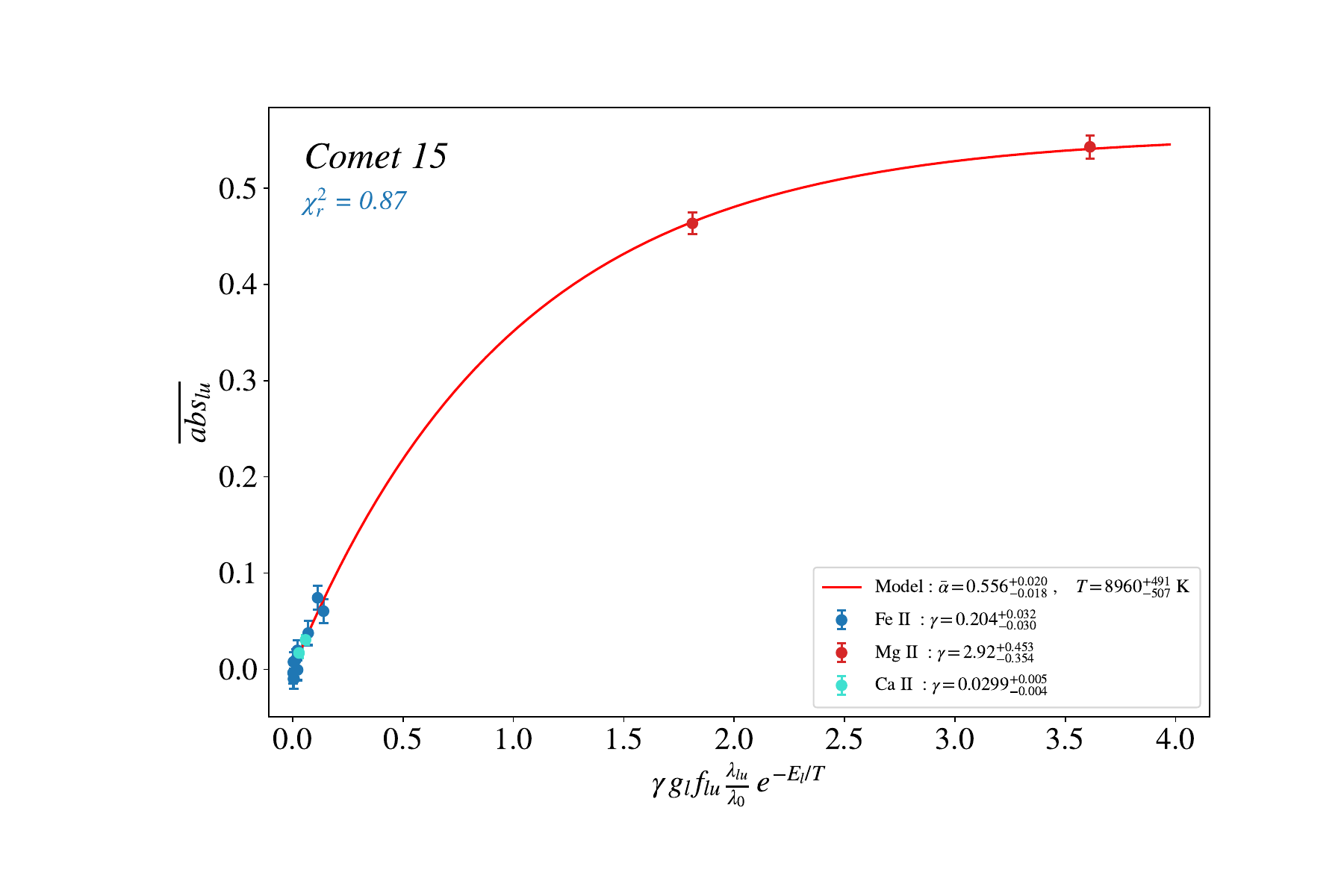}   
\end{figure*}
\small \textbf{Figure \ref{Fig. curves of growth all}}, continued.

\newpage

\begin{figure*}[h!]
\centering
    \includegraphics[scale = 0.38,     trim = 10 30 30 40,clip]{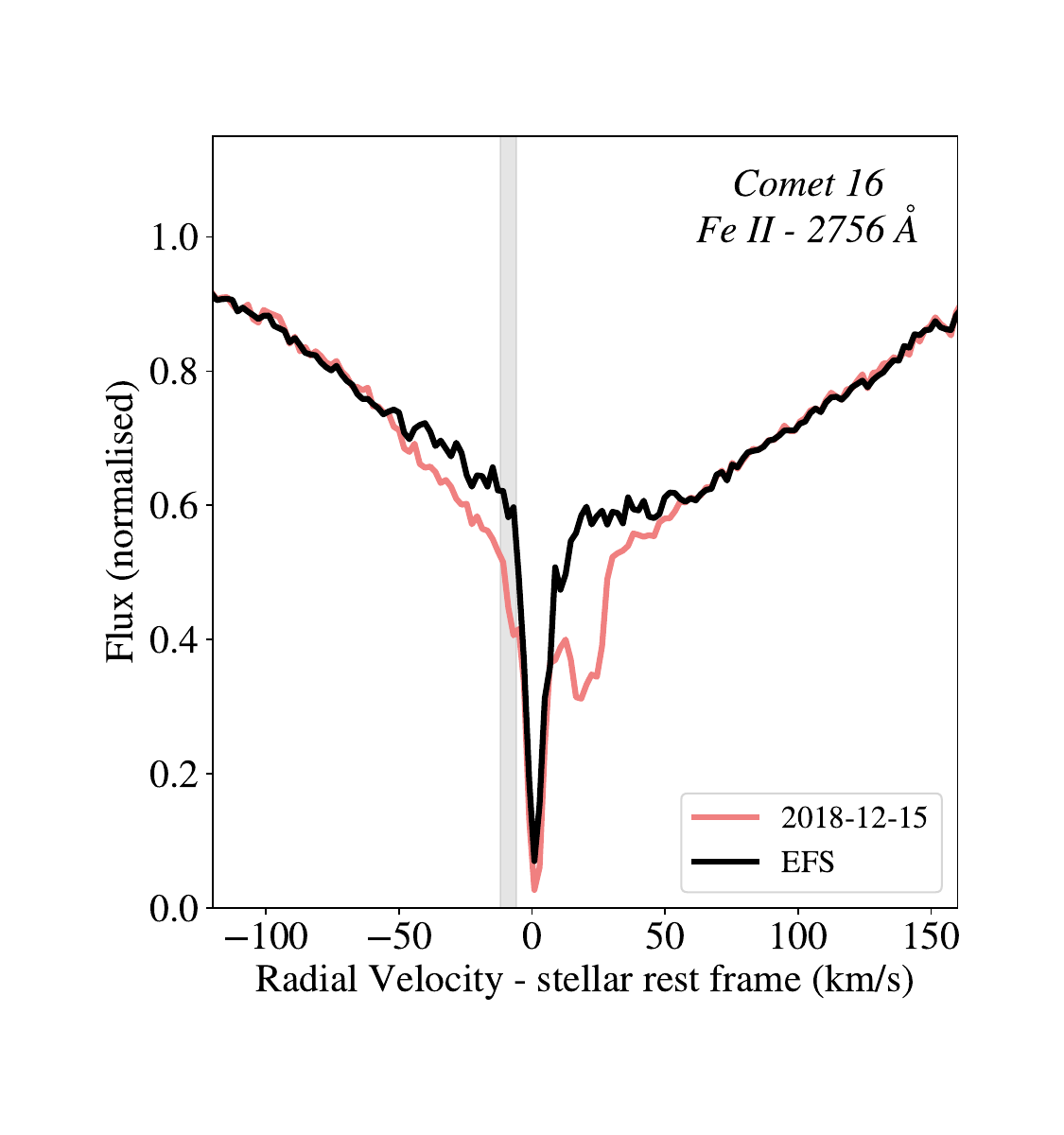}    
    \includegraphics[scale = 0.38,     trim = 85 30 80 40,clip]{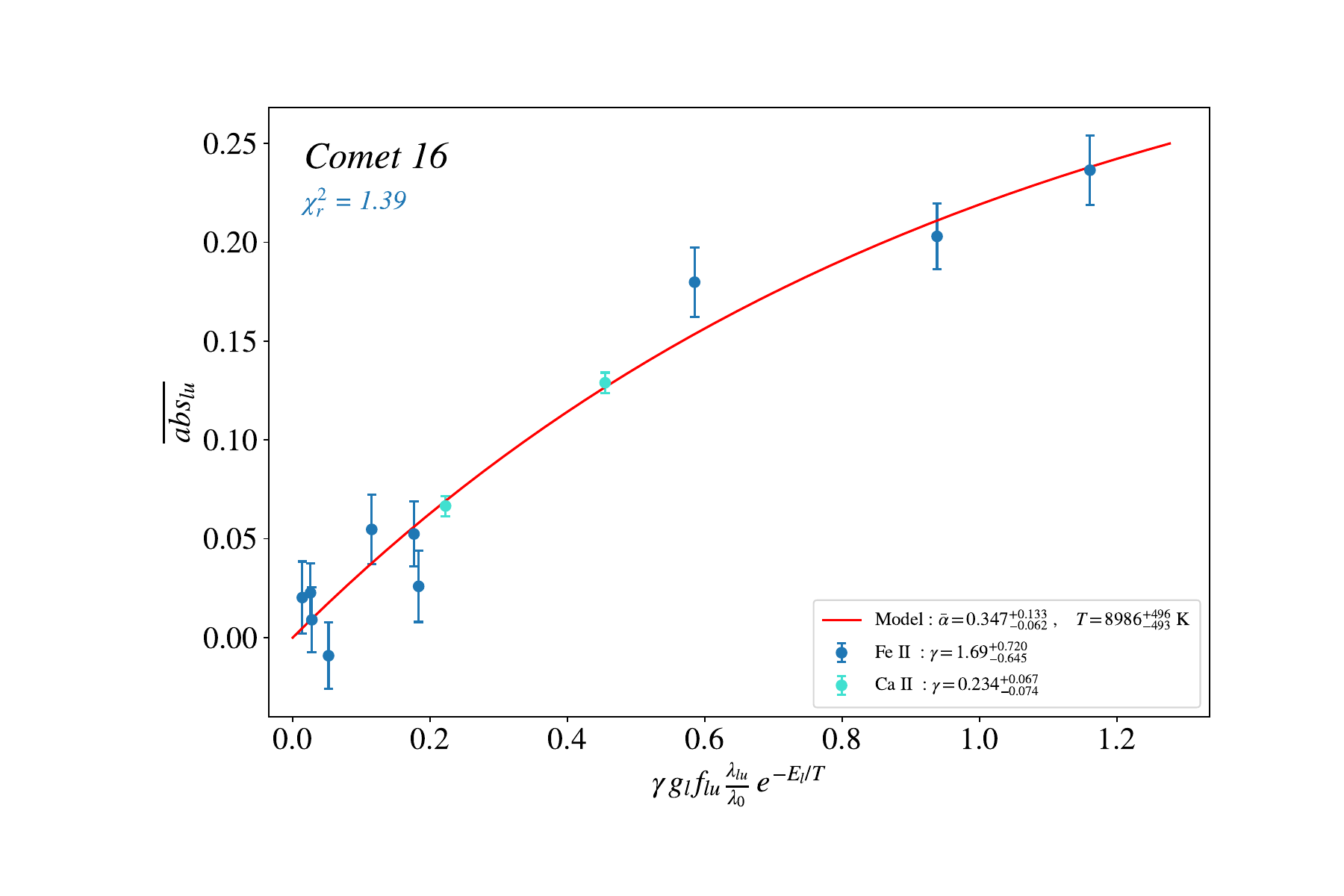}    
    
    \includegraphics[scale = 0.38,     trim = 10 30 30 40,clip]{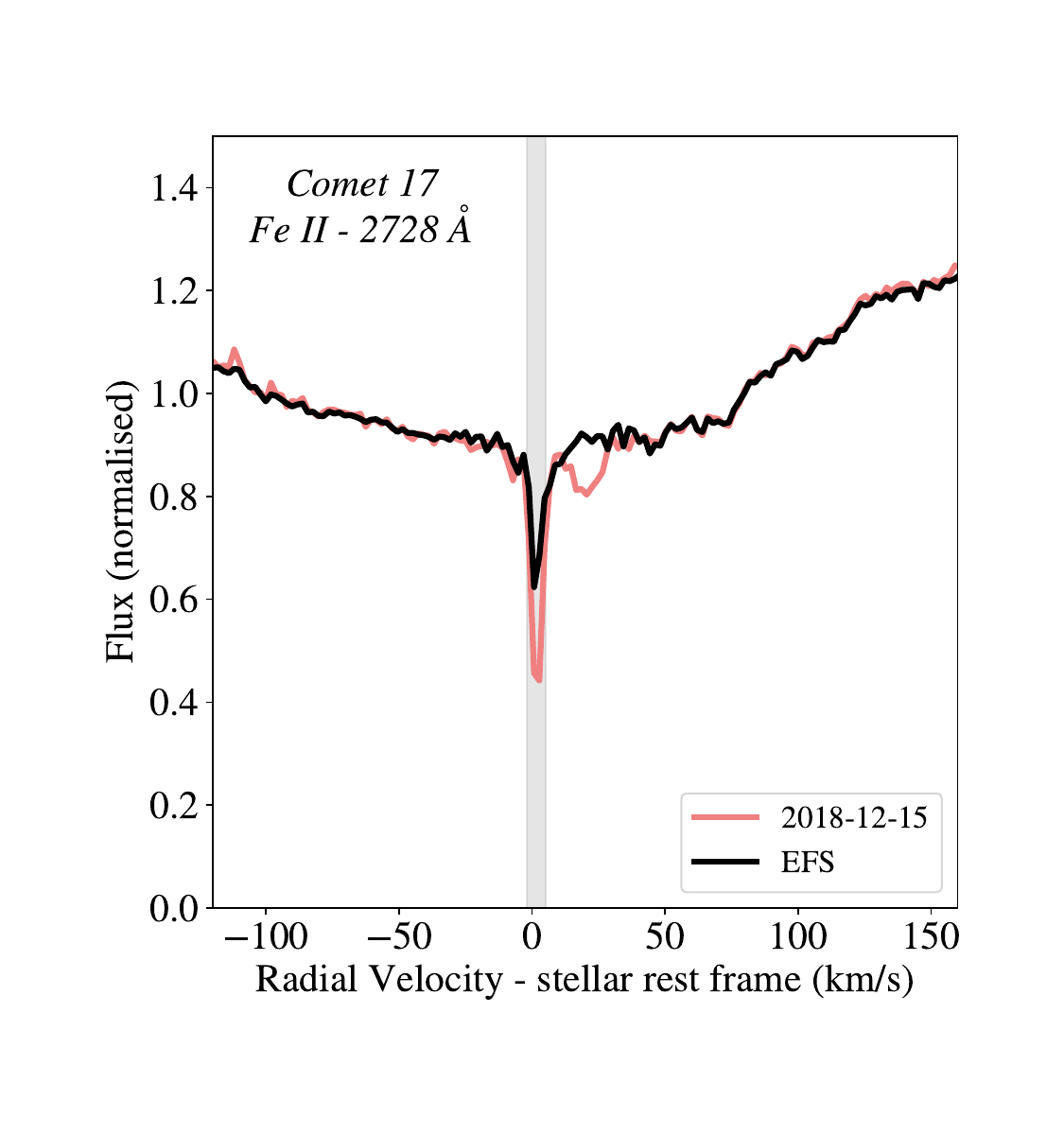}    
    \includegraphics[scale = 0.38,     trim = 85 30 80 40,clip]{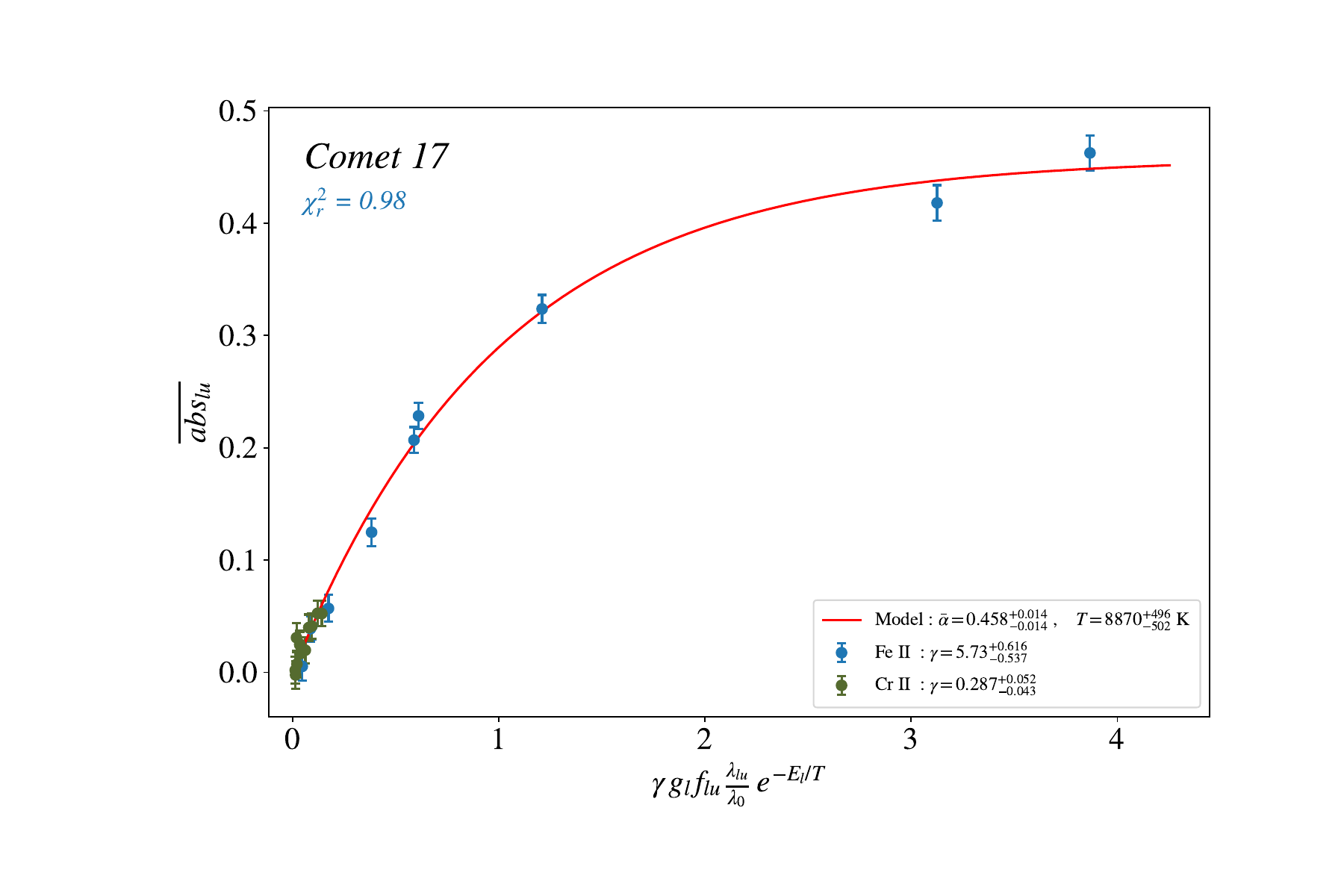}
    
    \includegraphics[scale = 0.38,     trim = 10 30 30 40,clip]{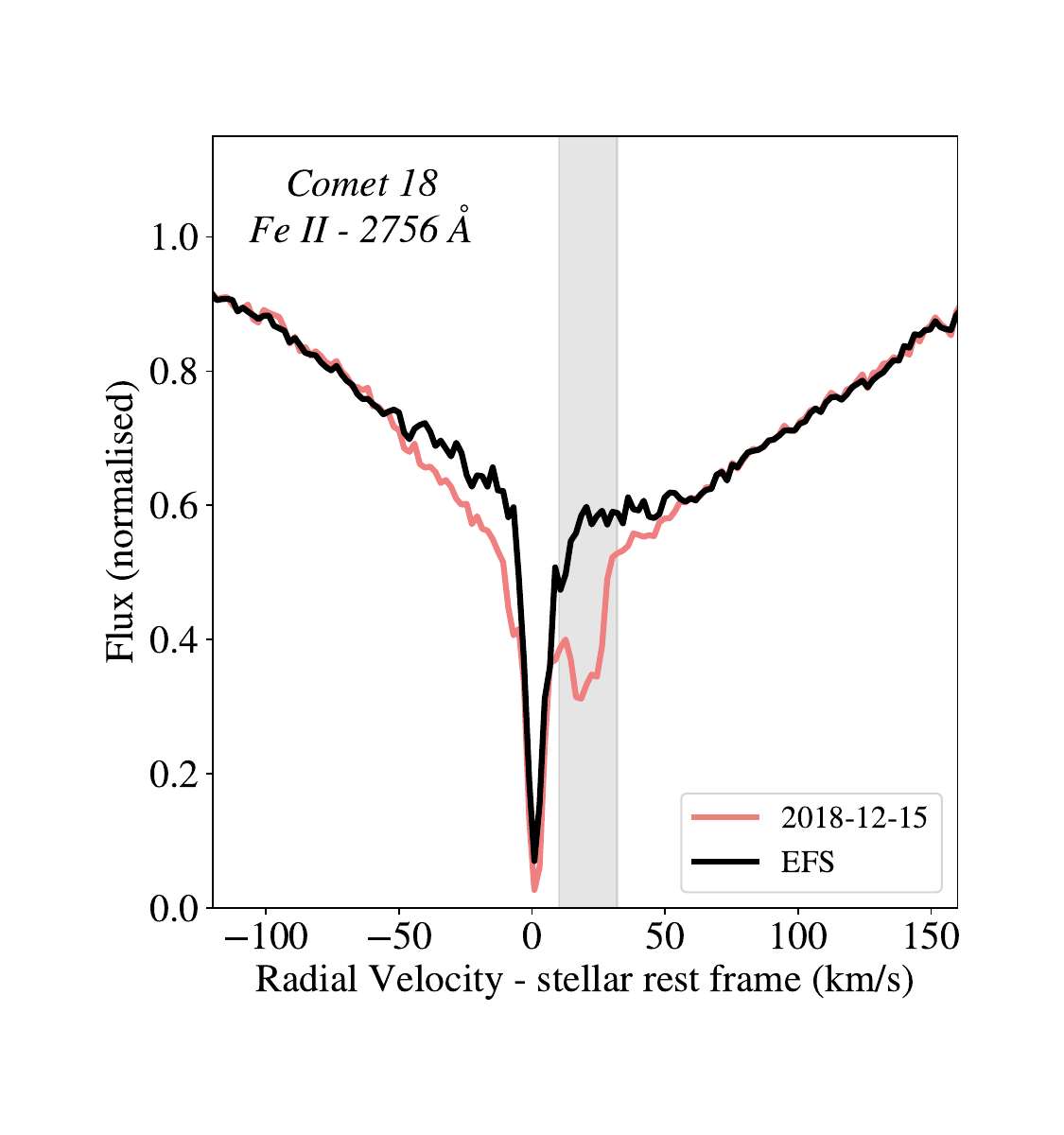}    
    \includegraphics[scale = 0.38,     trim = 85 30 80 40,clip]{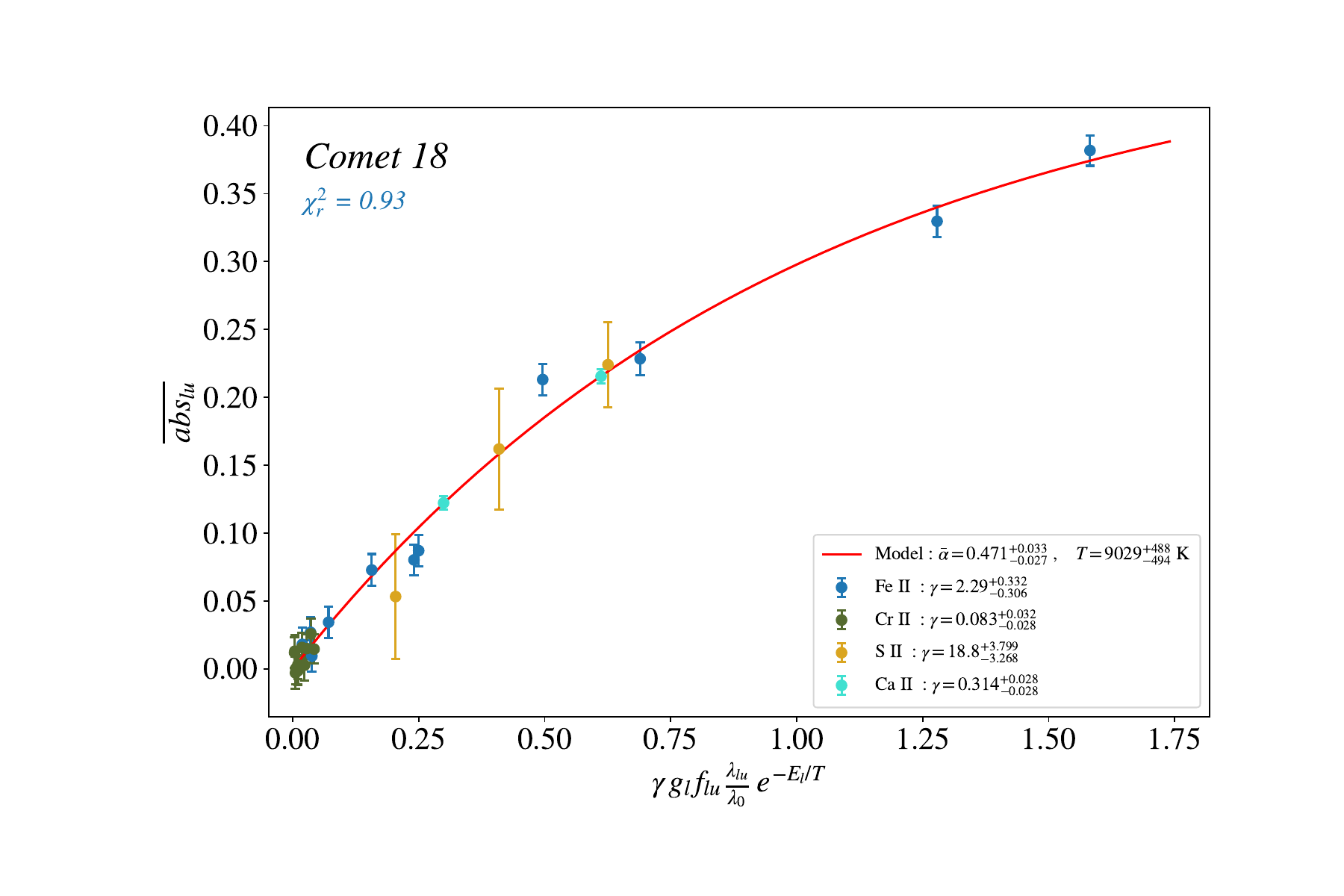}      
\end{figure*}
\small \textbf{Figure \ref{Fig. curves of growth all}}, continued.

\newpage

\begin{figure*}[h!]
\centering
    \includegraphics[scale = 0.38,     trim = 10 30 30 40,clip]{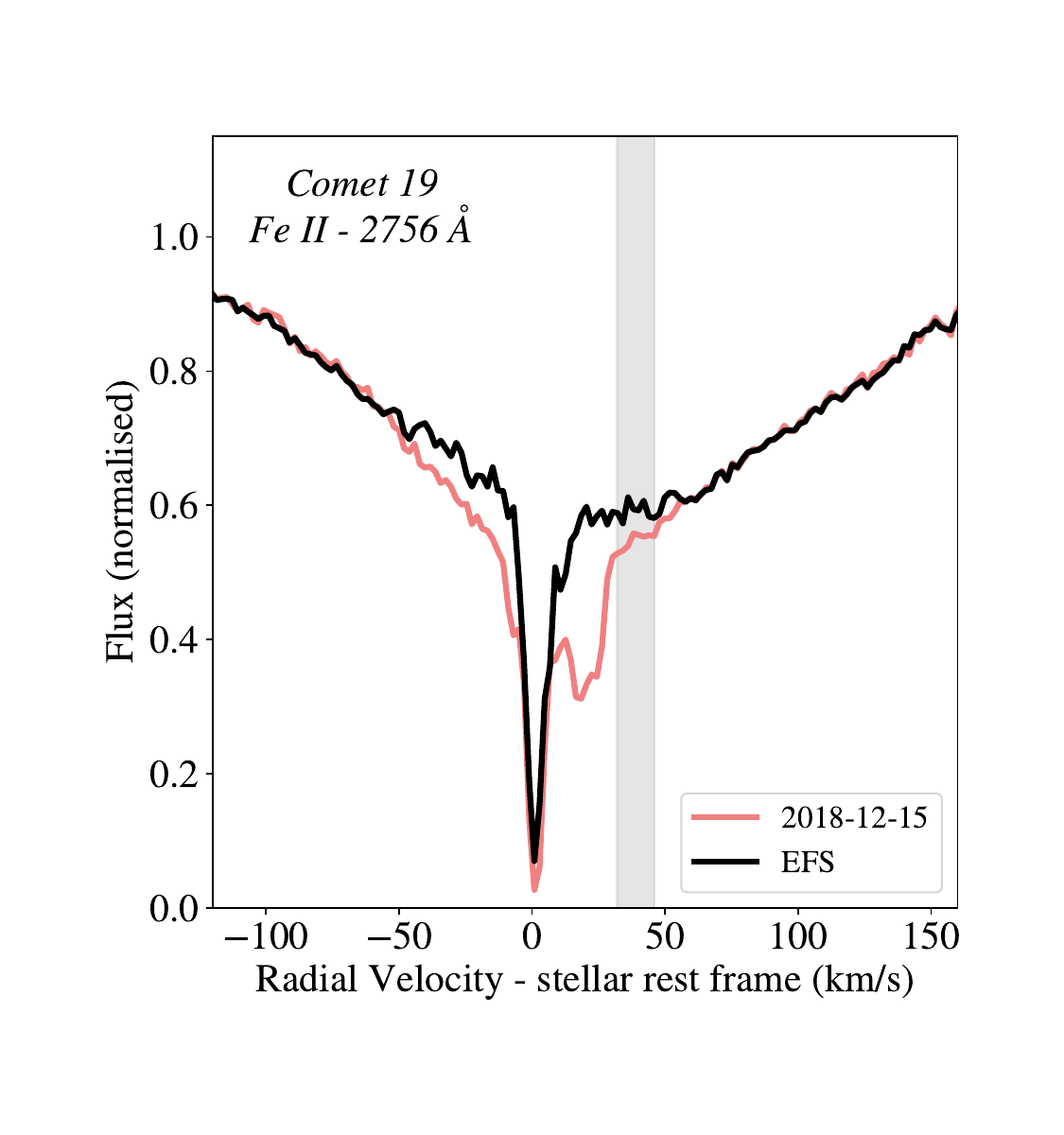}    
    \includegraphics[scale = 0.38,     trim = 85 30 80 40,clip]{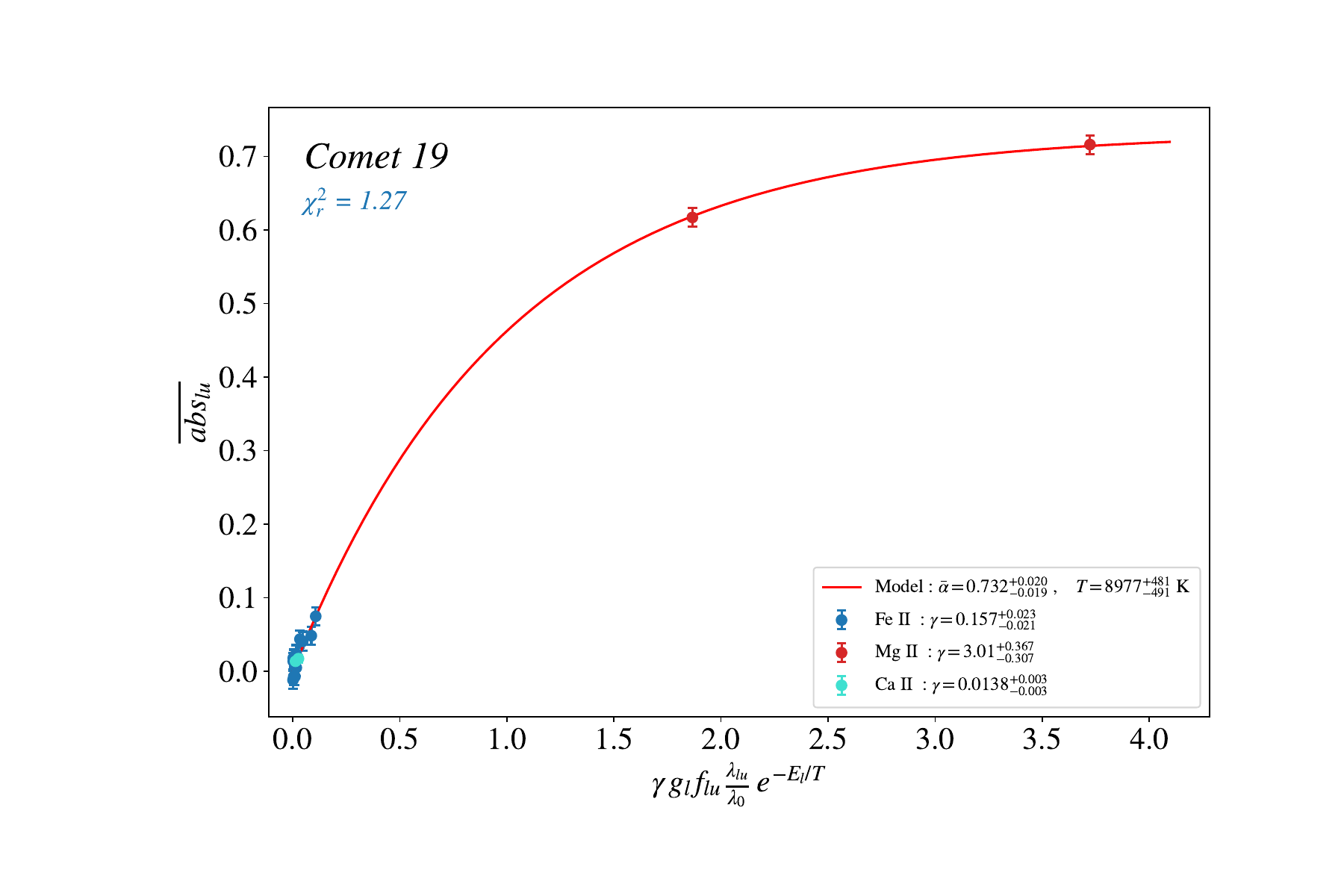}    
    
    \includegraphics[scale = 0.38,     trim = 10 30 30 40,clip]{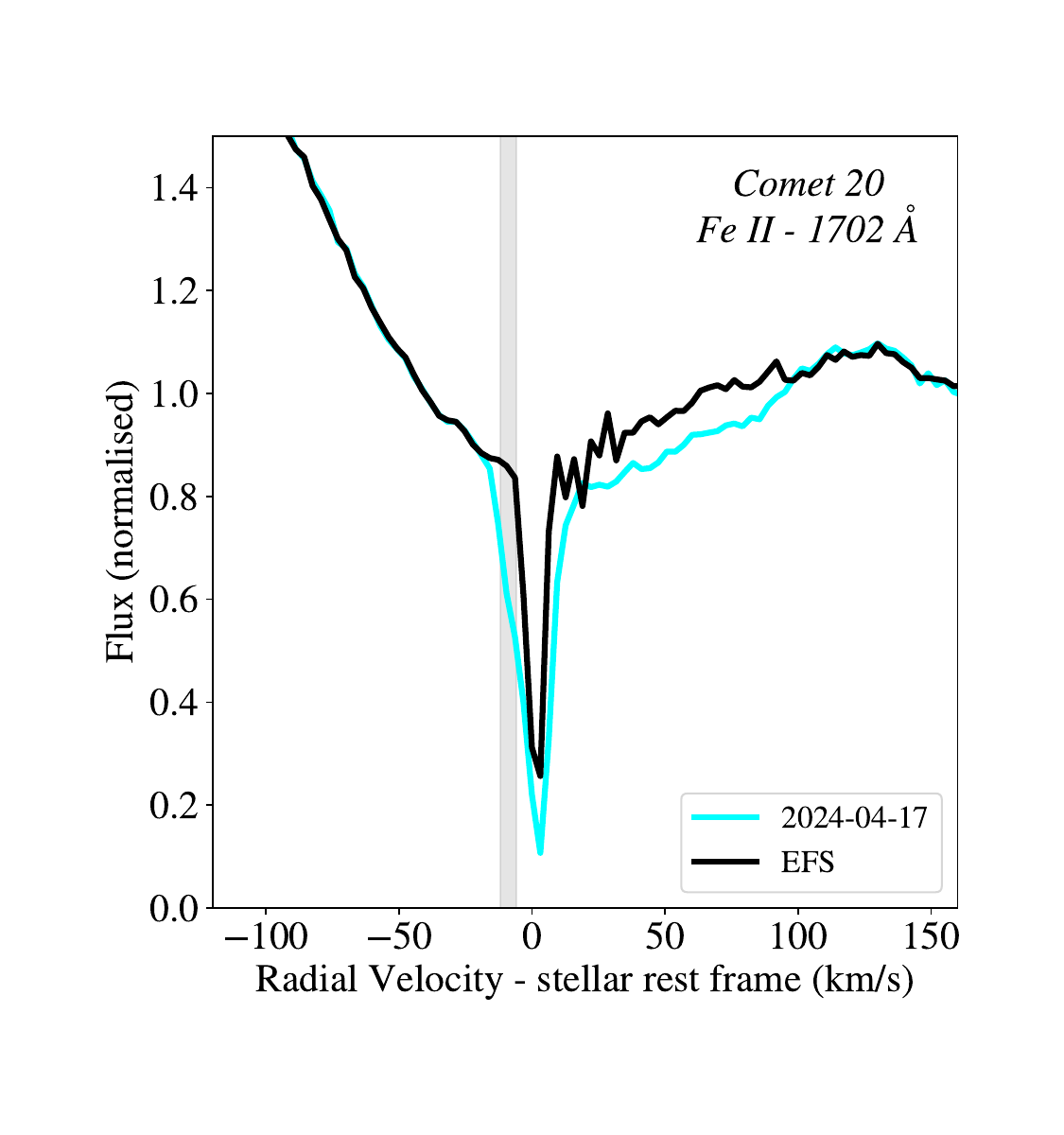}    
    \includegraphics[scale = 0.38,     trim = 85 30 80 40,clip]{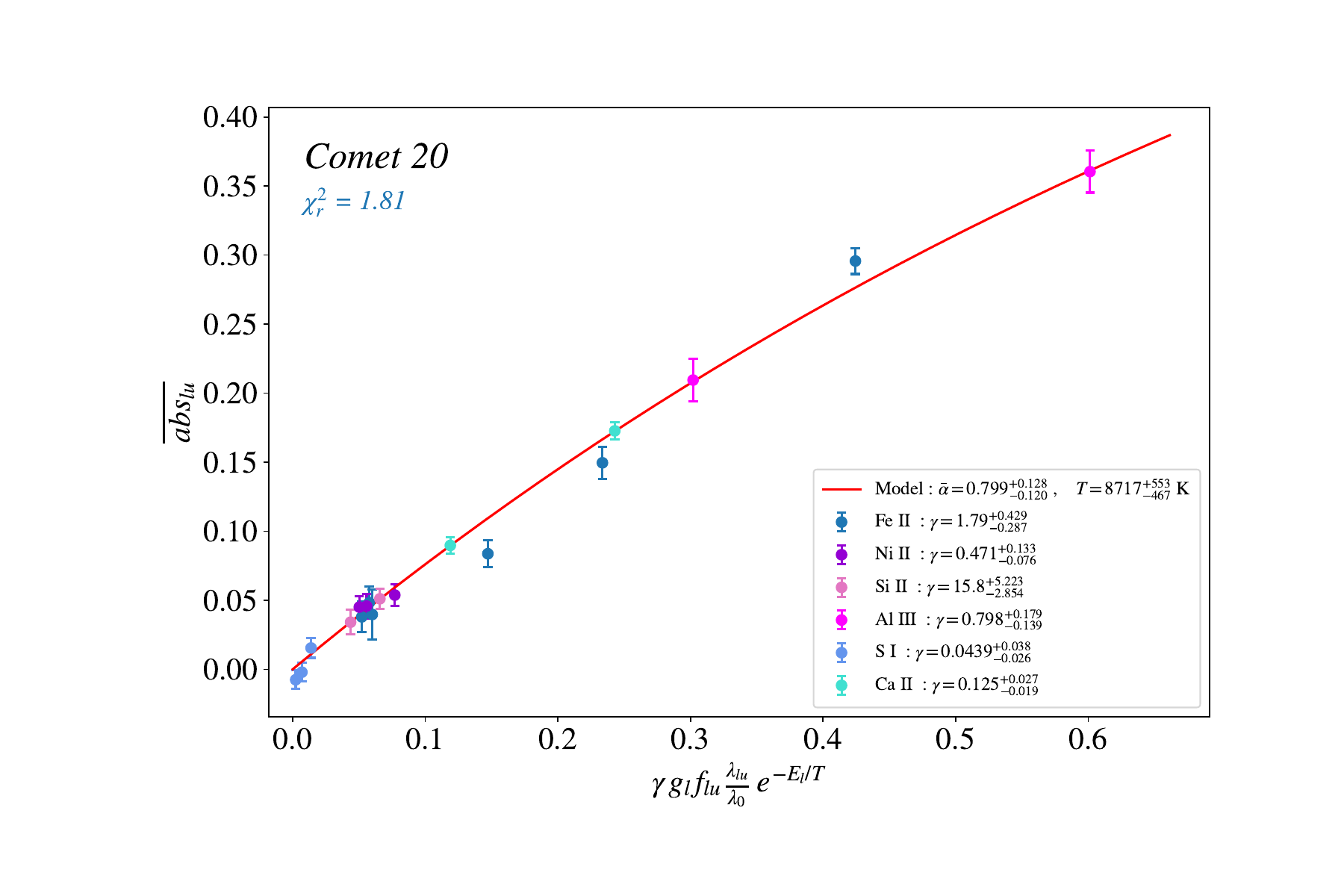}
    
    \includegraphics[scale = 0.38,     trim = 10 30 30 40,clip]{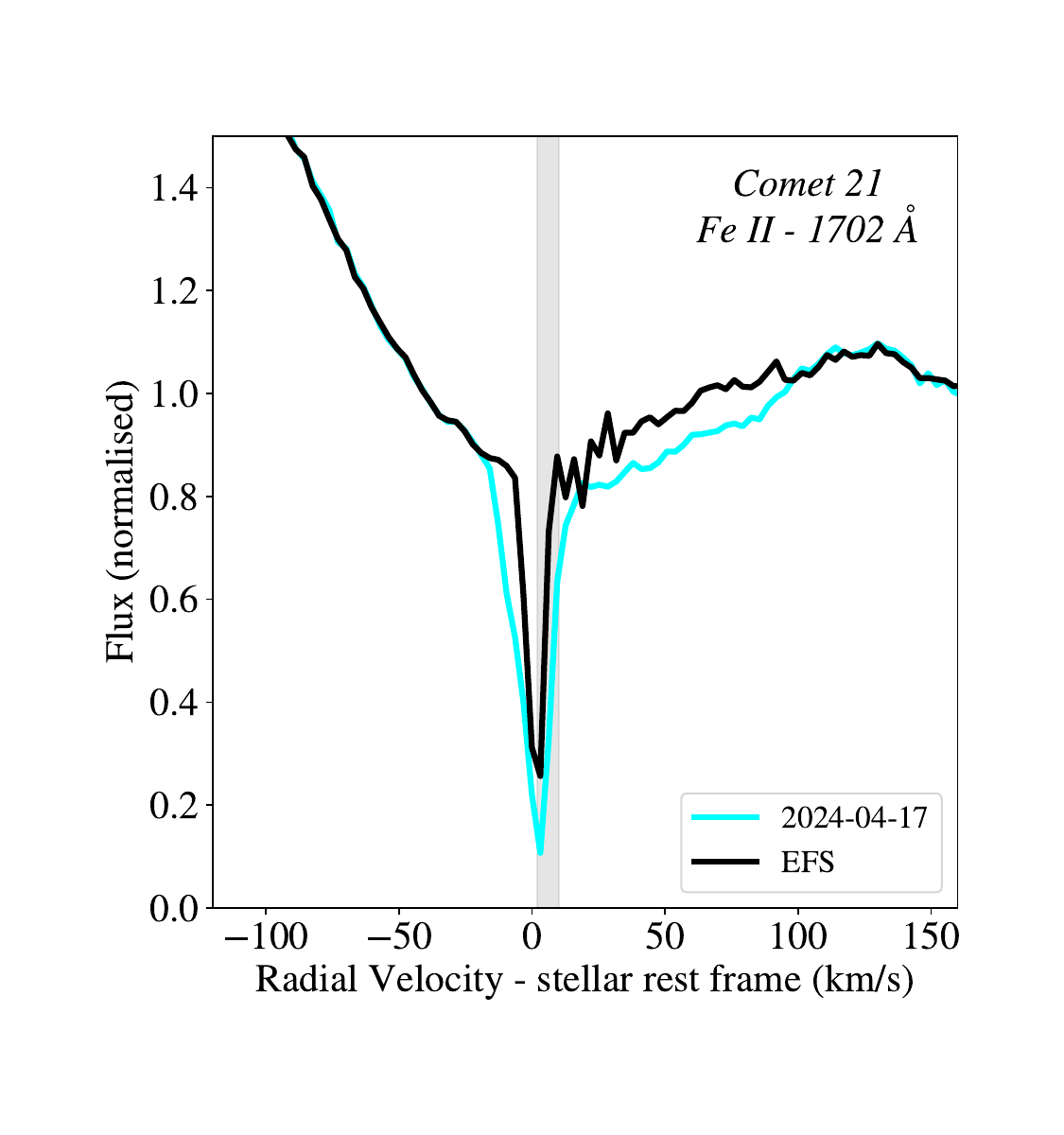}    
    \includegraphics[scale = 0.38,     trim = 85 30 80 40,clip]{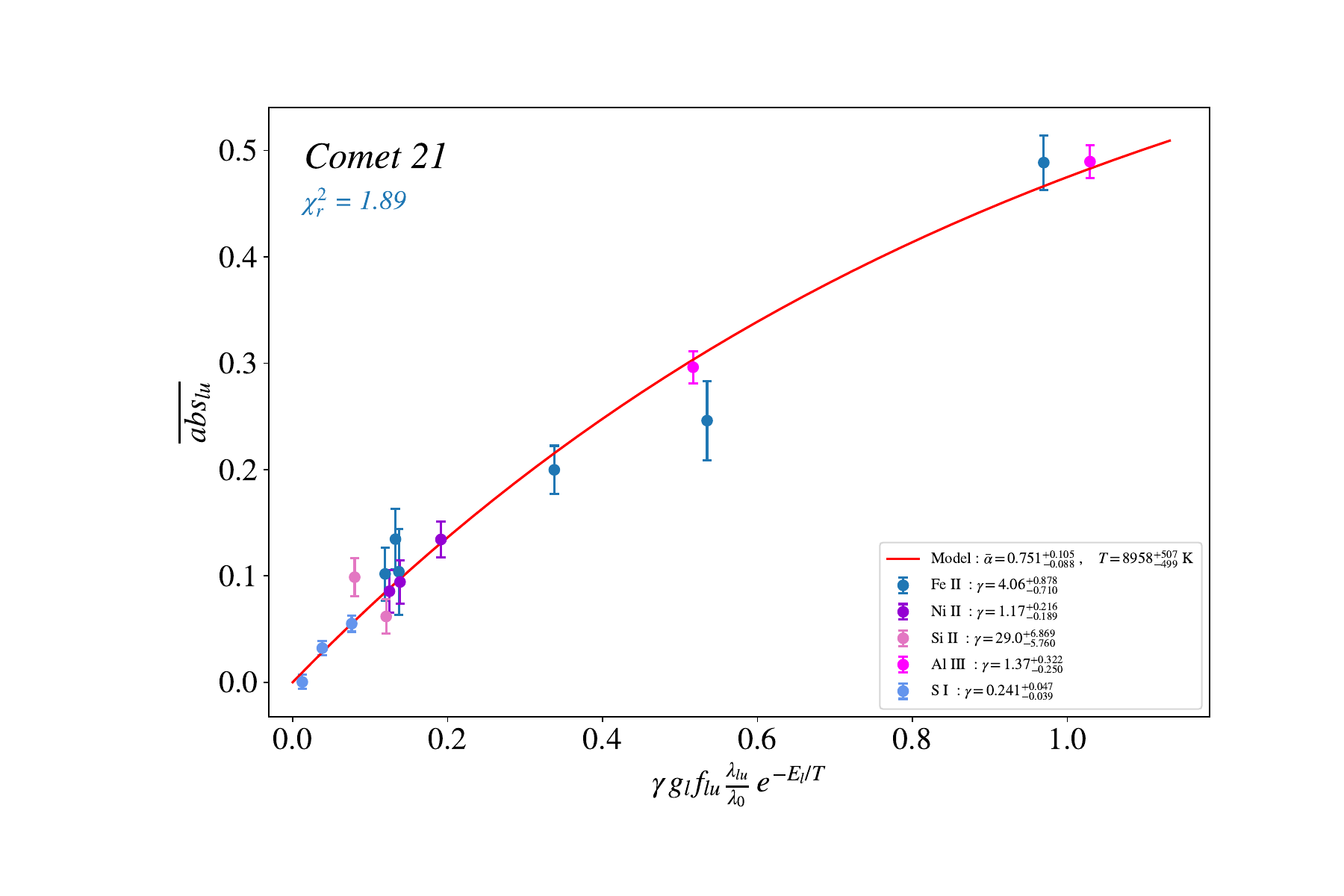}       
\end{figure*}
\small \textbf{Figure \ref{Fig. curves of growth all}}, continued.

\newpage

\begin{figure*}[h!]
\centering
    \includegraphics[scale = 0.38,     trim = 10 30 30 40,clip]{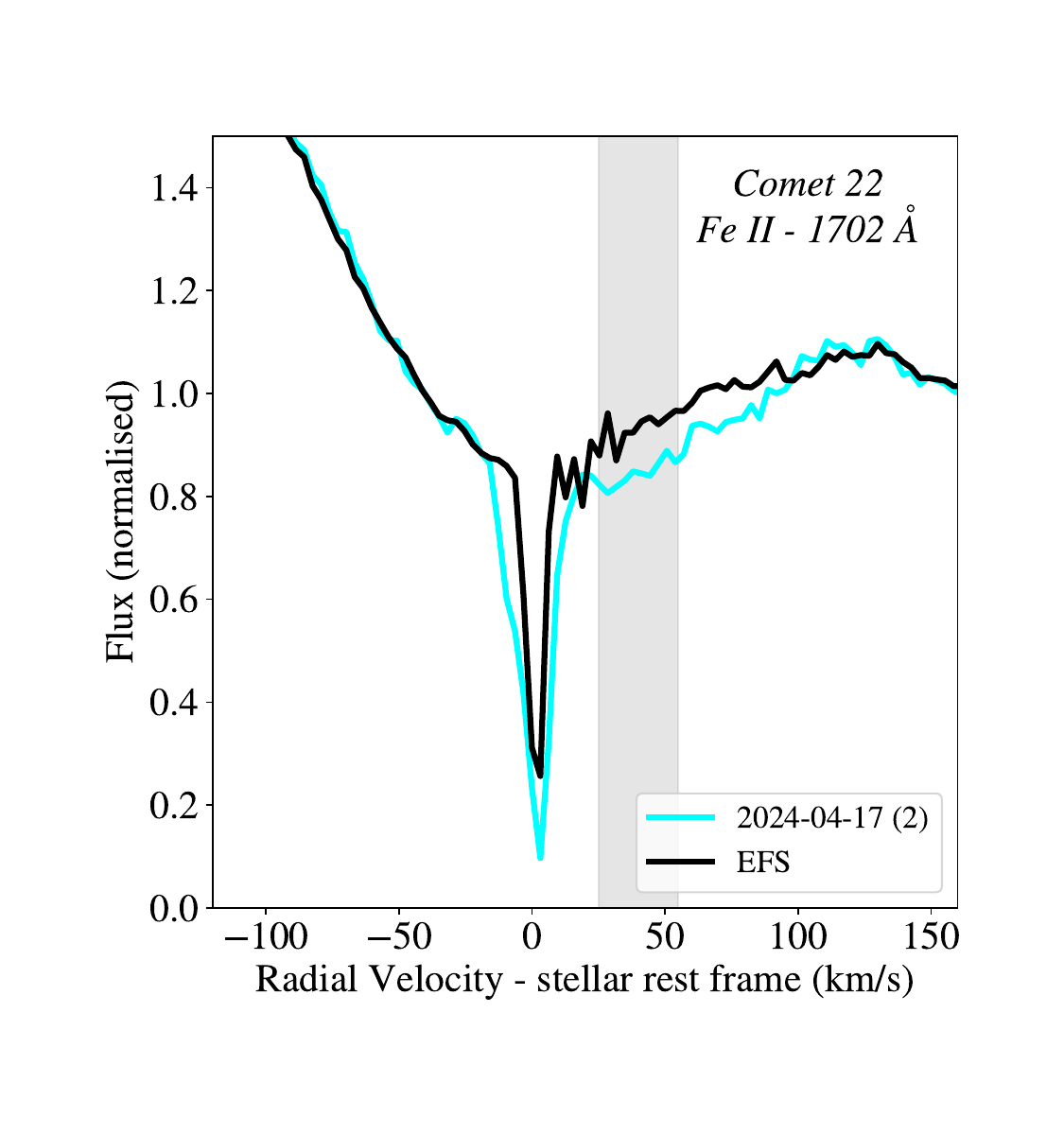}    
    \includegraphics[scale = 0.38,     trim = 85 30 80 40,clip]{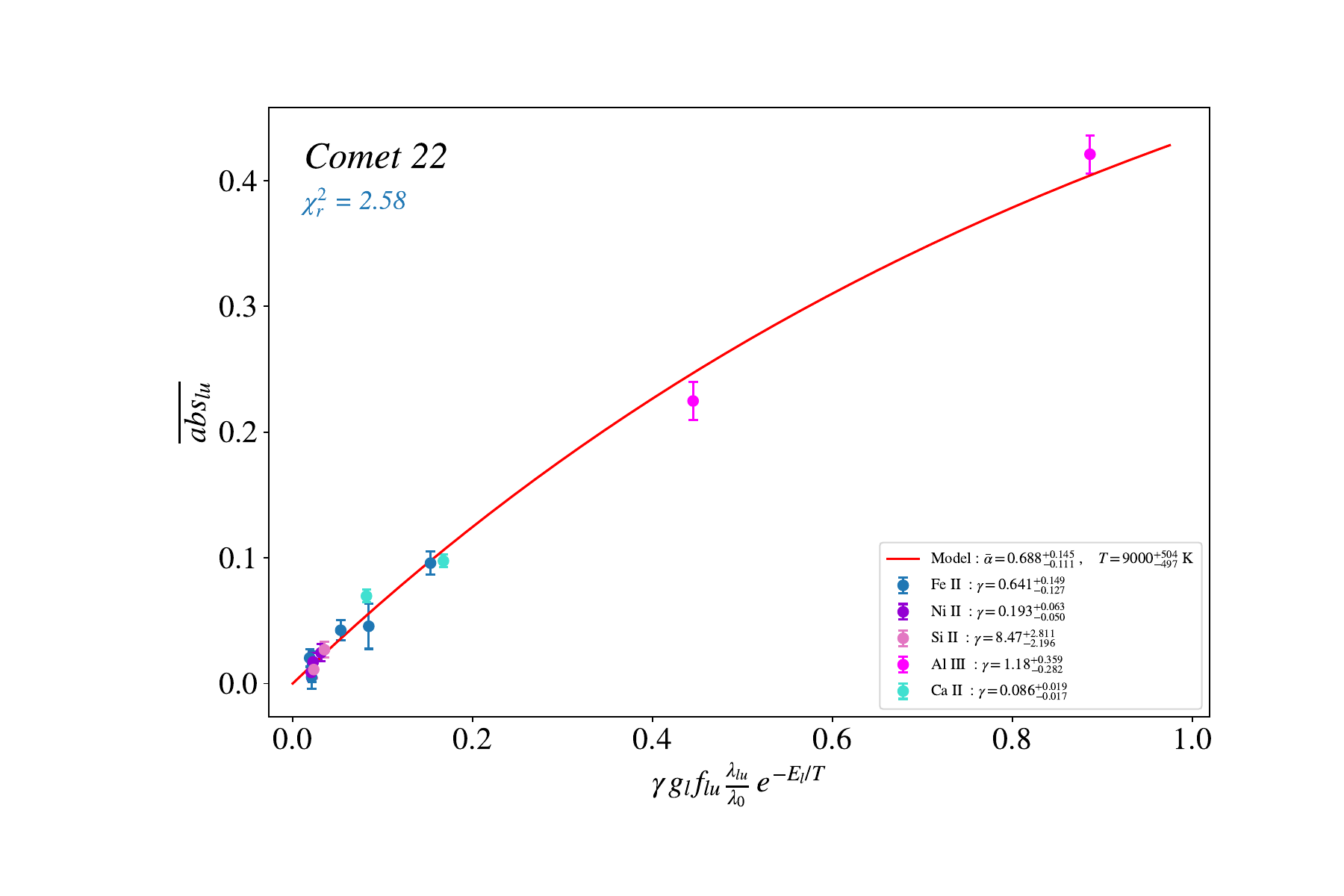}    
    
    \includegraphics[scale = 0.38,     trim = 10 30 30 40,clip]{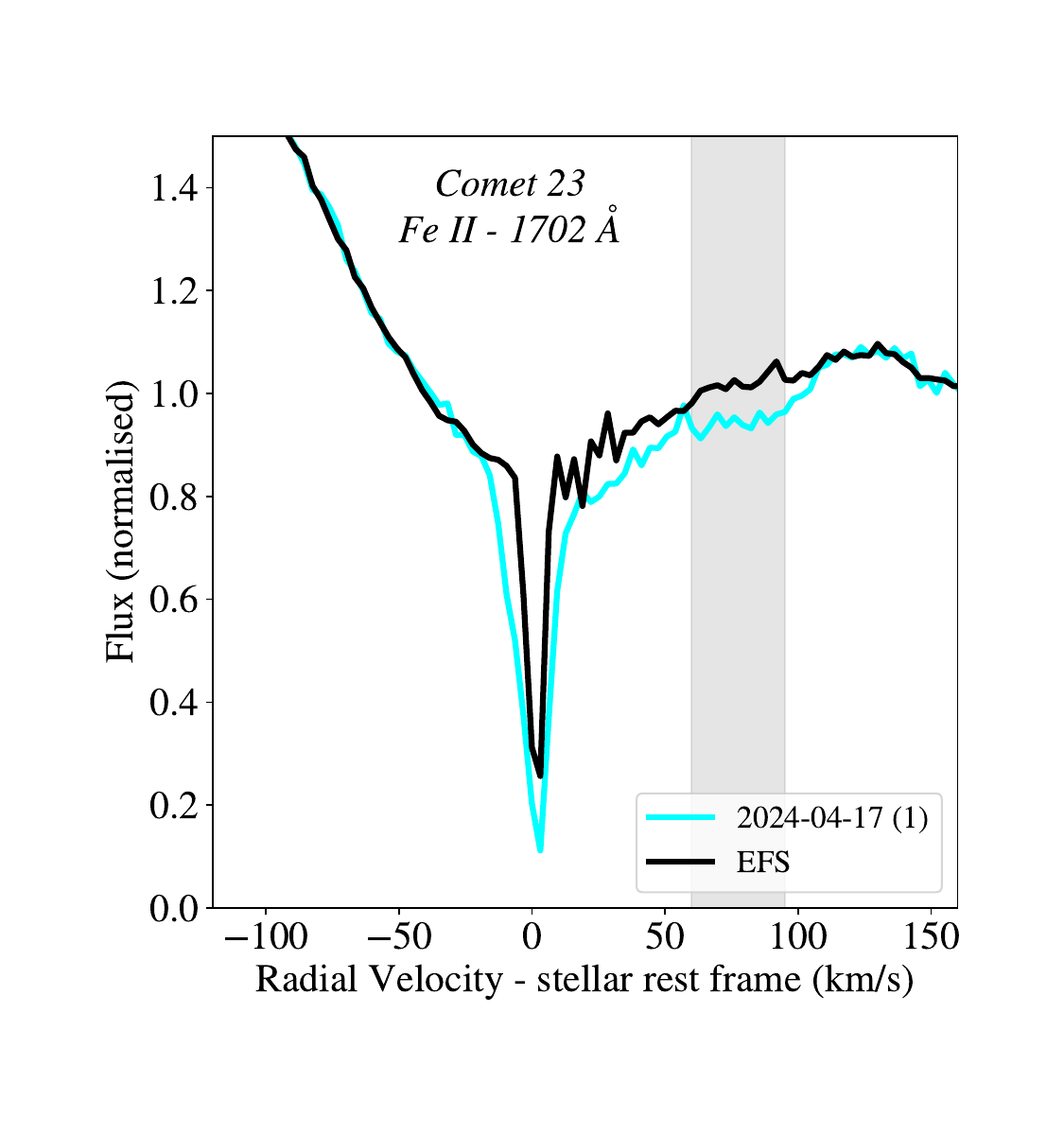}    
    \includegraphics[scale = 0.38,     trim = 85 30 80 40,clip]{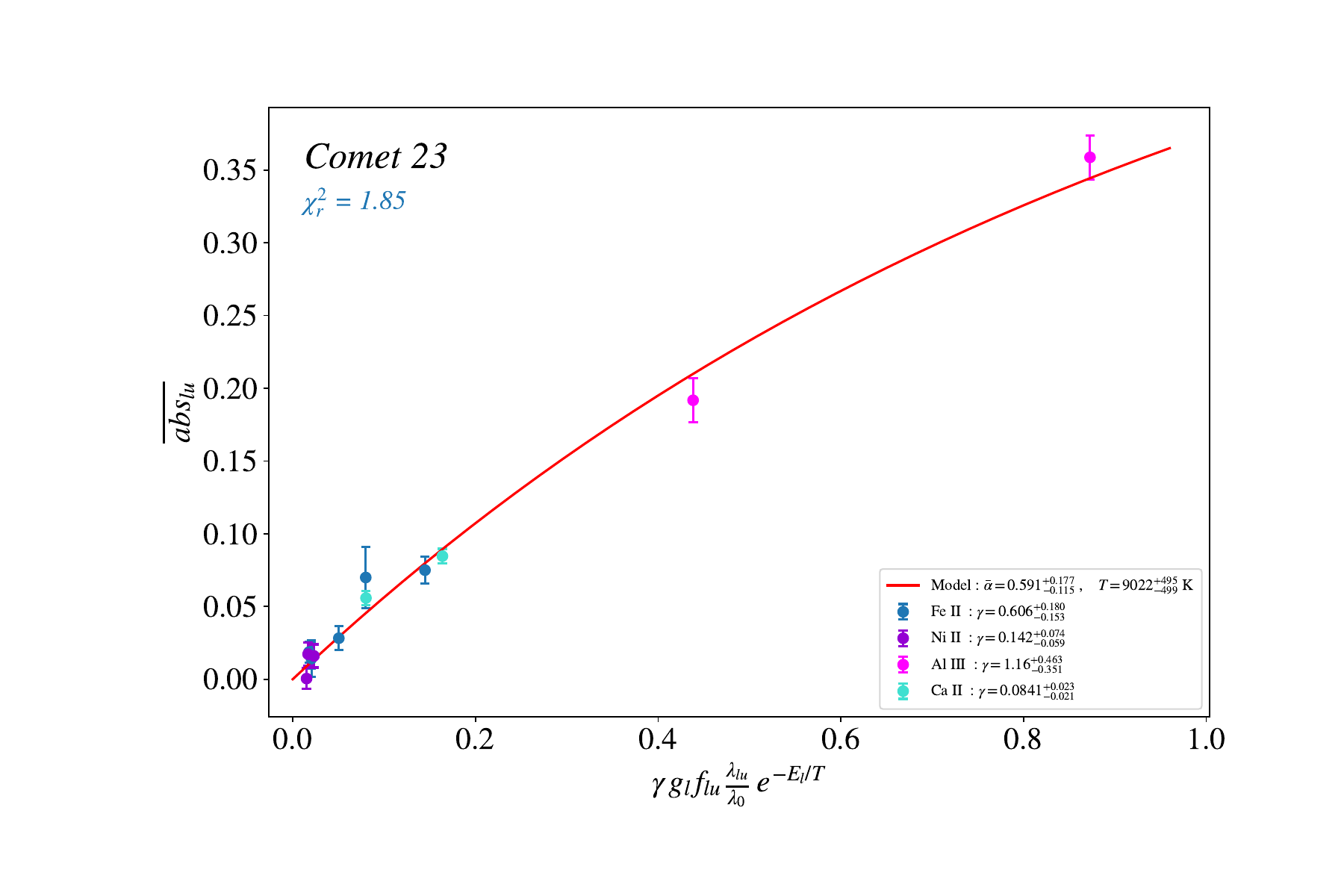}
    
    \includegraphics[scale = 0.38,     trim = 10 30 30 40,clip]{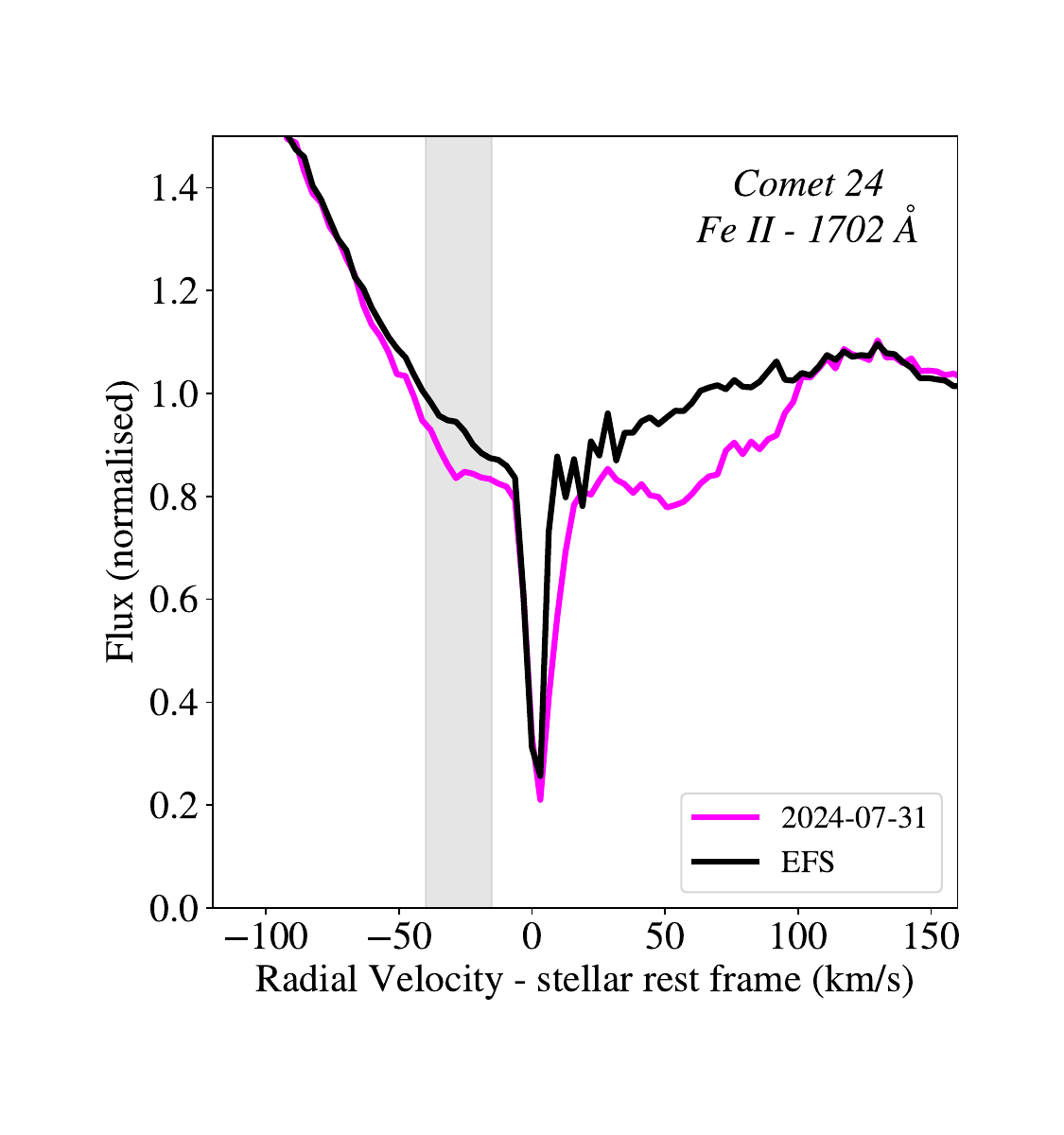}    
    \includegraphics[scale = 0.38,     trim = 85 30 80 40,clip]{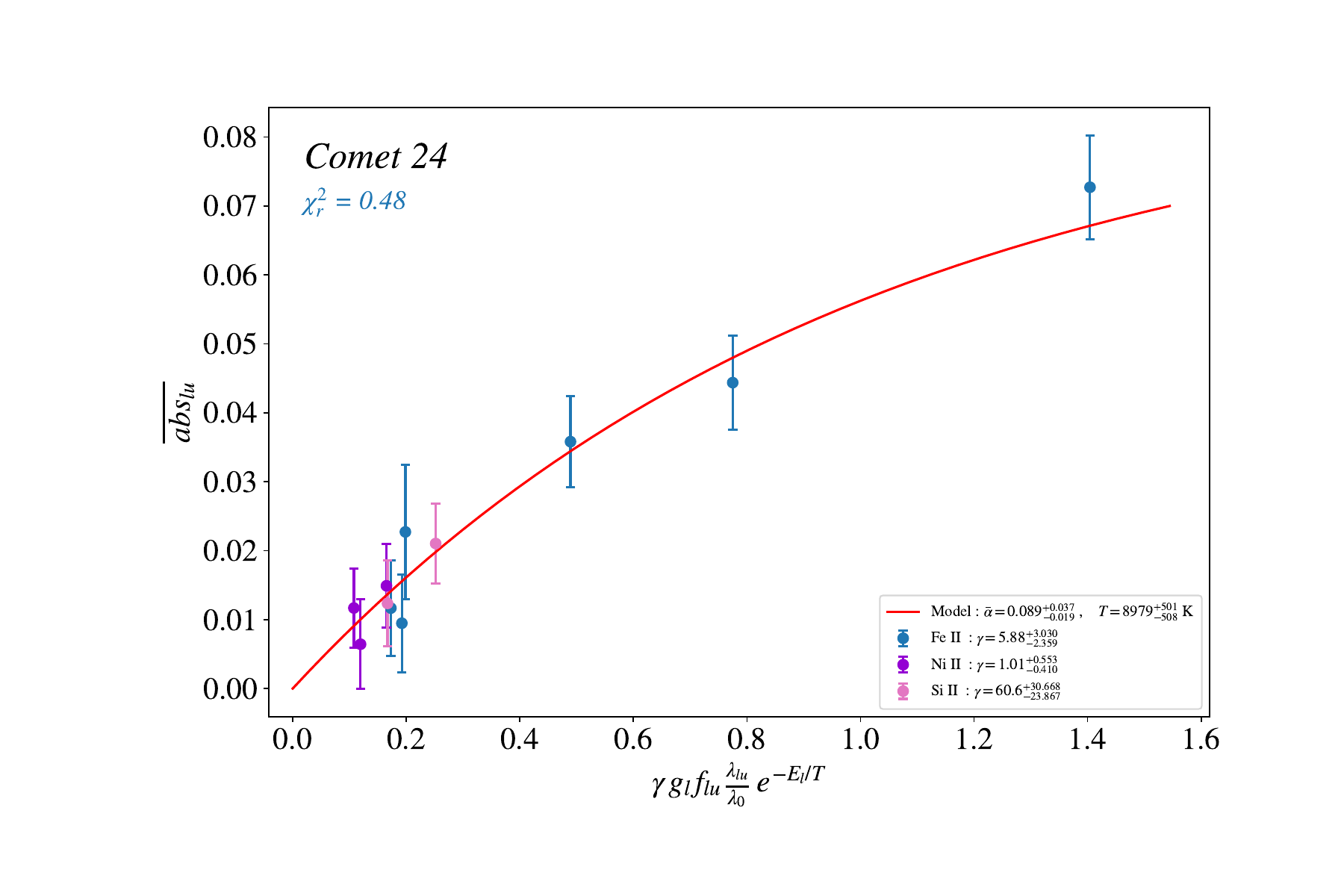}       
\end{figure*}
\small \textbf{Figure \ref{Fig. curves of growth all}}, continued.

\newpage

\begin{figure*}[h!]
\centering
    \includegraphics[scale = 0.38,     trim = 10 30 30 40,clip]{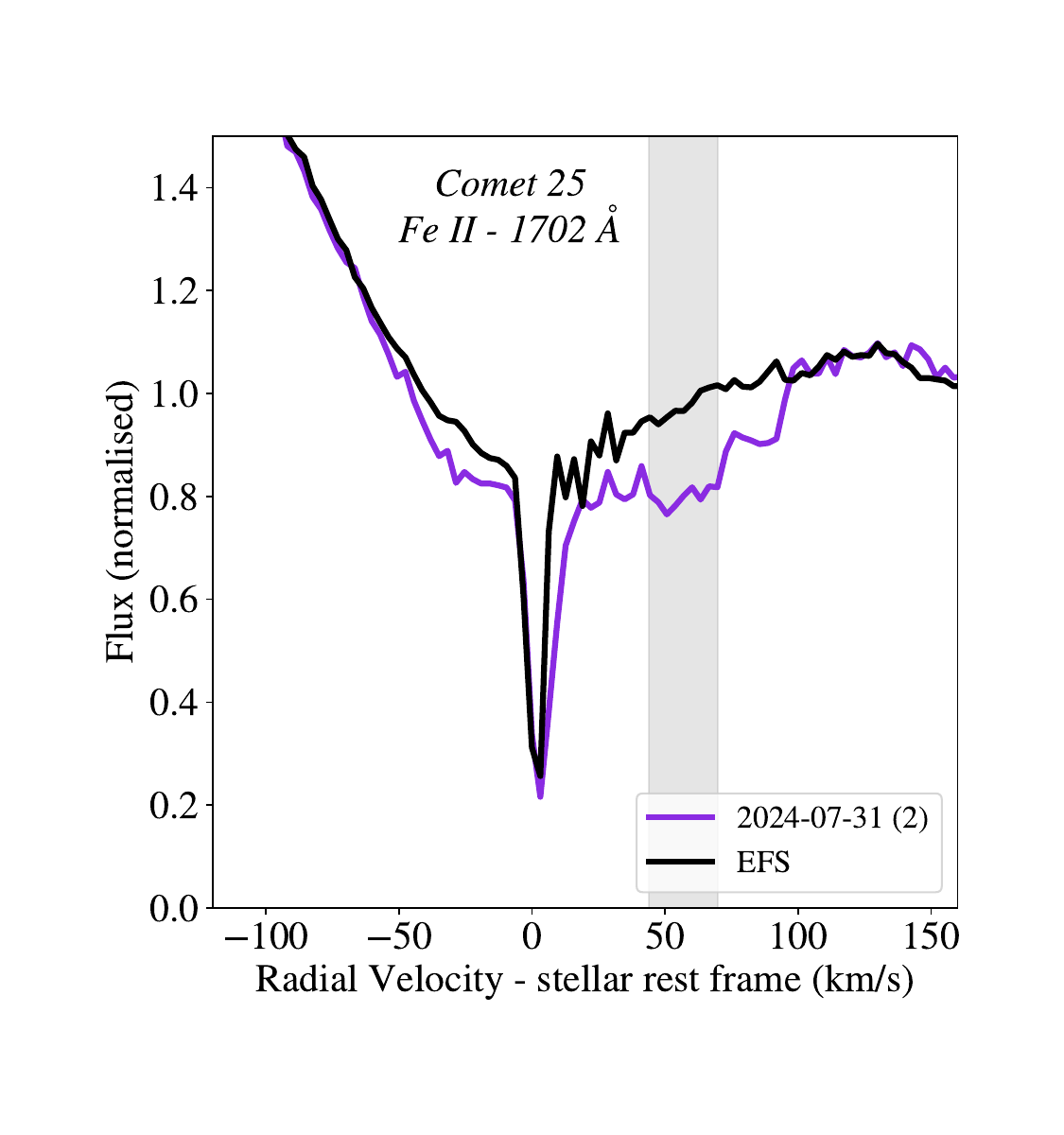}    
    \includegraphics[scale = 0.38,     trim = 85 30 80 40,clip]{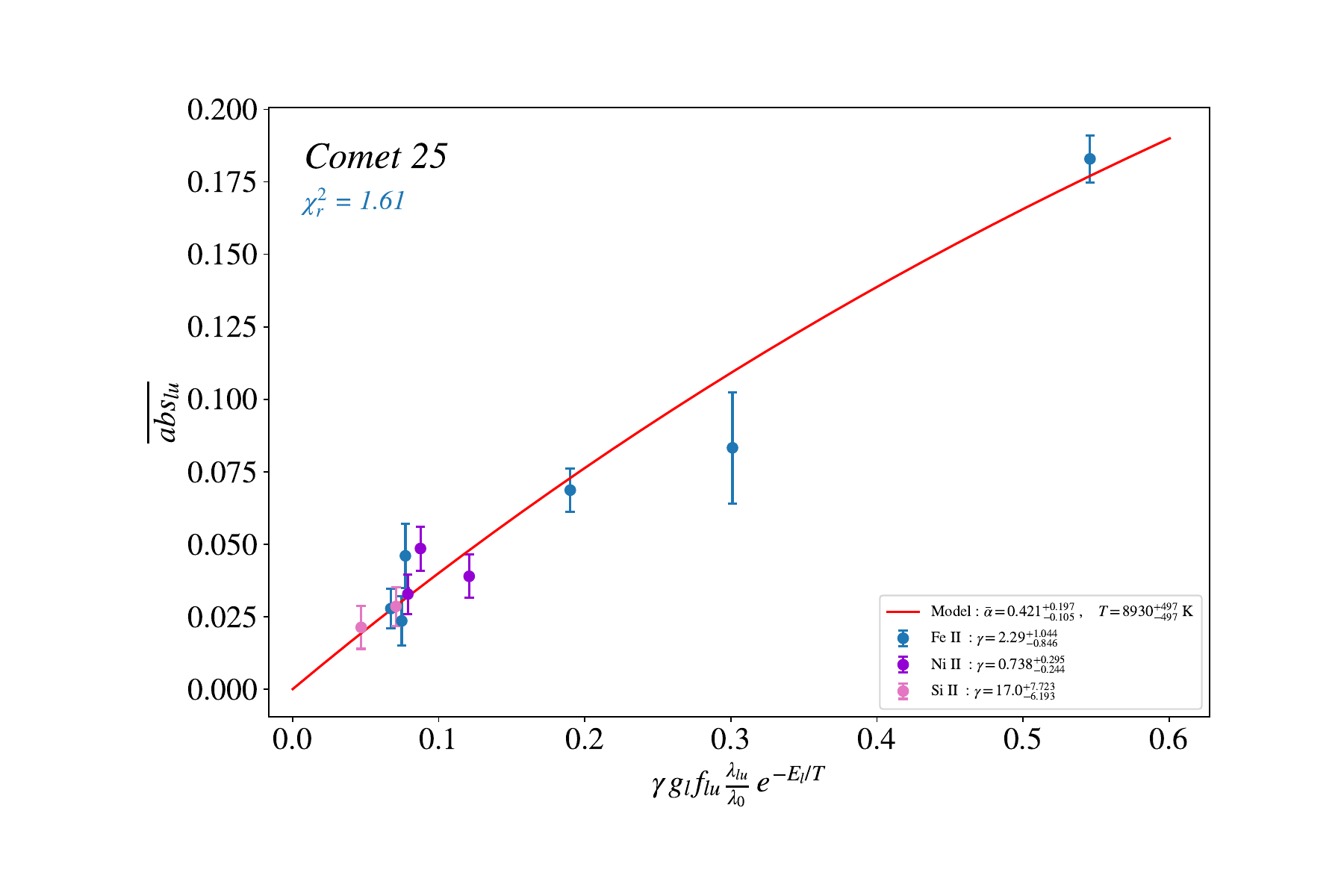}    
    
    \includegraphics[scale = 0.38,     trim = 10 30 30 40,clip]{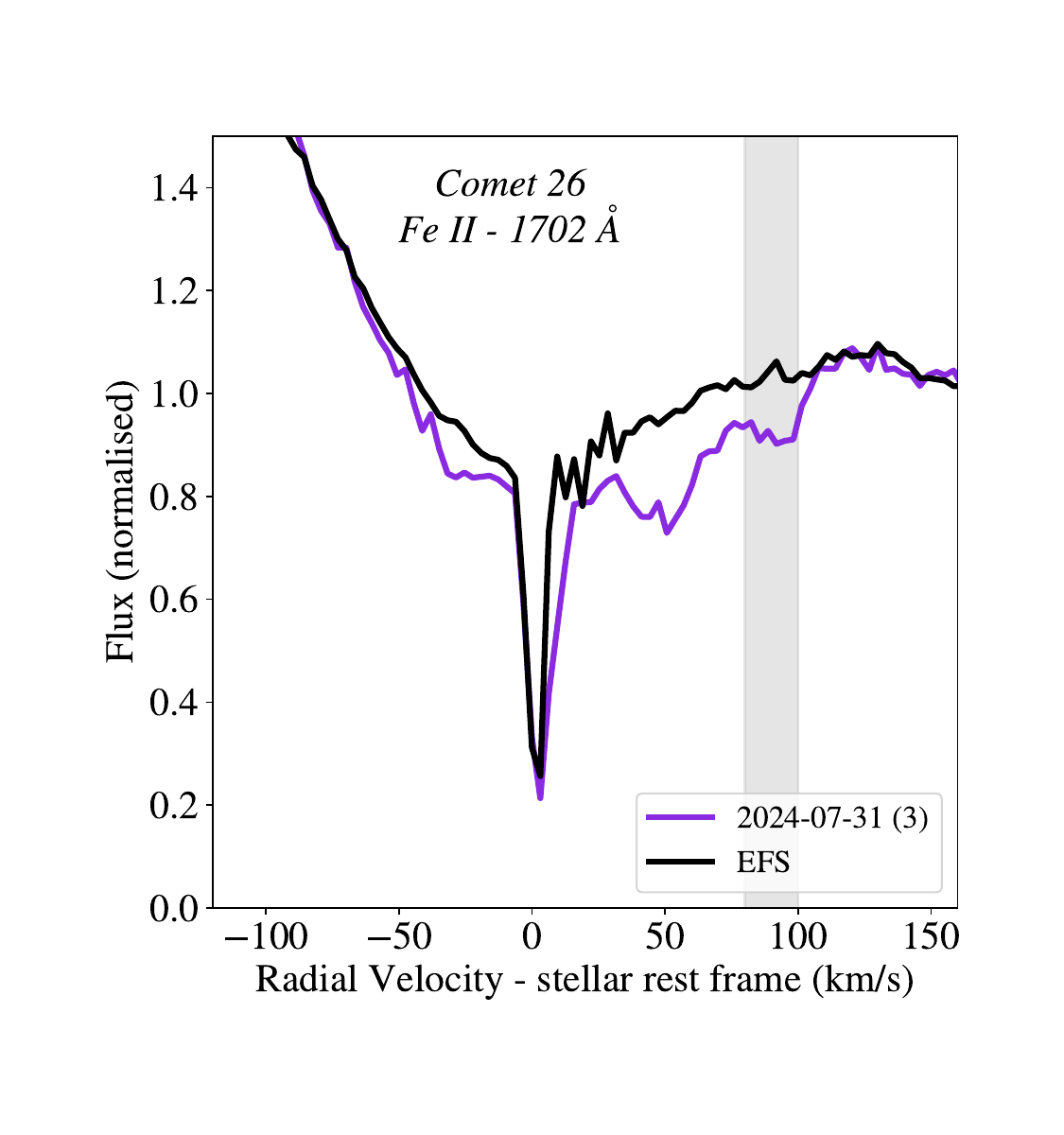}    
    \includegraphics[scale = 0.38,     trim = 85 30 80 40,clip]{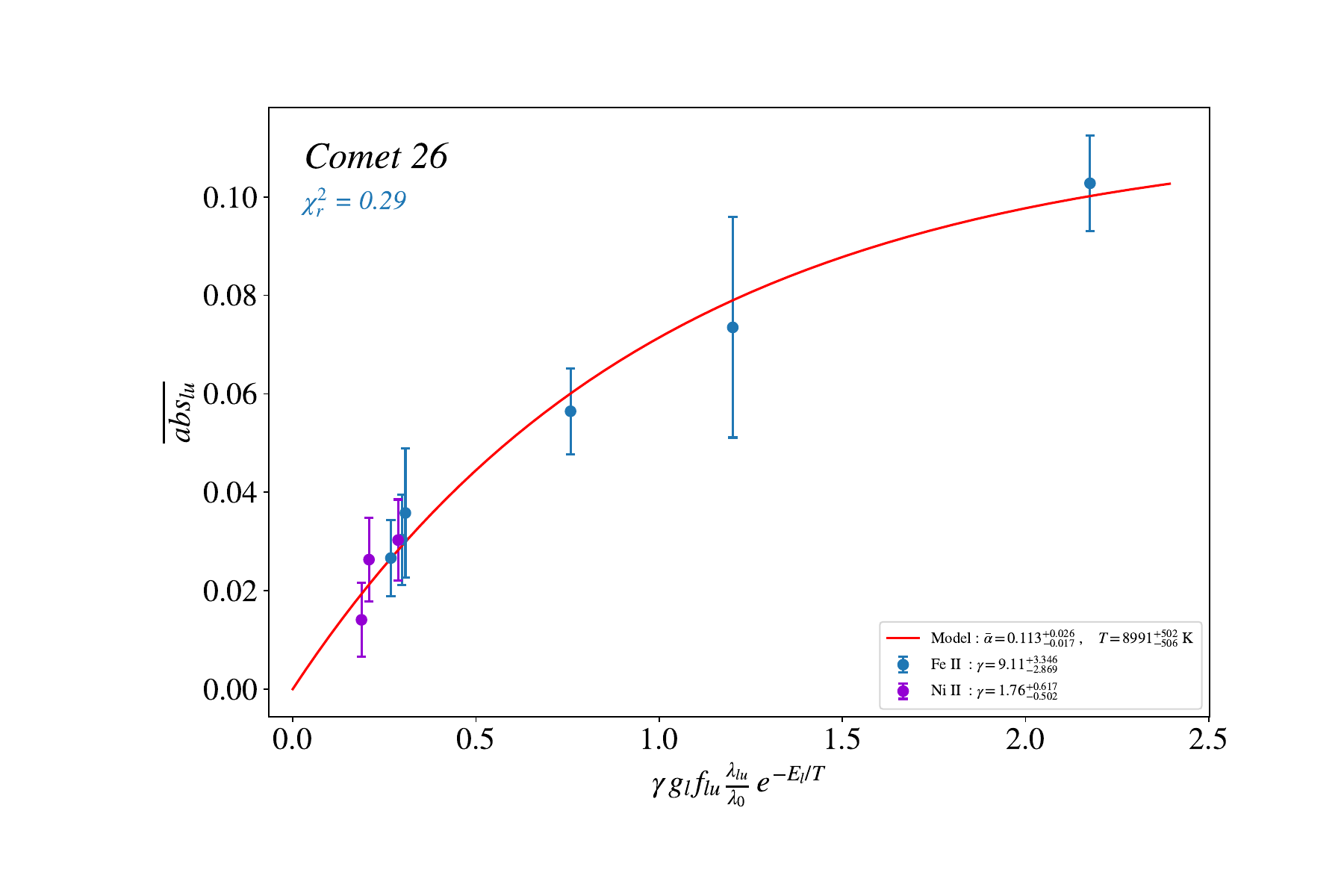}
    
    \includegraphics[scale = 0.38,     trim = 10 30 30 40,clip]{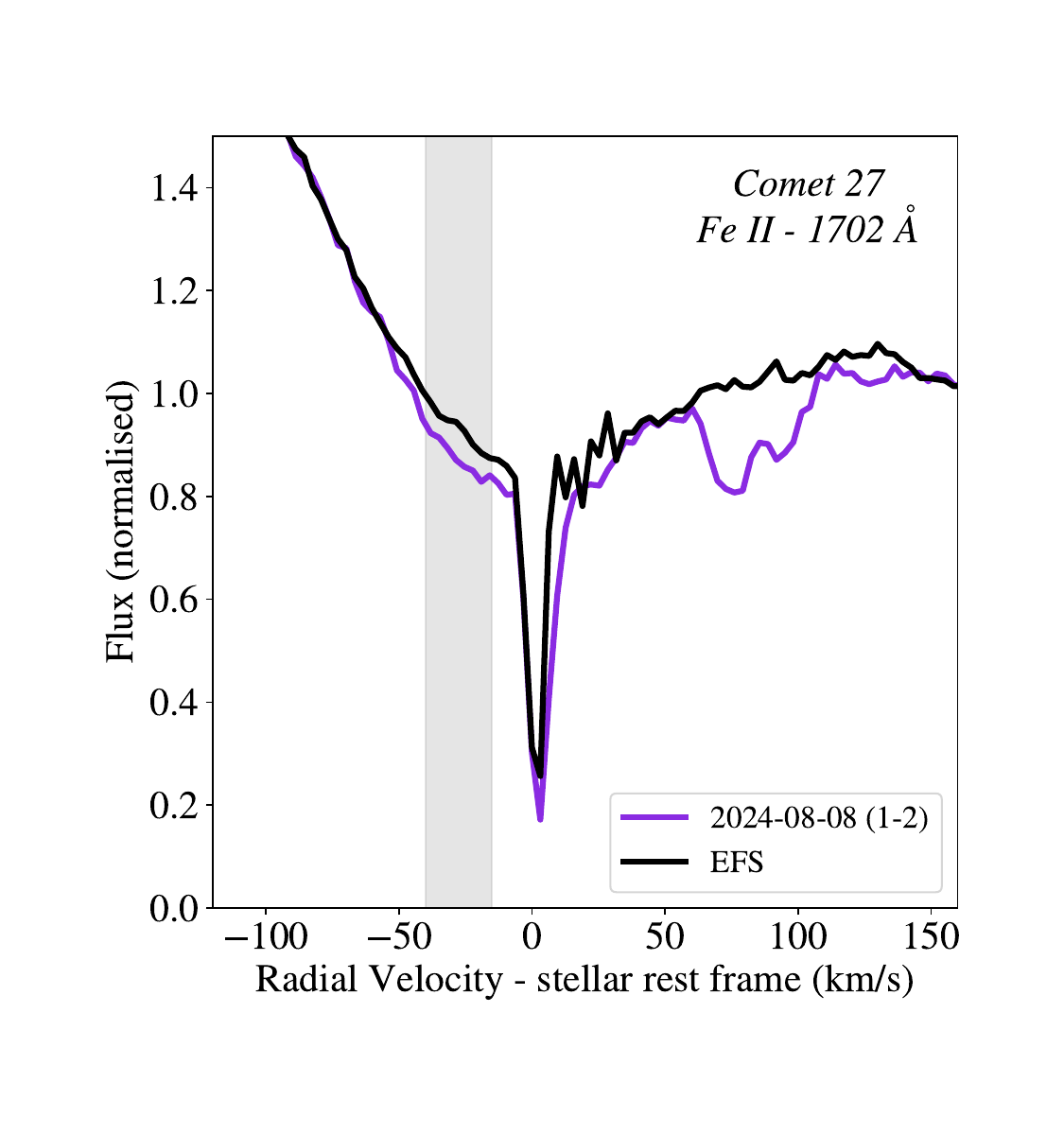}    
    \includegraphics[scale = 0.38,     trim = 85 30 80 40,clip]{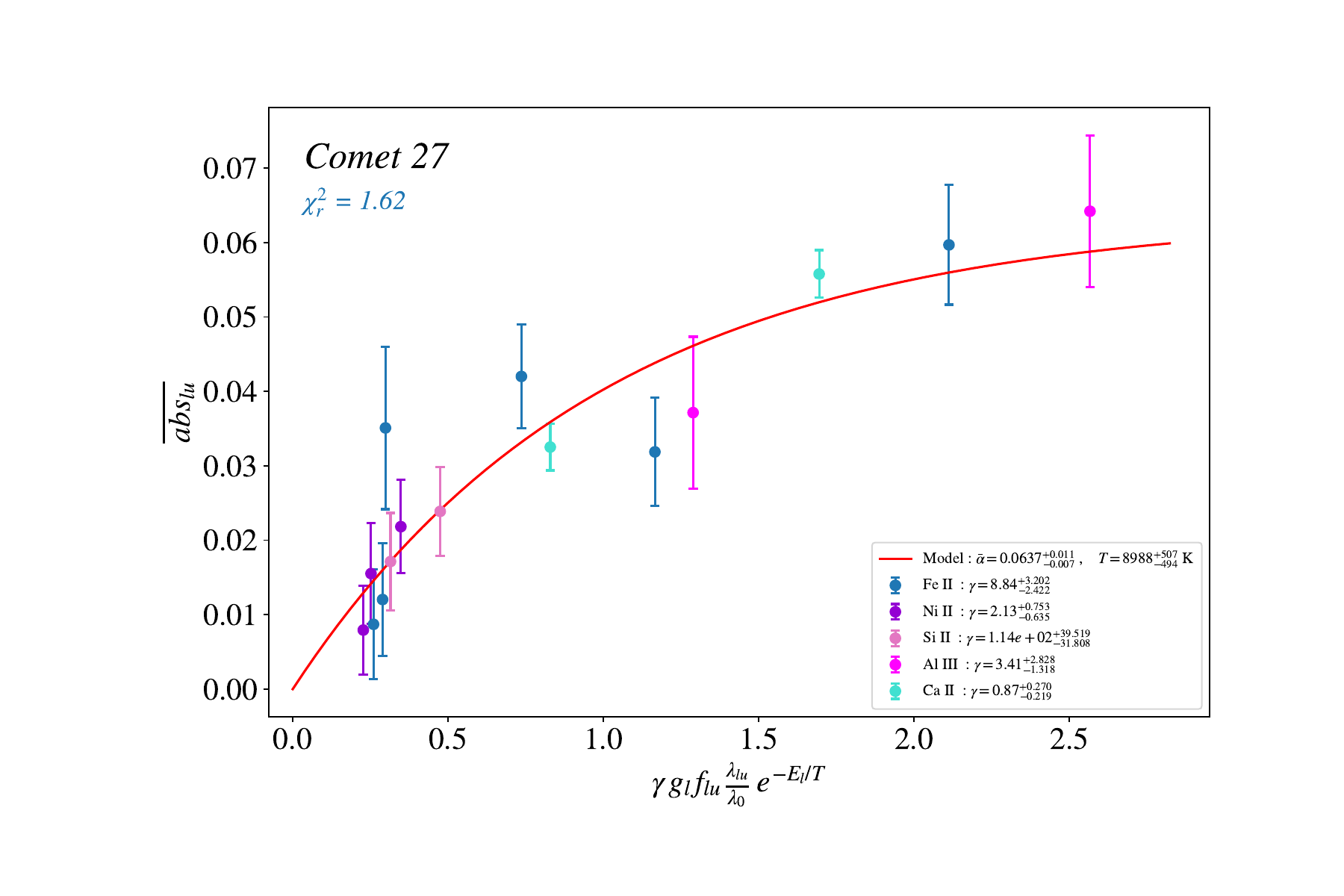}     
\end{figure*}
\small \textbf{Figure \ref{Fig. curves of growth all}}, continued.

\newpage

\begin{figure*}[h!]
\centering
    \includegraphics[scale = 0.38,     trim = 10 30 30 40,clip]{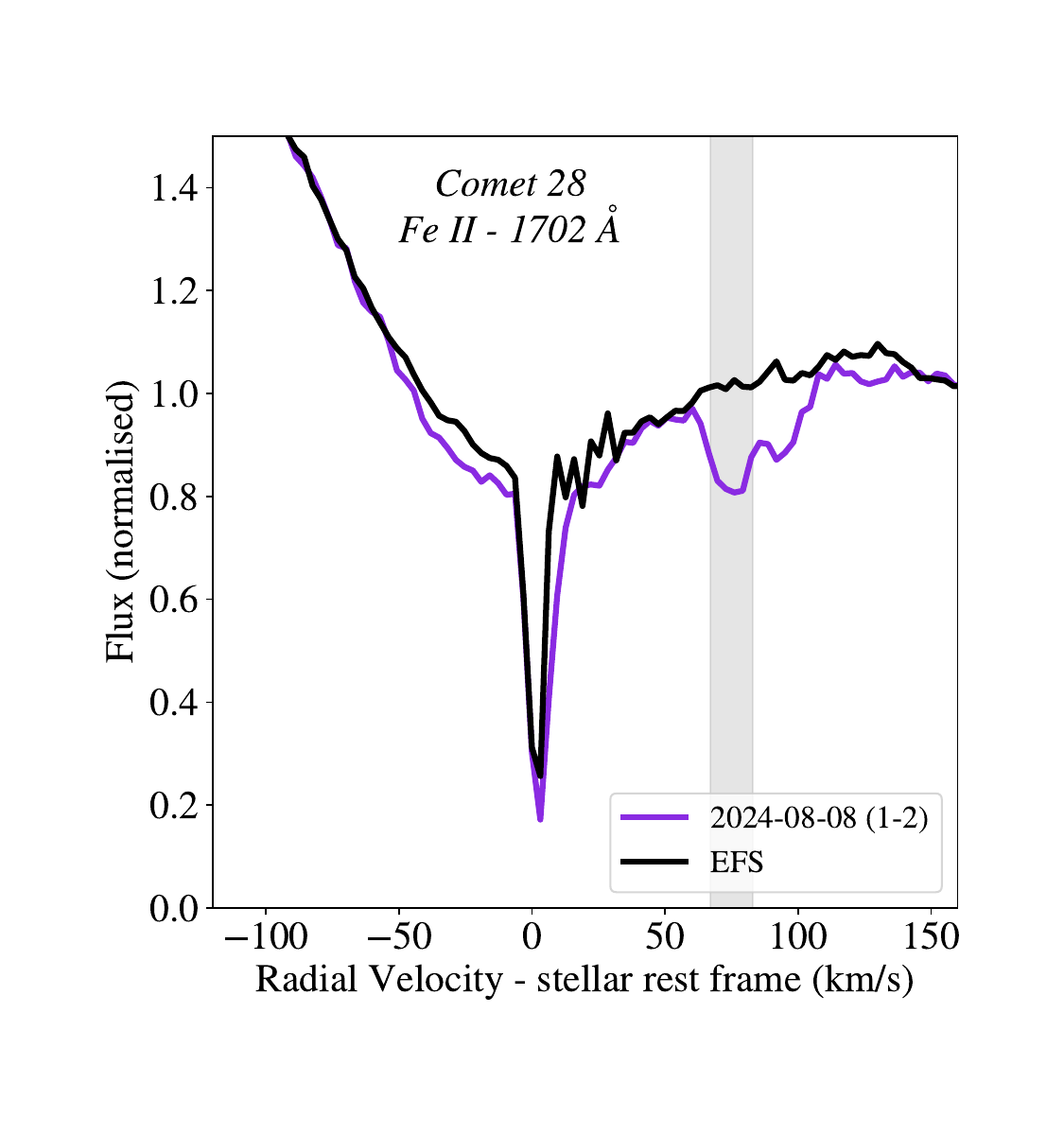}    
    \includegraphics[scale = 0.38,     trim = 85 30 80 40,clip]{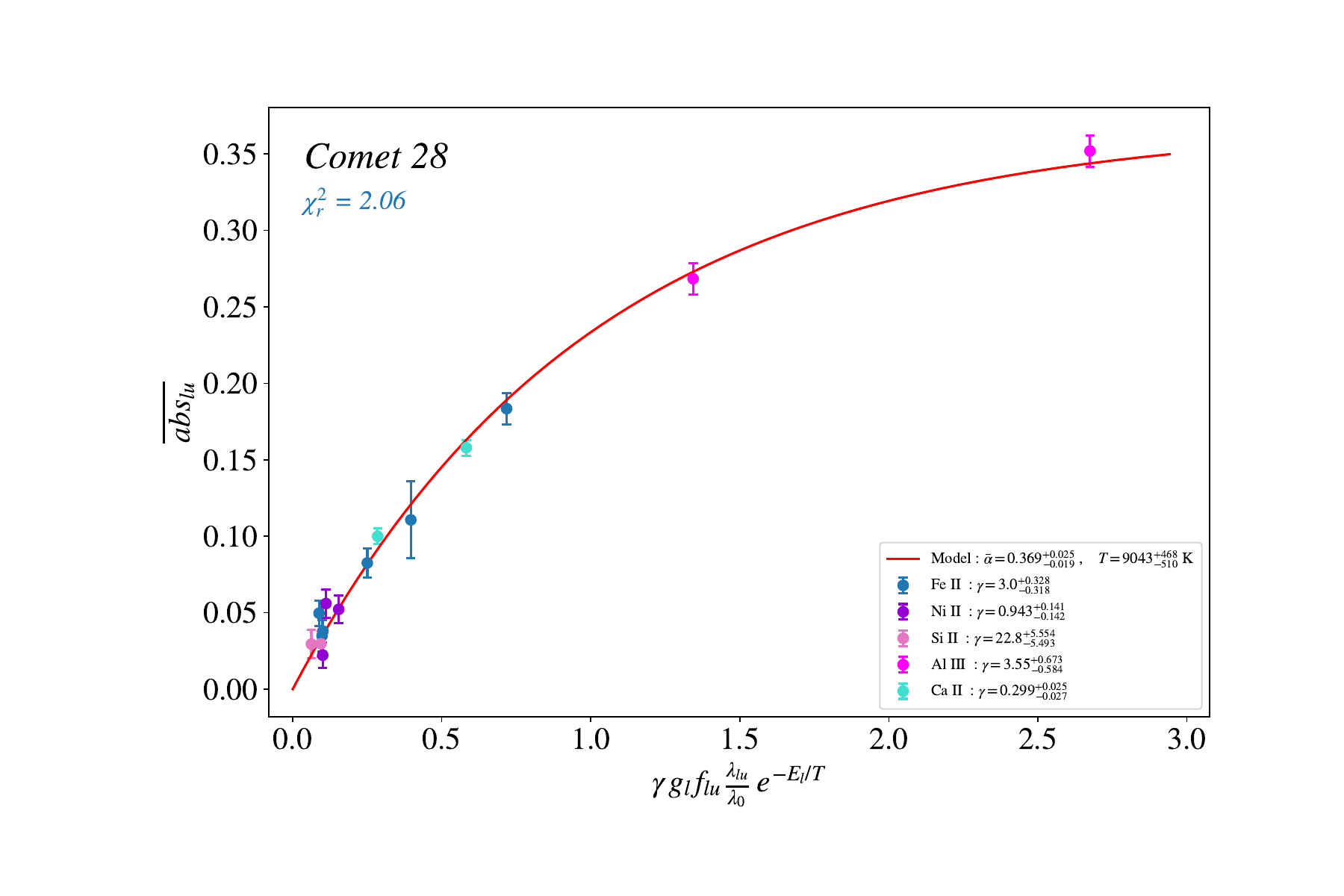}    
    
    \includegraphics[scale = 0.38,     trim = 10 30 30 40,clip]{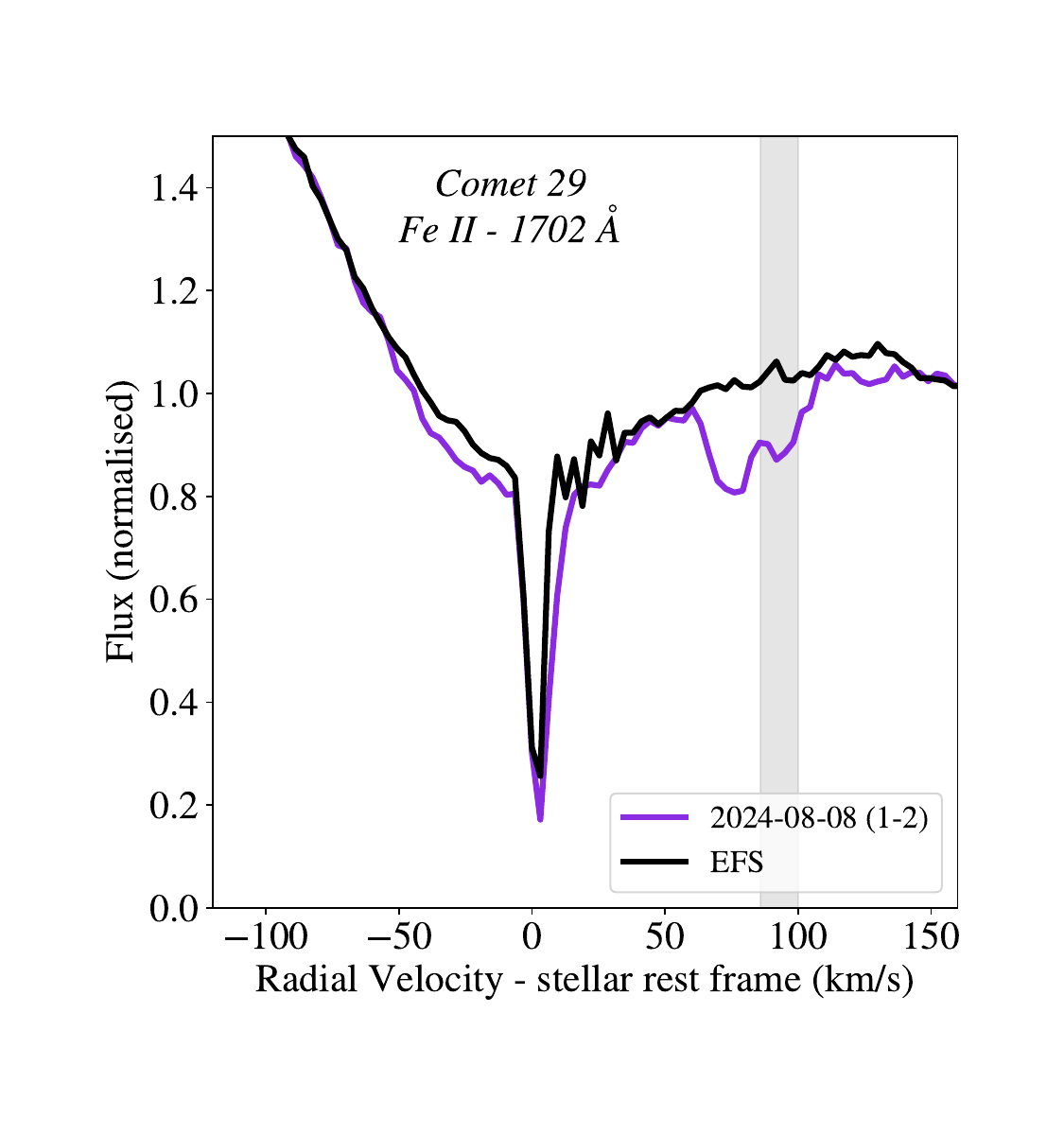}    
    \includegraphics[scale = 0.38,     trim = 85 30 80 40,clip]{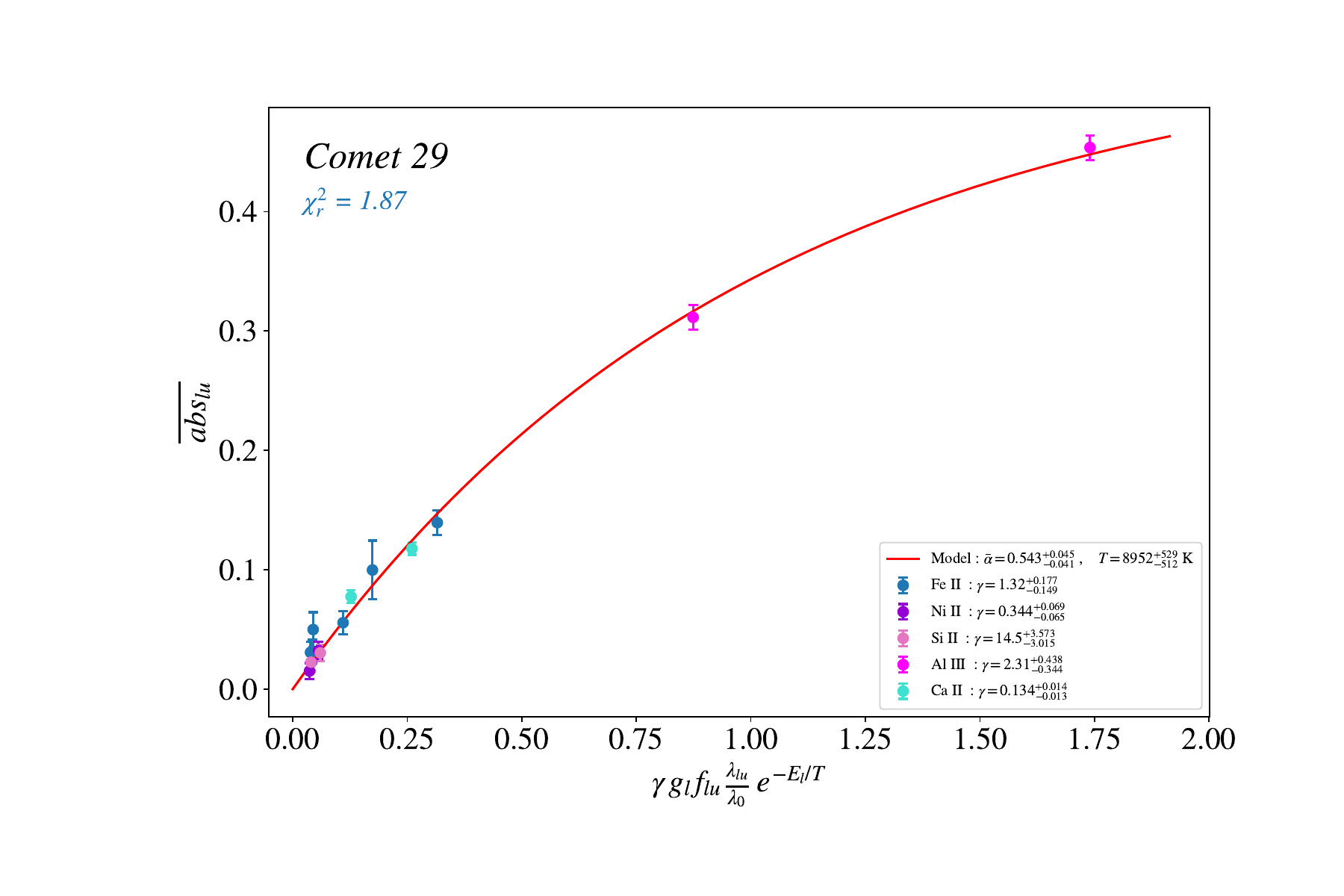}  
\end{figure*}
\small \textbf{Figure \ref{Fig. curves of growth all}}, continued.

\newpage
\onecolumn 
\section{Comet properties and composition}

\begin{table}[h!]
    \centering 
    \begin{threeparttable}
    \renewcommand{\arraystretch}{1.15}
    \caption{\small Fitted parameters of the cometary sample.}
    \begin{tabular}{ c c c c }          
            
            \cline{1-4}  
            \noalign{\smallskip}
            \cline{1-4}  
            \noalign{\smallskip}

            Label & $\overline{\alpha}\tnote{a}  $   &          $\log \left( \frac{ \text{N}_{\text{tot, \feii}} } {\Delta v} \right)$       &    $T_{\rm ex}$\tnote{b} \\
            \noalign{\smallskip}

                  &                                   &    ($\text{cm}^{-2} \ (\text{km} \ \text{s}^{-1})^{-1} $)                            &    $(\si{K})$            \\
                  
            \noalign{\smallskip}
            \cline{1-4} 
            \noalign{\smallskip}
            
    1 &  $0.58 \pm 0.01$ & $13.40 \pm 0.03 $ & $8900 \pm 300$ \\

    \noalign{\smallskip}

    2 &  $0.51\pm 0.02$ & $12.60_{-0.14}^{+0.22}$ & $10000 \pm 1400$ \\

    \noalign{\smallskip}

    3 &  $0.19\pm 0.04$ & $12.18 \pm 0.05$ & $10300_{-1400}^{+1800}$ \\

    4 &  $0.40\pm 0.02$ & $12.80 \pm 0.02$ & $9600 \pm 600$ \\

    5 &  $0.80\pm 0.02$ & $13.56 \pm 0.02$ & $9500 \pm 400$ \\

    6 &  $0.60\pm 0.02$ & $12.81_{-0.09}^{+0.13}$ & $7600 \pm 900$ \\

    7 & $0.73\pm 0.10$ & $13.23 \pm 0.05$ &   \\

    8 &  $0.81\pm 0.02$ & $13.08 \pm 0.02$ & $8400 \pm 400$  \\

    9 &  $0.84\pm 0.02$ & $13.49 \pm 0.02$ &   \\

    10 &  $0.84\pm 0.02$ & $13.52 \pm 0.02$ &   \\

    11 &  $0.46_{-0.11}^{+0.22}$ & $12.60 \pm 0.06$ &   \\

    12 &  $0.46\pm 0.03$ & $12.99 \pm 0.03$ &   \\

    13 &  $0.40\pm 0.04$ & $12.89 \pm 0.04$ &   \\

    14 &  $0.38\pm 0.02$ & $13.10 \pm 0.03$ &   \\

    15 &  $0.56\pm 0.02$ & $11.98 \pm 0.04$ &   \\

    16 &  $0.34_{-0.06}^{+0.12}$ & $12.70 \pm 0.08$ &   \\

    17 &  $0.46\pm 0.02$ & $13.34 \pm 0.03$ &   \\

    18 &  $0.48\pm 0.03$ & $12.96 \pm 0.03$ &   \\

    19 &  $0.73\pm 0.02$ & $11.98 \pm 0.06$ &   \\

    20 &  $0.83\pm 0.11$ & $13.08 \pm 0.03$ &   \\

    21 &  $0.76_{-0.09}^{+0.12}$ & $13.41 \pm 0.04$ &   \\

    22 & $0.70_{-0.12}^{+0.16}$ & $12.60 \pm 0.04$ &   \\

    23 & $0.60_{-0.12}^{+0.19}$ & $12.48 \pm 0.05$ &   \\

    24 &  $0.09_{-0.02}^{+0.04}$ & $12.63 \pm 0.09$ &   \\

    25 &  $0.42_{-0.11}^{+0.21}$ & $12.91 \pm 0.05$ &   \\

    26 &  $0.11_{-0.02}^{+0.03}$ & $12.94 \pm 0.09$ &   \\

    27 &  $0.06 \pm 0.01$ & $12.67 \pm 0.10$ &   \\

    28 &  $0.37 \pm 0.02$ & $12.98 \pm 0.03$ &   \\

    29 & $0.55 \pm 0.04$ & $12.79 \pm 0.04$ &   \\

    \noalign{\smallskip}
    \cline{1-4}  
    \noalign{\smallskip}

    \end{tabular} 
        
    \begin{tablenotes}
      \item[a] \small Replaced by $\overline{\alpha_{t}}$ for comets fitted with the two-component model (comets 1, 2 and 6).
      \item[b] \small As measured from the curve of growth fit. For comets were the excitation temperature could not be estimated (due to short wavelength coverage), a gaussian prior of $9000 \pm 500 \ \si{K}$ was assumed.  
    \end{tablenotes}
    
\label{Tab. comets properties}

\end{threeparttable}
\end{table}

\renewcommand{\arraystretch}{0.95} 
\begin{sidewaystable}
\centering
\captionsetup{width=.92\textwidth}
\caption{\small Measured composition of each studied exocomet.}
\begin{tabularx}{0.908\textwidth}{ c c c c c c c c c c c c }
    \hline 
    \noalign{\smallskip}
    \hline 
    \noalign{\smallskip}
    Comet & S$^0$/\feplus & \caplus/\feplus & \mgplus/\feplus & \mnplus/\feplus & \siplus/\feplus & \crplus/\feplus & \coplus/\feplus & \znplus/\feplus & \niplus/\feplus & \splus/\feplus & \aldeuxplus/\feplus \\
    & $(\cdot 10^{-3}$) & $(\cdot 10^{-3}$) & & $(\cdot 10^{-3}$) &  & $(\cdot 10^{-2}$)  & $(\cdot 10^{-3}$) & $(\cdot 10^{-3}$) & $(\cdot 10^{-2}$) & & $(\cdot 10^{-2}$)  \\
    \noalign{\smallskip}
    \hline
    \noalign{\medskip}
        
    1 & $0.8 \pm 2.0$ & $0.45 \pm 0.04$   &                 & $4.6 \pm 0.7$ & $1.19 \pm 0.16$ & $0.95 \pm 0.15$ & $4.1 \pm 1.4$ & $1.5 \pm 0.5$ & $8.1 \pm 0.5$  &                 & $5.5 \pm 0.4$   \\
    2 &               & $0.64 \pm 0.27$  & $0.61 \pm 0.14$ &               &                 &                 &               &               &                &                 &                 \\
    \noalign{\smallskip}
    \hdashline
    \noalign{\smallskip}
    3 &               &                 &                 &               &                 &                 &               &               & $8.2 \pm 3.2$  &                 &                 \\
    4 &               &                 &                 & $7.8 \pm 2.1$ &                 &                 &               &               & $6.4 \pm 0.9$  &                 & $1.5 \pm 0.2$   \\
    5 & $8.1 \pm 1.7$ &                 &                 & $7.7 \pm 0.7$ & $0.97 \pm 0.10$ & $1.15 \pm 0.10$ &               &               & $6.6 \pm 0.8$  &                 & $0.83 \pm 0.04$ \\
    6 &               &                 & $0.60 \pm 0.11$ &               &                 &                 &               &               &                &                 & $20 \pm 2.6$  \\
    \noalign{\smallskip}
    \hdashline
    \noalign{\smallskip}
    7 &               &                 &                 &               & $1.55 \pm 0.26$ &                 &               &               & $8.9 \pm 1.7$  &                 & $2.1 \pm 0.3$   \\
    \noalign{\smallskip}
    \hdashline
    \noalign{\smallskip}
    8 &               &                 &                 & $5.8 \pm 1.0$ &                 &                 &               &               & $10.6 \pm 2.7$ &                 &                 \\
    \noalign{\smallskip}
    \hdashline
    \noalign{\smallskip}
    9 &               &                 &                 &               &                 & $1.02 \pm 0.11$ &               &               &                &                 &                 \\
    \noalign{\smallskip}
    \hdashline
    \noalign{\smallskip}
    10&               &                 &                 &               &                 & $1.27 \pm 0.10$ &               &               &                &                 &                 \\
    \noalign{\smallskip}
    \hdashline
    \noalign{\smallskip}
    11&               &                 &                 &               &                 &                 &               &               &                & $0.50 \pm 0.20$ &                 \\
    12&               &                 &                 &               &                 & $1.43 \pm 0.29$ &               &               &                & $2.25 \pm 0.32$ &                 \\
    13&               &                 &                 &               &                 & $1.07 \pm 0.33$ &               &               &                & $3.08 \pm 0.55$ &                 \\
    14&               &                 &                 &               &                 & $1.11 \pm 0.22$ &               &               &                & $1.71 \pm 0.27$ &                 \\
    \noalign{\smallskip}
    \hdashline
    \noalign{\smallskip}
    15&               & $0.76 \pm 0.14$ & $0.48 \pm 0.07$ &               &                 &                 &               &               &                &                 &                 \\
    16&               & $0.72 \pm 0.10$ &                 &               &                 &                 &               &               &                &                 &                 \\
    17&               &                 &                 &               &                 & $1.20 \pm 0.15$ &               &               &                &                 &                 \\    
    18&               & $0.72 \pm 0.07$ &                 &               &                 & $0.90 \pm 0.30$ &               &               &                & $0.68 \pm 0.16$ &                 \\
    19&               & $0.46 \pm 0.12$ & $0.64 \pm 0.10$ &               &                 &                 &               &               &                &                 &                 \\
    \noalign{\smallskip}
    \hdashline
    \noalign{\smallskip}
    
    20& $3.6 \pm 2.2$ & $0.36 \pm 0.03$ &                 &               & $0.85 \pm 0.12$ &                 &               &               & $7.4 \pm 0.8$  &                 & $1.5 \pm 0.1$   \\
    21& $9.5 \pm 1.3$ &                 &                 &               & $0.68 \pm 0.13$ &                 &               &               & $8.1 \pm 1.1$  &                 & $1.1 \pm 0.1$   \\
    22&               & $0.70 \pm 0.07$ &                 &               & $1.27 \pm 0.30$ &                 &               &               &    $8.7 \pm 2.1$  &                 & $6.0 \pm 0.7$   \\
    23&               & $0.73 \pm 0.09$ &                 &               &                 &                 &               &               & $6.9 \pm 2.8$  &                 & $6.2 \pm 0.9$   \\
    \noalign{\smallskip}
    \hdashline
    \noalign{\smallskip}
    24&               &                 &                 &               & $1.00 \pm 0.32$ &                 &               &               & $5.2 \pm 1.9$  &                 &                 \\
    25&               &                 &                 &               & $0.71 \pm 0.16$ &                 &               &               & $9.2 \pm 1.3$  &                 &                 \\    
    26&               &                 &                 &               &                 &                 &               &               & $5.7 \pm 1.6$  &                 &                 \\
    \noalign{\smallskip}
    \hdashline
    \noalign{\smallskip}
    27&               & $0.52 \pm 0.10$ &                 &               & $1.23 \pm 0.40$ &                 &               &               & $6.9 \pm 2.3$  &                 & $1.3 \pm 0.5$   \\
    28&               & $0.52 \pm 0.04$ &                 &               & $0.72 \pm 0.17$ &                 &               &               & $8.9 \pm 1.3$  &                 & $3.0 \pm 0.4$   \\
    29&               & $0.53 \pm 0.05$ &                 &               & $1.02 \pm 0.22$ &                 &               &               & $7.4 \pm 1.4$  &                 & $5.0 \pm 0.6$   \\

    \noalign{\smallskip}
    \hline
\end{tabularx} 
\label{Tab. abundances}
\end{sidewaystable}

\FloatBarrier

\newpage

\section{List of the studied lines}

\vspace{0.2 cm}

\begin{table}[h!]
    \centering 
    \begin{threeparttable}

    \caption{\small List of all spectral lines used in the study, grouped by series.}
    \renewcommand{\arraystretch}{1.15}
    \begin{tabular}{ c c c c c c }          
            
    \cline{1-6}
    \noalign{\smallskip}
    \cline{1-6} 
    
            Series & $\lambda_{lu}$\tnote{a} & $E_l$\tnote{a} & $E_l/k_B$ & $A_{ul}$\tnote{a} & $g_l f_{lu}$ \\
            &  (\A) & ($\si{cm^{-1}}$) & ($\si{K}$) & ($10^7 \  \si{s^{-1}}$) &  \\
            
            \cline{1-6} 
            \noalign{\smallskip}

                                   & 1250.58 & 0 & 0 & 5.13 & 0.024 \\
            \Sii\ $\lambda$1250\A & 1253.81 & 0 & 0 & 5.12 & 0.048 \\
                                   & 1259.52 & 0 & 0 & 5.1 & 0.073 \\
            
            \noalign{\smallskip}
            \cline{1-6} 
            \noalign{\smallskip}

                                    & 1608.45 & 0    & 0    & 19.1 & 0.59\\
                                    & 1612.81 & 1873 & 2694 & 6.60 & 0.31\\
                                    & 1618.47 & 385  & 554  & 5.53 & 0.174\\
                                    & 1621.68 & 385  & 554  & 13.2 & 0.31\\
                                    & 1625.52 & 2430 & 3497 & 4.00 & 0.158\\
                                    & 1629.16 & 668  & 961  & 7.20 & 0.172\\
                                    & 1631.13 & 668  & 961  & 6.93 & 0.111\\
                                    & 1633.91 & 2838 & 4083 & 3.90 & 0.125\\
                                    & 1634.35 & 863  & 1241 & 3.21 & 0.077\\
                                    & 1635.40 & 7955 & 11446& 23.0 & 0.55\\
            \feii\ $\lambda$1600\A & 1636.33 & 863  & 1241 & 9.63 & 0.155\\
                                    & 1637.40 & 1873 & 2694 & 3.60 & 0.116\\
                                    & 1639.40 & 977  & 1406 & 6.85 & 0.110\\
                                    & 1640.15 & 3117 & 4485 & 5.90 & 0.143\\
                                    & 1641.76 & 8392 & 12074& 18.0 & 0.29\\
                                    & 1643.58 & 2430 & 3497 & 4.40 & 0.107\\
                                    & 1646.18 & 8680 & 12490& 14.0 & 0.114\\
                                    & 1647.16 & 8392 & 12074& 5.00 & 0.122\\
                                    & 1650.70 & 8847 & 12729& 14.0 & 0.114\\
                                    & 1659.48 & 2430 & 3497 & 8.00 & 0.198\\
                                    & 1663.22 & 1873 & 2694 & 6.70 & 0.111\\
                                    
            \noalign{\smallskip}
            \cline{1-6} 
            \noalign{\smallskip}

                                                     & 1696.80 & 1873 & 2694 & 1.70 & 0.073 \\
                                                     & 1702.05 & 1873 & 2694 & 10.0 & 0.52 \\
            \multirow{2}{*}{\feii\ $\lambda$1700\A} & 1708.62 & 2430 & 3497 & 2.00 & 0.070 \\
                                                     & 1713.00 & 2430 & 3497 & 7.10 & 0.31 \\
                                                     & 1720.61 & 2838 & 4083 & 5.90 & 0.21 \\
                                                     & 1726.39 & 3117 & 4485 & 3.20 & 0.086 \\
            
            \noalign{\smallskip}
            \cline{1-6} 
            \noalign{\smallskip}
            
                                             & 1709.61 & 0 & 0 & 7.25 & 0.191 \\
            \niii\ $\lambda$1700\A\tnote{b} & 1741.56 & 0 & 0 & 9.48 & 0.259 \\
                                             & 1751.91 & 0 & 0 & 4.55 & 0.168 \\

            \noalign{\smallskip}
            \cline{1-6} 
            \noalign{\smallskip}
            
            \multirow{2}{*}{\siii\ $\lambda$1800\A\tnote{c}} & 1808.01 & 0 & 0 & 0.213 & 0.0042 \\
                                                     & 1816.93 & 287 & 413 & 0.222 & 0.0066 \\
            
            \noalign{\smallskip}
            \cline{1-6} 
            \noalign{\smallskip}
                        
                                  & 1807.31 & 0 & 0 & 32.7 & 0.48 \\
            \Si\ $\lambda$1800\A & 1820.34 & 396 & 570 & 17.1 & 0.26 \\
                                  & 1826.25 & 573 & 824 & 5.64 & 0.085 \\

            \noalign{\smallskip}
            \cline{1-6} 
            \noalign{\smallskip}
            
        \end{tabular} 
        
    \begin{tablenotes}
      \item[a] \small Unless otherwise indicated, the level energies and line parameters ($\lambda_{lu}$, $A_{ul}$, $f_{lu}$) were collected from the NIST database (\cite{NIST_ASD}). The uncertainties are typically of the order of 10 \%.
      \item[b] \small Line parameters obtained from \cite{Boisse2019}.
      \item[c] \small Line parameters obtained from \cite{Bergeson1993}.
    \end{tablenotes}
    
\label{Tab. list lines}

\end{threeparttable}
\end{table}

\newpage

\centering
\small \textbf{Table \ref{Tab. list lines}}, continued.

\begin{table}[h!]
    \centering 
    \begin{threeparttable}
    \renewcommand{\arraystretch}{1.15}
    \begin{tabular}{ c c c c c c }          
            
    \cline{1-6}
    \noalign{\smallskip}
    \cline{1-6} 
            
            Series & $\lambda_{lu}$ & $E_l$ & $E_l/k_B$ & $A_{ul}$\tnote{a} & $g_l f_{lu}$ \\
            &  (\A) & ($\si{cm^{-1}}$) & ($\si{K}$) & ($10^7 \  \si{s^{-1}}$) &  \\

            \cline{1-6} 
            \noalign{\smallskip}
            
            \multirow{2}{*}{\aliii\ $\lambda$1850\A}  & 1854.72 & 0 & 0 & 54.4 & 1.12 \\
                                                       & 1862.79 & 0 & 0 & 53.6 & 0.56 \\

            \noalign{\smallskip}
            \cline{1-6} 
            \noalign{\smallskip}
            
            \multirow{2}{*}{\znii\ $\lambda$2000\A} & 2026.14 & 0 & 0 & 40.7 & 1.00 \\
                                                     & 2062.66 & 0 & 0 & 38.6 & 0.49 \\
            
            \noalign{\smallskip}
            \cline{1-6} 
            \noalign{\smallskip}

                                   & 2056.26 & 0 & 0 & 12.2 & 0.62 \\
            \crii\ $\lambda$2000\A & 2062.24 & 0 & 0 & 11.9 & 0.46 \\
                                   & 2066.16 & 0 & 0 & 12.0 & 0.31 \\

            \noalign{\smallskip}
            \cline{1-6} 
            \noalign{\smallskip}
        
                                                     & 2166.23 & 8394 & 12077 & 24.0 & 1.69 \\
                                                     & 2169.77 & 9330 & 13424 & 15.8 & 0.89 \\
                                                     & 2175.35 & 9330 & 13424 & 14.3 & 1.01 \\
                                                     & 2175.82 & 10116 & 14555 & 17.7 & 0.75 \\
                                                     & 2185.29 & 10664 & 15343 & 29.0 & 0.83 \\
                                                     & 2202.09 & 10664 & 15343 & 13.0 & 0.57 \\
                                                     & 2207.41 & 10116 & 14555 & 16.6 & 0.97 \\
                                                     & 2211.07 & 9330 & 13424 & 3.9 & 0.29 \\
                                                     & 2217.17 & 8394 & 12077 & 34.0 & 3.01 \\
                                                     & 2223.64 & 8394 & 12077 & 9.8 & 0.73 \\
            \multirow{2}{*}{\niii\ $\lambda$2200\A} & 2225.56 & 9330 & 13424 & 16.5 & 0.98 \\
                                                     & 2227.02 & 10116 & 14555 & 13.0 & 0.58 \\
                                                     & 2254.55 & 10664 & 15343 & 19.8 & 0.91 \\
                                                     & 2265.16 & 10116 & 14555 & 14.3 & 0.88 \\
                                                     & 2270.91 & 9330 & 13424 & 15.6 & 1.21 \\
                                                     & 2279.47 & 13550 & 19496 & 28.0 & 1.31 \\
                                                     & 2287.79 & 14996 & 21576 & 28.0 & 0.88 \\
                                                     & 2297.26 & 13550 & 19496 & 19.8 & 1.25 \\
                                                     & 2298.2 & 10664 & 15343 & 30.0 & 0.48 \\
                                                     & 2298.97 & 14996 & 21576 & 28.0 & 1.33 \\
                                                     & 2303.7 & 9330 & 13424 & 29.0 & 1.38 \\
                                                     & 2316.75 & 8394 & 12077 & 28.8 & 1.85 \\

            \noalign{\smallskip}
            \cline{1-6} 
            \noalign{\smallskip}

                                    & 2249.88 & 0 & 0 & 0.3 & 0.0182 \\
                                    & 2251.63 & 668 & 961 & 0.319 & 0.0146 \\
                                    & 2253.82 & 385 & 554 & 0.44 & 0.027 \\
            \feii\ $\lambda$2250\A & 2260.78 & 0 & 0 & 0.318 & 0.024 \\
                                    & 2261.56 & 863 & 1241 & 0.216 & 0.0099 \\
                                    & 2268.29 & 668 & 961 & 0.369 & 0.023 \\
                                    & 2280.62 & 385 & 554 & 0.449 & 0.035\\

            \noalign{\smallskip}
            \cline{1-6} 
            \noalign{\smallskip}

                                    & 2286.86 & 3350 & 4820 & 33.0 & 3.36 \\
                                    & 2308.57 & 4028 & 5796 & 26.0 & 2.29 \\
            \coii\ $\lambda$2300\A & 2312.32 & 4560 & 6561 & 28.0 & 2.02 \\
                                    & 2314.77 & 4950 & 7122 & 28.0 & 1.57 \\
                                    & 2315.69 & 5205 & 7489 & 27.0 & 1.09 \\

            \noalign{\smallskip}
            \cline{1-6} 
            \noalign{\smallskip}

        \end{tabular} 
        
\end{threeparttable}
\end{table}

\newpage

\centering
\small \textbf{Table \ref{Tab. list lines}}, continued.

\begin{table}[h!]
    \centering 
    \begin{threeparttable}
    \renewcommand{\arraystretch}{1.15}
    \begin{tabular}{ c c c c c c }          
            
    \cline{1-6}
    \noalign{\smallskip}
    \cline{1-6} 
            
            Series & $\lambda_{lu}$ & $E_l$ & $E_l/k_B$ & $A_{ul}$\tnote{a} & $g_l f_{lu}$ \\
            &  (\A) & ($\si{cm^{-1}}$) & ($\si{K}$) & ($10^7 \  \si{s^{-1}}$) &  \\

            \cline{1-6} 
            \noalign{\smallskip}

                                                     & 2328.11 & 667 & 960 & 6.6 & 0.22 \\
                                                     & 2332.02 & 1873 & 2694 & 3.17 & 0.21 \\
                                                     & 2333.51 & 385 & 554 & 13.1 & 0.64 \\
                                                     & 2338.73 & 863 & 1241 & 11.3 & 0.37 \\
                                                     & 2344.21 & 0 & 0 & 17.3 & 1.14 \\
                                                     & 2345.0 & 977 & 1406 & 9.27 & 0.31 \\
                                                     & 2348.83 & 1873 & 2694 & 6.5 & 0.43 \\
                                                     & 2349.02 & 668 & 961 & 11.5 & 0.57 \\
                                                     & 2355.61 & 2838 & 4083 & 2.67 & 0.089 \\
                                                     & 2359.83 & 863 & 1241 & 5.0 & 0.25 \\
                                                     & 2360.72 & 1873 & 2694 & 3.59 & 0.30 \\
                                                     & 2361.02 & 2430 & 3497 & 6.2 & 0.31 \\
                                                     & 2362.74 & 2430 & 3497 & 1.41 & 0.094 \\
                                                     & 2365.55 & 385 & 554 & 5.9 & 0.49 \\
                                                     & 2367.32 & 2838 & 4083 & 1.01 & 0.051 \\
            \multirow{2}{*}{\feii\ $\lambda$2400\A}  & 2369.32 & 2838 & 4083 & 6.06 & 0.20 \\
                                                     & 2371.22 & 3117 & 4485 & 1.73 & 0.058 \\
                                                     & 2374.46 & 0 & 0 & 4.25 & 0.36 \\
                                                     & 2375.92 & 3117 & 4485 & 9.8 & 0.166 \\ 
                                                     & 2380.00 & 2430 & 3497 & 2.73 & 0.185 \\
                                                     & 2381.49 & 668 & 961 & 3.1 & 0.21 \\
                                                     & 2382.76 & 0 & 0 & 31.3 & 3.20 \\
                                                     & 2385.11 & 3117 & 4485 & 3.2 & 0.109 \\
                                                     & 2389.36 & 385 & 554 & 10.5 & 0.72 \\
                                                     & 2392.21 & 2430 & 3497 & 0.377 & 0.032 \\
                                                     & 2396.35 & 385 & 554 & 25.9 & 2.23 \\
                                                     & 2399.97 & 668 & 961 & 13.9 & 0.72 \\
                                                     & 2405.62 & 668 & 961 & 19.6 & 1.36 \\
                                                     & 2407.39 & 863 & 1241 & 16.1 & 0.56 \\
                                                     & 2411.25 & 863 & 1241 & 15.5 & 0.81 \\
                                                     & 2411.80 & 977 & 1406 & 23.7 & 0.41 \\
                                                     & 2414.04 & 977 & 1406 & 10.2 & 0.36 \\
            
            \noalign{\smallskip}
            \cline{1-6} 
            \noalign{\smallskip}

                                                     & 2395.25 & 13550 & 19496 & 17.0 & 1.46 \\
            \multirow{2}{*}{\niii\ $\lambda$2400\A} & 2416.87 & 14996 & 21576 & 21.0 & 1.47 \\
                                                     & 2438.63 & 13550 & 19496 & 5.4 & 0.48 \\
                                                     & 2511.63 & 13550 & 19496 & 5.8 & 0.55 \\

            \noalign{\smallskip}
            \cline{1-6} 
            \noalign{\smallskip}

                                    & 2576.88 & 0 & 0 & 28.0 & 2.51 \\
            \mnii\ $\lambda$2600\A & 2594.50\tnote{d} & 0 & 0 & 27.6 & 1.95 \\
                                    & 2606.46 & 0 & 0 & 26.9 & 1.37 \\

            \noalign{\smallskip}
            \cline{1-6} 
            \noalign{\smallskip}

                                    & 2563.30 & 7955 & 11446 & 17.9 & 1.06 \\
                                    & 2564.24 & 8392 & 12074 & 15.1 & 0.60 \\
                                    & 2567.68 & 8680 & 12490 & 11.0 & 0.22 \\
            \feii\ $\lambda$2600\A & 2578.69 & 8847 & 12729 & 12.0 & 0.24 \\
                                    & 2583.36 & 8680 & 12490 & 8.8 & 0.35 \\
                                    & 2592.31 & 8392 & 12074 & 5.7 & 0.35 \\
                                    & 2594.50\tnote{d} & 8847 & 12729 & 1.63 & 0.066 \\
           \noalign{\smallskip}
            \cline{1-6} 
            \noalign{\smallskip}

        \end{tabular} 
        
    \begin{tablenotes}
      \item[d] The optical depths of the 2594.50 \A\ \mnii\ and 2594.50 \A\ \feii\ lines were added up together in the fits, the two lines being separated by 800 m/s only.
    \end{tablenotes}
    
\end{threeparttable}
\end{table}

\newpage

\centering
\small \textbf{Table \ref{Tab. list lines}}, continued.

\begin{table}[h!]
    \centering 
    \begin{threeparttable}
    \renewcommand{\arraystretch}{1.15}
    \begin{tabular}{ c c c c c c }          
            
    \cline{1-6}
    \noalign{\smallskip}
    \cline{1-6} 
            
            Series & $\lambda_{lu}$ & $E_l$ & $E_l/k_B$ & $A_{ul}$\tnote{a} & $g_l f_{lu}$ \\
            &  (\A) & ($\si{cm^{-1}}$) & ($\si{K}$) & ($10^7 \  \si{s^{-1}}$) &  \\

            \cline{1-6} 
            \noalign{\smallskip}

                                    & 2586.65 & 0 & 0 & 8.94 & 0.72 \\
                                    & 2599.15 & 385 & 554 & 14.3 & 0.87 \\
                                    & 2600.17 & 0 & 0 & 23.5 & 2.38 \\
                                    & 2607.87 & 668 & 961 & 17.3 & 0.71 \\
                                    & 2612.65 & 385 & 554 & 12.0 & 0.98 \\
            \feii\ $\lambda$2600\A & 2614.6 & 863 & 1241 & 21.2 & 0.44 \\
            (continued)             & 2618.4 & 668 & 961 & 4.88 & 0.30 \\
                                    & 2622.45 & 977 & 1406 & 5.6 & 0.115 \\
                                    & 2626.45 & 385 & 554 & 3.52 & 0.36 \\
                                    & 2629.08 & 977 & 1406 & 8.74 & 0.36 \\
                                    & 2631.83 & 863 & 1241 & 8.16 & 0.51 \\
                                    & 2632.11 & 668 & 961 & 6.29 & 0.52 \\

            \noalign{\smallskip}
            \cline{1-6} 
            \noalign{\smallskip}

                                                     & 2669.50 & 12033 & 17313 & 14.5 & 0.31 \\
                                                     & 2672.60 & 12148 & 17478 & 10.9 & 0.47 \\
                                                     & 2673.62 & 12304 & 17703 & 5.74 & 0.37 \\
                                                     & 2677.95\tnote{e} & 12304 & 17703 & 11.9 & 1.02 \\
                                                     & 2677.95\tnote{e} & 12496 & 17980 & 20.9 & 2.25 \\
                                                     & 2679.59 & 12033 & 17313 & 8.02 & 0.52 \\
            \multirow{2}{*}{\crii\ $\lambda$2700\A\tnote{f}} & 2767.35 & 12496 & 17980 & 22.3 & 2.05 \\
                                                     & 2836.47 & 12496 & 17980 & 25.0 & 3.62 \\
                                                     & 2844.08 & 12304 & 17703 & 18.9 & 2.29 \\
                                                     & 2850.67 & 12148 & 17478 & 15.2 & 1.48 \\
                                                     & 2856.51 & 12033 & 17313 & 11.3 & 0.83 \\
                                                     & 2863.41 & 12304 & 17703 & 8.66 & 0.85 \\
                                                     & 2865.95 & 12148 & 17478 & 11.1 & 0.82 \\

            \noalign{\smallskip}
            \cline{1-6} 
            \noalign{\smallskip}

                                                     & 2715.22 & 7955 & 11446 & 5.7 & 0.38 \\
                                                     & 2725.69 & 8392 & 12074 & 0.96 & 0.064 \\
                                                     & 2728.35 & 8392 & 12074 & 9.38 & 0.42 \\
                                                     & 2731.54 & 8680 & 12490 & 2.79 & 0.125 \\
                                                     & 2737.78 & 8680 & 12490 & 12.2 & 0.27 \\
            \multirow{2}{*}{\feii\ $\lambda$2750\A} & 2740.36 & 7955 & 11446 & 22.1 & 1.99 \\
                                                     & 2744.01 & 8847 & 12729 & 19.7 & 0.89 \\
                                                     & 2747.3 & 8680 & 12490 & 20.5 & 1.39 \\
                                                     & 2747.79 & 8392 & 12074 & 16.9 & 1.15 \\
                                                     & 2756.55 & 7955 & 11446 & 21.5 & 2.45 \\
                                                     & 2762.66 & 8847 & 12729 & 1.38 & 0.063 \\
                                                     & 2769.75 & 8680 & 12490 & 0.48 & 0.033 \\

            \noalign{\smallskip}
            \cline{1-6} 
            \noalign{\smallskip}

            \multirow{2}{*}{\mgii\ $\lambda$2800\A} & 2796.35 & 0 & 0 & 26.0 & 1.22 \\
                                                     & 2803.53 & 0 & 0 & 25.7 & 0.61 \\
                                                     
            \noalign{\smallskip}
            \cline{1-6} 
            \noalign{\smallskip}

            \multirow{2}{*}{\caii\ $\lambda$3900\A} & 3934.77 & 0 & 0 & 14.7 & 1.36 \\
                                                     & 3969.59 & 0 & 0 & 14.0 & 0.66 \\

            \noalign{\smallskip}
            \cline{1-6} 
            \noalign{\smallskip}

        \end{tabular} 
        
    \begin{tablenotes}
      \item[e] The two \crii\ lines at 2677.95 \A\ were considered as a single line, with an effective $gf$-value of 3.27.
      \item[f] Line parameters obtained from \cite{Nilsson2006}.
    \end{tablenotes}
    
\end{threeparttable}
\end{table}

\end{appendix}
\end{document}